\documentclass[12pt]{article}

\usepackage{graphicx}

\def\square{\kern1pt\vbox{\hrule height 1.2pt\hbox{\vrule width 1.2pt\hskip 3pt
\vbox{\vskip 6pt}\hskip 3pt\vrule width 0.6pt}\hrule height 0.6pt}\kern1pt}

\def\dlim{\mathrel{\raise.8ex\hbox{${\scriptscriptstyle D =
4}$\kern-1.5em\lower1ex\hbox{$\longrightarrow$}}}}
\def\ltwid{\mathrel{\raise.3ex\hbox{$<$\kern-.75em\lower1ex\hbox{$\sim$}}}}
\def \be{\begin{equation}}
\def \ee{\end{equation}}
\def \bea{\begin{eqnarray}}
\def \eea{\end{eqnarray}}

\def \del{\partial}

\def \f{\frac}

\def \g{\gamma}

\def \ka{\kappa}
\def \d{\delta}
\def \D{\Delta}

\begin{document}

\begin{titlepage}

\begin{flushright}
ITP-UU-12/08 \\ SPIN-12/07
\end{flushright}

\vspace{2cm}

\begin{center}
{\bf Quantum Gravitational Effects on Massive Fermions \\during
Inflation I}
\end{center}

\vspace{.5cm}

\begin{center}
S. P. Miao\dag
\end{center}

\vspace{.5cm}

\begin{center}
\it{Institute for Theoretical Physics, Spinoza Institute, University of Utrecht
\\Luevenlaan 4, Postbus 80.195, 3508TD Utrecht, NETHERLANDS }
\end{center}

\vspace{1cm}

\begin{center}
ABSTRACT
\end{center}
We compute the one loop graviton contribution to the self-energy of
a very light fermion on a locally de Sitter background. This result
can be used to study the effect that a small mass has on the
propagation of fermions through the sea of infrared gravitons
generated by inflation. We employ dimensional regularization and
obtain a fully renormalized result by absorbing all divergences with
BPHZ counterterms. An interesting technical aspect of this
computation is the need for two noninvariant counterterms owing to
the breaking of de Sitter invariance by our gauge condition.

\vspace{2cm}

\begin{flushleft}
\end{flushleft}

\begin{flushleft}
$^{\dagger}$ e-mail: s.miao@uu.nl
\end{flushleft}

\end{titlepage}

\section{Introduction}
In this paper we compute and renormalize the one loop quantum
grvaitational corrections to the self-energy of very light
fermions on a locally de Sitter background. The physical
motivation for this exercise is to facilitate a later study
of how inflationary gravitons affect fermions and, in
particular, the contrast between the case of exactly
massless fermions and those with a small mass. Nonzero mass
introduces two competing effects: it changes how fermions
propagate and it also alters how they interact with gravity.
The first of these changes tends to suppress the effects of
inflationary gravitons because it makes the fermion wave
function oscillate so that interactions at different times
tend to cancel. However, the new interaction enhances the
effect of inflationary gravitons because it does not fall
off with time.

The current work can be seen as complementing two previous
studies of massless fermions on de Sitter. In both cases the
technique was to compute the one loop fermion self-energy
$-i[\mbox{}_i \Sigma_j](x;x')$ and then use it to solve the
quantum-corrected Dirac equation for fermion mode functions,
\begin{equation}
\sqrt{-g} \, i\hspace{-.1cm}\not{\hspace{-.15cm} \mathcal{D}}_{ij}
\psi_{j}(x) - \int d^4x' \, \Bigl[\mbox{}_i \Sigma_j \Bigr](x;x') \,
\psi_{j}(x') = 0 \; . \label{Diraceq}
\end{equation}
The first model results from Yukawa coupling the fermion
to a massless, minimally coupled (MMC) scalar on a nondynamical
de Sitter background \cite{PW}. The second model consists of the
fermion with dynamical gravity on de Sitter background \cite{MW1}.
Powers of the inflationary scale factor $a = e^{H t}$ are
crucial for understanding the results in both cases. The
self-energy from the $\phi \overline{\psi} \psi$ interaction
of the first model grows like $a \ln(a)$ relative to the
classical term. The resulting fermion mode functions behave as
if they had a growing mass. The interactions of the second
model all possess derivatives --- for example, $\partial h
\overline{\psi} \psi$ --- which limit the induced self-energy
to grow no faster than $\ln(a)$ relative to the classical term.
The resulting fermion mode functions behave as if they had a
growing field strength, which could be understood as the
random walk that fermions take under buffeting from the sea of
inflationary gravitons \cite{MW1,MW3}. Although the effect
from gravitons is smaller than that from massless scalars,
{\it it is universal}, independent of assumptions about
the existence or couplings of unnaturally light scalars.
It is even conceivable that the graviton effect might,
in a more complicated model, lead to baryogenesis
during inflation.

What we expect for massive fermions in dynamical gravity is
that the absence of derivatives in the $a m h \overline{\Psi}
\Psi$\footnote{In section 2 we re-scale fermion fields and
the metric: $\Psi=a^{\frac{D-1}{2}}\psi\,;\,
g_{\mu\nu}=a^2\tilde{g}_{\mu\nu}$.}
interaction will cause the self-energy to grow like
$a \ln(a)$ relative to the classical kinetic term, and like
$\ln(a)$ relative to the classical mass term.
\begin{equation}
\sqrt{-\tilde{g}} \, \Bigl[i\hspace{-.1cm}\not{\hspace{-.15cm}
\mathcal{D}}-am \Bigr]_{ij}\Psi_{j}(x) - \int d^4x' \,
\Bigl[\mbox{}_i \Sigma_j \Bigr](x;x') \,
\Psi_{j}(x') = 0 \; . \label{Diraceq1}
\end{equation}
When the classical mass is large (relative to the Hubble parameter)
we expect at most a small enhancement of the fermion field
strength. When the classical mass is small, classical dynamics
are mostly controlled by the kinetic term and we expect the
quantum correction to have a much larger proportional impact.
One might intuitively expect the crossover to come for fermion
masses near the Hubble parameter. However, we shall specialize
to the case of very light fermions, both because this is where
the largest effects should occur, and because expanding in the
fermion mass makes an enormous simplification in the computation.

This work also deserves a place in the growing list of studies
of quantum infrared effects during inflation. Among these are:
\begin{itemize}
\item{The effects of self-interacting, MMC scalars on
nondynamical de Sitter background \cite{OW1,OW2,BOW,KO};}
\item{The effects of a charged, MMC scalar on nondynamical de
Sitter background \cite{PTW1,PTW3};}
\item{The effects of a nonlinear sigma model on nondynamical
de Sitter background \cite{KK1,KK2};}
\item{The effects of a MMC scalar on gravitons on de Sitter
background \cite{PW1,PW2};}
\item{The effects of gravitons on a MMC scalar on de Sitter
background \cite{KW1}; and}
\item{The effects of gravitons on interacting conformal matter
on de Sitter background \cite{KK3,KK4}.}
\end{itemize}
It should also be noted that the series of leading infrared
logarithms can be summed for scalar potential models using
the stochastic technique of Starobinksy and Yokoyama \cite{SY}.
The same resummation can be achieved for Yukawa theory \cite{MW2},
and for scalar QED \cite{PTW2}, but it has so far not been accomplished
for either the nonlinear sigma model \cite{KK1,KK2}, or for
quantum gravity \cite{RPW3}. Each fully renormalized quantum
gravitational result is an important piece of ``data'' in
the search for such a resummation.

Although Dirac $+$ Einstein is not perturbatively renormalizable
\cite{DVN}, ultraviolet divergences can always be absorbed in the BPHZ
sense \cite{BP,H,Z1,Z2}. A widespread misconception exists that no
valid quantum predictions can be extracted from such a theory.
This is not true: while nonrenormalizability does preclude being able
to compute {\it everything}, that is not the same thing as being able
to compute {\it nothing}. The problem with a nonrenormalizable
theory is that no physical principle fixes the finite parts of the
escalating series of BPHZ counterterms needed to absorb ultraviolet
divergences, order-by-order in perturbation theory. Hence any
prediction of the theory that can be changed by adjusting the finite
parts of these counterterms is essentially arbitrary. However, loops
of massless particles make nonlocal contributions to the effective
action that can never be affected by local counterterms. These
nonlocal contributions typically dominate the infrared. Further,
they cannot be affected by whatever modification of ultraviolet
physics ultimately results in a completely consistent formalism. As
long as the eventual fix introduces no new massless particles, and
does not disturb the low energy couplings of the existing ones, the
far infrared predictions of a BPHZ-renormalized quantum theory will
agree with those of its fully consistent descendant.

It is worth mentioning the many studies which have exploited
this basic facet of low energy effective field theory.
The oldest example is the solution of the infrared problem
in quantum electrodynamics by Bloch and Nordsieck \cite{BN},
long before that theory's renormalizability was suspected.
Weinberg \cite{SW} was able to achieve a similar resolution
for quantum gravity with zero cosmological constant. The same
principle was at work in the Fermi theory computation of the long
range force due to loops of massless neutrinos by Feinberg and
Sucher \cite{FS,HS}. Matter which is not supersymmetric generates
nonrenormalizable corrections to the graviton propagator at one
loop, but this did not prevent the computation of photon, massless
neutrino and massless, conformally coupled scalar loop corrections
to the long range gravitational force \cite{CDH,CD,DMC1,DL}.
The same principles of low energy effective field theory
have been applied to compute graviton loop corrections to
the long range force by Donoghue \cite{JFD1,JFD2} and many others
\cite{MV,HL,ABS,oldKK1,oldKK2}.

That summarizes why the exercise we have undertaken is both valid
and interesting. The necessary Feynman rules are given in sections 2-4.
Because some of these are the same as for the previous study of
massless fermions \cite{MW1, MW3}, we merely present the relevant
old results and reserve discussion for the new features associated
with a nonzero fermion mass. Section 2 covers the fermionic sector
which gives the fermion propagator and the interactions. Section 3
presents the graviton propagator. The BPHZ counterterms
necessary for our computations are carefully analyzed in section 4.
In section 5 we evaluate the contributions from diagrams involving
a single 4-point interaction. In section 6 we evaluate the more
difficult contributions which involve two 3-point interactions.
Renormalization is accomplished in section 7, and our conclusions
are given in section 8.

\section{Feynman Rules for Massive Dirac }

In this section we derive the Feynman rules to facilitate the
computation we are going to perform. To obtain three-point vertices,
four-point vertices and the massive fermion propagator we first start
with the Lagrangian of massive Dirac,
\begin{equation}
\mathcal{L}_{\rm Dirac} \equiv \overline{\psi} e^{\mu}_{~b}
\gamma^{b} i \mathcal{D}_{\mu} \psi \sqrt{-g}-m\overline{\psi}\psi
\sqrt{-g} \; . \label{Dirac}
\end{equation}
Here the vierbein $e^{\mu}_{~b}$ consists of coordinate indices expressed
by Greek letters and Lorentz indices denoted by Latin letters; and
$\mathcal{D}_{\mu}$ is the covariant derivative which is formed by
the spin connection and the Dirac Lorentz representation matrices,
\begin{eqnarray}
&&\mathcal{D}_{\mu} \equiv \partial_{\mu} + \frac{i}2 A_{\mu cd}
J^{cd} \; , \\ && A_{\mu c d} = e^{\nu}_{~c}
\Bigl( e_{\nu d, \mu} - \Gamma^{\rho}_{~\mu\nu} e_{\rho d}\Bigr)\; \;
,\;\; J^{bc} \equiv \frac{i}4 \Bigl(\gamma^b \gamma^c - \gamma^c
\gamma^b\Bigr).
\end{eqnarray}
Because our locally de Sitter background is conformally flat it is
useful to rescale the vierbein by an arbitrary function of spacetime
$a(x)$,
\begin{equation}
e_{\beta b} \equiv a \, \widetilde{e}_{\beta b}\;\;\;\;
\Longrightarrow \;\;\;\;g_{\mu\nu} = a^2 \, \widetilde{g}_{\mu\nu}.
\label{confg}
\end{equation}
Hence one can express the old connections in terms of the ones formed
from the rescaled fields,
\begin{eqnarray}
\Gamma^{\rho}_{~\mu\nu} & = & a^{-1} \Bigl(\delta^{\rho}_{~\mu} \,
a_{,\nu} \!+\!  \delta^{\rho}_{~\nu} \, a_{,\mu} \!-\!
\widetilde{g}^{\rho\sigma} \, a_{,\sigma} \,
\widetilde{g}_{\mu\nu}\Bigr) + \widetilde{\Gamma}^{\rho}_{
~\mu\nu} \label{confG} \; \\
A_{\mu cd} & = &-a^{-1} \Bigl(\widetilde{e}^{\nu}_{~c} \,
\widetilde{e}_{\mu d} \!-\!  \widetilde{e}^{\nu}_{~d} \,
\widetilde{e}_{\mu c} \Bigr) a_{,\nu} + \widetilde{A}_{\mu cd} \; ,
\end{eqnarray}
and re-defined fermion fields $\Psi \equiv a^{\frac{D-1}2} \psi$
to simplify the Lagrangian,
\begin{eqnarray}
\mathcal{L}_{\rm Dirac} = \overline{\Psi} \,
\widetilde{e}^{\mu}_{~b} \, \gamma^b \, i
\widetilde{\mathcal{D}}_{\mu} \Psi
\sqrt{-\widetilde{g}}-am\overline{\Psi}\Psi\sqrt{-\widetilde{g}}
\; , \label{Diract}
\end{eqnarray}
where $\widetilde{\mathcal{D}}_{\mu} \equiv \partial_{\mu} \!+\!
\frac{i}2 \widetilde{A}_{\mu cd} J^{cd}$.

We perturb the metric as,
\begin{equation}
\widetilde{g}_{\mu\nu} \equiv \eta_{\mu\nu} + \kappa h_{\mu\nu}
\qquad {\rm with} \qquad \kappa^2 = 16 \pi G \; .
\end{equation}
We then fix the local Lorentz gauge freedom by imposing symmetric gauge
($e_{\beta b} = e_{b \beta}$), and solve for the vierbein in terms
of the graviton \cite{RPW1},
\begin{equation}
\widetilde{e}[\widetilde{g}]_{\beta b} \equiv
\Bigl(\sqrt{\widetilde{g} \eta^{-1}} \, \Bigr)_{\!\beta}^{~\gamma}
\, \eta_{\gamma b} = \eta_{\beta b} + \frac12 \kappa h_{\beta b} -
\frac18 \kappa^2 h_{\beta}^{~\gamma} h_{\gamma b} + \dots
\end{equation}
At this stage there is no more point in distinguishing
between Latin letters for local Lorentz indices and Greek letters for
vector indices. Other conventions are that graviton indices are raised
and lowered with the Lorentz metric
($h^{\mu}_{~\nu} \equiv \eta^{\mu\rho} h_{\rho\nu}$, $h^{\mu\nu}
\equiv \eta^{\mu\rho} \eta^{\nu\sigma} h_{\rho\sigma}$) and that the
trace of the graviton field is $h \equiv \eta^{\mu\nu} h_{\mu\nu}$.
We also employ the usual Dirac ``slash'' notation,
\begin{equation}
\hspace{-.1cm}\not{\hspace{-.05cm} V}_{ij} \equiv V_{\mu}
\gamma^{\mu}_{ij} \; .
\end{equation}
Therefore one can expand all familiar operators accordingly in powers of
graviton field and obtain the perturbed, conformally rescaled Dirac
Lagrangian,
\begin{eqnarray}
\lefteqn{\mathcal{L}_{\rm Dirac} = \overline{\Psi}\Bigl[ i
\hspace{-.1cm}\not{ \hspace{-.1cm} \partial}-am\Bigr] \Psi +
\frac{\kappa}2 \overline{\Psi} \Bigl[ h i
\hspace{-.1cm}\not{\hspace{-.1cm}
\partial}  \!-\! h^{\mu\nu}  \gamma_{\mu} i
\partial_{\nu} \!-\! h_{\mu\rho , \sigma}
\gamma^{\mu} J^{\rho\sigma}
-am h\Bigr]\Psi } \nonumber \\
& & \hspace{0.8cm}+ \kappa^2 \Biggl\{ \Bigl[\frac18 h^2 \!-\!
\frac14 h^{\rho\sigma} h_{\rho\sigma}\Bigr] \overline{\Psi} i
\hspace{-.1cm}\not{\hspace{-.1cm}
\partial} \Psi \!+\! \Bigl[-\frac14 h h^{\mu\nu} \!+\! \frac38 h^{\mu\rho}
h_{\rho}^{~\nu}\Bigr] \overline{\Psi} \gamma_{\mu} i \partial_{\nu}
\Psi \nonumber\\
&& \hspace{0.8cm}+ \Bigl[-\frac14 h h_{\mu \rho , \sigma} + \frac18
h^{\nu}_{~\rho} h_{\nu \sigma , \mu} +
\frac14 (h^{\nu}_{~\mu} h_{\nu\rho})_{,\sigma} \!+\! \frac14
h^{\nu}_{~ \sigma} h_{\mu\rho ,\nu}\Bigr] \overline{\Psi}
\gamma^{\mu} J^{\rho\sigma}\Psi \nonumber\\
&&\hspace{4.8cm} -am\Bigl[\frac{1}{8}h^2
-\frac{1}{4}h^{\alpha\beta}h_{\alpha\beta}
\Bigr]\overline{\Psi}\Psi\Biggr\} + \mathcal{O}(\kappa^3).
\label{Dexp}
\end{eqnarray}
From the quadratic operator we see that the rescaled
massive fermion propagator can be connected to the solution
of Candelas and Raine \cite{CR, MW2} up to some powers of
scale factors,
\begin{eqnarray}
i S[m](x;x')_{C.R.}=(aa')^{-\frac{D-1}{2}}i S[m](x;x')\,.
\end{eqnarray}
Therefore the conformally re-scaled fermion propagator is,
\begin{eqnarray}
\lefteqn{ i S[m](x;x') = \frac{H^{D-2}}{(4\pi)^{\frac{D}2}}
\Gamma\Bigl( \frac{D}2 \!-\! 1\Bigr)(aa')^{\frac{D-1}{2}} \Bigl(a i
\hspace{-.08cm} \not{\hspace{-.15cm} \mathcal{D}} \; \frac1{\sqrt{a
a'}} \!+\! \sqrt{\frac{a}{a'}} m \, I \Bigr)}
\nonumber \\
& & \hspace{-.7cm} \times \Biggl\{ \frac{\Gamma(\frac{D}2 \!-\! 1
\!+\! i \frac{m}{H}) \Gamma(\frac{D}2 \!-\! i
\frac{m}{H})}{\Gamma(\frac{D}2 \!-\! 1) \Gamma(\frac{D}2)} \mbox{}_2
F_1\Bigl(\frac{D}2 \!-\! 1 \!+\! i \frac{m}{H}, \frac{D}2 \!-\! i
\frac{m}{H}; \frac{D}2 ;1 \!-\! \frac{y}4\Bigr) \Bigl(\frac{
I \!-\! \gamma^0}{2}\Bigr) \nonumber \\
& & \hspace{-.7cm} + \frac{\Gamma(\frac{D}2 \!-\! 1 \!-\! i
\frac{m}{H}) \Gamma(\frac{D}2 \!+\! i \frac{m}{H})}{\Gamma(\frac{D}2
\!-\! 1) \Gamma(\frac{D}2)} \mbox{}_2 F_1\Bigl(\frac{D}2 \!-\! 1
\!-\! i \frac{m}{H}, \frac{D}2 \!+\! i \frac{m}{H}; \frac{D}2 ;1
\!-\! \frac{y}4\Bigr) \Bigl(\frac{ I \!+\! \gamma^0}{2}\Bigr)
\Biggr\} . \qquad \label{fullp}
\end{eqnarray}
Here $ i \hspace{-.08cm} \not{\hspace{-.15cm}
\mathcal{D}} $ is just $ a^{-(\frac{D+1}2)} i \hspace{-.1cm}
\not{ \hspace{-.1cm} \partial} a^{(\frac{D-1}2)} $ and
a de Sitter invariant length function $y$ is formed by the following
function of the invariant length $\ell(x;x')$ between $x^{\mu}$ and
$x^{\prime \mu}$,
\begin{eqnarray}
y(x;x') & \equiv & 4 \sin^2\Bigl(\frac12 H \ell(x;x')\Bigr)
= a a' H^2 {\Delta x }^2(x;x') \; , \\
& = & a a' H^2 \Bigl( \Vert \vec{x} - \vec{x}'\Vert^2 - (\vert \eta
\!-\! \eta'\vert \!-\! i\delta)^2 \Bigr) \; . \label{fully}
\end{eqnarray}

It is useful to recast the solution (\ref{fullp}) using the
transformation formula for hypergeometric functions \cite{GR} and
then expand it in powers of $y$,

\begin{eqnarray}
\lefteqn{i S[m](x;x')= \frac{\Gamma(\frac{D}{2}\!-\!1) }
{4\pi^{\frac{D}{2}}}\Bigl[ i \hspace{-.08cm}
\not{\hspace{-.1cm}\partial} + a\,m \Bigr]
\frac{1}{\Delta x^{D-2}} }\nonumber\\ && \hspace{0.8cm}+
\frac{(H^2aa')^{\frac{D}{2}-1}}{(4\pi)^{\frac{D}{2}}}
\frac{\Gamma(\frac{D}{2}\!-\!1)\Gamma(2\!-\!\frac{D}{2})(i\frac{m}{H})}
{\Gamma(1\!+\!i\frac{m}{H})\Gamma(1\!-\!i\frac{m}{H})} \Biggl[ i
\hspace{-.08cm} \not{\hspace{-.1cm}\partial}
+\Bigl(\frac{D}{2}-1\Bigr)i Ha\gamma^0 + a\,m\Biggr] \nonumber\\
&& \hspace{0.7cm}\times\sum_{n=0}^{\infty}\Biggl\{
\frac{\Gamma(n\!+\!\frac{D}{2}\!-\!1\!+\!i\frac{m}{H})
\Gamma(n\!+\!\frac{D}{2}\!-\!1\!-\!i\frac{m}{H})}
{\Gamma(n\!+\!\frac{D}{2}\!-\!1)\Gamma(n\!+\!1)}
\Biggl[\frac{i\frac{m}{H}}{(n\!+\!\frac{D}{2}\!-\!1)} +
\gamma^0\Biggr]\Bigl(\frac{y}{4}\Bigr)^n \nonumber\\
& &\hspace{2cm} -\frac{\Gamma(n\!+\!1\!+\!i\frac{m}{H})
\Gamma(n\!+\!1\!-\!i\frac{m}{H})}
{\Gamma(n\!+\!1)\Gamma(n\!+\!3\!-\!\frac{D}{2})}
\Biggl[\frac{i\frac{m}{H}}{(n\!+\!1)}+\gamma^0\Biggr]\Bigl
(\frac{y}{4}\Bigr)^{n+2-\frac{D}{2}}\Biggr\}\,.\label{refullp}
\end{eqnarray}

Because we only endow fermions very small mass compared with
the Hubble parameter, for the computation purpose we simplify the
infinite series expansion by only keeping terms at order m,
\begin{eqnarray}
&& i\Bigl[\mbox{}_i S_j \Bigr](x;x') =
i\Bigl[\mbox{}_i S_j \Bigr]_{\rm cf}(x;x')+
i\Bigl[\mbox{}_i S_j \Bigr]_{\rm fm}(x;x')\nonumber\\
&&-\Bigl(\frac{m}{H}\Bigr)\frac{(H^2aa')^{\frac{D}{2}-1}}
{(4\pi)^{\frac{D}{2}}}\frac{\Gamma(\frac{D}{2}-1)
\Gamma(3-\frac{D}{2})}{(2-\frac{D}{2})}
\Bigl[\;\hspace{-.1cm}\not{\hspace{-.1cm}}\partial\gamma^0
+\Bigl(\frac{D}{2}-1\Bigr)Ha \Bigr]\nonumber\\
&&\sum_{n=0}^{\infty}\Biggl\{\frac{\Gamma(n+\frac{D}{2}-1)}
{\Gamma(n+1)}\Bigl(\frac{y}{4}\Bigr)^n
-\frac{\Gamma(n+1)}{\Gamma(n+3-\frac{D}{2})}
\Bigl(\frac{y}{4}\Bigr)^{n+2-\frac{D}{2}}\Biggr\}
+ \mathcal{O}(m^2).\label{mfprop}
\end{eqnarray}
The explicit expression for the first two terms in (\ref{mfprop})
are,
\begin{eqnarray}
&&i\Bigl[\mbox{}_i S_j \Bigr]_{\rm cf}(x;x') =
\frac{\Gamma(\frac{D}2\!-\!1)}{4\pi^{ \frac{D}2}} \, i
\hspace{-.1cm}\not{\hspace{-.1cm} \partial}_{ij} \frac1{\Delta
x^{D-2}}\,, \label{fprop}\\
&&i\Bigl[\mbox{}_i S_j \Bigr]_{\rm fm}(x;x')
=\frac{\Gamma(\frac{D}2\!-\!1)}{4\pi^{\frac{D}2}}
\frac{ma}{\Delta x^{D-2}}\,.
\end{eqnarray}
Here ``cf'' stands for ``conformal'' and ``fm'' stands for
``flat spacetime mass\footnote{ This term even at $a=1$ doesn't
stand for the full mass term in flat space. It is actually the most
singular term at order m.}.''
Even though the two infinite series expansions in (\ref{mfprop})
tend to cancel out with each other in $D=4$, the combinations are
still finite owing to the divergent factor $\frac{1}{(2-\frac{D}{2})}$.
In addition, they can not be reduced to an elementary function.
These facts complicate the computation.

We now represent the various interaction terms in (\ref{Dexp}) as
vertex operators acting on the fields. At order $\kappa$ the
interactions involve fields, $\overline{\Psi}_i$, $\Psi_j$ and
$h_{\alpha\beta}$, which we number ``1'', ``2'' and ``3'',
respectively. Each of the three interactions can be written as some
combination $V_{I ij}^{ \alpha\beta}$ of tensors, spinors and a
derivative operator acting on these fields. For example, the first
interaction is,
\begin{equation}
\frac{\kappa}2 h \overline{\Psi} i \hspace{-.1cm}\not{\hspace{-.1cm}
\partial} \Psi = \frac{\kappa}2 \eta^{\alpha \beta} i
\hspace{-.1cm}\not{\hspace{-.1cm}
\partial}_{2 ij} \times \overline{\Psi}_i \Psi_j h_{\alpha\beta} \equiv
V_{1ij}^{\alpha\beta} \times \overline{\Psi}_i \Psi_j
h_{\alpha\beta} \; .
\end{equation}
Hence the 3-point vertex operators are,
\begin{eqnarray}
&&V_{1ij}^{\alpha\beta} = \frac{\kappa}2 \eta^{\alpha \beta} i
\hspace{-.1cm} \not{\hspace{-.1cm} \partial}_{2 ij} \quad , \quad
V_{2ij}^{\alpha\beta} = -\frac{\kappa}2 \gamma^{(\alpha}_{ij}
i\partial_2^{\beta)}\nonumber\\ && V_{3ij}^{\alpha\beta} =
-\frac{\kappa}2 \Bigl(\gamma^{(\alpha} J^{\beta)\mu} \Bigr)_{ij}
\partial_{3 \mu} \quad , \quad V_{4ij}^{\alpha\beta}=
-\frac{\kappa}{2}am\eta^{\alpha\beta}I_{ij}\; . \label{3VO}
\end{eqnarray}
The order $\kappa^2$ interactions define 4-point vertex operators
$U_{I ij}^{ \alpha\beta\rho\sigma}$ similarly, for example,
\begin{equation}
\frac18 \kappa^2 h^2 \overline{\Psi}
i\hspace{-.1cm}\not{\hspace{-.1cm}
\partial} \Psi = \frac18 \kappa^2 \eta^{\alpha \beta} \eta^{\rho\sigma} i
\hspace{-.1cm} \not{\hspace{-.1cm} \partial}_{2 ij} \times
\overline{\Psi}_i \Psi_j h_{\alpha\beta} h_{\rho\sigma} \equiv
U_{1ij}^{\alpha\beta\rho\sigma} \times \overline{\Psi}_i \Psi_j
h_{\alpha\beta} h_{\rho\sigma} \; .
\end{equation}
The ten 4-point vertex operators are given in Table \ref{v4ops}.
Note that we do not bother to symmetrize upon the identical graviton
fields.

\begin{table}

\vbox{\tabskip=0pt \offinterlineskip
\def\tablerule{\noalign{\hrule}}
\halign to390pt {\strut#& \vrule#\tabskip=1em plus2em& \hfil#&
\vrule#& \hfil#\hfil& \vrule#& \hfil#& \vrule#& \hfil#\hfil&
\vrule#\tabskip=0pt\cr\tablerule
\omit&height4pt&\omit&&\omit&&\omit&&\omit&\cr &&\omit\hidewidth \#
&&\omit\hidewidth {\rm Vertex Operator}\hidewidth&& \omit\hidewidth
\#\hidewidth&& \omit\hidewidth {\rm Vertex Operator} \hidewidth&\cr
\omit&height4pt&\omit&&\omit&&\omit&&\omit&\cr \tablerule
\omit&height2pt&\omit&&\omit&&\omit&&\omit&\cr && 1 && $\frac18
\kappa^2 \eta^{\alpha\beta} \eta^{\rho\sigma} i \hspace{-.1cm}
\not{\hspace{-.1cm} \partial}_{2 ij}$ && 6 && $\frac18 \kappa^2
\eta^{\alpha\rho} (\gamma^{\mu} J^{\beta\sigma})_{ij}
\partial_{4\mu}$ &\cr
\omit&height2pt&\omit&&\omit&&\omit&&\omit&\cr \tablerule
\omit&height2pt&\omit&&\omit&&\omit&&\omit&\cr && 2 && $-\frac14
\kappa^2 \eta^{\alpha\rho} \eta^{\sigma\beta} i \hspace{-.1cm}
\not{\hspace{-.1cm} \partial}_{2 ij}$ && 7 && $\frac14 \kappa^2
\eta^{\alpha\rho} (\gamma^{\beta} J^{\sigma\mu})_{ij} (\partial_3 +
\partial_4)_{\mu}$ &\cr
\omit&height2pt&\omit&&\omit&&\omit&&\omit&\cr \tablerule
\omit&height2pt&\omit&&\omit&&\omit&&\omit&\cr && 3 && $-\frac14
\kappa^2 \eta^{\alpha\beta}\gamma^{\rho}_{ij} i\partial^{\sigma}_2$
&& 8 && $\frac14 \kappa^2 (\gamma^{\rho} J^{\sigma\alpha})_{ij}
\partial_4^{\beta}$ &\cr
\omit&height2pt&\omit&&\omit&&\omit&&\omit&\cr\tablerule
\omit&height2pt&\omit&&\omit&&\omit&&\omit&\cr && 4 && $\frac38
\kappa^2 \eta^{\alpha\rho} \gamma^{\beta}_{ij} i\partial^{\sigma}_2$
&& 9 && $-\frac{1}{8}\kappa^2
\eta^{\alpha\beta}\eta^{\rho\sigma}am$ &\cr
\omit&height2pt&\omit&&\omit&&\omit&&\omit&\cr \tablerule
\omit&height2pt&\omit&&\omit&&\omit&&\omit&\cr && 5 && $-\frac14
\kappa^2 \eta^{\alpha\beta} (\gamma^{\rho} J^{\sigma\mu})_{ij}$ &&
10&&$\frac{1}{4}\kappa^2\eta^{\alpha\rho}\eta^{\sigma\beta}am$&\cr
\omit&height2pt&\omit&&\omit&&\omit&&\omit&\cr\tablerule}}

\caption{Vertex operators $U_{I ij}^{\alpha\beta\rho\sigma}$
contracted into $\overline{\Psi}_i \Psi_j h_{\alpha\beta}
h_{\rho\sigma}$.}

\label{v4ops}

\end{table}

\section{Graviton Propagator}

In this section we briefly sketch how to obtain the graviton propagator
and presently give the explicit expression for it. The low energy
effective field theory for gravity is Einstein-Hilbert,
\begin{equation}
\mathcal{L}_{\rm E-H} \equiv \frac1{16\pi G} \Bigl( R -
(D\!-\!2) \Lambda\Bigr) \sqrt{-g} \; . \label{Einstein}
\end{equation}
We follow the same convention as the fermion sector to re-scale
the metric (\ref{confg}) and connections (\ref{confG}) for garvity
even though it is not conformally invariant. In order to obtain the
graviton propagator we also need to fix $a$. We work on the open
conformal coordinate patch of de Sitter, which implies,
\begin{equation}
ds^2 = a^2 \Bigl( -d\eta^2 + d\vec{x} \!\cdot\! d\vec{x}\Bigr)
\qquad {\rm where} \qquad a(\eta) = -\frac1{H\eta} \; ,
\end{equation}
and the $D$-dimensional Hubble constant is $H \equiv
\sqrt{\Lambda/(D\!-\!1)}$. Note that the conformal time $\eta$ runs
from $-\infty$ to zero. For this choice of scale factor we can
extract a surface term from the invariant Lagrangian and write it
in the form \cite{TW1},
\begin{eqnarray}
\lefteqn{\mathcal{L}_{\rm E-H} \!-\! {\rm Surface} =
{\scriptstyle (\frac{D}2 - 1)} H a^{D-1} \sqrt{-\widetilde{g}}
\widetilde{g}^{\rho\sigma} \widetilde{g}^{\mu \nu} h_{\rho\sigma
,\mu} h_{\nu 0} + a^{D-2} \sqrt{-\widetilde{g}}
\widetilde{g}^{\alpha\beta} \widetilde{g}^{\rho\sigma}
\widetilde{g}^{\mu\nu} } \nonumber \\
& & \hspace{2cm} \times \Bigl\{{\scriptstyle \frac12} h_{\alpha\rho
,\beta} h_{\sigma\mu ,\nu} \!-\! {\scriptstyle \frac12}
h_{\alpha\beta ,\rho} h_{\sigma\mu ,\nu} \!+\! {\scriptstyle
\frac14} h_{\alpha\beta ,\rho} h_{\mu\nu ,\sigma} \!-\!
{\scriptstyle \frac14} h_{\alpha\rho ,\mu} h_{\beta\sigma ,\nu}
\Bigr\} . \quad \label{Linv}
\end{eqnarray}

There has been a long controversy about the graviton propagator
in de Sitter space which is discussed in \cite{MTW1,MTW1a,MTW2,MTW3}.
It turns out that one cannot add a gauge fixing term which preserves
all de Sitter symmetries, so we break spatial special conformal
transformations\footnote{The de Sitter symmetry group in $D=4$ includes
3 spatial translations, 3 spatial rotations, 1 dilatation
and 3 spatial special conformal transformations.} with the term,
\begin{equation}
\mathcal{L}_{GF} = -\frac12 a^{D-2} \eta^{\mu\nu} F_{\mu} F_{\nu} \;
, \; F_{\mu} \equiv \eta^{\rho\sigma} \Bigl(h_{\mu\rho , \sigma} -
\frac12 h_{\rho \sigma , \mu} + (D \!-\! 2) H a h_{\mu \rho}
\delta^0_{\sigma} \Bigr) . \label{GR}
\end{equation}
The quadratic part of $\mathcal{L}_{\rm E-H} +
\mathcal{L}_{GF}$ can be partially integrated to take the form
$\frac12 h^{\mu\nu} D_{\mu\nu}^{~~\rho \sigma} h_{\rho\sigma}$.
Hence one can solve graviton propagator accordingly,
\begin{equation}
D_{\mu\nu}^{~~\rho\sigma} \times i\Bigl[{}_{\rho\sigma}
\Delta^{\alpha\beta} \Bigr](x;x') = \delta_{\mu}^{(\alpha}
\delta_{\nu}^{\beta)} i \delta^D(x-x') \; .
\end{equation}
For the more detail of solving this gauge-fixed graviton propagator
equation one can consult \cite{MW1,TW1}. Here we would like to
give the result without a derivation. The graviton propagator
in this gauge takes the form of a sum of constant index factors
times scalar propagators,
\begin{equation}
i\Bigl[{}_{\mu\nu} \Delta_{\rho\sigma}\Bigr](x;x') = \sum_{I=A,B,C}
\Bigl[{}_{\mu\nu} T^I_{\rho\sigma}\Bigr] i\Delta_I(x;x') \; ,
\label{gprop}
\end{equation}
where the index factors are,
\begin{eqnarray}
\Bigl[{}_{\mu\nu} T^A_{\rho\sigma}\Bigr] & = & 2 \,
\overline{\eta}_{\mu (\rho} \overline{\eta}_{\sigma) \nu} -
\frac2{D\!-\! 3} \overline{\eta}_{\mu\nu}
\overline{\eta}_{\rho \sigma} \; , \\
\Bigl[{}_{\mu\nu} T^B_{\rho\sigma}\Bigr] & = & -4 \delta^0_{(\mu}
\overline{\eta}_{\nu) (\rho} \delta^0_{\sigma)} \; , \\
\Bigl[{}_{\mu\nu} T^C_{\rho\sigma}\Bigr] & = & \frac2{(D \!-\!2) (D
\!-\!3)} \Bigl[(D \!-\!3) \delta^0_{\mu} \delta^0_{\nu} +
\overline{\eta}_{\mu\nu}\Bigr] \Bigl[(D \!-\!3) \delta^0_{\rho}
\delta^0_{\sigma} + \overline{\eta}_{\rho \sigma}\Bigr] \; .
\end{eqnarray}
\begin{equation}
\overline{\eta}_{\mu\nu} \equiv \eta_{\mu\nu} + \delta^0_{\mu}
\delta^0_{\nu} \qquad {\rm and} \qquad \overline{\delta}^{\mu}_{\nu}
\equiv \delta^{\mu}_{\nu} - \delta_0^{\mu} \delta^0_{\nu} \; .
\end{equation}
The most singular term for each scalar propagator is the propagator for
a massless, conformally coupled scalar \cite{BD},
\begin{equation}
{i\Delta}_{\rm cf}(x;x') = \frac{H^{D-2}}{(4\pi)^{\frac{D}2}}
\Gamma\Bigl( \frac{D}2 \!-\! 1\Bigr)
\Bigl(\frac4{y}\Bigr)^{\frac{D}2-1} \; .
\end{equation}
The three scalar propagators are,
\begin{eqnarray}
\lefteqn{i \Delta_A(x;x') =  i \Delta_{\rm cf}(x;x') } \nonumber \\
& & + \frac{H^{D-2}}{(4\pi)^{\frac{D}2}} \frac{\Gamma(D \!-\!
1)}{\Gamma( \frac{D}2)} \left\{\! \frac{D}{D\!-\! 4}
\frac{\Gamma^2(\frac{D}2)}{\Gamma(D \!-\! 1)}
\Bigl(\frac4{y}\Bigr)^{\frac{D}2 -2} \!\!\!\!\!\! - \pi
\cot\Bigl(\frac{\pi}2 D\Bigr) + \ln(a a') \!\right\} \nonumber \\
& & + \frac{H^{D-2}}{(4\pi)^{\frac{D}2}} \! \sum_{n=1}^{\infty}\!
\left\{\! \frac1{n} \frac{\Gamma(n \!+\! D \!-\! 1)}{\Gamma(n \!+\!
\frac{D}2)} \Bigl(\frac{y}4 \Bigr)^n \!\!\!\! - \frac1{n \!-\!
\frac{D}2 \!+\! 2} \frac{\Gamma(n \!+\!  \frac{D}2 \!+\!
1)}{\Gamma(n \!+\! 2)} \Bigl(\frac{y}4 \Bigr)^{n - \frac{D}2 +2}
\!\right\} \! . \quad \label{DeltaA}
\end{eqnarray}
\begin{eqnarray}
\lefteqn{i \Delta_B(x;x') =  i \Delta_{\rm cf}(x;x') -
\frac{H^{D-2}}{(4 \pi)^{\frac{D}2}} \! \sum_{n=0}^{\infty}\!
\left\{\!  \frac{\Gamma(n \!+\! D \!-\! 2)}{\Gamma(n \!+\!
\frac{D}2)} \Bigl(\frac{y}4 \Bigr)^n \right. }
\nonumber \\
& & \hspace{6.5cm} \left. - \frac{\Gamma(n \!+\!
\frac{D}2)}{\Gamma(n \!+\! 2)} \Bigl( \frac{y}4 \Bigr)^{n -
\frac{D}2 +2} \!\right\} \! , \qquad
\label{DeltaB} \\
\lefteqn{i \Delta_C(x;x') =  i \Delta_{\rm cf}(x;x') +
\frac{H^{D-2}}{(4\pi)^{\frac{D}2}} \! \sum_{n=0}^{\infty} \left\{\!
(n\!+\!1) \frac{\Gamma(n \!+\! D \!-\! 3)}{\Gamma(n \!+\!
\frac{D}2)}
\Bigl(\frac{y}4 \Bigr)^n \right. } \nonumber \\
& & \hspace{4.5cm} \left. - \Bigl(n \!-\! \frac{D}2 \!+\!  3\Bigr)
\frac{ \Gamma(n \!+\! \frac{D}2 \!-\! 1)}{\Gamma(n \!+\! 2)}
\Bigl(\frac{y}4 \Bigr)^{n - \frac{D}2 +2} \!\right\} \! . \qquad
\label{DeltaC}
\end{eqnarray}
These expressions might seem daunting but they are actually simple
to use because the infinite sums vanish in $D=4$, and each term in
these sums goes like a positive power of $y(x;x')$. This means the
infinite sums can only contribute when multiplied by a divergent
term, and even then only a small number of terms can contribute.
Note also that the $B$-type and $C$-type propagators agree with the
conformal propagator in $D=4$.

\section{Counterterm Analysis}

In this section we would deal with the local counterterms we must add,
order-by-order in perturbation theory, to absorb divergences in the
sense of BPHZ renormalization. The particular counterterms which
renormalize the ferm\-i\-on self-energy must obviously involve a
single $\overline{\psi}$ and a single $\psi$.
At one loop order the superficial degree of divergence (S.D.D.)
of quantum gravitational contributions to the fermion self-energy
is three, so the necessary counterterms can involve zero, one, two
or three derivatives. These derivatives can either act upon the
fermi fields or upon the metric, in which case they must be organized
into curvatures or derivatives of curvatures. We close with a discussion
of possible noninvariant counterterms.

All one loop corrections from quantum gravity must carry a factor of
$\kappa^2 \sim {\rm mass}^{-2}$. There will be additional dimensions
associated with derivatives and with the various fields, and the
balance must be struck using the renormalized fermion mass, $m$.
For the purpose of our computation, we only focus the counterterms
at order m. Because S.D.D. is three, the possible expressions at
order m must consist of one mass and two derivatives which can either
act upon the fermions or else on the metric to produce curvatures,
\begin{equation}
\kappa^2 m \overline{\psi} (i \hspace{-.1cm} \not{\hspace{-.15cm}
\mathcal{D}})^2 \psi \sqrt{-g} \quad , \quad \kappa^2 m R
\overline{\psi} \psi \sqrt{-g} \; . \label{two}
\end{equation}
We then specialize the above expressions from the general background to
de Sitter. Hence the invariant counter-Lagrangian we require at order m
is,
\begin{eqnarray}
\lefteqn{\Delta \mathcal{L}_{\rm inv} = \lambda_1 \kappa^2 m
\overline{\psi} (i \hspace{-.1cm} \not{\hspace{-.15cm}
\mathcal{D}})^2 \psi \sqrt{-g} + \lambda_2 \kappa^2 m R
\overline{\psi} \psi \sqrt{-g} \; , \label{invctms} } \\
& & \longrightarrow \lambda_1 \kappa^2 \overline{\Psi} m \Bigl(
i\hspace{-.1cm} \not{\hspace{-.1cm} \partial} a^{-1}
i\hspace{-.1cm} \not{\hspace{-.1cm} \partial} \Bigr)\Psi
+ \lambda_2 \kappa^2(D-1)DH^2 ma\overline{\Psi} \Psi \; . \qquad
\end{eqnarray}
Here $\lambda_1$ and $\lambda_2$ are $D$-dependent constants which
are dimensionless. The associated vertex operators are,
\begin{eqnarray}
C_{1 ij} & \equiv & \lambda_1 \kappa^2 \Bigl(
\frac{m}{a}\partial^2 + mH\gamma^{0}\hspace{-.1cm}
\not{\hspace{-.1cm} \partial} \Bigr) = \lambda_1 \kappa^2 m
\hspace{-.1cm}\not{\hspace{-.1cm} \partial}
a^{-1} \hspace{-.1cm}\not{\hspace{-.1cm} \partial}\; , \label{C1} \\
C_{2 ij} & \equiv & \lambda_2 D(D-1) \kappa^2 H^2 ma \; . \label{C2}
\end{eqnarray}
$C_1$ is the higher derivative counterterm. It will renormalize the
most singular terms --- coming from the $i\Delta_{\rm cf}$ part
of the graviton propagator ---which are unimportant because
they are suppressed by powers of the scale factor. The other
vertex operator, $C_2$, is a sort of dimensionful field strength
renormalization in de Sitter background. It will renormalize
the less singular contributions which derive physically from
inflationary particle production.

Because our gauge fixing functional (necessarily) breaks de Sitter
invariance, it is also necessary to consider noninvariant counterterms.
These noninvariant counterterms must respect the symmetries of the gauge
condition, which are homogeneity, isotropy and dilatation invariance.
As one loop counterterms, they should also contain a factor of
$\kappa^2$, multiplied by a spinor differential operator with the
dimension of mass-cubed, involving no more than three derivatives and
acting between $\overline{\Psi}$ and $\Psi$. As the only dimensionful
constant in our problem, powers of $H$ must be used to make up whatever
dimensions are not supplied by derivatives. Homogeneity implies that
the spinor differential operator cannot depend upon the spatial coordinate
$x^i$. Similarly, isotropy requires that any spatial derivative operators
$\partial_i$ must either be contracted into $\gamma^i$ or another
spatial derivative. Owing to the identity,
\begin{equation}
(\gamma^i \partial_i)^2 = - \nabla^2 \; ,
\end{equation}
we can think of all spatial derivatives as contracted into
$\gamma^i$. Although the temporal derivative is not required to be
multiplied by $\gamma^0$ we lose nothing by doing so provided
additional dependence upon $\gamma^0$ is allowed.

The final residual symmetry is dilatation invariance. It has the
crucial consequence that derivative operators can only appear in
the form $a^{-1}\partial_{\mu}$. In addition the entire counterterm
must have an overall factor of $a$, and there can be no other dependence
upon $\eta$. So the most general order m counterterm consistent with
our gauge condition takes the following form,
\begin{equation}
\Delta \mathcal{L}_{\rm non} = \kappa^2 H^2 ma \overline{\Psi}
\mathcal{S}\Bigl((H a)^{-1} \gamma^0 \partial_0, (H a)^{-1} \gamma^i
\partial_i\Bigr) \Psi \; , \label{noninv}
\end{equation}
where the spinor function $\mathcal{S}(b,c)$ is at most a second
order polynomial function of its arguments , and it may involve
$\gamma^0$ in an arbitrary way.

Three more principles constrain the order m noninvariant
counterterms. The first of these principles is that the
fermion self-energy at order m involves only even powers of
gamma matrices. This follows because the three-point vertices,
the four-point vertices and the fermion propagator all consist of an
even number of $\gamma$'s at order $\rm{m}^1$ and an odd number of
$\gamma$'s at order $\rm {m}^0$. The diagram which consists of
one 4-point vertex possesses an even number of gamma matrices at order m.
The contribution from any diagram with two 3-point vertices consists of
three factors involving gamma matrices: one factor from the fermion
propagator and one factor from each of the two vertices. At order m
such a product consists of one even and two odd factors, so it contains
an even number of gamma matrices. This principle fixes the dependence upon
$\gamma^0$ and allows us to express the spinor differential operator
in terms of just six constants $\beta_i$,
\begin{eqnarray}
&& \kappa^2 H^2 ma \mathcal{S}\Bigl((H a)^{-1} \gamma^0
\partial_0, (H a)^{-1} \gamma^i \partial_i\Bigr)
\nonumber \\ && =\kappa^2 ma \Biggl\{
\beta_1 (a^{-1} \gamma^0 \partial_0)^2
+ \beta_2 \Bigl[(a^{-1} \gamma^0 \partial_0)
(a^{-1} \gamma^i \partial_i)\Bigr]
+ \beta_3 (a^{-1} \gamma^i \partial_i)^2  \nonumber \\
&& \hspace{3.5cm} +H\gamma^{0} \Biggl( \beta_4
(a^{-1} \gamma^0 \partial_0)
+ \beta_5 (a^{-1} \gamma^i \partial_i) \Biggr)+
H^2 \beta_{6} \Biggr\} . \qquad \label{expS}
\end{eqnarray}
In this expansion, but for the rest of this section only, we define
noncommuting factors within square brackets to be symmetrically
ordered, for example,
\begin{eqnarray}
\Bigl[ ( a^{-1}\gamma^0\partial_0 )
( a^{-1}\gamma^i\partial_i ) \Bigr] \equiv
\frac12 ( a^{-1}\gamma^0\partial_0 )
(a^{-1}\gamma^i\partial_i)
+ \frac12 ( a^{-1}\gamma^i\partial_i )
(a^{-1}\gamma^0\partial_0) \; . \qquad
\end{eqnarray}

The second principle is that our gauge condition (\ref{GR}) becomes
Poincar\'e invariant in the flat space limit of $H \rightarrow 0$,
where the conformal time is $\eta = -e^{-Ht}/H$ with $t$ held fixed.
In that limit only the three quadratic terms of (\ref{expS}) survive,
\begin{eqnarray}
&& \lim_{H \rightarrow 0} \kappa^2 H^2 ma \mathcal{S}
\Bigl((Ha)^{-1} \gamma^0 \partial_0,
(H a)^{-1} \gamma^i \partial_i\Bigr)
\nonumber \\ && =\kappa^2 ma \Biggl\{
\beta_1 (a^{-1} \gamma^0 \partial_0)^2
+ \beta_2 \Bigl[(a^{-1} \gamma^0 \partial_0)
(a^{-1} \gamma^i \partial_i)\Bigr]
+ \beta_3 (a^{-1} \gamma^i \partial_i)^2 \Biggr\}
\qquad.
\end{eqnarray}
Because the entire theory is Poincar\'e invariant in that limit,
these three terms must sum to a term proportional to $(\gamma^{\mu}
\partial_{\mu})^2$, which implies,
\begin{equation}
\beta_1 = \frac12 \beta_2 = \beta_3 \; .
\end{equation}
But in that case the three quadratic terms sum to give
(\ref{C1}),
\begin{eqnarray}
\kappa^2 ma  \Biggl\{(a^{-1} \gamma^0 \partial_0)^2
+ 2\Bigl[(a^{-1} \gamma^0 \partial_0)
(a^{-1} \gamma^i \partial_i)\Bigr]
+ (a^{-1} \gamma^i \partial_i)^2 \Biggr\}
=\kappa^2 m \hspace{-.1cm}\not{\hspace{-.1cm} \partial}
a^{-1} \hspace{-.1cm}\not{\hspace{-.1cm} \partial}
\; . \qquad
\end{eqnarray}
Because it is the same as one of the invariant
counterterms, it need not be included in $\mathcal{S}$.
Besides, the final term in (\ref{expS}) recovers the other
invariant counterterm (\ref{C2}). So the two remaining
noninvariant counterterms we need to consider in (\ref{expS}) are,
\begin{equation}
\Delta \mathcal{L}_{non}=
\overline{\Psi}\Biggl\{\kappa^2maH\gamma^0\Bigl[
\beta_4(a^{-1} \gamma^0 \partial_0)
+ \beta_5 (a^{-1} \gamma^i \partial_i)\Bigr] \Biggr\}\Psi
\; .\label{nictm}
\end{equation}
However, these two terms are not independent of the last term
in (\ref{C1}). Therefore we could chose any four independent counterterm
operators we need for this computation,
\begin{eqnarray}
&& \alpha_1\kappa^2\frac{m}{a}\partial^2\;\;,\;\;
\alpha_4\kappa^2H^{2} ma \,, \label{inv12} \\
&& \alpha_2\kappa^{2} mH\partial_0\;\;,\;\;
\alpha_3\kappa^2 mH\gamma^0 \hspace{-.1cm}
\not{\hspace{-.1cm}\overline{\partial}}\;.\label{noninv34}
\end{eqnarray}

\section{Contributions from the 4-Point Vertices}
In this section the contributions from 4-point vertex operators of
Table~\ref{v4ops} is evaluated. The generic diagram topology is depicted
in Fig.~1. The analytic form is,
\begin{equation}
-i\Bigl[\mbox{}_i \Sigma_j^{\rm 4pt}\Bigr](x;x') = \sum_{I=1}^{10} i
U^{\alpha\beta \rho\sigma}_{Iij} \, i\Bigl[\mbox{}_{\alpha\beta}
\Delta_{\rho\sigma} \Bigr](x;x') \, \delta^D(x\!-\!x') \; .
\label{4ptloop}
\end{equation}
\begin{figure}
\begin{center}
\includegraphics[width=4.0cm]{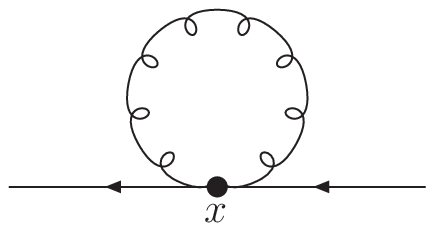}
\\{\rm Fig.~1: Contribution from 4-point vertices.}
\label{Fig1}
\end{center}
\end{figure}
And the generic contraction for each of the vertex operators in
Table~\ref{v4ops} is given in Table~\ref{4con}.

From an inspection of the generic contractions in Table~\ref{4con}
it is obvious that we must work out how the three index factors
$[\mbox{}_{\alpha \beta} T^I_{\rho \sigma}]$ which make up the
graviton propagator contract into $\eta^{\alpha\beta}$ and
$\eta^{\alpha\rho}$. For the $A$-type and $B$-type index factors the
various contractions give,

\begin{table}

\vbox{\tabskip=0pt \offinterlineskip
\def\tablerule{\noalign{\hrule}}
\halign to390pt {\strut#& \vrule#\tabskip=1em plus2em& \hfil#\hfil&
\vrule#& \hfil#\hfil& \vrule#\tabskip=0pt\cr
\tablerule
\omit&height4pt&\omit&&\omit&\cr
&&$\!\!\!\!{\rm I}\!\!\!\!$ &&
$\!\!\!\! i [\mbox{}_{\alpha\beta} \Delta_{\rho\sigma}](x;x') \, i
U_I^{\alpha\beta\rho\sigma} \, \delta^D(x\!-\!x') \!\!\!\!$ & \cr
\omit&height4pt&\omit&&\omit&\cr
\tablerule
\omit&height2pt&\omit&&\omit&\cr
&& 1 && $-\frac18 \kappa^2 \, i
[\mbox{}^{\alpha}_{~\alpha} \Delta^{\rho}_{~\rho}](x;x) \,
\hspace{-.1cm} \not{\hspace{-.1cm} \partial}\,\delta^D(x\!-\!x')$&\cr
\omit&height2pt&\omit&&\omit&\cr
\tablerule
\omit&height2pt&\omit&&\omit&\cr
&& 2 && $\frac14 \kappa^2 \, i
[\mbox{}^{\alpha\beta} \Delta_{\alpha\beta}](x;x) \, \hspace{-.1cm}
\not{\hspace{-.1cm} \partial} \, \delta^D(x\!-\!x')$ & \cr
\omit&height2pt&\omit&&\omit&\cr
\tablerule
\omit&height2pt&\omit&&\omit&\cr
&& 3 && $\frac14 \kappa^2 \, i
[\mbox{}^{\alpha}_{~\alpha} \Delta_{\rho\sigma}](x;x) \,
\gamma^{\rho} \partial^{\sigma} \, \delta^D(x\!-\!x')$ & \cr
\omit&height2pt&\omit&&\omit&\cr
\tablerule
\omit&height2pt&\omit&&\omit&\cr
&& 4 && $-\frac38 \kappa^2 \, i
[\mbox{}^{\alpha}_{~\beta} \Delta_{\alpha\sigma}](x;x) \,
\gamma^{\beta} \partial^{\sigma} \, \delta^D(x\!-\!x')$ & \cr
\omit&height2pt&\omit&&\omit&\cr
\tablerule
\omit&height2pt&\omit&&\omit&\cr
&& 5 && $-\frac{i}4 \kappa^2 \,
\partial_{\mu}' i [\mbox{}^{\alpha}_{~\alpha}
\Delta_{\rho\sigma}](x;x') \, \gamma^{\rho} J^{\sigma \mu} \,
\delta^D(x\!-\!x')$ & \cr
\omit&height2pt&\omit&&\omit&\cr
\tablerule
\omit&height2pt&\omit&&\omit&\cr
&& 6 && $\frac{i}8\kappa^2 \, \partial_{\mu}'
i [\mbox{}^{\alpha}_{~\beta}\Delta_{\alpha\sigma}](x;x')
\, \gamma^{\mu} J^{\beta\sigma} \,\delta^D(x\!-\!x')$ & \cr
\omit&height2pt&\omit&&\omit&\cr
\tablerule
\omit&height2pt&\omit&&\omit&\cr
&& 7 && $\frac{i}4\kappa^2 \, \partial_{\mu} i [\mbox{}^{\alpha}_{~\beta}
\Delta_{\alpha\sigma}](x;x) \, \gamma^{\beta} J^{\sigma\mu} \,
\delta^D(x\!-\!x')$ & \cr
\omit&height2pt&\omit&&\omit&\cr
\tablerule
\omit&height2pt&\omit&&\omit&\cr
&& 8 && $\frac{i}4\kappa^2 \, \partial^{\prime \beta}
i [\mbox{}_{\alpha \beta}\Delta_{\rho\sigma}](x;x') \, \gamma^{\rho}
J^{\sigma \alpha} \,\delta^D(x\!-\!x')$ & \cr
\omit&height2pt&\omit&&\omit&\cr
\tablerule
\omit&height2pt&\omit&&\omit&\cr
&& 9 && $-\frac{i}8\kappa^2 \,am i [\mbox{}^{\alpha}_{~\alpha}
\Delta^{\rho}_{~\rho}](x;x) \, \delta^D(x\!-\!x')$ & \cr
\omit&height2pt&\omit&&\omit&\cr
\tablerule
\omit&height2pt&\omit&&\omit&\cr
&& 10 && $\frac{i}4\kappa^2 \,am i [\mbox{}^{\alpha\beta}
\Delta_{\alpha\beta}](x;x)\, \delta^D(x\!-\!x')$ & \cr
\omit&height2pt&\omit&&\omit&\cr
\tablerule}}

\caption{Generic 4-point contractions}

\label{4con}

\end{table}

\begin{eqnarray}
\eta^{\alpha\beta} \, \Bigl[{}_{\alpha\beta} T^A_{\rho\sigma}\Bigr]
= - \Bigl(\frac4{D\!-\!3}\Bigr) \, \overline{\eta}_{\rho\sigma} & ,
& \eta^{\alpha\rho} \, \Bigl[{}_{\alpha\beta} T^A_{\rho\sigma}\Bigr]
= \Bigl(D\!-\!\frac2{D\!-\!3}\Bigr) \,\overline{\eta}_{\beta\sigma} \; , \\
\eta^{\alpha\beta} \, \Bigl[{}_{\alpha\beta} T^B_{\rho\sigma}\Bigr]
= 0 & , & \eta^{\alpha\rho} \, \Bigl[{}_{\alpha\beta}
T^B_{\rho\sigma} \Bigr] = -(D \!-\! 1) \, \delta^0_{\beta}
\delta^0_{\sigma} + \overline{\eta}_{ \beta\sigma} \; ,
\end{eqnarray}
For the $C$-type index factor they are,
\begin{eqnarray}
\eta^{\alpha\beta} \, \Bigl[{}_{\alpha\beta} T^C_{\rho\sigma}\Bigr]
& = & \Bigl(\frac4{D-2}\Bigr) \, \delta^0_{\rho} \delta^0_{\sigma} +
\frac4{(D\!-\!2)(D\!-\!3)} \, \overline{\eta}_{\rho\sigma} \; , \nonumber \\
\eta^{\alpha\rho} \, \Bigl[{}_{\alpha\beta} T^C_{\rho\sigma}\Bigr] &
= & -2 \Bigl(\frac{D\!-\!3}{D\!-\!2}\Bigr) \, \delta^0_{\beta}
\delta^0_{\sigma} \!+\! \frac2{(D\!-\!2) (D\!-\!3)} \,
\overline{\eta}_{\beta\sigma} \; . \qquad
\end{eqnarray}
At order m we actually only require double contractions. For the $A$-type
index factor these are,
\begin{eqnarray}
\eta^{\alpha\beta} \eta^{\rho\sigma} \, \Bigl[{}_{\alpha\beta}
T^A_{\rho\sigma}\Bigr] & = & -4 \Bigl(\frac{D\!-\!1}{D\!-\!3}\Bigr)
\; ,
\nonumber \\
\eta^{\alpha\rho} \eta^{\beta\sigma} \, \Bigl[{}_{\alpha\beta}
T^A_{\rho\sigma}\Bigr] & = & D (D\!-\!1) - 2
\Bigl(\frac{D\!-\!1}{D\!-\!3}\Bigr) \; .
\end{eqnarray}
The double contractions of the $B$-type and $C$-type index factors
are,
\begin{eqnarray}
\eta^{\alpha\beta} \eta^{\rho\sigma} \, \Bigl[{}_{\alpha\beta}
T^B_{\rho\sigma}\Bigr] = 0 & , & \eta^{\alpha\rho}
\eta^{\beta\sigma} \,
\Bigl[{}_{\alpha\beta} T^B_{\rho\sigma} \Bigr] = 2 (D \!-\! 1) \; , \\
\eta^{\alpha\beta} \eta^{\rho\sigma} \, \Bigl[{}_{\alpha\beta}
T^C_{\rho\sigma}\Bigr] = \frac8{(D\!-\!2)(D\!-\!3)} & , &
\eta^{\alpha\rho} \eta^{\beta\sigma} \, \Bigl[{}_{\alpha\beta}
T^C_{\rho\sigma}\Bigr] = 2 \frac{(D^2 \!-\! 5D \!+\!
8)}{(D\!-\!2)(D\!-\!3)} \; . \qquad
\end{eqnarray}

Table~\ref{4props} was generated from Table~\ref{4con} by expanding
the graviton propagator in terms of index factors,
\begin{equation}
i\Bigl[{}_{\alpha\beta} \Delta_{\rho\sigma}\Bigr](x;x') =
\Bigl[{}_{\alpha\beta} T^A_{\rho\sigma}\Bigr] i\Delta_A(x;x') +
\Bigl[{}_{\alpha\beta} T^B_{\rho\sigma}\Bigr] i\Delta_B(x;x') +
\Bigl[{}_{\alpha\beta} T^C_{\rho\sigma}\Bigr] i\Delta_C(x;x') \; .
\end{equation}
We then perform the relevant contractions using the previous
identities.
\begin{table}

\vbox{\tabskip=0pt \offinterlineskip
\def\tablerule{\noalign{\hrule}}
\halign to390pt {\strut#& \vrule#\tabskip=1em plus2em& \hfil#\hfil&
\vrule#& \hfil#\hfil& \vrule#& \hfil#\hfil& \vrule#\tabskip=0pt\cr
\tablerule \omit&height4pt&\omit&&\omit&&\omit&\cr &&$\!\!\!\!{\rm
I}\!\!\!\!$ && $\!\!\!\!{\rm J}\!\!\!\!$ && $\!\!\!\! i
[\mbox{}_{\alpha\beta} T^J_{\rho\sigma}]\, i\Delta_J(x;x') \, i
U_I^{\alpha\beta\rho\sigma} \, \delta^D(x\!-\!x') \!\!\!\!$ & \cr
\omit&height4pt&\omit&&\omit&&\omit&\cr
\tablerule
\omit&height2pt&\omit&&\omit&&\omit&\cr && 9 && A && $\!\!\!\!
\frac{i}{2}\kappa^2\frac{(D-1)}{(D-3)} \; am \; i\Delta_A(x;x) \,
\delta^D(x\!-\!x') \!\!\!\!$ & \cr
\omit&height2pt&\omit&&\omit&&\omit&\cr
\tablerule
\omit&height2pt&\omit&&\omit&&\omit&\cr && 9 && C && $\!\!\!\!
-\frac{i}{(D-2)(D-3)}\kappa^2 \, am \, i\Delta_C(x;x) \,
\delta^D(x\!-\!x') \!\!\!\!$ & \cr
\omit&height2pt&\omit&&\omit&&\omit&\cr
\tablerule
\omit&height2pt&\omit&&\omit&&\omit&\cr && 10 && A && $\!\!\!\!
\frac{i}{4}\kappa^2[D(D-1)-2\frac{(D-1)}{(D-3)}] \; am \;
i\Delta_A(x;x) \, \delta^D(x\!-\!x') \!\!\!\!$ & \cr
\omit&height2pt&\omit&&\omit&&\omit&\cr
\tablerule
\omit&height2pt&\omit&&\omit&&\omit&\cr && 10 && B &&
$\frac{i}{2}\kappa^2(D-1) \, am \, i\Delta_{B}(x;x) \, \delta^D
(x\!-\!x')$ & \cr \omit&height2pt&\omit&&\omit&&\omit&\cr
\tablerule
\omit&height2pt&\omit&&\omit&&\omit&\cr && 10 && C && $\!\!\!\!
\frac{i}{2}\kappa^2 \frac{(D^2-5D+8)}{(D-2)(D-3)} \, am \,
i\Delta_C(x;x) \, \delta^D(x\!-\!x') \!\!\!\!$ & \cr
\omit&height2pt&\omit&&\omit&&\omit&\cr \tablerule}}

\caption{4-point contribution from each part of the graviton
propagator at order $m^1$. The vertices 1-8 could only give the contribution
at order $m^0$.}

\label{4props}

\end{table}

From Table~\ref{4props} it is apparent that we require the
coincidence limits on each of the scalar propagators.
For the $A$-type propagator these are,
\begin{eqnarray}
\lim_{x' \rightarrow x} \, {i\Delta}_A(x;x') & = &
\frac{H^{D-2}}{(4\pi)^{ \frac{D}2}}
\frac{\Gamma(D-1)}{\Gamma(\frac{D}2)} \left\{-\pi \cot\Bigl(
\frac{\pi}2 D \Bigr) + 2 \ln(a) \right\} . \label{Acoin}
\end{eqnarray}
The analogous coincidence limits for the $B$-type propagator are
actually finite in $D=4$ dimensions,
\begin{eqnarray}
\lim_{x' \rightarrow x} \, {i\Delta}_B(x;x') & = &
\frac{H^{D-2}}{(4\pi)^{\frac{D}2}} \frac{\Gamma(D-1)}
{\Gamma(\frac{D}2)}\times -\frac1{D\!-\!2}\,.
\end{eqnarray}
The same is true for the coincidence limits of the $C$-type
propagator,
\begin{eqnarray}
\lim_{x' \rightarrow x} \, {i\Delta}_C(x;x') & = &
\frac{H^{D-2}}{(4\pi)^{ \frac{D}2}}
\frac{\Gamma(D-1)}{\Gamma(\frac{D}2)}\times \frac1{(D\!-\!2)
(D\!-\!3)} \; .
\end{eqnarray}
We apply the various coincidence limits to each contraction in
Table~\ref{4props} and present the order m, 4-point contributions
in Table~\ref{4fin-mass}. The total summation for this
local contributions is quite simple,
\begin{eqnarray}
&&-i\Bigl[\Sigma^{\rm 4pt}\Bigr](x;x')\!=\!
\frac{i\kappa^2\!H^2 ma}{2^{D+1}\pi^{\frac{D}{2}}}
\frac{H^{D-4}}{(D\!-\!4)}\frac{-\Gamma(D\!+\!1)}
{\Gamma(\frac{D}{2})}\delta^D(x\!-\!x')\nonumber\\
&&\hspace{4cm}+\frac{i\kappa^2H^2}{16\pi^2}ma
\Bigl[12\ln a \!-\!1\Bigr]\delta^4\!(x\!-\!x')\;.\label{tot4pt}
\end{eqnarray}
\begin{table}
\vbox{\tabskip=0pt \offinterlineskip
\def\tablerule{\noalign{\hrule}}
\halign to390pt {\strut#& \vrule#\tabskip=1em plus2em& \hfil#&
\vrule#& \hfil#& \vrule#& \hfil#\hfil& \vrule#\tabskip=0pt\cr
\tablerule \omit&height4pt&\omit&&\omit&&\omit&\cr &&\omit\hidewidth
{\rm I} &&\omit\hidewidth {\rm J} \hidewidth&& \omit\hidewidth
$i\,am\, \delta^D (x\!-\!x')$ \hidewidth&\cr
\omit&height4pt&\omit&&\omit&&\omit&\cr
\tablerule
\omit&height2pt&\omit&&\omit&&\omit&\cr && 9 && A &&
$-\frac{(D-1)}{(D-3)}A $ & \cr
\omit&height2pt&\omit&&\omit&&\omit&\cr
\tablerule
\omit&height2pt&\omit&&\omit&&\omit&\cr && 9 && B && $0$ & \cr
\omit&height2pt&\omit&&\omit&&\omit&\cr
\tablerule
\omit&height2pt&\omit&&\omit&&\omit&\cr && 9 && C &&
$-\frac{1}{(D-2)^2(D-3)^2}$ & \cr
\omit&height2pt&\omit&&\omit&&\omit&\cr
\tablerule
\omit&height2pt&\omit&&\omit&&\omit&\cr && 10 && A &&
$-\frac{1}{2}[D(D-1)-\frac{2(D-1)}{(D-3)}]A$ & \cr
\omit&height2pt&\omit&&\omit&&\omit&\cr
\tablerule
\omit&height2pt&\omit&&\omit&&\omit&\cr && 10 && B &&
$-\frac{1}{2}\frac{(D-1)}{(D-2)}$ & \cr
\omit&height2pt&\omit&&\omit&&\omit&\cr
\tablerule
\omit&height2pt&\omit&&\omit&&\omit&\cr && 10 && C &&
$\frac{1}{2}\frac{(D^2-5D+8)}{(D-2)^2(D-3)^2}$ & \cr
\omit&height2pt&\omit&&\omit&&\omit&\cr \tablerule}}

\caption{The 4-point contributions at order m.
All contributions are multiplied by
$\frac{\kappa^2 H^{D-2}}{(4 \pi)^{\frac{D}2}}
\frac{\Gamma(D-1)}{ \Gamma(\frac{D}2)}$. Here $A \equiv
 \frac{\pi}{2}\cot(\frac{D\pi}{2})\!-\! \ln(a)$.}

\label{4fin-mass}

\end{table}
\section{Contributions from the 3-Point Vertices}

In this section we work out the contributions from two 3-point
vertex operators. The generic diagram topology is depicted in
Fig.~2. The analytic form is,
\begin{equation}
-i\Bigl[\mbox{}_i \Sigma_j^{\rm 3pt}\Bigr](x;x') = \sum_{I=1}^4
iV^{\alpha\beta }_{Iik}(x) \, i\Bigl[\mbox{}_k S_{\ell}\Bigr](x;x')
\sum_{J=1}^4 iV^{\rho\sigma }_{J\ell j}(x') \,
i\Bigl[\mbox{}_{\alpha\beta} \Delta_{\rho\sigma}\Bigr](x;x') \; .
\label{3ptloop}
\end{equation}

\begin{figure}
\begin{center}
\includegraphics[width=5.0cm]{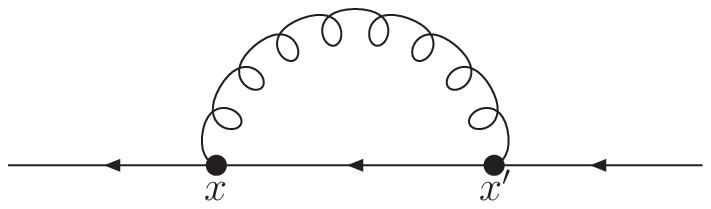}
\\{\rm Fig.~2: Contribution from two 3-point vertices.}
\label{Fig2}
\end{center}
\end{figure}

\begin{table}

\vbox{\tabskip=0pt \offinterlineskip
\def\tablerule{\noalign{\hrule}}
\halign to390pt {\strut#& \vrule#\tabskip=1em plus2em& \hfil#&
\vrule#& \hfil#& \vrule#& \hfil#\hfil& \vrule#\tabskip=0pt\cr
\tablerule \omit&height4pt&\omit&&\omit&&\omit&\cr &&\omit\hidewidth
{\rm I} &&\omit\hidewidth {\rm J} \hidewidth&& \omit\hidewidth
$iV_I^{\alpha\beta}(x) \, i[S](x;x') \, i V_J^{\rho\sigma}(x') \,
i[\mbox{}_{\alpha\beta} \Delta_{\rho\sigma}](x;x')$ \hidewidth&\cr
\omit&height4pt&\omit&&\omit&&\omit&\cr
\tablerule
\omit&height2pt&\omit&&\omit&&\omit&\cr
&& 1 && 1 && $-\frac14
\kappa^2 \partial_{\mu}^{\prime}\{\,\hspace{-.1cm}
\not{\hspace{-.1cm}\partial}\, i[S](x;x') \, \gamma^{\mu}
i[^{\alpha}_{~\alpha}\Delta^{\rho}_{~\rho}](x;x')\}$ & \cr
\omit&height2pt&\omit&&\omit&&\omit&\cr
\tablerule
\omit&height2pt&\omit&&\omit&&\omit&\cr
&& 1 && 2 && $\frac14\kappa^2 \partial^{\prime\rho} \{ \,\hspace{-.1cm}
\not{\hspace{-.1cm}\partial} \, i[S](x;x') \, \gamma^{\sigma} \,
i[^{\alpha}_{~\alpha}\Delta_{\rho\sigma}](x;x')\}$ & \cr
\omit&height2pt&\omit&&\omit&&\omit&\cr
\tablerule
\omit&height2pt&\omit&&\omit&&\omit&\cr
&& 1 && 3 && $\frac{1}{4} i\kappa^2 \hspace{-.1cm} \not{\hspace{-.1cm}
\partial}\, i[S](x;x') \, \gamma^{\rho} J^{\sigma \mu} \,
\partial^{\prime}_{\mu} i[^{\alpha}_{~\alpha}\Delta_{\rho\sigma}](x;x')$ & \cr
\omit&height2pt&\omit&&\omit&&\omit&\cr
\tablerule
\omit&height2pt&\omit&&\omit&&\omit&\cr
&& 1 && 4 && $\frac{1}{4} i\kappa^2 am \hspace{-.1cm} \not{\hspace{-.1cm}
\partial} i[S](x;x') \, i[^{\alpha}_{~\alpha}\Delta^{\rho}_{~\rho}](x;x')$ & \cr
\omit&height2pt&\omit&&\omit&&\omit&\cr
\tablerule
\omit&height2pt&\omit&&\omit&&\omit&\cr
&& 2 && 1 && $\frac14\kappa^2 \partial^{\prime}_{\mu} \{ \gamma^{\alpha}
\partial^{\beta} \, i[S](x;x') \, \gamma^{\mu} \, i[\mbox{}_{\alpha\beta}
\Delta^{\rho}_{~\rho}](x;x') \}$ & \cr
\omit&height2pt&\omit&&\omit&&\omit&\cr
\tablerule
\omit&height2pt&\omit&&\omit&&\omit&\cr
&& 2 && 2 && $-\frac14
\kappa^2 \partial^{\prime \rho} \{ \gamma^{\alpha}
\partial^{\beta} \, i[S](x;x') \, \gamma^{\sigma} \, i[\mbox{}_{\alpha\beta}
\Delta_{\rho\sigma}](x;x') \}$ & \cr
\omit&height2pt&\omit&&\omit&&\omit&\cr
\tablerule
\omit&height2pt&\omit&&\omit&&\omit&\cr
&& 2 && 3 && $-\frac14i\kappa^2 \, \gamma^{\alpha} \partial^{\beta}
\,i[S](x;x') \,\gamma^{\rho} J^{\sigma\mu} \partial^{\prime}_{\mu} \,
i[\mbox{}_{\alpha\beta} \Delta_{\rho\sigma}](x;x')$ & \cr
\omit&height2pt&\omit&&\omit&&\omit&\cr
\tablerule
\omit&height2pt&\omit&&\omit&&\omit&\cr
&& 2 && 4 && $-\frac14
i\kappa^2 am \, \gamma^{\alpha} \partial^{\beta} \, i[S](x;x') \,
i[\mbox{}_{\alpha\beta} \Delta^{\rho}_{~\rho}](x;x')$ & \cr
\omit&height2pt&\omit&&\omit&&\omit&\cr
\tablerule
\omit&height2pt&\omit&&\omit&&\omit&\cr
&& 3 && 1 && $-\frac14 i\kappa^2 \partial^{\prime}_{\nu}
\{ \gamma^{\alpha} J^{\beta\mu} \,i[S](x;x') \, \gamma^{\nu}
\partial_{\mu} \, i[\mbox{}_{\alpha\beta} \Delta^{\rho}_{~\rho}](x;x') \}$ & \cr
\omit&height2pt&\omit&&\omit&&\omit&\cr
\tablerule
\omit&height2pt&\omit&&\omit&&\omit&\cr
&& 3 && 2 && $\frac14 i\kappa^2 \partial^{\prime \rho}
\{ \gamma^{\alpha} J^{\beta\mu} \,i[S](x;x') \, \gamma^{\sigma}
\partial_{\mu} \,i[\mbox{}_{\alpha\beta} \Delta_{\rho\sigma}](x;x') \}$ & \cr
\omit&height2pt&\omit&&\omit&&\omit&\cr
\tablerule
\omit&height2pt&\omit&&\omit&&\omit&\cr
&& 3 && 3 && $-\frac14\kappa^2 \, \gamma^{\alpha} J^{\beta\mu} \,
i[S](x;x') \,\gamma^{\rho} J^{\sigma\nu} \partial_{\mu} \partial^{\prime}_{\nu}
\, i[\mbox{}_{\alpha\beta} \Delta_{\rho\sigma}](x;x')$ & \cr
\omit&height2pt&\omit&&\omit&&\omit&\cr
\tablerule
\omit&height2pt&\omit&&\omit&&\omit&\cr
&& 3 && 4 && $-\frac14\kappa^2 am \,\gamma^{\alpha}J^{\beta\mu}
\, i[S](x;x')\,\partial_{\mu}\, i[\mbox{}_{\alpha\beta}
\Delta^{\rho}_{~\rho}](x;x')$ & \cr
\omit&height2pt&\omit&&\omit&&\omit&\cr
\tablerule
\omit&height2pt&\omit&&\omit&&\omit&\cr
&& 4 && 1 && $-\frac14i\kappa^2\partial^{\prime}_{\mu}\{am\,i[S](x;x')\,
\gamma^{\mu}\,i[\mbox{}^{\alpha}_{~\alpha}\Delta^{\rho}_{~\rho}](x;x')\}$ & \cr
\omit&height2pt&\omit&&\omit&&\omit&\cr
\tablerule
\omit&height2pt&\omit&&\omit&&\omit&\cr
&& 4 && 2 && $ \frac14i\kappa^2\partial^{\prime\rho}\{am\,i[S](x;x')\,
\gamma^{\sigma}\, i[\mbox{}^{\alpha}_{~\alpha}\Delta_{\rho\sigma}](x;x')\}$ & \cr
\omit&height2pt&\omit&&\omit&&\omit&\cr
\tablerule
\omit&height2pt&\omit&&\omit&&\omit&\cr
&& 4 && 3 && $ -\frac14\kappa^2 am \,i[S](x;x')\,\gamma^{\rho}J^{\sigma\mu}\,
\partial_{\mu}\,i[\mbox{}^{\alpha}_{~\alpha}\Delta_{\rho\sigma}](x;x')$&\cr \omit&height2pt&\omit&&\omit&&\omit&\cr
\tablerule
\omit&height2pt&\omit&&\omit&&\omit&\cr
&& 4 && 4 && $ -\frac14\kappa^2 a^2m^2\,i[S](x;x')\,
i[\mbox{}^{\alpha}_{~\alpha}\Delta^{\rho}_{~\rho}](x;x')$&\cr
\omit&height2pt&\omit&&\omit&&\omit&\cr \tablerule}}

\caption{Generic Contributions from the 3-Point Vertices.}

\label{gen3}

\end{table}

Because there are four 3-point vertex operators in (\ref{3VO}), there
are sixteen vertex products in (\ref{3ptloop}). We label each
contribution by the numbers on its vertex pair, for example,
\begin{equation}
\Bigl[I\!\!-\!\!J\Bigr] \equiv iV_I^{\alpha\beta}(x) \times
i\Bigl[S\Bigr](x;x') \times i V_J^{\rho\sigma}(x') \times
i\Bigl[\mbox{}_{ \alpha\beta} \Delta_{\rho\sigma}\Bigr](x;x') \; .
\end{equation}
Table \ref{gen3} gives the generic reductions, before decomposing
the graviton propagator\footnote{We would not consider the 4-4
contraction because it is an order $\rm{m}^2$ contribution.}.
Most of these reductions are straightforward but one subtlety
deserve mention, that is, derivatives on external lines must be
partially integrated back on the entire diagram. This happens
whenever the second vertex is $J\!=\!1$ or $J\!=\!2$, for example,
\begin{eqnarray}
\Bigl[2\!\!-\!\!2\Bigr] & \equiv & -\frac{i\kappa}2 \gamma^{\alpha}
i
\partial^{\beta} \times i \Bigl[S\Bigr](x;x') \times -\frac{i\kappa}2
\gamma^{\rho} i \partial^{\prime \sigma}_{\rm ext} \times
i\Bigl[\mbox{}_{
\alpha\beta} \Delta_{\rho\sigma}\Bigr](x;x') \; , \qquad \\
& = & -\frac{\kappa^2}4 \partial^{\prime \sigma} \Biggl\{
\gamma^{\alpha}
\partial^{\beta} \, i\Bigl[S\Bigr](x;x') \, \gamma^{\rho} \, i \Bigl[
\mbox{}_{\alpha\beta} \Delta_{\rho\sigma} \Bigr](x;x') \Biggr\} .
\end{eqnarray}
Another simplification we might use for later contractions is that
the Dirac slash of the conformal part of fermion propagator gives
a delta function,
\begin{equation}
i \hspace{-.1cm} \not{\hspace{-.1cm} \partial}
i \Bigl[S\Bigr]_{\rm{cf}}(x;x')
= i \delta^D(x-x') \; . \label{fpeqn}
\end{equation}
\subsection{Conformal Contributions}

The key to accomplishing a tractable reduction of the diagrams of Fig.~2
is that the first term of each of the scalar propagators
$i\Delta_I(x;x')$ is the conformal propagator $i\Delta_{\rm
cf}(x;x)$. The sum of the three index factors also gives a simple
tensor, so it is very convenient to write the graviton propagator in
the form,
\begin{eqnarray}
\lefteqn{i\Bigl[{}_{\mu\nu} \Delta_{\rho\sigma}\Bigr](x;x') =
\Bigl[2 \eta_{\mu (\rho} \eta_{\sigma) \nu} - \frac2{D\!-\!2}
\eta_{\mu\nu}
\eta_{\rho\sigma}\Bigr] i\Delta_{\rm cf}(x;x') } \nonumber \\
& & \hspace{6cm} + \sum_{I=A,B,C} \Bigl[\mbox{}_{\mu\nu}
T^I_{\rho\sigma} \Bigr] \, i{\delta \! \Delta}_I(x;x') \; , \qquad
\end{eqnarray}
where $i{\delta \! \Delta}_I(x;x') \equiv i\Delta_I(x;x') -
i\Delta_{\rm cf}(x;x')$. In this subsection we evaluate the
contribution to (\ref{3ptloop}) using the 3-point vertex operators
(\ref{3VO}) and the fermion propagator (\ref{mfprop}) but only the
conformal part of the graviton propagator,
\begin{equation}
i\Bigl[{}_{\mu\nu} \Delta_{\rho\sigma}\Bigr](x;x') \longrightarrow
\Bigl[2 \eta_{\mu (\rho} \eta_{\sigma) \nu} - \frac2{D\!-\!2}
\eta_{\mu\nu} \eta_{\rho\sigma}\Bigr] i\Delta_{\rm cf}(x;x') \equiv
\Bigl[\mbox{}_{\alpha \beta} T^{\rm cf}_{\rho\sigma}\Bigr]
i\Delta_{\rm cf}(x;x) \; . \label{cfpart}
\end{equation}

We carry out the reduction in three stages. In the first stage the
conformal part (\ref{cfpart}) of the graviton propagator is
substituted into the generic results from Table \ref{gen3} and the
contractions are performed. We also make use of the following gamma
matrix identities,
\begin{equation}
\gamma^{\rho}J^{\beta\mu}+\gamma^{\beta}J^{\rho\mu}=\frac{i}{2}
\Bigl(\gamma^{\rho}\eta^{\beta\mu}+\gamma^{\beta}\eta^{\rho\mu}
\Bigr)-i\gamma^{\mu}\eta^{\rho\beta}\;\; ,\;\; \gamma_{\alpha}
J^{\alpha \mu} = -\frac{i}2 (D\!-\!1) \gamma^{\mu}  .
\end{equation}
At this stage we do not act any derivatives on the fermion propagator.
The results of these reductions are presented in Table \ref{Dcfcon}.
The conformal tensor factor $[{}_{\alpha\beta} T^{\rm
cf}_{\rho\sigma}]$ consists of three distinct terms, and the
factors of $\gamma^{\alpha} J^{\beta \mu}$ in Table~\ref{gen3} can
contribute different terms with a distinct structure, so we have
sometimes broken up the result for a given vertex pair into parts.
These parts are distinguished in Table~\ref{Dcfcon} and subsequently
by subscripts taken from the lower case Latin letters.

\begin{table}

\vbox{\tabskip=0pt \offinterlineskip
\def\tablerule{\noalign{\hrule}}
\halign to390pt {\strut#& \vrule#\tabskip=1em plus2em& \hfil#\hfil&
\vrule#& \hfil#\hfil& \vrule#& \hfil#\hfil& \vrule#& \hfil#\hfil&
\vrule#\tabskip=0pt\cr
\tablerule
\omit&height4pt&\omit&&\omit&&\omit&&\omit&\cr &&$\!\!\!\!{\rm
I}\!\!\!\!$ && $\!\!\!\!{\rm J}\!\!\!\!$ && $\!\!\!\!{\rm sub}
\!\!\!\!$ && $\!\!\!\!iV_I^{\alpha\beta}(x) \, i[S](x;x') \, i
V_J^{\rho \sigma}(x') \, [\mbox{}_{\alpha\beta} T^{\rm
cf}_{\rho\sigma}] \, i\Delta_{\rm cf}(x;x') \!\!\!\!$ & \cr
\omit&height4pt&\omit&&\omit&&\omit&&\omit&\cr
\tablerule
\omit&height2pt&\omit&&\omit&&\omit&&\omit&\cr
&& 1 && 1 && \omit &&$\frac{D}{D-2} \kappa^2 \partial_{\mu}'
\{\not{\hspace{-.1cm}\partial}i[S](x;x')\gamma^{\mu}i
\Delta _{\rm{cf}}(x;x')\}$ &\cr
\omit&height2pt&\omit&&\omit&&\omit&&\omit&\cr
\tablerule
\omit&height2pt&\omit&&\omit&&\omit&&\omit&\cr
&& 1 && 2 && \omit &&$-\frac{1}{D-2} \kappa^2 \partial_{\mu}'
\{\not{\hspace{-.1cm}\partial}i[S](x;x')\gamma^{\mu}i
\Delta _{\rm{cf}}(x;x')\}$ & \cr
\omit&height2pt&\omit&&\omit&&\omit&&\omit&\cr
\tablerule
\omit&height2pt&\omit&&\omit&&\omit&&\omit&\cr
&& 1 && 3&& \omit && $-\frac{(D-1)}{2(D-2)}\kappa^2
\not{\hspace{-.1cm}\partial}i[S](x;x')\not{\hspace{-.1cm}\partial'}
i\Delta_{\rm{cf}}(x;x')$ &\cr
\omit&height2pt&\omit&&\omit&&\omit&&\omit&\cr
\tablerule
\omit&height2pt&\omit&&\omit&&\omit&&\omit&\cr
&& 1 && 4 && \omit &&$-\frac{D}{D-2}\kappa^2 iam
\not{\hspace{-.1cm} \partial}i[S](x;x')\,i\Delta_{\rm cf}(x;x')$ &\cr
\omit&height2pt&\omit&&\omit&&\omit&&\omit&\cr
\tablerule
\omit&height2pt&\omit&&\omit&&\omit&&\omit&\cr
&& 2 && 1 && \omit &&$-\frac{1}{D-2}\kappa^2\partial_{\mu}'
\{\not{\hspace{-.1cm}\partial}i[S](x;x')\gamma^{\mu}
i\Delta _{\rm{cf}}(x;x')\}$ &\cr
\omit&height2pt&\omit&&\omit&&\omit&&\omit&\cr
\tablerule
\omit&height2pt&\omit&&\omit&&\omit&&\omit&\cr
&& 2 && 2 && a &&$-\frac{1}{4}\kappa^2\not{\hspace{-.1cm}\partial}'
\{ \partial_{\mu}i[S](x;x')\gamma^{\mu}i\Delta_{\rm{cf}}(x;x')\}$ &\cr
\omit&height2pt&\omit&&\omit&&\omit&&\omit&\cr
\tablerule
\omit&height2pt&\omit&&\omit&&\omit&&\omit&\cr
&& 2 && 2 && b &&$-\frac{1}{4}\kappa^2\partial^{\prime}_{\mu}
\{\partial^{\mu}\gamma^{\beta}i[S](x;x')\gamma_{\beta}
i\Delta_{\rm cf}(x;x')\} $ &\cr
\omit&height2pt&\omit&&\omit&&\omit&&\omit&\cr
\tablerule
\omit&height2pt&\omit&&\omit&&\omit&&\omit&\cr
&& 2 && 2 && c &&$\frac{1}{2(D-2)}\kappa^2\partial_{\mu}'
\{ \not{\hspace{-.1cm}\partial}i[S](x;x')\gamma^{\mu}
i\Delta_{\rm cf}(x;x') \}$ &\cr
\omit&height2pt&\omit&&\omit&&\omit&&\omit&\cr
\tablerule
\omit&height2pt&\omit&&\omit&&\omit&&\omit&\cr
&& 2 && 3 && a &&$\frac18 \kappa^2\gamma^{\beta}\partial^{\mu}i[S](x;x')
\gamma_{\beta}\partial_{\mu}'i\Delta_{\rm cf}(x;x')$ &\cr
\omit&height2pt&\omit&&\omit&&\omit&&\omit&\cr
\tablerule
\omit&height2pt&\omit&&\omit&&\omit&&\omit&\cr
&& 2 && 3 && b &&$\frac{1}{8}\kappa^2\not{\hspace{-.1cm}\partial}'
i\Delta_{\rm cf}(x;x')\partial_{\mu}i[S](x;x')\gamma^{\mu}$ &\cr
\omit&height2pt&\omit&&\omit&&\omit&&\omit&\cr
\tablerule
\omit&height2pt&\omit&&\omit&&\omit&&\omit&\cr
&& 2 && 3 && c &&$\frac{1}{4(D-2)}\kappa^2\not{\hspace{-.1cm}\partial}
i[S](x;x')\not{\hspace{-.1cm}\partial}'i\Delta_{\rm cf}(x;x')$ &\cr
\omit&height2pt&\omit&&\omit&&\omit&&\omit&\cr
\tablerule
\omit&height2pt&\omit&&\omit&&\omit&&\omit&\cr
&& 2 && 4 && \omit &&$\frac{1}{(D-2)}\kappa^2 iam\not{\hspace{-.1cm}\partial}
i[S](x;x')i\Delta_{\rm cf}(x;x')$ &\cr
\omit&height2pt&\omit&&\omit&&\omit&&\omit&\cr
\tablerule
\omit&height2pt&\omit&&\omit&&\omit&&\omit&\cr
&& 3 && 1 && \omit &&$\frac{(D-1)}{2(D-2)}\kappa^2\partial_{\mu}'
\{\not{\hspace{-.1cm}\partial}i\Delta_{\rm cf}(x;x')
i[S](x;x')\gamma^{\mu}\}$ &\cr
\omit&height2pt&\omit&&\omit&&\omit&&\omit&\cr
\tablerule
\omit&height2pt&\omit&&\omit&&\omit&&\omit&\cr
&& 3 && 2 && a &&$-\frac{1}{8}\kappa^2\not{\hspace{-.1cm}\partial}'
\{ i[S](x;x')\not{\hspace{-.1cm}\partial}i\Delta_{\rm cf}(x;x')\}$ &\cr
\omit&height2pt&\omit&&\omit&&\omit&&\omit&\cr
\tablerule
\omit&height2pt&\omit&&\omit&&\omit&&\omit&\cr
&& 3 && 2 && b &&$-\frac{1}{4(D-2)}\kappa^2 \partial_{\mu}'\{
\not{\hspace{-.1cm}\partial} i\Delta_{\rm cf}(x;x')
i[S](x;x') \gamma^{\mu} \}$ &\cr
\omit&height2pt&\omit&&\omit&&\omit&&\omit&\cr
\tablerule
\omit&height2pt&\omit&&\omit&&\omit&&\omit&\cr
&& 3 && 2 && c &&$-\frac{1}{8}\kappa^2\partial_{\mu}'\{ \gamma^{\beta}
i[S](x;x')\gamma_{\beta}\partial^{\mu}i\Delta_{\rm cf}(x;x')\}$ &\cr
\omit&height2pt&\omit&&\omit&&\omit&&\omit&\cr
\tablerule
\omit&height2pt&\omit&&\omit&&\omit&&\omit&\cr
&& 3 && 3 && a &&$\frac{1}{16}\kappa^2\gamma^{\beta}i[S](x;x')
\gamma_{\beta}\partial^{\mu}\partial_{\mu}'i\Delta_{\rm cf}(x;x')$ &\cr
\omit&height2pt&\omit&&\omit&&\omit&&\omit&\cr
\tablerule
\omit&height2pt&\omit&&\omit&&\omit&&\omit&\cr
&& 3 && 3 && b &&$\frac{1}{16}\kappa^2\gamma^{\mu}i[S](x;x')
\partial_{\mu}'\!\!\not{\hspace{-.1cm}\partial}i\Delta_{\rm cf}(x;x')$ &\cr
\omit&height2pt&\omit&&\omit&&\omit&&\omit&\cr
\tablerule
\omit&height2pt&\omit&&\omit&&\omit&&\omit&\cr
&& 3 && 3 && c &&$-\frac{(2D-3)}{8(D-2)}\kappa^2\gamma^{\mu}i[S](x;x')
\partial_{\mu}\!\!\not{\hspace{-.1cm}\partial}'i\Delta_{\rm cf}(x;x')$ &\cr
\omit&height2pt&\omit&&\omit&&\omit&&\omit&\cr
\tablerule
\omit&height2pt&\omit&&\omit&&\omit&&\omit&\cr
&& 3 && 4 && \omit &&$-\frac{(D-1)}{2(D-2)}\kappa^2 iam
\not{\hspace{-.1cm}\partial}i\Delta_{\rm cf}(x;x')i[S](x;x')$ &\cr
\omit&height2pt&\omit&&\omit&&\omit&&\omit&\cr
\tablerule
\omit&height2pt&\omit&&\omit&&\omit&&\omit&\cr
&& 4 && 1 && \omit &&$\frac{D}{(D-2)}\kappa^2 iam\,\partial_{\mu}'
\{i[S](x;x')\gamma^{\mu}i\Delta_{\rm cf}(x;x')\}$ &\cr
\omit&height2pt&\omit&&\omit&&\omit&&\omit&\cr
\tablerule
\omit&height2pt&\omit&&\omit&&\omit&&\omit&\cr
&& 4 && 2 && \omit &&$-\frac{1}{(D-2)}\kappa^2 iam\,\partial_{\mu}'
\{i[S](x;x')\gamma^{\mu}i\Delta_{\rm cf}(x;x')\}$ &\cr
\omit&height2pt&\omit&&\omit&&\omit&&\omit&\cr
\tablerule
\omit&height2pt&\omit&&\omit&&\omit&&\omit&\cr
&& 4 && 3 && \omit &&$-\frac{(D-1)}{2(D-2)}\kappa^2 iam \,i[S](x;x')
\not{\hspace{-.1cm}\partial}i\Delta_{\rm cf}(x;x')$ &\cr
\omit&height2pt&\omit&&\omit&&\omit&&\omit&\cr
\tablerule
\omit&height2pt&\omit&&\omit&&\omit&&\omit&\cr
&& 4 && 4 && \omit &&$\frac{D}{(D-2)}\kappa^2 ia^{2}m^{2}\,
i[S](x;x')\Delta_{\rm cf}(x;x')$ &\cr
\omit&height2pt&\omit&&\omit&&\omit&&\omit&\cr
\tablerule}}
\caption{Contractions from the $i\Delta_{\rm cf}$ part of the
Graviton Propagator.}

\label{Dcfcon}

\end{table}

In the second stage we substitute the conformal part of the graviton
propagator,
\begin{eqnarray}
i\Delta_{\rm cf}(x;x') & = & \frac{\Gamma(\frac{D}2 \!-\!1)}{4
\pi^{\frac{D}2}} \frac{(a a')^{1-\frac{D}2}}{\Delta x^{D-2}} \; ,
\end{eqnarray}
and decompose the fermion propagator (\ref{mfprop}) into the
conformal part, the flat spacetime mass term, $n=0$ part and $n\geq 1$
part of the infinite series expansion\footnote{We will explain why
we separate the $n=0$ part from the rest of the infinite series
expansion in a later paragraph.}. In the final stage we act
the derivatives. We start from the most singular contribution in
Table~\ref{Dcfcon}, which substitutes the conformal parts of the
fermion propagator into the contraction 1-4, 2-4, 3-4, 4-1, 4-2 and 4-3
\footnote{The contraction 4-4 is an order $\rm{m}^2$ contribution.}.
The contraction 1-4 and 2-4 vanish owing to the equation (\ref{fpeqn})
and owing to the zero contribution from D powers of the coordinate
separation in Dimensional regularization. We also must remember that
$[\Sigma](x;x')$ will be used inside an integral in the quantum-corrected
Dirac equation (\ref{Diraceq1}). For that purpose the most singular term
at $x^{\prime \mu} \!=\! x^{\mu}$ is quadratically divergent in
$D\!=\!4$ dimensions. Hence we first conveniently\footnote{Some
individual term is easier written as a derivative with respect to
$x^{\mu}$ acting upon a less singular coordinate separation
than taking the derivative directly.} employ the following identities
to express the rest of them as a less singular form,
\begin{eqnarray}
\frac{1}{\Delta x^{2D-2}}=\frac{\partial^2}{2(D-2)^2}
\frac{1}{\Delta x^{2D-4}}\,\,\,\,,\,\,\,\,
\frac{\Delta x_{\mu}}{\Delta x^{2D-2}}=\frac{-\partial_{\mu}}{2(D-2)}
\frac{1}{\Delta x^{2D-4}}\,.\label{id1}
\end{eqnarray}

\begin{table}

\vbox{\tabskip=0pt \offinterlineskip
\def\tablerule{\noalign{\hrule}}
\halign to390pt {\strut#& \vrule#\tabskip=1em plus2em& \hfil#\hfil&
\vrule#& \hfil#\hfil& \vrule#& \hfil#\hfil& \vrule#& \hfil#\hfil&
\vrule#& \hfil#\hfil& \vrule#\tabskip=0pt\cr
\tablerule
\omit&height4pt&\omit&&\omit&&\omit&&\omit&&\omit&\cr
&& \omit && \omit && Coefficient && Coefficient && Coefficient &\cr
\omit&height4pt&\omit&&\omit&&\omit&&\omit&&\omit&\cr
\tablerule
\omit&height4pt&\omit&&\omit&&\omit&&\omit&&\omit&\cr
&& I && J && $\partial^2 \frac{1}{\Delta x^{2D-4}}$
&&$Ha\gamma^0 \!\!\not{\hspace{-.1cm}\partial}\frac{1}{\Delta x^{2D-4}}$
&&$Ha'\!\!\not{\hspace{-.1cm}\partial}\gamma^0\!\frac{1}{\Delta x^{2D-4}}$&\cr
\omit&height4pt&\omit&&\omit&&\omit&&\omit&&\omit&\cr
\tablerule
\omit&height2pt&\omit&&\omit&&\omit&&\omit&&\omit&\cr
&& 1 && 4 && $0$ && $0$ && $0$ & \cr
\omit&height2pt&\omit&&\omit&&\omit&&\omit&&\omit&\cr
\tablerule
\omit&height2pt&\omit&&\omit&&\omit&&\omit&&\omit&\cr
&& 2 && 4 && $0$ && $0$ && $0$ & \cr
\omit&height2pt&\omit&&\omit&&\omit&&\omit&&\omit&\cr
\tablerule
\omit&height2pt&\omit&&\omit&&\omit&&\omit&&\omit&\cr
&& 3 && 4 && $-\frac{1}{4}\frac{(D-1)}{(D-2)^2}$
&& $-\frac{1}{8}\frac{(D-1)}{(D-2)}$ && $0$ & \cr
\omit&height2pt&\omit&&\omit&&\omit&&\omit&&\omit&\cr
\tablerule
\omit&height2pt&\omit&&\omit&&\omit&&\omit&&\omit&\cr
&& 4 && 1 && $-\frac{1}{2}\frac{D}{(D-2)^2}$ && $0$
&& $\frac{1}{4}\frac{D}{(D-2)}$ & \cr
\omit&height2pt&\omit&&\omit&&\omit&&\omit&&\omit&\cr
\tablerule
\omit&height2pt&\omit&&\omit&&\omit&&\omit&&\omit&\cr
&& 4 && 2 && $\frac{1}{2}\frac{1}{(D-2)^2}$ && $0$
&& $-\frac{1}{4}\frac{1}{(D-2)}$ & \cr
\omit&height2pt&\omit&&\omit&&\omit&&\omit&&\omit&\cr
\tablerule
\omit&height2pt&\omit&&\omit&&\omit&&\omit&&\omit&\cr
&& 4 && 3 && $-\frac{1}{4}\frac{(D-1)}{(D-2)^2}$ && $\frac{1}{8}
\frac{(D-1)}{(D-2)}$ && $0$ & \cr
\omit&height2pt&\omit&&\omit&&\omit&&\omit&&\omit&\cr
\tablerule}}

\caption{$i\Delta_{\rm cf}(x;x')\times i[S]_{\rm cf}(x;x')$.
All contributions are multiplied by $\frac{\kappa^2}{8 \pi^D}
\,\Gamma(\frac{D}2)\Gamma(\frac{D}2 \!-\! 1) ma (aa')^{1-\frac{D}2}$.}

\label{cfcfmost}

\end{table}

The individual result is quoted in Table~\ref{cfcfmost} and
collected all terms of this class,
\begin{eqnarray}
&&\frac{\kappa^{2}ma(aa')^{1-\frac{D}{2}}}{8\pi^D}\Gamma(\frac{D}2)
\Gamma(\frac{D}2 \!-\! 1)\nonumber\\
&&\hspace{.3cm}\times\frac{(D-1)}{(D-2)}\Biggl\{\frac{-1}{(D-2)}\partial^2
- \frac{H}{4}(a-a')\partial_0 - \frac{Ha'}{4}
\gamma^0\!\not{\hspace{-.1cm}\overline{\partial}}\Biggr\}
\frac{1}{\Delta x^{2D-4}}\,. \label{most}
\end{eqnarray}
The expression (\ref{most}) is still logarithmically divergent in $D=4$ after
pulling out various derivatives. To further renormalize this divergence
we extract derivatives with respect to the coordinate $x^{\mu}$ again,
which can of course be taken outside the integral in (\ref{Diraceq1})
to give a less singular integrand,
\begin{eqnarray}
\frac{1}{\Delta x^{2D-4}}=\frac{1}{2(D-3)(D-4)}\partial^2
\frac{1}{\Delta x^{2D-6}}\,.\label{ds}
\end{eqnarray}
Expression (\ref{ds}) is integrable in four dimensions and we could
take $D\!=\!4$ except for the explicit factor of $1/(D\!-\!4)$.
Of course that is how ultraviolet divergences manifest in dimensional
regularization. We can segregate the divergence on a local term by
employing a simple representation for a delta function,
\begin{eqnarray}
\lefteqn{\frac{\partial^2}{D\!-\!4} \, \Bigl(\frac1{\Delta
x^{2D-6}}\Bigr) = \frac{\partial^2}{D\!-\!4} \,
\Biggl\{\frac1{\Delta x^{2D-6}} \!-\! \frac{\mu^{D-4}}{\Delta
x^{D-2}} \Biggr\} \!+\! \frac{i 4 \pi^{\frac{D}2}
\mu^{D-4}}{\Gamma(\frac{D}2 \!-\! 1)} \,
\frac{\delta^D(x\!-\!x')}{D\!-\!4}\;,} \nonumber\\
&& = -\frac{\partial^2}2 \Biggl\{\mu^{2D-8} \frac{\ln(\mu^2
\Delta x^2)}{ \Delta x^2} \!+\! O(D\!-\!4) \Biggr\} + \frac{i 4
\pi^{\frac{D}2} \mu^{D-4}}{ \Gamma(\frac{D}2 \!-\! 1)} \,
\frac{\delta^D(x\!-\!x')}{D\!-\!4} \; . \label{ds1}\qquad
\end{eqnarray}
After substituting (\ref{ds}) and (\ref{ds1}) into (\ref{most}) one can get,
\begin{eqnarray}
&&\frac{3\kappa^2}{64\pi^4}\Biggl\{ \frac{1}{2}\frac{m}{a'}\partial^2
+ \frac{mH}{4}\Bigl( \frac{a}{a'}-1 \Bigr)\partial_0 +\frac{mH}{4}
\gamma^0\!\not{\hspace{-.1cm}\overline{\partial}} \Biggr\} \partial^2
\Biggl[ \frac{\ln(\mu^2\Delta x^2)}{\Delta x^2}\Biggr]
+\frac{\kappa^2}{8\pi^{\frac{D}{2}}}\Gamma(\frac{D}{2})
\frac{(aa')^{1-\frac{D}{2}}}{(D-2)}\nonumber\\
&&\hspace{-.3cm}\frac{(D-1)}{(D-3)}\frac{\mu^{D-4}}{(D-4)}
\Biggl\{ \frac{-2}{(D-2)}\frac{m}{a'}\partial^2 -\frac{mH}{2}
\Bigl(\frac{a}{a'}-1\Bigr)\partial_0 -\frac{mH}{2}\gamma^0\!
\not{\hspace{-.1cm}\overline{\partial}} \Biggr\}i\delta^D(x-x')\,.
\label{2ndcon}
\end{eqnarray}
To reach the same expressions as the counterterms we mentioned in
the suction 3 we make use of the following identities,
\begin{eqnarray}
&&\frac{1}{a'}\partial^2\delta^D(x-x')=\Bigl\{\frac{1}{a}\partial^2
+ 2H\partial_0 \Bigr\}\delta^D(x-x')\,,\nonumber\\
&&\frac{1}{a'}\partial_0\delta^D(x-x')=\Bigl\{\frac{1}{a}
\partial_0 -H\Bigr\}\delta^D(x-x')\,,\nonumber\\
&&\ln(a')\partial_0\delta^D(x-x')=\Bigl\{\ln(a)\partial_0
+ Ha\Bigr\}\delta^D(x-x')\,,\nonumber\\
&&\frac{\ln(a')}{a'}\partial_0\delta^D(x-x')=\Bigl\{\frac{\ln(a)}{a}
\partial_0+H\Bigl(1-\ln(a)\Bigr)\Bigr\}\delta^D(x-x')\,,\nonumber\\
&&\frac{\ln(a')}{a'}\partial^2\delta^D(x-x')=\Bigl\{\frac{\ln(a)}{a}
\partial^2 +2H\Bigl(\ln(a)-1\Bigr)\partial_0 +H^{2}a\Bigr\}
\delta^D(x-x')\,.\nonumber\\ \label{identity}
\end{eqnarray}
After applying (\ref{identity}) to (\ref{2ndcon}) and expanding out
$(aa')^{1-\frac{D}{2}}$ we get the total of this most singular class
which is consistent with our counterterm convention,
\begin{eqnarray}
&&-i\Bigl[\Sigma^{\rm cfcf}\Bigr]\!(x;x')\!=\!
\frac{i\kappa^2}{16\pi^{\frac{D}{2}}}
\frac{\Gamma(\frac{D}{2}\!\!-\!1)(D\!\!-\!1)\mu^{D\!-\!4}}
{(D\!\!-\!3)(D\!\!-\!4)}\Biggl\{\frac{-2}{(D\!\!-\!2)}
\frac{m}{a}\partial^2\!-\!\frac{4mH}{(D\!\!-\!2)}
\partial_0\nonumber\\
&&\,\,\,\,-\frac{1}{2}mH\gamma^0\!\!
\not{\hspace{-.1cm}\overline{\partial}}
\!+\!\frac{1}{2}mH^{2}\!a\Biggr\}\delta^{D}\!(x\!\!-\!\!x')
\!+\!\frac{i\kappa^2}{8\pi^2}\Biggl\{\ln(a)
\Biggl[\frac{3}{2}\frac{m}{a}\partial^2
\!+\!3mH\partial_0\nonumber\\
&&\hspace{2cm}+\frac{3}{4}mH\gamma^0\!\!
\not{\hspace{-.1cm}\overline{\partial}}
\!-\!\frac{3}{4}mH^{2}\!a\Biggr]\!-\!\frac{3}{2}mH\partial_0
\!+\!\frac{3}{4}mH^{2}\!a\Biggr\}\delta^4\!(x\!\!-\!\!x')\nonumber\\
&&\hspace{.5cm}+\frac{\kappa^2}{64\pi^2}\Biggl\{\frac{3}{2}\frac{m}{a'}
\partial^2\!+\!\frac{3}{4}mH(\frac{a}{a'}\!-\!1)\partial_0
\!+\!\frac{3}{4}mH\gamma^0\!\!\not{\hspace{-.1cm}\overline{\partial}}
\Biggr\}\partial^2\Biggl[\frac{\ln(\mu^2\Delta x^2)}{\Delta x^2}\Biggr]
\,.\label{re-cfcf}
\end{eqnarray}

\begin{table}

\vbox{\tabskip=0pt \offinterlineskip
\def\tablerule{\noalign{\hrule}}
\halign to390pt {\strut#& \vrule#\tabskip=1em plus2em& \hfil#\hfil&
\vrule#& \hfil#\hfil& \vrule#& \hfil#\hfil& \vrule#& \hfil#\hfil&
\vrule#& \hfil#\hfil& \vrule#& \hfil#\hfil& \vrule#&
\hfil#\hfil& \vrule#\tabskip=0pt\cr
\tablerule
\omit&height4pt&\omit&&\omit&&\omit&&\omit&&\omit&&\omit&&\omit&\cr
&&$\!\!\!\!\!(\rm I\!\!-\!\!J)_{\rm sub}\!\!\!\!\!\!\!\!$
&&$\!\!\!\!\!\!\partial^2\frac{1}{\Delta x^{2D-4}}\!\!\!\!\!\!\!$
&&$\!\!\!\!\!\!{\scriptstyle Ha}\partial_0\frac{1}{\Delta x^{2D-4}}
\!\!\!\!\!\!\!$&&$\!\!\!\!\!\!\gamma^0\!\!
\not{\hspace{-.1cm}\overline{\partial}}\frac{Ha}{\Delta x^{2D-4}}\!\!\!\!\!\!\!$
&&$\!\!\!\!\!\!\partial_0\frac{Ha'}{\Delta x^{2D-4}}\!\!\!\!\!\!\!$&&$ \!\!\!\!\!\!\gamma^0\!\!\not{\hspace{-.1cm}\overline{\partial}}
\frac{Ha'}{\Delta x^{2D-4}}\!\!\!\!\!\!\!\!$
&&$\!\!\!\!\!\frac{H^2 aa'}{\Delta x^{2D-4}}\!\!\!\!\!$& \cr
\omit&height4pt&\omit&&\omit&&\omit&&\omit&&\omit&&\omit&&\omit&\cr
\tablerule
\omit&height4pt&\omit&&\omit&&\omit&&\omit&&\omit&&\omit&&\omit&\cr
&&$(\rm 1\!\!-\!\!1)$
&&$\!\!\!\!\!\!\frac{D}{(D-2)^2}\!\!\!\!\!\!$
&&$\!\!\!\!\!\!\frac{-2D}{(D-2)^2}\!\!\!\!\!\!$
&&$\!\!\!\!\!\!\frac{-2D}{(D-2)^2}\!\!\!\!\!\!$
&&$\!\!\!\!\!\!\frac{-D}{2(D-2)}\!\!\!\!\!\!$
&&$\!\!\!\!\!\!\frac{D}{2(D-2)}\!\!\!\!\!\!$
&&$\!\!\!\!\frac{-D}{(D-2)}\!\!\!\!$& \cr
\omit&height4pt&\omit&&\omit&&\omit&&\omit&&\omit&&\omit&&\omit&\cr
\tablerule
\omit&height4pt&\omit&&\omit&&\omit&&\omit&&\omit&&\omit&&\omit&\cr
&&$(\rm 1\!\!-\!\!2)$
&&$\!\!\!\!\!\!\frac{-1}{(D-2)^2}\!\!\!\!\!\!$
&&$\!\!\!\!\!\!\frac{2}{(D-2)^2}\!\!\!\!\!\!$
&&$\!\!\!\!\!\!\frac{2}{(D-2)^2}\!\!\!\!\!\!$
&&$\!\!\!\!\!\!\frac{1}{2(D-2)}\!\!\!\!\!\!$
&&$\!\!\!\!\!\!\frac{-1}{2(D-2)}\!\!\!\!\!\!$
&&$\!\!\!\!\frac{1}{(D-2)}\!\!\!\!$& \cr
\omit&height4pt&\omit&&\omit&&\omit&&\omit&&\omit&&\omit&&\omit&\cr
\tablerule
\omit&height4pt&\omit&&\omit&&\omit&&\omit&&\omit&&\omit&&\omit&\cr
&&$(\rm 1\!\!-\!\!3)$
&&$\!\!\!\!\!\!\frac{-(D-1)}{2(D-2)^2}\!\!\!\!\!\!$
&&$\!\!\!\!\!\!\frac{(D-1)}{2(D-2)^2}\!\!\!\!\!\!$
&&$\!\!\!\!\!\!\frac{(D-1)}{2(D-2)^2}\!\!\!\!\!\!$
&&$\!\!\!\!\!\!\frac{(D-1)}{4(D-2)}\!\!\!\!\!\!$
&&$\!\!\!\!\!\!\frac{-(D-1)}{4(D-2)}\!\!\!\!\!\!$
&&$\!\!\!\!\frac{(D-1)}{2(D-2)}\!\!\!\!$& \cr
\omit&height4pt&\omit&&\omit&&\omit&&\omit&&\omit&&\omit&&\omit&\cr
\tablerule
\omit&height4pt&\omit&&\omit&&\omit&&\omit&&\omit&&\omit&&\omit&\cr
&&$(\rm 2\!\!-\!\!1)$
&&$\!\!\!\!\!\!\frac{-1}{(D-2)^2}\!\!\!\!\!\!$
&&$\!\!\!\!\!\!\frac{2}{(D-2)^2}\!\!\!\!\!\!$
&&$\!\!\!\!\!\!\frac{2}{(D-2)^2}\!\!\!\!\!\!$
&&$\!\!\!\!\!\!\frac{1}{2(D-2)}\!\!\!\!\!\!$
&&$\!\!\!\!\!\!\frac{-1}{2(D-2)}\!\!\!\!\!\!$
&&$\!\!\!\!\frac{1}{(D-2)}\!\!\!\!$& \cr
\omit&height4pt&\omit&&\omit&&\omit&&\omit&&\omit&&\omit&&\omit&\cr
\tablerule
\omit&height4pt&\omit&&\omit&&\omit&&\omit&&\omit&&\omit&&\omit&\cr
&&$(\rm 2\!\!-\!\!2)_{\rm a}$
&&$\!\!\!\!\!\!\frac{-1}{4(D-2)}\!\!\!\!\!\!$
&&$\!\!\!\!\!\!\frac{1}{2(D-2)}\!\!\!\!\!\!$
&&$\!\!\!\!\!\!\frac{-1}{2(D-2)}\!\!\!\!\!\!$
&&$\!\!\!\!\!\!\frac{1}{8}\!\!\!\!\!\!$
&&$\!\!\!\!\!\!\frac{1}{8}\!\!\!\!\!\!$
&&$\!\!\!\!\frac{1}{4}\!\!\!\!$& \cr
\omit&height4pt&\omit&&\omit&&\omit&&\omit&&\omit&&\omit&&\omit&\cr
\tablerule
\omit&height4pt&\omit&&\omit&&\omit&&\omit&&\omit&&\omit&&\omit&\cr
&&$(\rm 2\!\!-\!\!2)_{\rm b}$
&&$\!\!\!\!\!\!\frac{-D}{4(D-2)}\!\!\!\!\!\!$
&&$\!\!\!\!\!\!\frac{D}{2(D-2)}\!\!\!\!\!\!$
&&$\!\!\!\!\!\!0\!\!\!\!\!\!$
&&$\!\!\!\!\!\!\frac{D}{8}\!\!\!\!\!\!$
&&$\!\!\!\!\!\!0\!\!\!\!\!\!$
&&$\!\!\!\!\frac{D}{4}\!\!\!\!$& \cr
\omit&height4pt&\omit&&\omit&&\omit&&\omit&&\omit&&\omit&&\omit&\cr
\tablerule
\omit&height4pt&\omit&&\omit&&\omit&&\omit&&\omit&&\omit&&\omit&\cr
&&$(\rm 2\!\!-\!\!2)_{\rm c}$
&&$\!\!\!\!\!\!\frac{1}{2(D-2)^2}\!\!\!\!\!\!$
&&$\!\!\!\!\!\!\frac{-1}{(D-2)^2}\!\!\!\!\!\!$
&&$\!\!\!\!\!\!\frac{-1}{(D-2)^2}\!\!\!\!\!\!$
&&$\!\!\!\!\!\!\frac{-1}{4(D-2)}\!\!\!\!\!\!$
&&$\!\!\!\!\!\!\frac{1}{4(D-2)}\!\!\!\!\!\!$
&&$\!\!\!\!\frac{-1}{2(D-2)}\!\!\!\!$& \cr
\omit&height4pt&\omit&&\omit&&\omit&&\omit&&\omit&&\omit&&\omit&\cr
\tablerule
\omit&height4pt&\omit&&\omit&&\omit&&\omit&&\omit&&\omit&&\omit&\cr
&&$(\rm 2\!\!-\!\!3)_{\rm a}$
&&$\!\!\!\!\!\!\frac{D}{8(D-2)}\!\!\!\!\!\!$
&&$\!\!\!\!\!\!\frac{-D}{8(D-2)}\!\!\!\!\!\!$
&&$\!\!\!\!\!\!0\!\!\!\!\!\!$
&&$\!\!\!\!\!\!\frac{-D}{16}\!\!\!\!\!\!$
&&$\!\!\!\!\!\!0\!\!\!\!\!\!$
&&$\!\!\!\!\frac{-D}{8}\!\!\!\!$& \cr
\omit&height4pt&\omit&&\omit&&\omit&&\omit&&\omit&&\omit&&\omit&\cr
\tablerule
\omit&height4pt&\omit&&\omit&&\omit&&\omit&&\omit&&\omit&&\omit&\cr
&&$(\rm 2\!\!-\!\!3)_{\rm b}$
&&$\!\!\!\!\!\!\frac{1}{4(D-2)^2}\!\!\!\!\!\!$
&&$\!\!\!\!\!\!\frac{-1}{4(D-2)^2}\!\!\!\!\!\!$
&&$\!\!\!\!\!\!\frac{-1}{4(D-2)^2}\!\!\!\!\!\!$
&&$\!\!\!\!\!\!\frac{-1}{8(D-2)}\!\!\!\!\!\!$
&&$\!\!\!\!\!\!\frac{1}{8(D-2)}\!\!\!\!\!\!$
&&$\!\!\!\!\frac{-1}{4(D-2)}\!\!\!\!$& \cr
\omit&height4pt&\omit&&\omit&&\omit&&\omit&&\omit&&\omit&&\omit&\cr
\tablerule
\omit&height4pt&\omit&&\omit&&\omit&&\omit&&\omit&&\omit&&\omit&\cr
&&$(\rm 2\!\!-\!\!3)_{\rm c}$
&&$\!\!\!\!\!\!\frac{1}{8(D-2)}\!\!\!\!\!\!$
&&$\!\!\!\!\!\!\frac{-1}{8(D-2)}\!\!\!\!\!\!$
&&$\!\!\!\!\!\!\frac{1}{8(D-2)}\!\!\!\!\!\!$
&&$\!\!\!\!\!\!\frac{-1}{16}\!\!\!\!\!\!$
&&$\!\!\!\!\!\!\frac{-1}{16}\!\!\!\!\!\!$
&&$\!\!\!\!\frac{-1}{8}\!\!\!\!$& \cr
\omit&height4pt&\omit&&\omit&&\omit&&\omit&&\omit&&\omit&&\omit&\cr
\tablerule
\omit&height4pt&\omit&&\omit&&\omit&&\omit&&\omit&&\omit&&\omit&\cr
&&$\!\!\!\!\!\!(\rm \!3\!\!-\!\!1\!)\!\!
+\!\!(\rm \!3\!\!-\!\!2\!)_{\rm b}\!\!\!\!\!\!$
&&$\!\!\!\!\!\!\frac{(2D-3)}{4(D-2)^2}\!\!\!\!\!\!$
&&$\!\!\!\!\!\!\frac{(2D-3)}{4(D-2)}\!\!\!\!\!\!$
&&$\!\!\!\!\!\!\frac{(2D-3)}{4(D-2)}\!\!\!\!\!\!$
&&$\!\!\!\!\!\!\frac{-(2D-3)}{8(D-2)}\!\!\!\!\!\!$
&&$\!\!\!\!\!\!\frac{(2D-3)}{8(D-2)}\!\!\!\!\!\!$
&&$\!\!\!\!\frac{(2D-3)}{8}\!\!\!\!$& \cr
\omit&height4pt&\omit&&\omit&&\omit&&\omit&&\omit&&\omit&&\omit&\cr
\tablerule
\omit&height4pt&\omit&&\omit&&\omit&&\omit&&\omit&&\omit&&\omit&\cr
&&$(\rm 3\!\!-\!\!2)_{\rm a}$
&&$\!\!\!\!\!\!\frac{-1}{8(D-2)}\!\!\!\!\!\!$
&&$\!\!\!\!\!\!\frac{-1}{8}\!\!\!\!\!\!$
&&$\!\!\!\!\!\!\frac{1}{8}\!\!\!\!\!\!$
&&$\!\!\!\!\!\!\frac{1}{16(D-2)}\!\!\!\!\!\!$
&&$\!\!\!\!\!\!\frac{1}{16(D-2)}\!\!\!\!\!\!$
&&$\!\!\!\!\frac{-1}{16}\!\!\!\!$& \cr
\omit&height4pt&\omit&&\omit&&\omit&&\omit&&\omit&&\omit&&\omit&\cr
\tablerule
\omit&height4pt&\omit&&\omit&&\omit&&\omit&&\omit&&\omit&&\omit&\cr
&&$(\rm 3\!\!-\!\!2)_{\rm c}$
&&$\!\!\!\!\!\!\frac{-D}{8(D-2)}\!\!\!\!\!\!$
&&$\!\!\!\!\!\!\frac{-D}{8}\!\!\!\!\!\!$
&&$\!\!\!\!\!\!0\!\!\!\!\!\!$
&&$\!\!\!\!\!\!\frac{D}{16}\!\!\!\!\!\!$
&&$\!\!\!\!\!\!0\!\!\!\!\!\!$
&&$\!\!\!\!\frac{-D(D-2)}{16}\!\!\!\!$& \cr
\omit&height4pt&\omit&&\omit&&\omit&&\omit&&\omit&&\omit&&\omit&\cr
\tablerule
\omit&height4pt&\omit&&\omit&&\omit&&\omit&&\omit&&\omit&&\omit&\cr
&&$(\rm 3\!\!-\!\!3)_{\rm a}$
&&$\!\!\!\!\!\!0\!\!\!\!\!\!$
&&$\!\!\!\!\!\!\frac{D}{32}\!\!\!\!\!\!$
&&$\!\!\!\!\!\!0\!\!\!\!\!\!$
&&$\!\!\!\!\!\!\frac{-D}{32}\!\!\!\!\!\!$
&&$\!\!\!\!\!\!0\!\!\!\!\!\!$
&&$\!\!\!\!\frac{D(D-2)}{32}\!\!\!\!$& \cr
\omit&height4pt&\omit&&\omit&&\omit&&\omit&&\omit&&\omit&&\omit&\cr
\tablerule
\omit&height4pt&\omit&&\omit&&\omit&&\omit&&\omit&&\omit&&\omit&\cr
&&$(\rm 3\!\!-\!\!2)_{\rm b}$
&&$\!\!\!\!\!\!0\!\!\!\!\!\!$
&&$\!\!\!\!\!\!\frac{1}{32}\!\!\!\!\!\!$
&&$\!\!\!\!\!\!\frac{-1}{32}\!\!\!\!\!\!$
&&$\!\!\!\!\!\!\frac{-1}{32}\!\!\!\!\!\!$
&&$\!\!\!\!\!\!\frac{-1}{32}\!\!\!\!\!\!$
&&$\!\!\!\!\frac{(D-2)}{32}\!\!\!\!$& \cr
\omit&height4pt&\omit&&\omit&&\omit&&\omit&&\omit&&\omit&&\omit&\cr
\tablerule
\omit&height4pt&\omit&&\omit&&\omit&&\omit&&\omit&&\omit&&\omit&\cr
&&$(\rm 3\!\!-\!\!3)_{\rm c}$
&&$\!\!\!\!\!\!0\!\!\!\!\!\!$
&&$\!\!\!\!\!\!\frac{-(2D-3)}{16(D-2)}\!\!\!\!\!\!$
&&$\!\!\!\!\!\!\frac{-(2D-3)}{16(D-2)}\!\!\!\!\!\!$
&&$\!\!\!\!\!\!\frac{(2D-3)}{16(D-2)}\!\!\!\!\!\!$
&&$\!\!\!\!\!\!\frac{-(2D-3)}{16(D-2)}\!\!\!\!\!\!$
&&$\!\!\!\!\frac{-(2D-3)}{16}\!\!\!\!$& \cr
\omit&height4pt&\omit&&\omit&&\omit&&\omit&&\omit&&\omit&&\omit&\cr
\tablerule}}

\caption{$i\Delta_{\rm cf}\times i[S]_{\rm fm}$. Note that all contributions
are multiplied by the factor $\frac{\kappa^2}{16\pi^D}
\Gamma(\frac{D}2)\Gamma(\frac{D}{2}-1)ma(aa')^{1-\frac{D}2}$.}

\label{cffm}
\end{table}

A less singular contribution comes from the flat spacetime mass term
of the fermion propagator. Note that the contraction (3-3) involves
two derivatives acting upon the conformal graviton propagator, which
would produce a delta function,
\begin{eqnarray}
\partial^{\mu}\partial'_{\mu}i\Delta_{\rm cf}=\frac{-\Gamma(\frac{D}{2})}
{4\pi^{\frac{D}{2}}}\frac{(D\!-\!2)}{(aa')^{\frac{D}{2}-\!1}}\Biggl[
\frac{H^2\!aa'\!\Delta\eta^2}{\Delta x^D}\!+\!\frac{H^2\!aa'}
{2\Delta x^{D-2}}\!+\!\frac{i2\pi^{\frac{D}{2}}}{\Gamma(\frac{D}{2})}
\delta^D\!(x\!-\!x')\Biggr].\label{2dcf}
\end{eqnarray}
This delta function would give zero when it is multiplied by D powers
of the coordinate separation, which occurs in this case. Next, in
order to facilitate our computation, we make use of (\ref{id1}) and
the following identity,
\begin{eqnarray}
\frac{H^2aa'\Delta\eta^2}{\Delta x^{2D-2}}=\frac{Ha\!-\!Ha'}
{2(D\!-\!2)}\partial_0\frac{1}{\Delta x^{2D-4}}\,,\label{id2}
\end{eqnarray}
and breaks up $Ha(a')\!\!\not{\!\partial}\gamma^0$ into
$Ha(a')\partial_0$ and $Ha(a')\gamma^0\!\!\not{\!\bar{\partial}}$.
The intermediate results are summarized in Table~\ref{cffm}.
When all terms in Table~\ref{cffm} are summed, we employ (\ref{ds})
for making the expression integrable in $D=4$ and also segregate
the ultraviolet divergences into the local terms using (\ref{ds1}).
The total result we get from this contribution is,
\begin{eqnarray}
&&-i\Bigl[\Sigma^{\rm cffm}\Bigr]\!(x;x')\!=\!
\frac{i\kappa^2\!mH}{16\pi^{\frac{D}{2}}}
\frac{2\Gamma(\frac{D}{2})\mu^{D\!-\!4}}{(D\!\!-\!3)(D\!\!-\!4)}
\Biggl\{(b_2\!+\!b_3)\partial_0\!+\!(b_{2a}\!+\!b_{3a})\gamma^{0}\!\!
\not{\hspace{-.1cm}\overline{\partial}}\nonumber\\
&&+(b_4\!-\!b_2)H\!a\Biggr\}\delta^{D}\!(x\!\!-\!\!x')
\!+\!\frac{i\kappa^{2}\!mH}{64\pi^2}\Biggl\{\ln(a)
\Biggl[\frac{1}{4}\gamma^{0}\!\!\not{\hspace{-.1cm}\partial}
\!-\!\frac{1}{2}H\!a\Biggr]\!+\!\frac{1}{8}Ha\Biggr\}
\delta^4\!(x\!\!-\!\!x')\nonumber\\
&&+\frac{\kappa^{2}\!mH}{64\pi^4}\Biggl\{\Biggl[\frac{1}{8}
\frac{a}{a'}\partial_0\!-\!\frac{3}{32}\partial_0
\!+\!\frac{9}{16}\frac{a}{a'}\gamma^0\!\!
\not{\hspace{-.1cm}\overline{\partial}}\!-\!\frac{17}{32}
\gamma^0\!\!\not{\hspace{-.1cm}\overline{\partial}}
\!+\!\frac{1}{4}Ha\Biggr]\partial^2\Biggl[
\frac{\ln(\mu^{2}\Delta x^2)}{\Delta x^2}\Biggr]\nonumber\\
&&\hspace{6.2cm}+\Biggl[-\frac{3}{8}\frac{1}{Ha'}\partial^4\!+\!
\frac{Ha}{4}\partial^2_0\Biggr]\!\frac{1}{\Delta x^2}
\Biggr\}\,.\label{re-cffm}
\end{eqnarray}
$b_{2}, b_{2a}, b_{3}, b_{3a}$ and $b_{4}$ are D dimension-dependent
coefficients,
\begin{eqnarray}
&&b_{2}=-\frac{(D-1)(D-5)}{8(D-2)^2}(D-4)-\frac{(3D-4)}{32(D-2)}\,;\nonumber\\
&&b_{2a}=\frac{(D-4)}{8(D-2)^2}+\frac{(-3D+5)}{2(D-2)^2}
+\frac{(6D-17)}{16(D-2)}+\frac{3}{32}\,;\nonumber\\
&&b_{3}=\frac{(4D-7)}{32(D-2)}(D-4)+\frac{(2D-5)}{16(D-2)}\,;\nonumber\\
&&b_{3a}=\frac{(D-4)}{32(D-2)}+\frac{(6D-7)}{16(D-2)}\,;\nonumber\\
&&b_{4}=\Biggl[\frac{-D+6}{32}+\frac{1}{8(D-2)}\Biggr](D-4)
-\frac{1}{16}\,.\label{cffmcoef}
\end{eqnarray}

\begin{table}

\vbox{\tabskip=0pt \offinterlineskip
\def\tablerule{\noalign{\hrule}}
\halign to390pt {\strut#& \vrule#\tabskip=1em plus2em& \hfil#\hfil&
\vrule#& \hfil#\hfil& \vrule#& \hfil#\hfil& \vrule#& \hfil#\hfil&
\vrule#& \hfil#\hfil& \vrule#&\hfil#\hfil& \vrule#\tabskip=0pt\cr
\tablerule
\omit&height4pt&\omit&&\omit&&\omit&&\omit&&\omit&&\omit&\cr
&&$\!\!\!\!(\rm I\!\!-\!\!J)_{\rm sub}\!\!\!\!$
&&$\!\!\!\!\partial_0\frac{1}{\Delta x^{2D-4}}\!\!\!\!\!$
&&$\!\!\!\!\!\!\gamma^0\!\!\not{\hspace{-.1cm}\overline{\partial}}
\frac{1}{\Delta x^{2D-4}}\!\!\!\!\!\!$
&&$\!\!\!\!\frac{Ha}{\Delta x^{2D-4}}\!\!\!\!$
&&$\!\!\!\!\!\!\frac{Ha'}{\Delta x^{2D-4}}\!\!\!\!\!\!$
&&$\!\!\!\!\frac{H^{2}aa'\gamma^0\gamma^k\Delta x_k}
{\Delta x^{2D-4}}\!\!\!$& \cr
\omit&height4pt&\omit&&\omit&&\omit&&\omit&&\omit&&\omit&\cr
\tablerule
\omit&height4pt&\omit&&\omit&&\omit&&\omit&&\omit&&\omit&\cr
&&$(\rm 1\!\!-\!\!1)$
&&$\!\!\!\!\frac{4D}{(D-2)^2}\!\!\!\!$
&&$\!\!\!\!\frac{4D}{(D-2)^2}\!\!\!\!$
&&$\!\!\!\!0\!\!\!\!$
&&$\!\!\!\!0\!\!\!\!$
&&$\!\!\!\!0\!\!\!\!$& \cr
\omit&height4pt&\omit&&\omit&&\omit&&\omit&&\omit&&\omit&\cr
\tablerule
\omit&height4pt&\omit&&\omit&&\omit&&\omit&&\omit&&\omit&\cr
&&$(\rm 1\!\!-\!\!2)$
&&$\!\!\!\!\frac{-4}{(D-2)^2}\!\!\!\!$
&&$\!\!\!\!\frac{-4}{(D-2)^2}\!\!\!\!$
&&$\!\!\!\!0\!\!\!\!$
&&$\!\!\!\!0\!\!\!\!$
&&$\!\!\!\!0\!\!\!\!$& \cr
\omit&height4pt&\omit&&\omit&&\omit&&\omit&&\omit&&\omit&\cr
\tablerule
\omit&height4pt&\omit&&\omit&&\omit&&\omit&&\omit&&\omit&\cr
&&$(\rm 1\!\!-\!\!3)$
&&$\!\!\!\!\frac{-(D-1)}{(D-2)^2}\!\!\!\!$
&&$\!\!\!\!\frac{-(D-1)}{(D-2)^2}\!\!\!\!$
&&$\!\!\!\!\frac{-(D-1)(D-3)}{(D-2)^2}\!\!\!\!$
&&$\!\!\!\!0\!\!\!\!$
&&$\!\!\!\!0\!\!\!\!$& \cr
\omit&height4pt&\omit&&\omit&&\omit&&\omit&&\omit&&\omit&\cr
\tablerule
\omit&height4pt&\omit&&\omit&&\omit&&\omit&&\omit&&\omit&\cr
&&$(\rm 2\!\!-\!\!1)$
&&$\!\!\!\!\frac{-4}{(D-2)^2}\!\!\!\!$
&&$\!\!\!\!\frac{-4}{(D-2)^2}\!\!\!\!$
&&$\!\!\!\!0\!\!\!\!$
&&$\!\!\!\!0\!\!\!\!$
&&$\!\!\!\!0\!\!\!\!$& \cr
\omit&height4pt&\omit&&\omit&&\omit&&\omit&&\omit&&\omit&\cr
\tablerule
\omit&height4pt&\omit&&\omit&&\omit&&\omit&&\omit&&\omit&\cr
&&$(\rm 2\!\!-\!\!2)_{\rm a}$
&&$\!\!\!\!\frac{(D-3)}{2(D-2)}\!\!\!\!$
&&$\!\!\!\!0\!\!\!\!$
&&$\!\!\!\!\frac{(2D-9)}{2(D-2)}\!\!\!\!$
&&$\!\!\!\!0\!\!\!\!$
&&$\!\!\!\!0\!\!\!\!$& \cr
\omit&height4pt&\omit&&\omit&&\omit&&\omit&&\omit&&\omit&\cr
\tablerule
\omit&height4pt&\omit&&\omit&&\omit&&\omit&&\omit&&\omit&\cr
&&$(\rm 2\!\!-\!\!2)_{\rm b}$
&&$\!\!\!\!\frac{(D-5)}{2}\!\!\!\!$
&&$\!\!\!\!0\!\!\!\!$
&&$\!\!\!\!\frac{D(D-4)}{2(D-2)}\!\!\!\!$
&&$\!\!\!\!0\!\!\!\!$
&&$\!\!\!\!0\!\!\!\!$& \cr
\omit&height4pt&\omit&&\omit&&\omit&&\omit&&\omit&&\omit&\cr
\tablerule
\omit&height4pt&\omit&&\omit&&\omit&&\omit&&\omit&&\omit&\cr
&&$(\rm 2\!\!-\!\!2)_{\rm c}$
&&$\!\!\!\!\frac{2}{(D-2)^2}\!\!\!\!$
&&$\!\!\!\!\frac{2}{(D-2)^2}\!\!\!\!$
&&$\!\!\!\!0\!\!\!\!$
&&$\!\!\!\!0\!\!\!\!$
&&$\!\!\!\!0\!\!\!\!$& \cr
\omit&height4pt&\omit&&\omit&&\omit&&\omit&&\omit&&\omit&\cr
\tablerule
\omit&height4pt&\omit&&\omit&&\omit&&\omit&&\omit&&\omit&\cr
&&$(\rm 2\!\!-\!\!3)_{\rm a}$
&&$\!\!\!\!\frac{-(D-3)}{4}\!\!\!\!$
&&$\!\!\!\!0\!\!\!\!$
&&$\!\!\!\!\frac{-1}{4}\!\!\!\!$
&&$\!\!\!\!\frac{-(D-2)}{4}\!\!\!\!$
&&$\!\!\!\!0\!\!\!\!$& \cr
\omit&height4pt&\omit&&\omit&&\omit&&\omit&&\omit&&\omit&\cr
\tablerule
\omit&height4pt&\omit&&\omit&&\omit&&\omit&&\omit&&\omit&\cr
&&$(\rm 2\!\!-\!\!3)_{\rm b}$
&&$\!\!\!\!\frac{1}{2(D-2)^2}\!\!\!\!$
&&$\!\!\!\!\frac{1}{2(D-2)^2}\!\!\!\!$
&&$\!\!\!\!\frac{(D-3)}{2(D-2)^2}\!\!\!\!$
&&$\!\!\!\!0\!\!\!\!$
&&$\!\!\!\!0\!\!\!\!$& \cr
\omit&height4pt&\omit&&\omit&&\omit&&\omit&&\omit&&\omit&\cr
\tablerule
\omit&height4pt&\omit&&\omit&&\omit&&\omit&&\omit&&\omit&\cr
&&$(\rm 2\!\!-\!\!3)_{\rm c}$
&&$\!\!\!\!\frac{-1}{4}\!\!\!\!$
&&$\!\!\!\!0\!\!\!\!$
&&$\!\!\!\!\frac{-1}{4}\!\!\!\!$
&&$\!\!\!\!\frac{-3}{8}\!\!\!\!$
&&$\!\!\!\!\frac{-1}{4}\!\!\!\!$& \cr
\omit&height4pt&\omit&&\omit&&\omit&&\omit&&\omit&&\omit&\cr
\tablerule
\omit&height4pt&\omit&&\omit&&\omit&&\omit&&\omit&&\omit&\cr
&&$\!\!\!\!(\rm 3\!\!-\!\!1)\!\!+\!\!(3\!\!-\!\!2)_{\rm b}\!\!\!\!\!\!\!$
&&$\!\!\!\!\!\!\frac{-(2D-3)}{2(D-2)}\!\!\!\!\!\!$
&&$\!\!\!\!\!\!\frac{-(2D-3)}{2(D-2)}\!\!\!\!\!\!$
&&$\!\!\!\!\!\!\!\!\frac{-(D-5)(2D-3)}{4(D-2)}\!\!\!\!\!\!\!\!$
&&$\!\!\!\!\!\!\frac{-(2D-3)}{4(D-2)}\!\!\!\!\!\!$
&&$\!\!\!\!\!\!\frac{(2D-3)}{4(D-2)}\!\!\!\!\!\!$& \cr
\omit&height4pt&\omit&&\omit&&\omit&&\omit&&\omit&&\omit&\cr
\tablerule
\omit&height4pt&\omit&&\omit&&\omit&&\omit&&\omit&&\omit&\cr
&&$(3\!\!-\!\!2)_{\rm a}$
&&$\!\!\!\!\frac{-1}{4(D-2)}\!\!\!\!$
&&$\!\!\!\!\frac{(D-3)}{4(D-2)}\!\!\!\!$
&&$\!\!\!\!\frac{-(D-3)}{8}\!\!\!\!$
&&$\!\!\!\!\frac{1}{8}\!\!\!\!$
&&$\!\!\!\!\frac{1}{8}\!\!\!\!$& \cr
\omit&height4pt&\omit&&\omit&&\omit&&\omit&&\omit&&\omit&\cr
\tablerule
\omit&height4pt&\omit&&\omit&&\omit&&\omit&&\omit&&\omit&\cr
&&$(3\!\!-\!\!2)_{\rm c}$
&&$\!\!\!\!\frac{D-3}{4}\!\!\!\!$
&&$\!\!\!\!0\!\!\!\!$
&&$\!\!\!\!\frac{-D}{8}\!\!\!\!$
&&$\!\!\!\!\frac{D}{8}\!\!\!\!$
&&$\!\!\!\!0\!\!\!\!$& \cr
\omit&height4pt&\omit&&\omit&&\omit&&\omit&&\omit&&\omit&\cr
\tablerule
\omit&height4pt&\omit&&\omit&&\omit&&\omit&&\omit&&\omit&\cr
&&$(3\!\!-\!\!3)_{\rm a}$
&&$\!\!\!\!0\!\!\!\!$
&&$\!\!\!\!0\!\!\!\!$
&&$\!\!\!\!\!\!\frac{-(D-2)(D-3)}{16}\!\!\!\!\!\!$
&&$\!\!\!\!\frac{(D-2)(D-3)}{16}\!\!\!\!$
&&$\!\!\!\!0\!\!\!\!$& \cr
\omit&height4pt&\omit&&\omit&&\omit&&\omit&&\omit&&\omit&\cr
\tablerule
\omit&height4pt&\omit&&\omit&&\omit&&\omit&&\omit&&\omit&\cr
&&$(3\!\!-\!\!3)_{\rm b}$
&&$\!\!\!\!0\!\!\!\!$
&&$\!\!\!\!\frac{-1}{8}\!\!\!\!$
&&$\!\!\!\!\frac{(D-2)}{32}\!\!\!\!$
&&$\!\!\!\!\frac{(3D-8)}{32}\!\!\!\!$
&&$\!\!\!\!\frac{D-2}{32}\!\!\!\!$& \cr
\omit&height4pt&\omit&&\omit&&\omit&&\omit&&\omit&&\omit&\cr
\tablerule
\omit&height4pt&\omit&&\omit&&\omit&&\omit&&\omit&&\omit&\cr
&&$(3\!\!-\!\!3)_{\rm c}$
&&$\!\!\!\!0\!\!\!\!$
&&$\!\!\!\!\frac{(2D-3)}{4(D-2)}\!\!\!\!$
&&$\!\!\!\!\frac{(2D-3)(3D-7)}{16(D-2)}\!\!\!\!$
&&$\!\!\!\!\frac{(2D-3)}{16}\!\!\!\!$
&&$\!\!\!\!\frac{-(2D-3)}{16}\!\!\!\!$& \cr
\omit&height4pt&\omit&&\omit&&\omit&&\omit&&\omit&&\omit&\cr
\tablerule}}

\caption{$i\Delta_{\rm cf}\times i[S]_{n=0}$. Note that all contributions
are multiplied by the factor $\frac{\kappa^{2}mH}{32\pi^D}
\Gamma(\frac{D}2)\Gamma(\frac{D}{2}-1)(aa')^{2-\frac{D}2}$.}

\label{cfn01}

\end{table}
At the next step we are going to consider contributions from the infinite
series expansion of the fermion propagator. Because the series carries at
least one power of mass we only need to consider diagrams which do not
originate from the mass term in the Lagrangian. Because the infinite series is
vastly more complicated than other parts of the fermion propagator it would
be desirable to carry out the computation in $D=4$ dimensions.
Whether or not it is legitimate for us to do this entirely depends on
whether this kind of contraction is integrable in four dimensions.
The dimensionality of the series of the fermion propagator is
$\frac{1}{\Delta x^{D-3-2n}}$ and the one from the conformal part of
the graviton propagator is $\frac{1}{\Delta x^{D-2}}$. Also remember
that all the terms in Table~\ref{gen3} which derive from two order $m^0$
vertices carry two derivatives . This means that the total dimensionality
in this class is $\frac{1}{\Delta x^{2D-3-2n}}$. Therefore we shall separate
the $n=0$ part, which working on an arbitrary $D$ dimension is necessary,
from the rest of the infinite series expansion, which is integrable
in four dimensions. Because $n=0$ part is not integrable in $D=4$
it worths mentioning its simplification from (\ref{mfprop}) by performing
$\frac{m}{H}$ expansion for gamma functions rather than
expanding it out around $D=4$,
\begin{eqnarray}
&&\Gamma\Bigl(\frac{D}{2}\!-\!1\pm i\frac{m}{H}\Bigr)\!=\!
\Gamma\Bigl(\frac{D}{2}\!-\!1\Bigr)\Bigl[1\pm i\frac{m}{H}
\psi(\frac{D}{2}\!-\!1)\Bigr]+\mathcal{O}(m^2)\,,\\
&&\Gamma\Bigl(1\pm i\frac{m}{H}\Bigr)\!=\!\Bigl[1\pm i\frac{m}{H}
\psi(1)\Bigr]+\mathcal{O}(m^2)\,.
\end{eqnarray}
Here $\psi$'s stand for digamma functions and they cancel out completely
at order m when one substitutes the above equations back to the
$n=0$ part of series,
\begin{eqnarray}
&&i\Bigl[S\Bigr](x;x')_{n=0}\!=\!\frac{H^{D-2}}{(4\pi)^{\frac{D}{2}}}
\frac{\Gamma(\frac{D}{2}\!-\!1)}{(2\!-\!\frac{D}{2})}(i\frac{m}{H})
(aa')^{\frac{D}{2}-1}\Bigl[i\!\not{\!\partial}\gamma^0\!+\!
i(\frac{D}{2}\!-\!1)Ha\Bigr]\nonumber\\
&&\times\Bigl[\Gamma(3\!-\!\frac{D}{2})\Gamma(\frac{D}{2}\!-\!1)
\!-\!(\frac{y}{4})^{2-\frac{D}{2}}\Bigr]\!=\!\Gamma(\frac{D}{2}\!-\!1)
\Biggl\{\frac{mHaa'}{16\pi^{\frac{D}{2}}}\Biggl[
\frac{2\gamma^{\nu}\gamma^{0}\Delta x_{\nu}}{\Delta x^{D-2}}\nonumber\\
&&+\frac{1}{(2\!-\!\frac{D}{2})}\frac{Ha}{\Delta x^{D-4}}\Biggr]
\!-\!\frac{mH^{D-3}(aa')^{\frac{D}{2}-1}}{(4\pi)^{\frac{D}{2}}}
\Gamma(\frac{D}{2})\Gamma(2\!-\!\frac{D}{2})Ha\Biggr\}.\label{n0mf}
\end{eqnarray}
\begin{table}

\vbox{\tabskip=0pt \offinterlineskip
\def\tablerule{\noalign{\hrule}}
\halign to390pt {\strut#& \vrule#\tabskip=1em plus2em& \hfil#\hfil&
\vrule#& \hfil#\hfil& \vrule#& \hfil#\hfil& \vrule#& \hfil#\hfil&
\vrule#& \hfil#\hfil& \vrule#&\hfil#\hfil& \vrule#\tabskip=0pt\cr
\tablerule
\omit&height4pt&\omit&&\omit&&\omit&&\omit&&\omit&&\omit&\cr
&&$\!\!\!\!(\rm I\!\!-\!\!J)_{\rm sub}\!\!\!\!$
&&$\!\!\!\!\!\!\!H\!a\partial^2_0\frac{1}{\Delta x^{2D-6}}\!\!\!\!\!\!\!\!$
&&$\!\!\!\!\!\!\partial^2_0\frac{Ha'}{\Delta x^{2D-6}}\!\!\!\!\!\!$
&&$\!\!\!\!\!\!H\!a\gamma^0\partial_0\!\!
\not{\hspace{-.1cm}\overline{\partial}}\!\!
\frac{1}{\Delta x^{2D-6}}\!\!\!\!\!\!$
&&$\!\!\!\!\!\!\!\!\gamma^0\partial_0\!\!
\not{\hspace{-.1cm}\overline{\partial}}\!\!
\frac{Ha'}{\Delta x^{2D-6}}\!\!\!\!\!\!\!\!$
&&$\!\!\!\frac{H^{2}a^{2}Ha'}
{\Delta x^{2D-6}}\!\!\!\!$& \cr
\omit&height4pt&\omit&&\omit&&\omit&&\omit&&\omit&&\omit&\cr
\tablerule
\omit&height4pt&\omit&&\omit&&\omit&&\omit&&\omit&&\omit&\cr
&&$(\rm 1\!\!-\!\!1)$
&&$\!\!\!\!\frac{2D}{(D-2)^2(D-3)}\!\!\!\!$
&&$\!\!\!\!0\!\!\!\!$
&&$\!\!\!\!\frac{2D}{(D-2)^2(D-3)}\!\!\!\!$
&&$\!\!\!\!0\!\!\!\!$
&&$\!\!\!\!\frac{2D}{(D-2)^2}\!\!\!\!$& \cr
\omit&height4pt&\omit&&\omit&&\omit&&\omit&&\omit&&\omit&\cr
\tablerule
\omit&height4pt&\omit&&\omit&&\omit&&\omit&&\omit&&\omit&\cr
&&$(\rm 1\!\!-\!\!2)$
&&$\!\!\!\!\frac{-2}{(D-2)^2(D-3)}\!\!\!\!$
&&$\!\!\!\!0\!\!\!\!$
&&$\!\!\!\!\frac{-2}{(D-2)^2(D-3)}\!\!\!\!$
&&$\!\!\!\!0\!\!\!\!$
&&$\!\!\!\!\frac{-2}{(D-2)^2}\!\!\!\!$& \cr
\omit&height4pt&\omit&&\omit&&\omit&&\omit&&\omit&&\omit&\cr
\tablerule
\omit&height4pt&\omit&&\omit&&\omit&&\omit&&\omit&&\omit&\cr
&&$(\rm 1\!\!-\!\!3)$
&&$\!\!\!\!\frac{-(D-1)}{2(D-2)^2(D-3)}\!\!\!\!$
&&$\!\!\!\!0\!\!\!\!$
&&$\!\!\!\!\frac{-(D-1)}{2(D-2)^2(D-3)}\!\!\!\!$
&&$\!\!\!\!0\!\!\!\!$
&&$\!\!\!\!0\!\!\!\!$& \cr
\omit&height4pt&\omit&&\omit&&\omit&&\omit&&\omit&&\omit&\cr
\tablerule
\omit&height4pt&\omit&&\omit&&\omit&&\omit&&\omit&&\omit&\cr
&&$(\rm 2\!\!-\!\!1)$
&&$\!\!\!\!\frac{-2}{(D-2)^2(D-3)}\!\!\!\!$
&&$\!\!\!\!0\!\!\!\!$
&&$\!\!\!\!\frac{-2}{(D-2)^2(D-3)}\!\!\!\!$
&&$\!\!\!\!0\!\!\!\!$
&&$\!\!\!\!\frac{-2}{(D-2)^2}\!\!\!\!$& \cr
\omit&height4pt&\omit&&\omit&&\omit&&\omit&&\omit&&\omit&\cr
\tablerule
\omit&height4pt&\omit&&\omit&&\omit&&\omit&&\omit&&\omit&\cr
&&$(\rm 2\!\!-\!\!2)_{\rm a}$
&&$\!\!\!\!0\!\!\!\!$
&&$\!\!\!\!0\!\!\!\!$
&&$\!\!\!\!0\!\!\!\!$
&&$\!\!\!\!0\!\!\!\!$
&&$\!\!\!\!\frac{-1}{2(D-2)}\!\!\!\!$& \cr
\omit&height4pt&\omit&&\omit&&\omit&&\omit&&\omit&&\omit&\cr
\tablerule
\omit&height4pt&\omit&&\omit&&\omit&&\omit&&\omit&&\omit&\cr
&&$(\rm 2\!\!-\!\!2)_{\rm b}$
&&$\!\!\!\!\frac{-1}{2(D-3)}\!\!\!\!$
&&$\!\!\!\!0\!\!\!\!$
&&$\!\!\!\!0\!\!\!\!$
&&$\!\!\!\!0\!\!\!\!$
&&$\!\!\!\!\frac{-D}{2(D-2)}\!\!\!\!$& \cr
\omit&height4pt&\omit&&\omit&&\omit&&\omit&&\omit&&\omit&\cr
\tablerule
\omit&height4pt&\omit&&\omit&&\omit&&\omit&&\omit&&\omit&\cr
&&$(\rm 2\!\!-\!\!2)_{\rm c}$
&&$\!\!\!\!\frac{1}{(D-2)^2(D-3)}\!\!\!\!$
&&$\!\!\!\!0\!\!\!\!$
&&$\!\!\!\!\frac{1}{(D-2)^2(D-3)}\!\!\!\!$
&&$\!\!\!\!0\!\!\!\!$
&&$\!\!\!\!\frac{1}{D-2}\!\!\!\!$& \cr
\omit&height4pt&\omit&&\omit&&\omit&&\omit&&\omit&&\omit&\cr
\tablerule
\omit&height4pt&\omit&&\omit&&\omit&&\omit&&\omit&&\omit&\cr
&&$(\rm 2\!\!-\!\!3)_{\rm a}$
&&$\!\!\!\!\frac{1}{8(D-3)}\!\!\!\!$
&&$\!\!\!\!0\!\!\!\!$
&&$\!\!\!\!0\!\!\!\!$
&&$\!\!\!\!0\!\!\!\!$
&&$\!\!\!\!0\!\!\!\!$& \cr
\omit&height4pt&\omit&&\omit&&\omit&&\omit&&\omit&&\omit&\cr
\tablerule
\omit&height4pt&\omit&&\omit&&\omit&&\omit&&\omit&&\omit&\cr
&&$(\rm 2\!\!-\!\!3)_{\rm b}$
&&$\!\!\!\!\frac{1}{4(D-2)^2(D-3)}\!\!\!\!$
&&$\!\!\!\!0\!\!\!\!$
&&$\!\!\!\!0\!\!\!\!$
&&$\!\!\!\!\!\!\frac{1}{4(D-2)^2(D-3)}\!\!\!\!\!\!$
&&$\!\!\!\!0\!\!\!\!$& \cr
\omit&height4pt&\omit&&\omit&&\omit&&\omit&&\omit&&\omit&\cr
\tablerule
\omit&height4pt&\omit&&\omit&&\omit&&\omit&&\omit&&\omit&\cr
&&$(\rm 2\!\!-\!\!3)_{\rm c}$
&&$\!\!\!\!0\!\!\!\!$
&&$\!\!\!\!\frac{1}{16(D-3)}\!\!\!\!$
&&$\!\!\!\!0\!\!\!\!$
&&$\!\!\!\!0\!\!\!\!$
&&$\!\!\!\!0\!\!\!\!$& \cr
\omit&height4pt&\omit&&\omit&&\omit&&\omit&&\omit&&\omit&\cr
\tablerule
\omit&height4pt&\omit&&\omit&&\omit&&\omit&&\omit&&\omit&\cr
&&$\!\!\!\!\!\!(\rm 3\!\!-\!\!1)\!\!+\!\!(3\!\!-\!\!2)_{\rm b}\!\!\!\!\!\!\!$
&&$\!\!\!\!\!\!\frac{-(2D-3)}{4(D-2)(D-3)}\!\!\!\!\!\!$
&&$\!\!\!\!\!\!0\!\!\!\!\!\!$
&&$\!\!\!\!\!\!\!\!\frac{-(2D-3)}{4(D-2)(D-3)}\!\!\!\!\!\!$
&&$\!\!\!\!\!\!0\!\!\!\!\!\!$
&&$\!\!\!\!\!\!\frac{-(2D-3)}{8(D-2)}\!\!\!\!\!\!$& \cr
\omit&height4pt&\omit&&\omit&&\omit&&\omit&&\omit&&\omit&\cr
\tablerule
\omit&height4pt&\omit&&\omit&&\omit&&\omit&&\omit&&\omit&\cr
&&$(3\!\!-\!\!2)_{\rm a}$
&&$\!\!\!\!0\!\!\!\!$
&&$\!\!\!\!0\!\!\!\!$
&&$\!\!\!\!0\!\!\!\!$
&&$\!\!\!\!0\!\!\!\!$
&&$\!\!\!\!\frac{1}{16}\!\!\!\!$& \cr
\omit&height4pt&\omit&&\omit&&\omit&&\omit&&\omit&&\omit&\cr
\tablerule
\omit&height4pt&\omit&&\omit&&\omit&&\omit&&\omit&&\omit&\cr
&&$(3\!\!-\!\!2)_{\rm c}$
&&$\!\!\!\!\frac{D-2}{8(D-3)}\!\!\!\!$
&&$\!\!\!\!0\!\!\!\!$
&&$\!\!\!\!0\!\!\!\!$
&&$\!\!\!\!0\!\!\!\!$
&&$\!\!\!\!\frac{D}{16}\!\!\!\!$& \cr
\omit&height4pt&\omit&&\omit&&\omit&&\omit&&\omit&&\omit&\cr
\tablerule
\omit&height4pt&\omit&&\omit&&\omit&&\omit&&\omit&&\omit&\cr
&&$(3\!\!-\!\!3)_{\rm a}$
&&$\!\!\!\!\frac{-(D-2)}{32(D-3)}\!\!\!\!$
&&$\!\!\!\!\frac{(D-2)}{32(D-3)}\!\!\!\!$
&&$\!\!\!\!\!\!0\!\!\!\!\!\!$
&&$\!\!\!\!0\!\!\!\!$
&&$\!\!\!\!0\!\!\!\!$& \cr
\omit&height4pt&\omit&&\omit&&\omit&&\omit&&\omit&&\omit&\cr
\tablerule
\omit&height4pt&\omit&&\omit&&\omit&&\omit&&\omit&&\omit&\cr
&&$(3\!\!-\!\!3)_{\rm b}$
&&$\!\!\!\!0\!\!\!\!$
&&$\!\!\!\!\frac{1}{32(D-3)}\!\!\!\!$
&&$\!\!\!\!0\!\!\!$
&&$\!\!\!\!\frac{1}{32(D-3)}\!\!\!\!$
&&$\!\!\!\!0\!\!\!\!$& \cr
\omit&height4pt&\omit&&\omit&&\omit&&\omit&&\omit&&\omit&\cr
\tablerule
\omit&height4pt&\omit&&\omit&&\omit&&\omit&&\omit&&\omit&\cr
&&$(3\!\!-\!\!3)_{\rm c}$
&&$\!\!\!\!\frac{(2D-3)}{16(D-2)(D-3)}\!\!\!\!$
&&$\!\!\!\!0\!\!\!\!$
&&$\!\!\!\!\frac{(2D-3)}{16(D-2)(D-3)}\!\!\!\!$
&&$\!\!\!\!0\!\!\!\!$
&&$\!\!\!\!0\!\!\!\!$& \cr
\omit&height4pt&\omit&&\omit&&\omit&&\omit&&\omit&&\omit&\cr
\tablerule}}

\caption{$i\Delta_{\rm cf}\times i[S]_{n=0}$. Note that all contributions
are multiplied by  the factor $\frac{\kappa^{2}mH}{32\pi^D}
\Gamma(\frac{D}2)\Gamma(\frac{D}{2}-1)(aa')^{2-\frac{D}2}$.}

\label{cfn02}

\end{table}
Note that the final two terms which have the same $D=4$ limit in
(\ref{n0mf}) tend to cancel out with each other. The first derivative
of (\ref{n0mf}) has the same pattern,
\begin{eqnarray}
&&\partial_{\mu}[S]_{n=0}\!=\!\Gamma(\frac{D}{2}\!-\!1)
\Biggl\{\frac{mHaa'}{8\pi^{\frac{D}{2}}}\Biggl[\frac{\gamma_{\mu}\gamma^0}
{\Delta x^{D-2}}\!-\!(D\!-\!2)
\frac{\gamma^{\nu}\!\gamma^0\!\Delta x_{\nu}\Delta x_{\mu}}{\Delta x^D}
+\frac{Ha\delta^0_{\mu}\gamma^{\nu}\!\gamma^0\!\Delta x_{\nu}}
{\Delta x^{D-2}}\nonumber\\
&&+\frac{Ha\Delta x_{\mu}}{\Delta x^{D-2}}\!-\!
\frac{2H^2\!a^2\!\delta^0_{\mu}}{(D\!-\!4)\Delta x^{D-4}}\Biggr]
-\frac{m(H^2aa')^{\frac{D}{2}-1}}{H(4\pi)^{\frac{D}{2}}}
\Gamma(\frac{D}{2}\!+\!1)\Gamma(2\!-\!\frac{D}{2})
H^2\!a^2\!\delta^0_{\mu}\Biggr\}.\nonumber\\ \label{n0mfd1}
\end{eqnarray}
The final two terms of equations (\ref{n0mf}) and (\ref{n0mfd1})
would give a non-zero contribution when they are multiplied by
the divergent term\footnote{One can consult the various gamma
function contractions with (\ref{n0mfd1}) in Appendix \ref{fermionprop}.}.

The results derived from this class are lengthy and we would tabulate
them separately based on their distinctive characteristics.
Some contractions would produce at least one $(D\!-\!4)$ factor.
One source of $(D\!-\!4)$ is from total derivatives acting upon
$(aa')^{2-\frac{D}{2}}$. This factor can arise when one power of $(aa')$
comes from the fermion propagator and the rest of it,
$(aa')^{1-\frac{D}{2}}$, originates from the conformal part of
the graviton propagator, i.e. (1-1), (1-2), (2-1), (2-2), (3-1), (3-2).
Another source of $(D\!-\!4)$ comes from the following peculiar
gamma function contraction,
\begin{eqnarray}
\gamma^{\beta}\gamma^{\nu}\gamma^0\gamma_{\beta}\!=\!(D\!-\!4)
\gamma^0\gamma^{\nu}\!+\!2(D\!-\!2)\eta^{0\nu}\,,
\end{eqnarray}
which occurs in the contractions $(2\!-\!2)_{\rm b}$, $(2\!-\!3)_{\rm a}$,
$(3\!-\!2)_{\rm c}$ and $(3\!-\!3)_{\rm a}$.
We summarized the terms without $(D\!-\!4)$\footnote{The contractions
$(1\!-\!3)$, $(2\!-\!3)_{\rm b}$, $(2\!-\!3)_{\rm c}$,
$(3\!-\!3)_{\rm b}$ and $(3\!-\!3)_{\rm c}$ produce no (D-4) factor.}
in Table~\ref{cfn01} and Table~\ref{cfn02} whereas the terms with
the $(D\!-\!4)$ factor are presented in Table~\ref{cfn03}.

\begin{table}

\vbox{\tabskip=0pt \offinterlineskip
\def\tablerule{\noalign{\hrule}}
\halign to390pt {\strut#& \vrule#\tabskip=1em plus2em& \hfil#\hfil&
\vrule#& \hfil#\hfil& \vrule#& \hfil#\hfil& \vrule#& \hfil#\hfil&
\vrule#& \hfil#\hfil& \vrule#&\hfil#\hfil& \vrule#\tabskip=0pt\cr
\tablerule
\omit&height4pt&\omit&&\omit&&\omit&&\omit&&\omit&&\omit&\cr
&&$\!\!\!\!(\rm I\!\!-\!\!J)_{\rm sub}\!\!\!\!$
&&$\!\!\!\!\partial_0\frac{1}{\Delta x^{2D-4}}\!\!\!\!\!$
&&$\!\!\!\!\!\!\gamma^0\!\!\not{\hspace{-.1cm}\overline{\partial}}
\frac{1}{\Delta x^{2D-4}}\!\!\!\!\!\!$
&&$\!\!\!\!\frac{Ha}{\Delta x^{2D-4}}\!\!\!\!$
&&$\!\!\!\!\!\!\frac{Ha'}{\Delta x^{2D-4}}\!\!\!\!\!\!$
&&$\!\!\!\!\frac{H^{2}aa'\gamma^0\gamma^k\Delta x_k}
{\Delta x^{2D-4}}\!\!\!$& \cr
\omit&height4pt&\omit&&\omit&&\omit&&\omit&&\omit&&\omit&\cr
\tablerule
\omit&height4pt&\omit&&\omit&&\omit&&\omit&&\omit&&\omit&\cr
&&$(\rm 1\!\!-\!\!1)$
&&$\!\!\!\!0\!\!\!\!$
&&$\!\!\!\!0\!\!\!\!$
&&$\!\!\!\!\frac{4D}{(D-2)^2}\!\!\!\!$
&&$\!\!\!\!0\!\!\!\!$
&&$\!\!\!\!0\!\!\!\!$& \cr
\omit&height4pt&\omit&&\omit&&\omit&&\omit&&\omit&&\omit&\cr
\tablerule
\omit&height4pt&\omit&&\omit&&\omit&&\omit&&\omit&&\omit&\cr
&&$(\rm 1\!\!-\!\!2)$
&&$\!\!\!\!0\!\!\!\!$
&&$\!\!\!\!0\!\!\!\!$
&&$\!\!\!\!\frac{-4}{(D-2)^2}\!\!\!\!$
&&$\!\!\!\!0\!\!\!\!$
&&$\!\!\!\!0\!\!\!\!$& \cr
\omit&height4pt&\omit&&\omit&&\omit&&\omit&&\omit&&\omit&\cr
\tablerule
\omit&height4pt&\omit&&\omit&&\omit&&\omit&&\omit&&\omit&\cr
&&$(\rm 1\!\!-\!\!3)$
&&$\!\!\!\!0\!\!\!\!$
&&$\!\!\!\!0\!\!\!\!$
&&$\!\!\!\!0\!\!\!\!$
&&$\!\!\!\!0\!\!\!\!$
&&$\!\!\!\!0\!\!\!\!$& \cr
\omit&height4pt&\omit&&\omit&&\omit&&\omit&&\omit&&\omit&\cr
\tablerule
\omit&height4pt&\omit&&\omit&&\omit&&\omit&&\omit&&\omit&\cr
&&$(\rm 2\!\!-\!\!1)$
&&$\!\!\!\!0\!\!\!\!$
&&$\!\!\!\!0\!\!\!\!$
&&$\!\!\!\!\frac{-4}{(D-2)^2}\!\!\!\!$
&&$\!\!\!\!0\!\!\!\!$
&&$\!\!\!\!0\!\!\!\!$& \cr
\omit&height4pt&\omit&&\omit&&\omit&&\omit&&\omit&&\omit&\cr
\tablerule
\omit&height4pt&\omit&&\omit&&\omit&&\omit&&\omit&&\omit&\cr
&&$(\rm 2\!\!-\!\!2)_{\rm a}$
&&$\!\!\!\!0\!\!\!\!$
&&$\!\!\!\!0\!\!\!\!$
&&$\!\!\!\!\frac{-1}{(D-2)}\!\!\!\!$
&&$\!\!\!\!\frac{3}{2(D-2)}\!\!\!\!$
&&$\!\!\!\!\frac{1}{(D-2)}\!\!\!\!$& \cr
\omit&height4pt&\omit&&\omit&&\omit&&\omit&&\omit&&\omit&\cr
\tablerule
\omit&height4pt&\omit&&\omit&&\omit&&\omit&&\omit&&\omit&\cr
&&$(\rm 2\!\!-\!\!2)_{\rm b}$
&&$\!\!\!\!\frac{-(D-5)}{2(D-2)}\!\!\!\!$
&&$\!\!\!\!\frac{-(D-5)}{2(D-2)}\!\!\!\!$
&&$\!\!\!\!\frac{-(3D-4)}{2(D-2)}\!\!\!\!$
&&$\!\!\!\!\frac{3}{2}\!\!\!\!$
&&$\!\!\!\!0\!\!\!\!$& \cr
\omit&height4pt&\omit&&\omit&&\omit&&\omit&&\omit&&\omit&\cr
\tablerule
\omit&height4pt&\omit&&\omit&&\omit&&\omit&&\omit&&\omit&\cr
&&$(\rm 2\!\!-\!\!2)_{\rm c}$
&&$\!\!\!\!0\!\!\!\!$
&&$\!\!\!\!0\!\!\!\!$
&&$\!\!\!\!\frac{2}{(D-2)^2}\!\!\!\!$
&&$\!\!\!\!0\!\!\!\!$
&&$\!\!\!\!0\!\!\!\!$& \cr
\omit&height4pt&\omit&&\omit&&\omit&&\omit&&\omit&&\omit&\cr
\tablerule
\omit&height4pt&\omit&&\omit&&\omit&&\omit&&\omit&&\omit&\cr
&&$(\rm 2\!\!-\!\!3)_{\rm a}$
&&$\!\!\!\!\frac{(D-3)}{4(D-2)}\!\!\!\!$
&&$\!\!\!\!\frac{(D-3)}{4(D-2)}\!\!\!\!$
&&$\!\!\!\!\frac{1}{4(D-2)}\!\!\!\!$
&&$\!\!\!\!\frac{3}{8}\!\!\!\!$
&&$\!\!\!\!\frac{1}{4}\!\!\!\!$& \cr
\omit&height4pt&\omit&&\omit&&\omit&&\omit&&\omit&&\omit&\cr
\tablerule
\omit&height4pt&\omit&&\omit&&\omit&&\omit&&\omit&&\omit&\cr
&&$(\rm 2\!\!-\!\!3)_{\rm b}$
&&$\!\!\!\!0\!\!\!\!$
&&$\!\!\!\!0\!\!\!\!$
&&$\!\!\!\!0\!\!\!\!$
&&$\!\!\!\!0\!\!\!\!$
&&$\!\!\!\!0\!\!\!\!$& \cr
\omit&height4pt&\omit&&\omit&&\omit&&\omit&&\omit&&\omit&\cr
\tablerule
\omit&height4pt&\omit&&\omit&&\omit&&\omit&&\omit&&\omit&\cr
&&$(\rm 2\!\!-\!\!3)_{\rm c}$
&&$\!\!\!\!0\!\!\!\!$
&&$\!\!\!\!0\!\!\!\!$
&&$\!\!\!\!0\!\!\!\!$
&&$\!\!\!\!0\!\!\!\!$
&&$\!\!\!\!0\!\!\!\!$& \cr
\omit&height4pt&\omit&&\omit&&\omit&&\omit&&\omit&&\omit&\cr
\tablerule
\omit&height4pt&\omit&&\omit&&\omit&&\omit&&\omit&&\omit&\cr
&&$\!\!\!\!(\rm 3\!\!-\!\!1)\!\!+\!\!(3\!\!-\!\!2)_{\rm b}\!\!\!\!\!\!$
&&$\!\!\!\!0\!\!\!\!$
&&$\!\!\!\!0\!\!\!\!$
&&$\!\!\!\!\frac{-(2D-3)}{4(D-2)}\!\!\!\!$
&&$\!\!\!\!\frac{-(2D-3)}{4(D-2)}\!\!\!\!$
&&$\!\!\!\!\frac{(2D-3)}{4(D-2)}\!\!\!\!$& \cr
\omit&height4pt&\omit&&\omit&&\omit&&\omit&&\omit&&\omit&\cr
\tablerule
\omit&height4pt&\omit&&\omit&&\omit&&\omit&&\omit&&\omit&\cr
&&$(\rm 3\!\!-\!\!2)_{\rm a}$
&&$\!\!\!\!0\!\!\!\!$
&&$\!\!\!\!0\!\!\!\!$
&&$\!\!\!\!\frac{-1}{8}\!\!\!\!$
&&$\!\!\!\!\frac{-(D-4)}{8(D-2)}\!\!\!\!$
&&$\!\!\!\!\frac{1}{8}\!\!\!\!$& \cr
\omit&height4pt&\omit&&\omit&&\omit&&\omit&&\omit&&\omit&\cr
\tablerule
\omit&height4pt&\omit&&\omit&&\omit&&\omit&&\omit&&\omit&\cr
&&$(\rm 3\!\!-\!\!2)_{\rm c}$
&&$\!\!\!\!\frac{-(D-3)}{4(D-2)}\!\!\!\!$
&&$\!\!\!\!\frac{-(D-3)}{4(D-2)}\!\!\!\!$
&&$\!\!\!\!\frac{-(D-2)}{4)}\!\!\!\!$
&&$\!\!\!\!\frac{(D-3)}{4}\!\!\!\!$
&&$\!\!\!\!0\!\!\!\!$& \cr
\omit&height4pt&\omit&&\omit&&\omit&&\omit&&\omit&&\omit&\cr
\tablerule
\omit&height4pt&\omit&&\omit&&\omit&&\omit&&\omit&&\omit&\cr
&&$(\rm 3\!\!-\!\!3)_{\rm a}$
&&$\!\!\!\!0\!\!\!\!$
&&$\!\!\!\!0\!\!\!\!$
&&$\!\!\!\!\frac{(D-3)}{16}\!\!\!\!$
&&$\!\!\!\!\frac{-(D-3)}{16}\!\!\!\!$
&&$\!\!\!\!\frac{-(D-2)}{16}\!\!\!\!$& \cr
\omit&height4pt&\omit&&\omit&&\omit&&\omit&&\omit&&\omit&\cr
\tablerule
\omit&height4pt&\omit&&\omit&&\omit&&\omit&&\omit&&\omit&\cr
&&$(\rm 3\!\!-\!\!3)_{\rm b}$
&&$\!\!\!\!0\!\!\!\!$
&&$\!\!\!\!0\!\!\!\!$
&&$\!\!\!\!0\!\!\!\!$
&&$\!\!\!\!0\!\!\!\!$
&&$\!\!\!\!0\!\!\!\!$& \cr
\omit&height4pt&\omit&&\omit&&\omit&&\omit&&\omit&&\omit&\cr
\tablerule
\omit&height4pt&\omit&&\omit&&\omit&&\omit&&\omit&&\omit&\cr
&&$(\rm 3\!\!-\!\!3)_{\rm c}$
&&$\!\!\!\!0\!\!\!\!$
&&$\!\!\!\!0\!\!\!\!$
&&$\!\!\!\!0\!\!\!\!$
&&$\!\!\!\!0\!\!\!\!$
&&$\!\!\!\!0\!\!\!\!$& \cr
\omit&height4pt&\omit&&\omit&&\omit&&\omit&&\omit&&\omit&\cr
\tablerule}}

\caption{$i\Delta_{\rm cf}\times i[S]_{n=0}$ with a $(D\!-\!4)$ pre-factor.
All contributions are multiplied by $\frac{\kappa^{2}mH}{32\pi^D}
\Gamma(\frac{D}2)\Gamma(\frac{D}{2}-1)(aa')^{2-\frac{D}2}\frac{(D-4)}{2}$.}

\label{cfn03}

\end{table}
After segregating the divergences into the local terms, the total result
from Table~\ref{cfn01} and Table~\ref{cfn02} is,
\begin{eqnarray}
&&-i\Bigl[\Sigma^{\rm cfn0\!-\!1}\Bigr]\!(x;\!x')\!=\!
\frac{i\kappa^2\!H^2}{32\pi^{\frac{D}{2}}}
\frac{2\Gamma(\frac{D}{2})}{(D\!\!-\!3)}\frac{\mu^{D\!-\!4}}
{(D\!\!-\!4)}\Biggl\{\!d_1\frac{m}{H}\gamma^0\!\!
\not{\hspace{-.1cm}\overline{\partial}}\!+\!
(d_2\!+\!d_3\!+\!d_4)ma\Biggr\}\nonumber\\
&&\times\delta^{D}\!(x\!\!-\!\!x')
\!+\!\frac{i\kappa^2\!H^2}{32\pi^2}\!\ln(a)
\Biggl[-2\!\frac{m}{H}\gamma^0\!\!
\not{\hspace{-.1cm}\overline{\partial}}\!+\!\frac{9}{4}ma
\Biggr]\delta^4\!(x\!\!-\!\!x')\!+\!\frac{\kappa^2\!H^2}{64\pi^4}
\Biggl\{\!\Bigl[-\frac{1}{2}\frac{m}{H}\gamma^0\!\!
\not{\hspace{-.1cm}\overline{\partial}}\nonumber\\
&&+\frac{9m(a\!+\!a')}{32}\Bigr]\partial^2\Bigl[
\frac{\ln(\mu^2\Delta x^2)}{\Delta x^2}\Bigr]
\!+\!\Bigl[\frac{-5}{16}\frac{m}{H}\partial_0
\!+\!\frac{3}{4}ma\!+\!\frac{13}{32}ma'\Bigr]
\partial^2\frac{1}{\Delta x^2}\nonumber\\
&&+\frac{9m(a\!+\!a')}{16}\partial^2_0
\!\frac{1}{\Delta x^2}\!+\!\frac{3m(5a\!+\!a')}{16}
\gamma^0\partial_0\!\!\not{\hspace{-.1cm}\overline{\partial}}
\frac{1}{\Delta x^2}\!-\!\frac{1}{4}mH\!aa'\gamma^0\!\!
\not{\hspace{-.1cm}\overline{\partial}}\!\frac{1}{\Delta x^2}
\Biggr\}\,.\label{n012}
\end{eqnarray}
Here $d_1,\; d_2,\; d_3$ and $d_4$ are dimension-dependent coefficients,
\begin{eqnarray}
&&d_1=\frac{-(D-8)(3D-4)}{8(D-2)^2};\nonumber\\
&&d_2=-\frac14-\frac{1}{(D-2)}+(D-4)\Bigl[\frac{-D}{16}
+\frac{9}{32}-\frac{3}{8(D-2)}\Bigr];\nonumber\\
&&d_3=\frac{(D-6)}{8}\Bigl[\frac12-\frac{(2D-3)}{(D-2)^2}\Bigr]
\;\;\;;\;\;\;d_4=-\frac{3}{16}(D-1)\,.\label{d1234}
\end{eqnarray}
Because the contributions from Table~\ref{cfn03} carries a factor of
$(D-4)$, they would survive only if they combine with a $\frac{1}{(D-4)}$
after making them integrable in $D=4$ dimensions. The result from
this part is finite so we give an expression for the $D=4$,
\begin{eqnarray}
-i\Bigl[\Sigma^{\rm cfn0\!-\!2}\Bigr](x;x')\!=\!
\frac{\kappa^{2}H^2}{64\pi^4}\Biggl\{\frac18\frac{m}{H}\gamma^0\!\!
\not{\hspace{-.1cm}\partial}-\frac{17}{32}ma+\frac{35}{32}ma'
\Biggr\}\partial^2\frac{1}{\Delta x^2}\;.\label{n03}
\end{eqnarray}

\begin{table}

\vbox{\tabskip=0pt \offinterlineskip
\def\tablerule{\noalign{\hrule}}
\halign to390pt {\strut#& \vrule#\tabskip=1em plus2em& \hfil#\hfil&
\vrule#& \hfil#\hfil& \vrule#& \hfil#\hfil& \vrule#& \hfil#\hfil&
\vrule#& \hfil#\hfil& \vrule#\tabskip=0pt\cr
\tablerule
\omit&height4pt&\omit&&\omit&&\omit&&\omit&&\omit&\cr
&&$\!\!\!\!(\rm I\!\!-\!\!J)_{\rm sub}\!\!\!\!$
&&$\!\!\!\!H^{2}\!a^2\partial_0\frac{1}{\Delta x^2}\!\!\!\!$
&&$\!\!\!\!\gamma^0\!\!\not{\hspace{-.1cm}\overline{\partial}}
\frac{H^{2}a'^2}{\Delta x^2}\!\!\!\!$
&&$\!\!\!\!\frac{H^{2}a^{2}Ha'}{\Delta x^2}\!\!\!\!$
&&$\!\!\!\!\!\!H\!a\partial^2\Bigl[\frac{\ln(\mu^2\Delta x^2)}
{\Delta x^2}\Bigr]\!\!\!\!\!\!$& \cr
\omit&height4pt&\omit&&\omit&&\omit&&\omit&&\omit&\cr
\tablerule
\omit&height4pt&\omit&&\omit&&\omit&&\omit&&\omit&\cr
&&$\!\!\!\!(\rm 1\!\!-\!\!1)\!\!\!\!$
&&$\!\!\!\!-2-4\ln(\frac{y}{4})\!\!\!\!$
&&$\!\!\!\!-2-4\ln(\frac{y}{4})\!\!\!\!$
&&$\!\!\!\!0\!\!\!\!$
&&$\!\!\!\!0\!\!\!\!$ &\cr
\omit&height4pt&\omit&&\omit&&\omit&&\omit&&\omit&\cr
\tablerule
\omit&height4pt&\omit&&\omit&&\omit&&\omit&&\omit&\cr
&&$\!\!\!\!(\rm 1\!\!-\!\!2)\!\!\!\!$
&&$\!\!\!\!\frac{1}{2}+\ln(\frac{y}{4})\!\!\!\!$
&&$\!\!\!\!\frac12+\ln(\frac{y}{4})\!\!\!\!$
&&$\!\!\!\!0\!\!\!\!$
&&$\!\!\!\!0\!\!\!\!$ &\cr
\omit&height4pt&\omit&&\omit&&\omit&&\omit&&\omit&\cr
\tablerule
\omit&height4pt&\omit&&\omit&&\omit&&\omit&&\omit&\cr
&&$\!\!\!\!(\rm 1\!\!-\!\!3)\!\!\!\!$
&&$\!\!\!\!\frac94+\frac32\ln(\frac{y}{4})\!\!\!\!$
&&$\!\!\!\!\frac94+\frac32\ln(\frac{y}{4})\!\!\!\!$
&&$\!\!\!\!\frac94+\frac32\ln(\frac{y}{4})\!\!\!\!$
&&$\!\!\!\!0\!\!\!\!$ &\cr
\omit&height4pt&\omit&&\omit&&\omit&&\omit&&\omit&\cr
\tablerule
\omit&height4pt&\omit&&\omit&&\omit&&\omit&&\omit&\cr
&&$\!\!\!\!(\rm 2\!\!-\!\!1)\!\!\!\!$
&&$\!\!\!\!\frac12+\ln(\frac{y}{4})\!\!\!\!$
&&$\!\!\!\!\frac12+\ln(\frac{y}{4})\!\!\!\!$
&&$\!\!\!\!0\!\!\!\!$
&&$\!\!\!\!0\!\!\!\!$ &\cr
\omit&height4pt&\omit&&\omit&&\omit&&\omit&&\omit&\cr
\tablerule
\omit&height4pt&\omit&&\omit&&\omit&&\omit&&\omit&\cr
&&$\!\!\!\!(\rm 2\!\!-\!\!2)_{\rm a}\!\!\!\!$
&&$\!\!\!\!\frac14+\frac12\ln(\frac{y}{4})\!\!\!\!$
&&$\!\!\!\!-\frac14-\frac12\ln(\frac{y}{4})\!\!\!\!$
&&$\!\!\!\!0\!\!\!\!$
&&$\!\!\!\!0\!\!\!\!$ &\cr
\omit&height4pt&\omit&&\omit&&\omit&&\omit&&\omit&\cr
\tablerule
\omit&height4pt&\omit&&\omit&&\omit&&\omit&&\omit&\cr
&&$\!\!\!\!(\rm 2\!\!-\!\!2)_{\rm b}\!\!\!\!$
&&$\!\!\!\!1+2\ln(\frac{y}{4})\!\!\!\!$
&&$\!\!\!\!0\!\!\!\!$
&&$\!\!\!\!0\!\!\!\!$
&&$\!\!\!\!0\!\!\!\!$ &\cr
\omit&height4pt&\omit&&\omit&&\omit&&\omit&&\omit&\cr
\tablerule
\omit&height4pt&\omit&&\omit&&\omit&&\omit&&\omit&\cr
&&$\!\!\!\!(\rm 2\!\!-\!\!2)_{\rm c}\!\!\!\!$
&&$\!\!\!\!-\frac14-\frac12\ln(\frac{y}{4})\!\!\!\!$
&&$\!\!\!\!-\frac14-\frac12\ln(\frac{y}{4})\!\!\!\!$
&&$\!\!\!\!0\!\!\!\!$
&&$\!\!\!\!0\!\!\!\!$ &\cr
\omit&height4pt&\omit&&\omit&&\omit&&\omit&&\omit&\cr
\tablerule
\omit&height4pt&\omit&&\omit&&\omit&&\omit&&\omit&\cr
&&$\!\!\!\!(\rm 2\!\!-\!\!3)_{\rm a}\!\!\!\!$
&&$\!\!\!\!-\frac32-\ln(\frac{y}{4})\!\!\!\!$
&&$\!\!\!\!0\!\!\!\!$
&&$\!\!\!\!-\frac32-\ln(\frac{y}{4})\!\!\!\!$
&&$\!\!\!\!0\!\!\!\!$ &\cr
\omit&height4pt&\omit&&\omit&&\omit&&\omit&&\omit&\cr
\tablerule
\omit&height4pt&\omit&&\omit&&\omit&&\omit&&\omit&\cr
&&$\!\!\!\!(\rm 2\!\!-\!\!3)_{\rm b}\!\!\!\!$
&&$\!\!\!\!-\frac38-\frac14\ln(\frac{y}{4})\!\!\!\!$
&&$\!\!\!\!-\frac38-\frac14\ln(\frac{y}{4})\!\!\!\!$
&&$\!\!\!\!-\frac38-\frac14\ln(\frac{y}{4})\!\!\!\!$
&&$\!\!\!\!0\!\!\!\!$ &\cr
\omit&height4pt&\omit&&\omit&&\omit&&\omit&&\omit&\cr
\tablerule
\omit&height4pt&\omit&&\omit&&\omit&&\omit&&\omit&\cr
&&$\!\!\!\!(\rm 2\!\!-\!\!3)_{\rm c}\!\!\!\!$
&&$\!\!\!\!-\frac38-\frac14\ln(\frac{y}{4})\!\!\!\!$
&&$\!\!\!\!\frac38+\frac14\ln(\frac{y}{4})\!\!\!\!$
&&$\!\!\!\!-\frac38-\frac14\ln(\frac{y}{4})\!\!\!\!$
&&$\!\!\!\!0\!\!\!\!$ &\cr
\omit&height4pt&\omit&&\omit&&\omit&&\omit&&\omit&\cr
\tablerule
\omit&height4pt&\omit&&\omit&&\omit&&\omit&&\omit&\cr
&&$\!\!\!\!\!(\rm 3\!\!-\!\!1)\!+\!(\rm 3\!\!-\!\!2)_{\rm b}\!\!\!\!\!\!$
&&$\!\!\!\!\frac58\ln(\frac{y}{4})\!\!\!\!$
&&$\!\!\!\!\frac58\ln(\frac{y}{4})\!\!\!\!$
&&$\!\!\!\!0\!\!\!\!$
&&$\!\!\!\!\frac58\!\!\!\!$ &\cr
\omit&height4pt&\omit&&\omit&&\omit&&\omit&&\omit&\cr
\tablerule
\omit&height4pt&\omit&&\omit&&\omit&&\omit&&\omit&\cr
&&$\!\!\!\!(\rm 3\!\!-\!\!2)_{\rm a}\!\!\!\!$
&&$\!\!\!\!-\frac18\ln(\frac{y}{4})\!\!\!\!$
&&$\!\!\!\!-\frac18\ln(\frac{y}{4})\!\!\!\!$
&&$\!\!\!\!0\!\!\!\!$
&&$\!\!\!\!-\frac18\!\!\!\!$ &\cr
\omit&height4pt&\omit&&\omit&&\omit&&\omit&&\omit&\cr
\tablerule
\omit&height4pt&\omit&&\omit&&\omit&&\omit&&\omit&\cr
&&$\!\!\!\!(\rm 3\!\!-\!\!2)_{\rm c}\!\!\!\!$
&&$\!\!\!\!-\frac12\ln(\frac{y}{4})\!\!\!\!$
&&$\!\!\!\!0\!\!\!\!$
&&$\!\!\!\!0\!\!\!\!$
&&$\!\!\!\!-\frac12\!\!\!\!$ &\cr
\omit&height4pt&\omit&&\omit&&\omit&&\omit&&\omit&\cr
\tablerule
\omit&height4pt&\omit&&\omit&&\omit&&\omit&&\omit&\cr
&&$\!\!\!\!{\rm Sub\!\!-\!\!total} \!\!\!\!$
&&$\!\!\!\!0\!\!\!\!$
&&$\!\!\!\!\frac34-\ln(\frac{y}{4})\!\!\!\!$
&&$\!\!\!\!0\!\!\!\!$
&&$\!\!\!\!0\!\!\!\!$ &\cr
\omit&height4pt&\omit&&\omit&&\omit&&\omit&&\omit&\cr
\tablerule}}

\caption{$i\Delta_{\rm cf}\times i[S]_{n=0}$ form the two terms
which tend to cancel. All contributions are multiplied by
$\frac{\kappa^{2}mH}{64\pi^4}$.}

\label{cfn04}

\end{table}
The contributions which originate from the final two terms of (\ref{n0mf})
and (\ref{n0mfd1}) tend to cancel. They could be taken special care first
before they are tabulated because most of contributions from
this class give finite results in $D=4$ dimensions except for
a few divergent terms from the double derivatives on $\Delta_{\rm cf}(x;x')$.
The strategy to deal with the finite part of the contributions is
to perform the $(D\!-\!4)$ expansion and make use of (\ref{ds1}) and the
following key identities,
\begin{eqnarray}
&&\partial_{\mu}\frac{1}{\Delta x^{D-2}}\!=\!\mu^{D-4}
\Bigl[1+\frac12(D\!-\!4)(1\!-\!\ln\Delta x^2)\Bigr]
\partial_{\mu}\frac{1}{\Delta x^2}\,,\label{id3}\\
&&\partial_{\mu}\frac{1}{\Delta x^{2D-6}}\!=\!\mu^{2D-8}
\Bigl[1+\frac12(D\!-\!4)(2-2\ln\Delta x^2)\Bigr]
\partial_{\mu}\frac{1}{\Delta x^2}\,,\label{id4}\\
&&\partial^2\frac{1}{\Delta x^{D-2}}\!=\!\frac{i4\pi^{\frac{D}{2}}}
{\Gamma(\frac{D}{2}\!-\!1)}\!\delta^D(x\!-\!x')\label{id5}\,.
\end{eqnarray}
Here we present one example from $(3\!-\!2)_{\rm c}$,
\begin{eqnarray}
&&\Gamma(\frac{D}{2})\Gamma(\frac{D}{2}\!-\!1)(\frac{m}{H})
\frac{\kappa^2}{8}\Biggl\{\frac{H^2(aa')^{2-\frac{D}{2}}}
{32\pi^D\frac12(D\!-\!4)}\Biggl[\frac{DHa}{2(D\!-\!3)}
\partial^2\!+\!\frac{D}{2}H^2a^2\partial_0\Biggr]
\frac{1}{\Delta x^{2D-6}}\nonumber\\
&&\hspace{1cm}-\frac{H^{D-2}}{2^{D+1}\pi^D}\Gamma(\frac{D}{2})
\frac{\Gamma(3\!-\!\frac{D}{2})}{\frac12(D\!-\!4)}\Biggl[
\frac{DHa}{(D\!-\!2)}\partial^2\!+\!\frac{D}{2}H^2a^2\partial_0
\Biggr]\frac{1}{\Delta x^{D-2}}\Biggr\}\,.\label{2-3c}
\end{eqnarray}
There are three distinctive contributions in (\ref{2-3c}).
We extract out the pre-factor
$\Gamma(\frac{D}{2})\Gamma(\frac{D}{2}\!-\!1)\frac{\kappa^2mH}{64\pi^D}$
to avoid the repeated and lengthy expressions. The first kind comes from
the two single derivative terms,
\begin{eqnarray}
&&\frac{\frac14\mu^{2D-8}}{\frac12(D\!-\!4)}\frac{DH^2a^2}{2}\Biggl\{
\Bigl[1\!+\!\frac12(D\!-\!4)(-\ln aa'\!+\!2\!-\!
2\ln\mu^2\!\Delta x^2)\Bigr]\!+\!\Bigl[-1\!+\!\frac12(D\!-\!4)\nonumber\\
&&\times(-\ln\frac{H^2}{4\mu^2}\!-\!1\!-\!1\!+\!\ln\mu^2\!\Delta x^2)
\Bigr]\Biggr\}\partial_0\frac{1}{\Delta x^2}\longrightarrow
\frac{DH^2a^2}{8}\Bigl[-\ln\frac{y}{4}\Bigr]\partial_0
\frac{1}{\Delta x^2}\,,
\end{eqnarray}
and we presented the temporal and spatial contributions separately
in Table~\ref{cfn04}. The second kind is the contributions which
produce a delta function originated from the two double derivative terms,
\begin{eqnarray}
&&\frac{\frac14\mu^{D-4}}{\frac12(D\!-\!4)}
\frac{DHa}{2}\Biggl\{\Bigl[1\!+\!\frac12(D\!-\!4)
(-\ln aa'\!-\!2)\Bigr]+\Bigl[-1\!+\!\frac12(D\!-\!4)
(-\ln\frac{H^2}{4\mu^2}\!-\!1\!+\!1)\Bigr]\Biggr\}\nonumber\\
&&\times\frac{i4\pi^{\frac{D}{2}}\delta^D(x\!-\!x')}
{\Gamma(\frac{D}{2}\!-\!1)}\longrightarrow -DHa\Bigl[
\ln\frac{Ha}{2\mu}+1\Bigr]\frac{i\pi^{\frac{D}{2}}\mu^{D-4}}
{\Gamma(\frac{D}{2}\!-\!1)}\delta^D\!(x\!-\!x')\,,
\end{eqnarray}
and the final results are displayed in the first column of Table~\ref{cfn06}.
The third kind is the residual term from
$\partial^2\frac{1}{\Delta x^{2D-6}}$,
\begin{eqnarray}
\frac{\mu^{2D-8}}{\frac12(D\!-\!4)}\times-\frac12(D\!-\!4)
\frac{DHa}{8}\partial^2\Bigl[\frac{\ln\mu^2\Delta x^2}
{\Delta x^2}\Bigr]\,,
\end{eqnarray}
and we give the results in the final column of Table~\ref{cfn04}.

\begin{table}

\vbox{\tabskip=0pt \offinterlineskip
\def\tablerule{\noalign{\hrule}}
\halign to390pt {\strut#& \vrule#\tabskip=1em plus2em& \hfil#\hfil&
\vrule#& \hfil#\hfil& \vrule#& \hfil#\hfil& \vrule#& \hfil#\hfil&
\vrule#& \hfil#\hfil& \vrule#&\hfil#\hfil& \vrule#\tabskip=0pt\cr
\tablerule
\omit&height4pt&\omit&&\omit&&\omit&&\omit&&\omit&&\omit&\cr
&&$\!\!\!\!(\rm I\!\!-\!\!J)_{\rm sub}\!\!\!\!$
&&$\!\!\!\!\frac{H^{2}a^2\Delta\eta}{\Delta x^4}\!\!\!\!$
&&$\!\!\!\!\frac{H^{2}aa'\Delta\eta}{\Delta x^4}\!\!\!\!\!\!$
&&$\!\!\!\!\frac{H^{2}a^{2}\gamma^0\gamma^k\Delta x_k}
{\Delta x^4}\!\!\!\!$
&&$\!\!\!\!\frac{H^{2}aa'\gamma^0\gamma^k\Delta x_k}
{\Delta x^4}\!\!\!\!$
&&$\!\!\!\!\frac{H^{2}a^{2}Ha'}{\Delta x^2}\!\!\!\!$& \cr
\omit&height4pt&\omit&&\omit&&\omit&&\omit&&\omit&&\omit&\cr
\tablerule
\omit&height4pt&\omit&&\omit&&\omit&&\omit&&\omit&&\omit&\cr
&&$(\rm 3\!\!-\!\!3)_{\rm a}$
&&$\!\!\!\!\frac12\!\!\!\!$
&&$\!\!\!\!-\frac12\!\!\!\!$
&&$\!\!\!\!0\!\!\!\!$
&&$\!\!\!\!0\!\!\!\!$
&&$\!\!\!\!\frac14\!\!\!\!$& \cr
\omit&height4pt&\omit&&\omit&&\omit&&\omit&&\omit&&\omit&\cr
\tablerule
\omit&height4pt&\omit&&\omit&&\omit&&\omit&&\omit&&\omit&\cr
&&$(\rm 3\!\!-\!\!3)_{\rm b}$
&&$\!\!\!\!\frac18\!\!\!\!$
&&$\!\!\!\!-\frac18\!\!\!\!$
&&$\!\!\!\!\frac18\!\!\!\!$
&&$\!\!\!\!\frac18\!\!\!\!$
&&$\!\!\!\!\frac1{16}\!\!\!\!$& \cr
\omit&height4pt&\omit&&\omit&&\omit&&\omit&&\omit&&\omit&\cr
\tablerule
\omit&height4pt&\omit&&\omit&&\omit&&\omit&&\omit&&\omit&\cr
&&$(\rm 3\!\!-\!\!3)_{\rm c}$
&&$\!\!\!\!-\frac58\!\!\!\!$
&&$\!\!\!\!\frac58\!\!\!\!$
&&$\!\!\!\!\frac58\!\!\!\!$
&&$\!\!\!\!\frac58\!\!\!\!$
&&$\!\!\!\!\frac{-5}{16}\!\!\!\!$& \cr
\omit&height4pt&\omit&&\omit&&\omit&&\omit&&\omit&&\omit&\cr
\tablerule
\omit&height4pt&\omit&&\omit&&\omit&&\omit&&\omit&&\omit&\cr
&&$\!\!\!\!{\rm Sub\!\!-\!\!total}\!\!\!\!$
&&$\!\!\!\!0\!\!\!\!$
&&$\!\!\!\!0\!\!\!\!$
&&$\!\!\!\!\frac34\!\!\!\!$
&&$\!\!\!\!\frac34\!\!\!\!$
&&$\!\!\!\!0\!\!\!\!$& \cr
\omit&height4pt&\omit&&\omit&&\omit&&\omit&&\omit&&\omit&\cr
\tablerule}}

\caption{$i\Delta_{\rm cf}\times i[S]_{n=0}$ from the two terms
which tend to cancel. All contributions are multiplied by
$\frac{\kappa^{2}mH}{64\pi^4}\times\Bigl(1+\ln(\frac{y}4)\Bigr)$.}

\label{cfn05}

\end{table}

\begin{table}

\vbox{\tabskip=0pt \offinterlineskip
\def\tablerule{\noalign{\hrule}}
\halign to390pt {\strut#& \vrule#\tabskip=1em plus2em& \hfil#\hfil&
\vrule#& \hfil#\hfil& \vrule#&\hfil#\hfil& \vrule#\tabskip=0pt\cr
\tablerule
\omit&height4pt&\omit&&\omit&&\omit&\cr
&&$\!\!\!\!(\rm I\!\!-\!\!J)_{\rm sub}\!\!\!\!$
&&$\!\!\!\!\frac{i\kappa^{2}H^2}{64\pi^2}ma\delta^4(x-x')\!\!\!\!$
&&$\!\!\!\!\!\!\frac{i\kappa^{2}H^{D-2}}{2^{D+2}\pi^{\frac{D}{2}}}
\Gamma(\frac{D}{2})\frac{\Gamma(3\!-\!\frac{D}{2})}
{\frac12(D\!-\!4)}ma\delta^{D}\!(x\!-\!x')\!\!\!\!\!\!$& \cr
\omit&height4pt&\omit&&\omit&&\omit&\cr
\tablerule
\omit&height4pt&\omit&&\omit&&\omit&\cr
&&$\!\!\!\!(\rm 3\!\!-\!\!1)\!\!+\!\!(3\!\!-\!\!2)_{\rm b}\!\!\!\!$
&&$\!\!\!\!5[1+\ln(\frac{Ha}{2\mu})]\!\!\!\!$
&&$\!\!\!\!0\!\!\!\!$& \cr
\omit&height4pt&\omit&&\omit&&\omit&\cr
\tablerule
\omit&height4pt&\omit&&\omit&&\omit&\cr
&&$\!\!\!\!(\rm 3\!\!-\!\!2)_{\rm a}\!\!\!\!$
&&$\!\!\!\!-[1+\ln(\frac{Ha}{2\mu})]\!\!\!\!$
&&$\!\!\!\!0\!\!\!\!$& \cr
\omit&height4pt&\omit&&\omit&&\omit&\cr
\tablerule
\omit&height4pt&\omit&&\omit&&\omit&\cr
&&$\!\!\!\!(\rm 3\!\!-\!\!2)_{\rm c}\!\!\!\!$
&&$\!\!\!\!-4[1+\ln(\frac{Ha}{2\mu})]\!\!\!\!$
&&$\!\!\!\!0\!\!\!\!$& \cr
\omit&height4pt&\omit&&\omit&&\omit&\cr
\tablerule
\omit&height4pt&\omit&&\omit&&\omit&\cr
&&$\!\!\!\!(\rm 3\!\!-\!\!3)_{\rm a}\!\!\!\!$
&&$\!\!\!\!0\!\!\!\!$
&&$\!\!\!\!\Gamma(\frac{D}{2})\times\frac{D}{4}\!\!\!\!$& \cr
\omit&height4pt&\omit&&\omit&&\omit&\cr
\tablerule
\omit&height4pt&\omit&&\omit&&\omit&\cr
&&$\!\!\!\!(\rm 3\!\!-\!\!3)_{\rm b}\!\!\!\!$
&&$\!\!\!\!0\!\!\!\!$
&&$\!\!\!\!\Gamma(\frac{D}{2})\times(1\!-\!\frac{D}{2})\!\!\!\!$& \cr
\omit&height4pt&\omit&&\omit&&\omit&\cr
\tablerule
\omit&height4pt&\omit&&\omit&&\omit&\cr
&&$\!\!\!\!(\rm 3\!\!-\!\!3)_{\rm c}\!\!\!\!$
&&$\!\!\!\!0\!\!\!\!$
&&$\!\!\!\!\Gamma(\frac{D}{2}\!-\!1)\times\frac{(2D-3)}{2}\!\!\!\!$& \cr
\omit&height4pt&\omit&&\omit&&\omit&\cr
\tablerule}}

\caption{$i\Delta_{\rm cf}\times i[S]_{\rm n=0}$. The delta function
contribution from the two terms which tend to cancel.}

\label{cfn06}

\end{table}

The final contribution in this category is from $(3\!-\!3)$ which
consists of some finite terms and local divergent terms.
Recall in (\ref{2dcf}) that two derivatives acting on the conformal part
of the graviton propagator would produce a delta function.
In general, it would be zero in dimensional regularization when it acts
on dimension-dependent power of the coordinate separation. However, the
last term with the divergent coefficient in (\ref{n0mf}) does not possess
any dimension-dependent power of $\Delta x^2$. As a result, when
$i[S]_{\rm n=0}$ are multiplied by $\partial .\partial\Delta_{\rm cf}$,
no any other terms in the calculation can be used to cancel this
particular divergent local term. Here we present $(3\!-\!3)_{\rm a}$
as an example,
\begin{eqnarray}
&&\Gamma(\frac{D}{2})\Gamma(\frac{D}{2}\!-\!1)\frac{-(D\!-\!2)}{16}
\frac{m}{H}\frac{\kappa^2}{2}\Biggl\{\frac{H^2(aa')^{2-\frac{D}{2}}}
{32\pi^D\frac12(D\!-\!4)}\Biggl[\frac{D\!H^3a^2a'\Delta\!\eta^2}
{\Delta x^{2D-4}}\!+\!\frac{D}{2}\frac{H^3a^2a'}
{\Delta x^{2D-6}}\Biggr]\nonumber\\
&&-\frac{H^{D-2}\Gamma(\frac{D}{2})}{2^{D+1}\pi^D}
\frac{\Gamma(3\!-\!\frac{D}{2})}{\frac12(D\!-\!4)}\Biggl[
\frac{D\!H^3a^2a'\Delta\!\eta^2}{\Delta x^D}\!+\!
\frac{D}{2}\frac{H^3a^2a'}{\Delta x^{D-2}}\!+\!
\frac{DHai2\pi^{\frac{D}{2}}}{\Gamma(\frac{D}{2})}
\delta^D\!(x\!-\!x')\Biggr]\Biggr\}.\nonumber\\
\end{eqnarray}
We employ the same trick to deal with the finite part, extract
out the prefactor $\Gamma(\frac{D}{2})
\Gamma(\frac{D}{2}\!-\!1)\frac{\kappa^2 mH}{64\pi^D}
\frac{-(D\!-\!2)}{16}$,
\begin{eqnarray}
&&\hspace{-1cm}\Biggl\{\Bigl[1\!+\!\frac12(D\!-\!4)(-\ln aa'\!-\!
2\ln\mu^2\Delta x^2)\Bigr]\!+\!\Bigl[-1\!+\!\frac12(D\!-\!4)
(-\ln\frac{H^2}{4\mu^2}\!-\!1\!+\!\ln\mu^2\Delta x^2)
\Bigr]\Biggr\}\nonumber\\
&&\hspace{5.4cm}\times\frac{\mu^{2D-8}}{\frac12(D\!-\!4)}\Biggl[
\frac{D\!H^3a^2a'\Delta\!\eta^2}{\Delta x^2}\!+\!
\frac{D}{2}\frac{H^3a^2a'}{\Delta x^2}\Biggr]\,,
\end{eqnarray}
and tabulate the result for $D=4$ in Table~\ref{cfn05}.
The divergent term can be read off directly and we present
it in the second column of Table~\ref{cfn05}. Finally we
enclose this sub-class by summing up all the terms from
Tables~\ref{cfn04}, \ref{cfn05} and \ref{cfn06},
\begin{eqnarray}
&&-i\Bigl[\Sigma^{\rm cfn0\!-\!3}\Bigr]\!(x;x')\!=\!
\frac{i\kappa^{2}\!H^{D-2}}{2^{D+1}\pi^{\frac{D}{2}}}
\frac{\Gamma(\frac{D}{2})\Gamma(\frac{D}{2}\!-\!1)
\Gamma(3\!-\!\frac{D}{2})(2D\!\!-\!3)}{2(D\!-\!4)}
ma\delta^D\!(x\!\!-\!\!x')\nonumber\\
&&\hspace{3cm}-\frac{i\kappa^2\!H^2}{32\pi^2}
\frac{ma}{4}\delta^4\!(x\!\!-\!\!x')
\!+\!\frac{\kappa^2\!H^2}{64\pi^4}\Biggl\{\Bigl[\frac38
\!-\!\frac{11}{8}\ln(\frac{y}{4})\Bigr]mH\!a^2\nonumber\\
&&\hspace{6cm}-\frac38\Bigl[1\!+\!\ln(\frac{y}{4})\Bigr]mH\!aa'
\Biggr\}\gamma^0\!\!\not{\hspace{-.1cm}\overline{\partial}}
\!\frac{1}{\Delta x^2}\,.\label{n0456}
\end{eqnarray}

The last computation in this sub-section involves the rest of the
infinite series expansion which are all integrable and hence we can
compute it in $D=4$ dimensions directly. The fermion propagator for the
$n\geq 1$ of the series in four dimensions we employed is,
\begin{eqnarray}
i[S]_{n\geq 1}\!=\!\! \frac{mHaa'}{16\pi^2}\!\!\sum_{n=1}^{\infty}
\Biggl\{\frac{2\gamma^{\nu}\!\gamma^0\!\Delta x_{\nu}}{\Delta x^2}\Bigl[
1\!\!+\!\!n\ln(\frac{y}{4})\Bigr]\!\!+\!\!Ha\Bigl[1\!\!+\!\!(n\!\!+\!\!1)
\ln(\frac{y}{4})\Bigr]\Biggr\}\Bigl(\frac{y}{4}\Bigr)^n,\label{4Dfpropn1}
\end{eqnarray}
and its derivative\footnote{One can find various gamma functions
contracted with (\ref{4Ddfpropn}) in Appendix \ref{fermionprop}} is,
\begin{eqnarray}
&&\partial_{\mu}i[S]_{n\geq 1}\!=\!\frac{mHaa'}{16\pi^2}
\Sigma_{n=1}^{\infty}(\frac{y}{4})^n\Biggl\{
\frac{4\gamma^{\nu}\gamma^0\!\Delta x_{\nu}\Delta x_{\mu}}{\Delta x^4}
\Bigl[(2n\!-\!1)\!+\!(n^2\!-\!n)\ln\frac{y}{4}\Bigr]\nonumber\\
&&+\frac{2\gamma_{\mu}\gamma^0}{\Delta x^2}
\Bigl[1\!+\!n\ln\frac{y}{4}\Bigr]
+\frac{2Ha(\delta^0_{\mu}\gamma^{\nu}\gamma^0\!\Delta x_{\nu}\!+\!
\Delta x_{\mu})}{\Delta x^2}\Bigl[(2n\!+\!1)\!+\!(n^2\!+\!n)
\ln\frac{y}{4}\Bigr]\nonumber\\
&&\hspace{4.5cm}+H^2a^2\!\delta^0_{\mu}\Bigl[(2n\!+\!3)
\!+\!(n^2\!+\!3n\!+\!2)\ln\frac{y}{4}\Bigr]\Biggr\}\,.\label{4Ddfpropn}
\end{eqnarray}

\begin{table}

\vbox{\tabskip=0pt \offinterlineskip
\def\tablerule{\noalign{\hrule}}
\halign to390pt {\strut#& \vrule#\tabskip=1em plus2em& \hfil#\hfil&
\vrule#& \hfil#\hfil& \vrule#& \hfil#\hfil& \vrule#& \hfil#\hfil&
\vrule#& \hfil#\hfil& \vrule#& \hfil#\hfil& \vrule#&
\hfil#\hfil& \vrule#\tabskip=0pt\cr
\tablerule
\omit&height4pt&\omit&&\omit&&\omit&&\omit&&\omit&&\omit&&\omit&\cr
&&$\!\!\!\!\!\scriptstyle{(\rm I\!-\!J)_{\rm sub}}\!\!\!\!\!\!\!\!$
&&$\!\!\!\!\!\!\frac{\gamma^0\gamma^k\Delta x_k}{\Delta x^6}\!\!\!\!\!\!\!$
&&$\!\!\!\!\!\!\frac{\Delta\eta}{\Delta x^6}\!\!\!\!\!\!\!$
&&$\!\!\!\!\!\!\scriptstyle{\frac{Ha\gamma^0\!\Delta\eta\gamma^k\!\Delta x_k}
{\Delta x^6}}\!\!\!\!\!\!\!\!\!\!\!$
&&$\!\!\!\!\!\!\scriptstyle{\frac{Ha'\gamma^0\!\Delta\eta\gamma^k\!\Delta x_k}
{\Delta x^6}}\!\!\!\!\!\!\!\!\!\!\!$
&&$ \!\!\!\!\!\!\frac{Ha\Delta\eta^2}{\Delta x^6}\!\!\!\!\!\!\!\!$
&&$\!\!\!\!\!\frac{Ha'\Delta\eta^2}{\Delta x^6}\!\!\!\!\!$& \cr
\omit&height4pt&\omit&&\omit&&\omit&&\omit&&\omit&&\omit&&\omit&\cr
\tablerule
\omit&height4pt&\omit&&\omit&&\omit&&\omit&&\omit&&\omit&&\omit&\cr
&&$\!\!\!\!\!\scriptstyle{(\rm 1\!-\!1)}\!\!\!\!\!\!\!\!$
&&$\!\!\!\!\!\!\scriptstyle{16(3n^2-2n-2)}\!\!\!\!\!\!\!\!\!\!\!$
&&$\!\!\!\!\!\!\scriptstyle{-16(3n^2-2n-2)}\!\!\!\!\!\!\!\!\!\!\!$
&&$\!\!\!\!\!\!\scriptstyle{16(3n^2-2n-2)}\!\!\!\!\!\!\!\!\!\!\!$
&&$\!\!\!\!\!\!\scriptstyle{0}\!\!\!\!\!\!\!$
&&$ \!\!\!\!\!\!\scriptstyle{-16(3n^2-2n-2)}\!\!\!\!\!\!\!\!\!\!$
&&$\!\!\!\!\!\scriptstyle{0}\!\!\!\!\!$& \cr
\omit&height4pt&\omit&&\omit&&\omit&&\omit&&\omit&&\omit&&\omit&\cr
\tablerule
\omit&height4pt&\omit&&\omit&&\omit&&\omit&&\omit&&\omit&&\omit&\cr
&&$\!\!\!\!\scriptstyle{(\rm 1\!-\!2)}\!\!\!\!\!\!\!\!$
&&$\!\!\!\!\!\!\scriptstyle{-4(3n^2-2n-2)}\!\!\!\!\!\!\!\!\!\!\!$
&&$\!\!\!\!\!\!\scriptstyle{4(3n^2-2n-2)}\!\!\!\!\!\!\!\!\!\!\!$
&&$\!\!\!\!\!\!\scriptstyle{-4(3n^2-2n-2)}\!\!\!\!\!\!\!\!\!\!\!$
&&$\!\!\!\!\!\!\scriptstyle{0}\!\!\!\!\!\!\!$
&&$ \!\!\!\!\!\!\scriptstyle{4(3n^2-2n-2)}\!\!\!\!\!\!\!\!\!\!$
&&$\!\!\!\!\!\scriptstyle{0}\!\!\!\!\!$& \cr
\omit&height4pt&\omit&&\omit&&\omit&&\omit&&\omit&&\omit&&\omit&\cr
\tablerule
\omit&height4pt&\omit&&\omit&&\omit&&\omit&&\omit&&\omit&&\omit&\cr
&&$\!\!\!\!\!\scriptstyle{(\rm 1\!-\!3)}\!\!\!\!\!\!\!\!$
&&$\!\!\!\!\!\!\scriptstyle{6(2n+1)}\!\!\!\!\!\!\!\!\!\!\!$
&&$\!\!\!\!\!\!\scriptstyle{-6(2n+1)}\!\!\!\!\!\!\!\!\!\!\!$
&&$\!\!\!\!\!\!\scriptstyle{6(2n+1)}\!\!\!\!\!\!\!\!\!\!\!$
&&$\!\!\!\!\!\!\scriptstyle{0}\!\!\!\!\!\!\!$
&&$ \!\!\!\!\!\!\scriptstyle{-6(2n+1)}\!\!\!\!\!\!\!\!\!\!$
&&$\!\!\!\!\!\scriptstyle{0}\!\!\!\!\!$& \cr
\omit&height4pt&\omit&&\omit&&\omit&&\omit&&\omit&&\omit&&\omit&\cr
\tablerule
\omit&height4pt&\omit&&\omit&&\omit&&\omit&&\omit&&\omit&&\omit&\cr
&&$\!\!\!\!\scriptstyle{(\rm 2\!-\!1)}\!\!\!\!\!\!\!\!$
&&$\!\!\!\!\!\!\scriptstyle{-4(3n^2-2n-2)}\!\!\!\!\!\!\!\!\!\!\!$
&&$\!\!\!\!\!\!\scriptstyle{4(3n^2-2n-2)}\!\!\!\!\!\!\!\!\!\!\!$
&&$\!\!\!\!\!\!\scriptstyle{-4(3n^2-2n-2)}\!\!\!\!\!\!\!\!\!\!\!$
&&$\!\!\!\!\!\!\scriptstyle{0}\!\!\!\!\!\!\!$
&&$ \!\!\!\!\!\!\scriptstyle{4(3n^2-2n-2)}\!\!\!\!\!\!\!\!\!\!$
&&$\!\!\!\!\!\scriptstyle{0}\!\!\!\!\!$& \cr
\omit&height4pt&\omit&&\omit&&\omit&&\omit&&\omit&&\omit&&\omit&\cr
\tablerule
\omit&height4pt&\omit&&\omit&&\omit&&\omit&&\omit&&\omit&&\omit&\cr
&&$\!\!\!\!\!\scriptstyle{(\rm 2\!-\!2)_{\rm a}}\!\!\!\!\!\!\!\!$
&&$\!\!\!\!\!\!\scriptstyle{-2(3n^2-2n-1)}\!\!\!\!\!\!\!\!\!\!\!$
&&$\!\!\!\!\!\!\scriptstyle{2(3n^2-2n+1)}\!\!\!\!\!\!\!\!\!\!\!$
&&$\!\!\!\!\!\!\scriptstyle{0}\!\!\!\!\!\!\!\!\!\!\!$
&&$\!\!\!\!\!\!\!\!\!\!\!\scriptstyle{2(3n^2-2n)}\!\!\!\!\!\!\!\!\!\!\!\!\!\!$
&&$ \!\!\!\!\!\!\scriptstyle{0}\!\!\!\!\!\!\!\!\!\!\!$
&&$\!\!\!\!\!\scriptstyle{-2(3n^2-2n)}\!\!\!\!\!\!\!$& \cr
\omit&height4pt&\omit&&\omit&&\omit&&\omit&&\omit&&\omit&&\omit&\cr
\tablerule
\omit&height4pt&\omit&&\omit&&\omit&&\omit&&\omit&&\omit&&\omit&\cr
&&$\!\!\!\!\!\scriptstyle{(\rm 2\!-\!2)_{\rm b}}\!\!\!\!\!\!\!\!$
&&$\!\!\!\!\!\!\scriptstyle{0}\!\!\!\!\!\!\!\!\!\!\!$
&&$\!\!\!\!\!\!\scriptstyle{4(6n^2-4n-1)}\!\!\!\!\!\!\!\!\!\!\!$
&&$\!\!\!\!\!\!\scriptstyle{0}\!\!\!\!\!\!\!\!\!\!\!$
&&$\!\!\!\!\!\!\!\!\!\!\!\scriptstyle{0}\!\!\!\!\!\!\!\!\!\!\!\!\!\!$
&&$ \!\!\!\!\!\!\scriptstyle{4(3n^2-2n-2)}\!\!\!\!\!\!\!\!\!\!\!$
&&$\!\!\!\!\!\scriptstyle{-4(3n^2-2n)}\!\!\!\!\!\!\!$& \cr
\omit&height4pt&\omit&&\omit&&\omit&&\omit&&\omit&&\omit&&\omit&\cr
\tablerule
\omit&height4pt&\omit&&\omit&&\omit&&\omit&&\omit&&\omit&&\omit&\cr
&&$\!\!\!\!\!\scriptstyle{(\rm 2\!-\!2)_{\rm c}}\!\!\!\!\!\!\!\!$
&&$\!\!\!\!\!\!\scriptstyle{2(3n^2-2n-2)}\!\!\!\!\!\!\!\!\!\!\!$
&&$\!\!\!\!\!\!\scriptstyle{-2(3n^2-2n-2)}\!\!\!\!\!\!\!\!\!\!\!$
&&$\!\!\!\!\!\!\scriptstyle{2(3n^2-2n-2)}\!\!\!\!\!\!\!\!\!\!\!$
&&$\!\!\!\!\!\!\!\!\!\!\!\scriptstyle{0}\!\!\!\!\!\!\!\!\!\!\!\!\!\!$
&&$ \!\!\!\!\!\!\scriptstyle{-2(3n^2-2n-2)}\!\!\!\!\!\!\!\!\!\!\!$
&&$\!\!\!\!\!\scriptstyle{0}\!\!\!\!\!\!\!$& \cr
\omit&height4pt&\omit&&\omit&&\omit&&\omit&&\omit&&\omit&&\omit&\cr
\tablerule
\omit&height4pt&\omit&&\omit&&\omit&&\omit&&\omit&&\omit&&\omit&\cr
&&$\!\!\!\!\!\scriptstyle{(\rm 2\!-\!3)_{\rm a}}\!\!\!\!\!\!\!\!$
&&$\!\!\!\!\!\!\scriptstyle{0}\!\!\!\!\!\!\!\!\!\!\!$
&&$\!\!\!\!\!\!\scriptstyle{2(4n-1)}\!\!\!\!\!\!\!\!\!\!\!$
&&$\!\!\!\!\!\!\scriptstyle{0}\!\!\!\!\!\!\!\!\!\!\!$
&&$\!\!\!\!\!\!\!\!\!\!\!\scriptstyle{0}\!\!\!\!\!\!\!\!\!\!\!\!\!\!$
&&$ \!\!\!\!\!\!\scriptstyle{2(2n+1)}\!\!\!\!\!\!\!\!\!\!\!$
&&$\!\!\!\!\!\scriptstyle{-2(2n-1)}\!\!\!\!\!\!\!$& \cr
\omit&height4pt&\omit&&\omit&&\omit&&\omit&&\omit&&\omit&&\omit&\cr
\tablerule
\omit&height4pt&\omit&&\omit&&\omit&&\omit&&\omit&&\omit&&\omit&\cr
&&$\!\!\!\!\!\scriptstyle{(\rm 2\!-\!3)_{\rm b}}\!\!\!\!\!\!\!\!$
&&$\!\!\!\!\!\!\scriptstyle{-(2n+1)}\!\!\!\!\!\!\!\!\!\!\!$
&&$\!\!\!\!\!\!\scriptstyle{(2n+1)}\!\!\!\!\!\!\!\!\!\!\!$
&&$\!\!\!\!\!\!\scriptstyle{-(2n+1)}\!\!\!\!\!\!\!\!\!\!\!$
&&$\!\!\!\!\!\!\!\!\!\!\!\scriptstyle{0}\!\!\!\!\!\!\!\!\!\!\!\!\!\!$
&&$ \!\!\!\!\!\!\scriptstyle{(2n+1)}\!\!\!\!\!\!\!\!\!\!\!$
&&$\!\!\!\!\!\scriptstyle{0}\!\!\!\!\!\!\!$& \cr
\omit&height4pt&\omit&&\omit&&\omit&&\omit&&\omit&&\omit&&\omit&\cr
\tablerule
\omit&height4pt&\omit&&\omit&&\omit&&\omit&&\omit&&\omit&&\omit&\cr
&&$\!\!\!\!\!\scriptstyle{(\rm 2\!-\!3)_{\rm c}}\!\!\!\!\!\!\!\!$
&&$\!\!\!\!\!\!\scriptstyle{-2n}\!\!\!\!\!\!\!\!\!\!\!$
&&$\!\!\!\!\!\!\scriptstyle{(2n-2)}\!\!\!\!\!\!\!\!\!\!\!$
&&$\!\!\!\!\!\!\scriptstyle{0}\!\!\!\!\!\!\!\!\!\!\!$
&&$\!\!\!\!\!\!\!\!\!\!\!\scriptstyle{(2n-1)}\!\!\!\!\!\!\!\!\!\!\!\!\!\!$
&&$ \!\!\!\!\!\!\scriptstyle{0}\!\!\!\!\!\!\!\!\!\!\!$
&&$\!\!\!\!\!\scriptstyle{-(2n-1)}\!\!\!\!\!\!\!$& \cr
\omit&height4pt&\omit&&\omit&&\omit&&\omit&&\omit&&\omit&&\omit&\cr
\tablerule
\omit&height4pt&\omit&&\omit&&\omit&&\omit&&\omit&&\omit&&\omit&\cr
&&$\!\!\!\!\!\scriptstyle{(\rm 3\!-\!1)}\!\!\!\!\!\!\!\!$
&&$\!\!\!\!\!\!\scriptstyle{-6(2n-2)}\!\!\!\!\!\!\!\!\!\!\!$
&&$\!\!\!\!\!\!\scriptstyle{6(2n-2)}\!\!\!\!\!\!\!\!\!\!\!$
&&$\!\!\!\!\!\!\scriptstyle{-6(2n-2)}\!\!\!\!\!\!\!\!\!\!\!$
&&$\!\!\!\!\!\!\!\!\!\!\!\scriptstyle{0}\!\!\!\!\!\!\!\!\!\!\!\!\!\!$
&&$ \!\!\!\!\!\!\scriptstyle{6(2n-2)}\!\!\!\!\!\!\!\!\!\!\!$
&&$\!\!\!\!\!\scriptstyle{0}\!\!\!\!\!\!\!$& \cr
\omit&height4pt&\omit&&\omit&&\omit&&\omit&&\omit&&\omit&&\omit&\cr
\tablerule
\omit&height4pt&\omit&&\omit&&\omit&&\omit&&\omit&&\omit&&\omit&\cr
&&$\!\!\!\!\!\scriptstyle{(\rm 3\!-\!2)_{\rm a}}\!\!\!\!\!\!\!\!$
&&$\!\!\!\!\!\!\scriptstyle{(2n-1)}\!\!\!\!\!\!\!\!\!\!\!$
&&$\!\!\!\!\!\!\scriptstyle{-(2n+1)}\!\!\!\!\!\!\!\!\!\!\!$
&&$\!\!\!\!\!\!\scriptstyle{0}\!\!\!\!\!\!\!\!\!\!\!$
&&$\!\!\!\!\!\!\!\!\!\!\!\scriptstyle{-2n}\!\!\!\!\!\!\!\!\!\!\!\!\!\!$
&&$ \!\!\!\!\!\!\scriptstyle{0}\!\!\!\!\!\!\!\!\!\!\!$
&&$\!\!\!\!\!\scriptstyle{2n}\!\!\!\!\!\!\!$& \cr
\omit&height4pt&\omit&&\omit&&\omit&&\omit&&\omit&&\omit&&\omit&\cr
\tablerule
\omit&height4pt&\omit&&\omit&&\omit&&\omit&&\omit&&\omit&&\omit&\cr
&&$\!\!\!\!\!\scriptstyle{(\rm 3\!-\!2)_{\rm b}}\!\!\!\!\!\!\!\!$
&&$\!\!\!\!\!\!\scriptstyle{(2n-2)}\!\!\!\!\!\!\!\!\!\!\!$
&&$\!\!\!\!\!\!\scriptstyle{-(2n-2)}\!\!\!\!\!\!\!\!\!\!\!$
&&$\!\!\!\!\!\!\scriptstyle{(2n-2)}\!\!\!\!\!\!\!\!\!\!\!$
&&$\!\!\!\!\!\!\!\!\!\!\!\scriptstyle{0}\!\!\!\!\!\!\!\!\!\!\!\!\!\!$
&&$ \!\!\!\!\!\!\scriptstyle{-(2n-2)}\!\!\!\!\!\!\!\!\!\!\!$
&&$\!\!\!\!\!\scriptstyle{0}\!\!\!\!\!\!\!$& \cr
\omit&height4pt&\omit&&\omit&&\omit&&\omit&&\omit&&\omit&&\omit&\cr
\tablerule
\omit&height4pt&\omit&&\omit&&\omit&&\omit&&\omit&&\omit&&\omit&\cr
&&$\!\!\!\!\!\scriptstyle{(\rm 3\!-\!2)_{\rm c}}\!\!\!\!\!\!\!\!$
&&$\!\!\!\!\!\!\scriptstyle{0}\!\!\!\!\!\!\!\!\!\!\!$
&&$\!\!\!\!\!\!\scriptstyle{-2(4n-1)}\!\!\!\!\!\!\!\!\!\!\!$
&&$\!\!\!\!\!\!\scriptstyle{0}\!\!\!\!\!\!\!\!\!\!\!$
&&$\!\!\!\!\!\!\!\!\!\!\!\scriptstyle{0}\!\!\!\!\!\!\!\!\!\!\!\!\!\!$
&&$ \!\!\!\!\!\!\scriptstyle{-2(2n-2)}\!\!\!\!\!\!\!\!\!\!\!$
&&$\!\!\!\!\!\scriptstyle{4n}\!\!\!\!\!\!\!$& \cr
\omit&height4pt&\omit&&\omit&&\omit&&\omit&&\omit&&\omit&&\omit&\cr
\tablerule
\omit&height4pt&\omit&&\omit&&\omit&&\omit&&\omit&&\omit&&\omit&\cr
&&$\!\!\!\!\!\scriptstyle{(\rm 3\!-\!3)_{\rm a}}\!\!\!\!\!\!\!\!$
&&$\!\!\!\!\!\!\scriptstyle{0}\!\!\!\!\!\!\!\!\!\!\!$
&&$\!\!\!\!\!\!\scriptstyle{0}\!\!\!\!\!\!\!\!\!\!\!$
&&$\!\!\!\!\!\!\scriptstyle{0}\!\!\!\!\!\!\!\!\!\!\!$
&&$\!\!\!\!\!\!\!\!\!\!\!\scriptstyle{0}\!\!\!\!\!\!\!\!\!\!\!\!\!\!$
&&$ \!\!\!\!\!\!\scriptstyle{-1}\!\!\!\!\!\!\!\!\!\!\!$
&&$\!\!\!\!\!\scriptstyle{+1}\!\!\!\!\!\!\!$& \cr
\omit&height4pt&\omit&&\omit&&\omit&&\omit&&\omit&&\omit&&\omit&\cr
\tablerule
\omit&height4pt&\omit&&\omit&&\omit&&\omit&&\omit&&\omit&&\omit&\cr
&&$\!\!\!\!\!\scriptstyle{(\rm 3\!-\!3)_{\rm b}}\!\!\!\!\!\!\!\!$
&&$\!\!\!\!\!\!\scriptstyle{1}\!\!\!\!\!\!\!\!\!\!\!$
&&$\!\!\!\!\!\!\scriptstyle{0}\!\!\!\!\!\!\!\!\!\!\!$
&&$\!\!\!\!\!\!\scriptstyle{0}\!\!\!\!\!\!\!\!\!\!\!$
&&$\!\!\!\!\!\!\!\!\!\!\!\scriptstyle{-\frac12}\!\!\!\!\!\!\!\!\!\!\!\!\!\!$
&&$ \!\!\!\!\!\!\scriptstyle{0}\!\!\!\!\!\!\!\!\!\!\!$
&&$\!\!\!\!\!\scriptstyle{\frac12}\!\!\!\!\!\!\!$& \cr
\omit&height4pt&\omit&&\omit&&\omit&&\omit&&\omit&&\omit&&\omit&\cr
\tablerule
\omit&height4pt&\omit&&\omit&&\omit&&\omit&&\omit&&\omit&&\omit&\cr
&&$\!\!\!\!\!\scriptstyle{(\rm 3\!-\!3)_{\rm c}}\!\!\!\!\!\!\!\!$
&&$\!\!\!\!\!\!\scriptstyle{-5}\!\!\!\!\!\!\!\!\!\!\!$
&&$\!\!\!\!\!\!\scriptstyle{0}\!\!\!\!\!\!\!\!\!\!\!$
&&$\!\!\!\!\!\!\scriptstyle{-\frac52}\!\!\!\!\!\!\!\!\!\!\!$
&&$\!\!\!\!\!\!\!\!\!\!\!\scriptstyle{0}\!\!\!\!\!\!\!\!\!\!\!\!\!\!$
&&$ \!\!\!\!\!\!\scriptstyle{\frac52}\!\!\!\!\!\!\!\!\!\!\!$
&&$\!\!\!\!\!\scriptstyle{0}\!\!\!\!\!\!\!$& \cr
\omit&height4pt&\omit&&\omit&&\omit&&\omit&&\omit&&\omit&&\omit&\cr
\tablerule
\omit&height4pt&\omit&&\omit&&\omit&&\omit&&\omit&&\omit&&\omit&\cr
&&$\!\!\!\!\!\scriptscriptstyle{\rm Total_{\rm sub}}\!\!\!\!\!\!\!\!\!\!\!\!$
&&$\!\!\!\!\!\!\scriptstyle{8(3n^2-2n-1)}\!\!\!\!\!\!\!\!\!\!\!$
&&$\!\!\!\!\!\!\scriptstyle{0}\!\!\!\!\!\!\!\!\!\!\!$
&&$\!\!\!\!\!\!\scriptstyle{5f'(n)}\!\!\!\!\!\!\!\!\!\!\!$
&&$\!\!\!\!\!\!\!\!\!\!\!\scriptstyle{f'(n)}\!\!\!\!\!\!\!$
&&$ \!\!\!\!\!\!\scriptstyle{-3f'(n)}\!\!\!\!\!\!\!\!\!\!\!$
&&$\!\!\!\!\!\scriptstyle{-3f'(n)}\!\!\!\!\!\!\!$& \cr
\omit&height4pt&\omit&&\omit&&\omit&&\omit&&\omit&&\omit&&\omit&\cr
\tablerule}}

\caption{$i\Delta_{\rm cf}\times i[S]_{n\geq 1}-{\rm I}$. All contributions
are multiplied by $\frac{\kappa^2 mH}{64\pi^4}\sum_{n\geq 1}^{\infty}
\Bigl(\frac{y}{4}\Bigr)^n$. Here $\scriptstyle{f'(n)=(6n^2-4n-\frac32)}$.}

\label{cfn11}

\end{table}
\begin{table}

\vbox{\tabskip=0pt \offinterlineskip
\def\tablerule{\noalign{\hrule}}
\halign to390pt {\strut#& \vrule#\tabskip=1em plus2em& \hfil#\hfil&
\vrule#& \hfil#\hfil& \vrule#& \hfil#\hfil& \vrule#& \hfil#\hfil&
\vrule#& \hfil#\hfil& \vrule#& \hfil#\hfil& \vrule#&
\hfil#\hfil& \vrule#\tabskip=0pt\cr
\tablerule
\omit&height4pt&\omit&&\omit&&\omit&&\omit&&\omit&&\omit&&\omit&\cr
&&$\!\!\!\!\!\scriptstyle{(\rm I\!-\!J)_{\rm sub}}\!\!\!\!\!\!\!\!$
&&$\!\!\!\!\!\!\frac{\gamma^0\gamma^k\Delta x_k}{\Delta x^6}\!\!\!\!\!\!\!$
&&$\!\!\!\!\!\!\frac{\Delta\eta}{\Delta x^6}\!\!\!\!\!\!\!$
&&$\!\!\!\!\!\!\scriptstyle{\frac{Ha\gamma^0\!\Delta\eta\gamma^k\!\Delta x_k}
{\Delta x^6}}\!\!\!\!\!\!\!\!\!\!\!$
&&$\!\!\!\!\!\!\scriptstyle{\frac{Ha'\gamma^0\!\Delta\eta\gamma^k\!\Delta x_k}
{\Delta x^6}}\!\!\!\!\!\!\!\!\!\!\!$
&&$ \!\!\!\!\!\!\frac{Ha\Delta\eta^2}{\Delta x^6}\!\!\!\!\!\!\!\!$
&&$\!\!\!\!\!\frac{Ha'\Delta\eta^2}{\Delta x^6}\!\!\!\!\!$& \cr
\omit&height4pt&\omit&&\omit&&\omit&&\omit&&\omit&&\omit&&\omit&\cr
\tablerule
\omit&height4pt&\omit&&\omit&&\omit&&\omit&&\omit&&\omit&&\omit&\cr
&&$\!\!\!\!\!\scriptstyle{(\rm 1\!-\!1)}\!\!\!\!\!\!\!\!$
&&$\!\!\!\!\!\!\scriptstyle{16(n^3-n^2-2n)}\!\!\!\!\!\!\!\!\!\!\!$
&&$\!\!\!\!\!\!\scriptstyle{-16(n^3-n^2-2n)}\!\!\!\!\!\!\!\!\!\!\!$
&&$\!\!\!\!\!\!\scriptstyle{16(n^3-n^2-2n)}\!\!\!\!\!\!\!\!\!\!\!$
&&$\!\!\!\!\!\!\scriptstyle{0}\!\!\!\!\!\!\!$
&&$ \!\!\!\!\!\!\scriptstyle{-16(n^3-n^2-2n)}\!\!\!\!\!\!\!\!\!\!$
&&$\!\!\!\!\!\scriptstyle{0}\!\!\!\!\!$& \cr
\omit&height4pt&\omit&&\omit&&\omit&&\omit&&\omit&&\omit&&\omit&\cr
\tablerule
\omit&height4pt&\omit&&\omit&&\omit&&\omit&&\omit&&\omit&&\omit&\cr
&&$\!\!\!\!\scriptstyle{(\rm 1\!-\!2)}\!\!\!\!\!\!\!\!$
&&$\!\!\!\!\!\!\scriptstyle{-4(n^3-n^2-2n)}\!\!\!\!\!\!\!\!\!\!\!$
&&$\!\!\!\!\!\!\scriptstyle{4(n^3-n^2-2n)}\!\!\!\!\!\!\!\!\!\!\!$
&&$\!\!\!\!\!\!\scriptstyle{-4(n^3-n^2-2n)}\!\!\!\!\!\!\!\!\!\!\!$
&&$\!\!\!\!\!\!\scriptstyle{0}\!\!\!\!\!\!\!$
&&$ \!\!\!\!\!\!\scriptstyle{4(n^3-n^2-2n)}\!\!\!\!\!\!\!\!\!\!$
&&$\!\!\!\!\!\scriptstyle{0}\!\!\!\!\!$& \cr
\omit&height4pt&\omit&&\omit&&\omit&&\omit&&\omit&&\omit&&\omit&\cr
\tablerule
\omit&height4pt&\omit&&\omit&&\omit&&\omit&&\omit&&\omit&&\omit&\cr
&&$\!\!\!\!\!\scriptstyle{(\rm 1\!-\!3)}\!\!\!\!\!\!\!\!$
&&$\!\!\!\!\!\!\scriptstyle{6(n^2+n)}\!\!\!\!\!\!\!\!\!\!\!$
&&$\!\!\!\!\!\!\scriptstyle{-6(n^2+n)}\!\!\!\!\!\!\!\!\!\!\!$
&&$\!\!\!\!\!\!\scriptstyle{6(n^2+n)}\!\!\!\!\!\!\!\!\!\!\!$
&&$\!\!\!\!\!\!\scriptstyle{0}\!\!\!\!\!\!\!$
&&$ \!\!\!\!\!\!\scriptstyle{-6(n^2+n)}\!\!\!\!\!\!\!\!\!\!$
&&$\!\!\!\!\!\scriptstyle{0}\!\!\!\!\!$& \cr
\omit&height4pt&\omit&&\omit&&\omit&&\omit&&\omit&&\omit&&\omit&\cr
\tablerule
\omit&height4pt&\omit&&\omit&&\omit&&\omit&&\omit&&\omit&&\omit&\cr
&&$\!\!\!\!\scriptstyle{(\rm 2\!-\!1)}\!\!\!\!\!\!\!\!$
&&$\!\!\!\!\!\!\scriptstyle{-4(n^3-n^2-2n)}\!\!\!\!\!\!\!\!\!\!\!$
&&$\!\!\!\!\!\!\scriptstyle{4(n^3-n^2-2n)}\!\!\!\!\!\!\!\!\!\!\!$
&&$\!\!\!\!\!\!\scriptstyle{-4(n^3-n^2-2n)}\!\!\!\!\!\!\!\!\!\!\!$
&&$\!\!\!\!\!\!\scriptstyle{0}\!\!\!\!\!\!\!$
&&$ \!\!\!\!\!\!\scriptstyle{4(n^3-n^2-2n)}\!\!\!\!\!\!\!\!\!\!$
&&$\!\!\!\!\!\scriptstyle{0}\!\!\!\!\!$& \cr
\omit&height4pt&\omit&&\omit&&\omit&&\omit&&\omit&&\omit&&\omit&\cr
\tablerule
\omit&height4pt&\omit&&\omit&&\omit&&\omit&&\omit&&\omit&&\omit&\cr
&&$\!\!\!\!\!\scriptstyle{(\rm 2\!-\!2)_{\rm a}}\!\!\!\!\!\!\!\!$
&&$\!\!\!\!\!\!\scriptstyle{-2(n^3-n^2-n)}\!\!\!\!\!\!\!\!\!\!\!$
&&$\!\!\!\!\!\!\scriptstyle{2(n^3-n^2+n)}\!\!\!\!\!\!\!\!\!\!\!$
&&$\!\!\!\!\!\!\scriptstyle{0}\!\!\!\!\!\!\!\!\!\!\!$
&&$\!\!\!\!\!\!\!\!\!\!\!\scriptstyle{2(n^3-n^2)}\!\!\!\!\!\!\!\!\!\!\!\!\!\!$
&&$ \!\!\!\!\!\!\scriptstyle{0}\!\!\!\!\!\!\!\!\!\!\!$
&&$\!\!\!\!\!\scriptstyle{-2(n^3-n^2)}\!\!\!\!\!\!\!$& \cr
\omit&height4pt&\omit&&\omit&&\omit&&\omit&&\omit&&\omit&&\omit&\cr
\tablerule
\omit&height4pt&\omit&&\omit&&\omit&&\omit&&\omit&&\omit&&\omit&\cr
&&$\!\!\!\!\!\scriptstyle{(\rm 2\!-\!2)_{\rm b}}\!\!\!\!\!\!\!\!$
&&$\!\!\!\!\!\!\scriptstyle{0}\!\!\!\!\!\!\!\!\!\!\!$
&&$\!\!\!\!\!\!\scriptstyle{4(2n^3-2n^2-n)}\!\!\!\!\!\!\!\!\!\!\!$
&&$\!\!\!\!\!\!\scriptstyle{0}\!\!\!\!\!\!\!\!\!\!\!$
&&$\!\!\!\!\!\!\!\!\!\!\!\scriptstyle{0}\!\!\!\!\!\!\!\!\!\!\!\!\!\!$
&&$ \!\!\!\!\!\!\scriptstyle{4(n^3-n^2-2n)}\!\!\!\!\!\!\!\!\!\!\!$
&&$\!\!\!\!\!\scriptstyle{-4(n^3-n^2)}\!\!\!\!\!\!\!$& \cr
\omit&height4pt&\omit&&\omit&&\omit&&\omit&&\omit&&\omit&&\omit&\cr
\tablerule
\omit&height4pt&\omit&&\omit&&\omit&&\omit&&\omit&&\omit&&\omit&\cr
&&$\!\!\!\!\!\scriptstyle{(\rm 2\!-\!2)_{\rm c}}\!\!\!\!\!\!\!\!$
&&$\!\!\!\!\!\!\scriptstyle{2(n^3-n^2-2n)}\!\!\!\!\!\!\!\!\!\!\!$
&&$\!\!\!\!\!\!\scriptstyle{-2(n^3-n^2-2n)}\!\!\!\!\!\!\!\!\!\!\!$
&&$\!\!\!\!\!\!\scriptstyle{2(n^3-n^2-2n)}\!\!\!\!\!\!\!\!\!\!\!$
&&$\!\!\!\!\!\!\!\!\!\!\!\scriptstyle{0}\!\!\!\!\!\!\!\!\!\!\!\!\!\!$
&&$ \!\!\!\!\!\!\scriptstyle{-2(n^3-n^2-2n)}\!\!\!\!\!\!\!\!\!\!\!$
&&$\!\!\!\!\!\scriptstyle{0}\!\!\!\!\!\!\!$& \cr
\omit&height4pt&\omit&&\omit&&\omit&&\omit&&\omit&&\omit&&\omit&\cr
\tablerule
\omit&height4pt&\omit&&\omit&&\omit&&\omit&&\omit&&\omit&&\omit&\cr
&&$\!\!\!\!\!\scriptstyle{(\rm 2\!-\!3)_{\rm a}}\!\!\!\!\!\!\!\!$
&&$\!\!\!\!\!\!\scriptstyle{0}\!\!\!\!\!\!\!\!\!\!\!$
&&$\!\!\!\!\!\!\scriptstyle{2(2n^2-n)}\!\!\!\!\!\!\!\!\!\!\!$
&&$\!\!\!\!\!\!\scriptstyle{0}\!\!\!\!\!\!\!\!\!\!\!$
&&$\!\!\!\!\!\!\!\!\!\!\!\scriptstyle{0}\!\!\!\!\!\!\!\!\!\!\!\!\!\!$
&&$ \!\!\!\!\!\!\scriptstyle{2(n^2+n)}\!\!\!\!\!\!\!\!\!\!\!$
&&$\!\!\!\!\!\scriptstyle{-2(n^2-n)}\!\!\!\!\!\!\!$& \cr
\omit&height4pt&\omit&&\omit&&\omit&&\omit&&\omit&&\omit&&\omit&\cr
\tablerule
\omit&height4pt&\omit&&\omit&&\omit&&\omit&&\omit&&\omit&&\omit&\cr
&&$\!\!\!\!\!\scriptstyle{(\rm 2\!-\!3)_{\rm b}}\!\!\!\!\!\!\!\!$
&&$\!\!\!\!\!\!\scriptstyle{-(n^2+n)}\!\!\!\!\!\!\!\!\!\!\!$
&&$\!\!\!\!\!\!\scriptstyle{(n^2+n)}\!\!\!\!\!\!\!\!\!\!\!$
&&$\!\!\!\!\!\!\scriptstyle{-(n^2+n)}\!\!\!\!\!\!\!\!\!\!\!$
&&$\!\!\!\!\!\!\!\!\!\!\!\scriptstyle{0}\!\!\!\!\!\!\!\!\!\!\!\!\!\!$
&&$ \!\!\!\!\!\!\scriptstyle{(n^2+n)}\!\!\!\!\!\!\!\!\!\!\!$
&&$\!\!\!\!\!\scriptstyle{0}\!\!\!\!\!\!\!$& \cr
\omit&height4pt&\omit&&\omit&&\omit&&\omit&&\omit&&\omit&&\omit&\cr
\tablerule
\omit&height4pt&\omit&&\omit&&\omit&&\omit&&\omit&&\omit&&\omit&\cr
&&$\!\!\!\!\!\scriptstyle{(\rm 2\!-\!3)_{\rm c}}\!\!\!\!\!\!\!\!$
&&$\!\!\!\!\!\!\scriptstyle{-n^2}\!\!\!\!\!\!\!\!\!\!\!$
&&$\!\!\!\!\!\!\scriptstyle{(n^2-2n)}\!\!\!\!\!\!\!\!\!\!\!$
&&$\!\!\!\!\!\!\scriptstyle{0}\!\!\!\!\!\!\!\!\!\!\!$
&&$\!\!\!\!\!\!\!\!\!\!\!\scriptstyle{(n^2-n)}\!\!\!\!\!\!\!\!\!\!\!\!\!\!$
&&$ \!\!\!\!\!\!\scriptstyle{0}\!\!\!\!\!\!\!\!\!\!\!$
&&$\!\!\!\!\!\scriptstyle{-(n^2-n)}\!\!\!\!\!\!\!$& \cr
\omit&height4pt&\omit&&\omit&&\omit&&\omit&&\omit&&\omit&&\omit&\cr
\tablerule
\omit&height4pt&\omit&&\omit&&\omit&&\omit&&\omit&&\omit&&\omit&\cr
&&$\!\!\!\!\!\scriptstyle{(\rm 3\!-\!1)}\!\!\!\!\!\!\!\!$
&&$\!\!\!\!\!\!\scriptstyle{-6(n^2-2n)}\!\!\!\!\!\!\!\!\!\!\!$
&&$\!\!\!\!\!\!\scriptstyle{6(n^2-2n)}\!\!\!\!\!\!\!\!\!\!\!$
&&$\!\!\!\!\!\!\scriptstyle{-6(n^2-2n)}\!\!\!\!\!\!\!\!\!\!\!$
&&$\!\!\!\!\!\!\!\!\!\!\!\scriptstyle{0}\!\!\!\!\!\!\!\!\!\!\!\!\!\!$
&&$ \!\!\!\!\!\!\scriptstyle{6(n^2-2n)}\!\!\!\!\!\!\!\!\!\!\!$
&&$\!\!\!\!\!\scriptstyle{0}\!\!\!\!\!\!\!$& \cr
\omit&height4pt&\omit&&\omit&&\omit&&\omit&&\omit&&\omit&&\omit&\cr
\tablerule
\omit&height4pt&\omit&&\omit&&\omit&&\omit&&\omit&&\omit&&\omit&\cr
&&$\!\!\!\!\!\scriptstyle{(\rm 3\!-\!2)_{\rm a}}\!\!\!\!\!\!\!\!$
&&$\!\!\!\!\!\!\scriptstyle{(n^2-n)}\!\!\!\!\!\!\!\!\!\!\!$
&&$\!\!\!\!\!\!\scriptstyle{-(n^2+n)}\!\!\!\!\!\!\!\!\!\!\!$
&&$\!\!\!\!\!\!\scriptstyle{0}\!\!\!\!\!\!\!\!\!\!\!$
&&$\!\!\!\!\!\!\!\!\!\!\!\scriptstyle{-n^2}\!\!\!\!\!\!\!\!\!\!\!\!\!\!$
&&$ \!\!\!\!\!\!\scriptstyle{0}\!\!\!\!\!\!\!\!\!\!\!$
&&$\!\!\!\!\!\scriptstyle{n^2}\!\!\!\!\!\!\!$& \cr
\omit&height4pt&\omit&&\omit&&\omit&&\omit&&\omit&&\omit&&\omit&\cr
\tablerule
\omit&height4pt&\omit&&\omit&&\omit&&\omit&&\omit&&\omit&&\omit&\cr
&&$\!\!\!\!\!\scriptstyle{(\rm 3\!-\!2)_{\rm b}}\!\!\!\!\!\!\!\!$
&&$\!\!\!\!\!\!\scriptstyle{(n^2-2n)}\!\!\!\!\!\!\!\!\!\!\!$
&&$\!\!\!\!\!\!\scriptstyle{-(n^2-2n)}\!\!\!\!\!\!\!\!\!\!\!$
&&$\!\!\!\!\!\!\scriptstyle{(n^2-2n)}\!\!\!\!\!\!\!\!\!\!\!$
&&$\!\!\!\!\!\!\!\!\!\!\!\scriptstyle{0}\!\!\!\!\!\!\!\!\!\!\!\!\!\!$
&&$ \!\!\!\!\!\!\scriptstyle{-(n^2-2n)}\!\!\!\!\!\!\!\!\!\!\!$
&&$\!\!\!\!\!\scriptstyle{0}\!\!\!\!\!\!\!$& \cr
\omit&height4pt&\omit&&\omit&&\omit&&\omit&&\omit&&\omit&&\omit&\cr
\tablerule
\omit&height4pt&\omit&&\omit&&\omit&&\omit&&\omit&&\omit&&\omit&\cr
&&$\!\!\!\!\!\scriptstyle{(\rm 3\!-\!2)_{\rm c}}\!\!\!\!\!\!\!\!$
&&$\!\!\!\!\!\!\scriptstyle{0}\!\!\!\!\!\!\!\!\!\!\!$
&&$\!\!\!\!\!\!\scriptstyle{-2(2n^2-n)}\!\!\!\!\!\!\!\!\!\!\!$
&&$\!\!\!\!\!\!\scriptstyle{0}\!\!\!\!\!\!\!\!\!\!\!$
&&$\!\!\!\!\!\!\!\!\!\!\!\scriptstyle{0}\!\!\!\!\!\!\!\!\!\!\!\!\!\!$
&&$ \!\!\!\!\!\!\scriptstyle{-2(n^2-2n)}\!\!\!\!\!\!\!\!\!\!\!$
&&$\!\!\!\!\!\scriptstyle{2n^2}\!\!\!\!\!\!\!$& \cr
\omit&height4pt&\omit&&\omit&&\omit&&\omit&&\omit&&\omit&&\omit&\cr
\tablerule
\omit&height4pt&\omit&&\omit&&\omit&&\omit&&\omit&&\omit&&\omit&\cr
&&$\!\!\!\!\!\scriptstyle{(\rm 3\!-\!3)_{\rm a}}\!\!\!\!\!\!\!\!$
&&$\!\!\!\!\!\!\scriptstyle{0}\!\!\!\!\!\!\!\!\!\!\!$
&&$\!\!\!\!\!\!\scriptstyle{0}\!\!\!\!\!\!\!\!\!\!\!$
&&$\!\!\!\!\!\!\scriptstyle{0}\!\!\!\!\!\!\!\!\!\!\!$
&&$\!\!\!\!\!\!\!\!\!\!\!\scriptstyle{0}\!\!\!\!\!\!\!\!\!\!\!\!\!\!$
&&$ \!\!\!\!\!\!\scriptstyle{-n}\!\!\!\!\!\!\!\!\!\!\!$
&&$\!\!\!\!\!\scriptstyle{+n}\!\!\!\!\!\!\!$& \cr
\omit&height4pt&\omit&&\omit&&\omit&&\omit&&\omit&&\omit&&\omit&\cr
\tablerule
\omit&height4pt&\omit&&\omit&&\omit&&\omit&&\omit&&\omit&&\omit&\cr
&&$\!\!\!\!\!\scriptstyle{(\rm 3\!-\!3)_{\rm b}}\!\!\!\!\!\!\!\!$
&&$\!\!\!\!\!\!\scriptstyle{n}\!\!\!\!\!\!\!\!\!\!\!$
&&$\!\!\!\!\!\!\scriptstyle{0}\!\!\!\!\!\!\!\!\!\!\!$
&&$\!\!\!\!\!\!\scriptstyle{0}\!\!\!\!\!\!\!\!\!\!\!$
&&$\!\!\!\!\!\!\!\!\!\!\!\scriptstyle{-\frac{1}{2}n}\!\!\!\!\!\!\!\!\!\!\!\!\!\!$
&&$ \!\!\!\!\!\!\scriptstyle{0}\!\!\!\!\!\!\!\!\!\!\!$
&&$\!\!\!\!\!\scriptstyle{\frac{1}{2}n}\!\!\!\!\!\!\!$& \cr
\omit&height4pt&\omit&&\omit&&\omit&&\omit&&\omit&&\omit&&\omit&\cr
\tablerule
\omit&height4pt&\omit&&\omit&&\omit&&\omit&&\omit&&\omit&&\omit&\cr
&&$\!\!\!\!\!\scriptstyle{(\rm 3\!-\!3)_{\rm c}}\!\!\!\!\!\!\!\!$
&&$\!\!\!\!\!\!\scriptstyle{-5n}\!\!\!\!\!\!\!\!\!\!\!$
&&$\!\!\!\!\!\!\scriptstyle{0}\!\!\!\!\!\!\!\!\!\!\!$
&&$\!\!\!\!\!\!\scriptstyle{-\frac{5}{2}n}\!\!\!\!\!\!\!\!\!\!\!$
&&$\!\!\!\!\!\!\!\!\!\!\!\scriptstyle{0}\!\!\!\!\!\!\!\!\!\!\!\!\!\!$
&&$ \!\!\!\!\!\!\scriptstyle{\frac{5}{2}n}\!\!\!\!\!\!\!\!\!\!\!$
&&$\!\!\!\!\!\scriptstyle{0}\!\!\!\!\!\!\!$& \cr
\omit&height4pt&\omit&&\omit&&\omit&&\omit&&\omit&&\omit&&\omit&\cr
\tablerule
\omit&height4pt&\omit&&\omit&&\omit&&\omit&&\omit&&\omit&&\omit&\cr
&&$\!\!\!\!\!\scriptscriptstyle{\rm Total_{\rm sub}}\!\!\!\!\!\!\!\!\!\!\!\!$
&&$\!\!\!\!\!\!\scriptstyle{8(n^3-n^2-n)}\!\!\!\!\!\!\!\!\!\!\!$
&&$\!\!\!\!\!\!\scriptstyle{0}\!\!\!\!\!\!\!\!\!\!\!$
&&$\!\!\!\!\!\!\scriptstyle{5f(n)}\!\!\!\!\!\!\!\!\!\!\!$
&&$\!\!\!\!\!\!\!\!\!\!\!\scriptstyle{f(n)}\!\!\!\!\!\!\!$
&&$ \!\!\!\!\!\!\scriptstyle{-3f(n)}\!\!\!\!\!\!\!\!\!\!\!$
&&$\!\!\!\!\!\scriptstyle{-3f(n)}\!\!\!\!\!\!\!$& \cr
\omit&height4pt&\omit&&\omit&&\omit&&\omit&&\omit&&\omit&&\omit&\cr
\tablerule}}

\caption{$i\Delta_{\rm cf}\times i[S]_{n\geq 1}-{\rm I'}$. All contributions
are multiplied by $\frac{\kappa^2 mH}{64\pi^4}\sum_{n\geq 1}^{\infty}
\Bigl(\frac{y}{4}\Bigr)^n\ln(\frac{y}{4})$.
Here $\scriptstyle{f(n)=2n^3-2n^2-\frac{3}{2}n}$.}

\label{cfn12}

\end{table}
One interesting pattern is that taking the derivative
of the coefficient of the logarithm term with respect to n
gives the coefficient for the term without logarithms.
Before we table the result from each length expression,
we present the result from the contraction (1-1),
\begin{eqnarray}
&&2\kappa^2\partial_{\mu}^{\prime}\{\not{\hspace{-.1cm}\partial}
i[S](x;x')\gamma^{\mu}i\Delta_{\rm cf}(x;x')\}=
\frac{\kappa^2 mH}{32\pi^4}\Biggl\{
8\Bigl[\frac{\scriptstyle{\gamma^0\gamma^{\mu}\Delta x_{\mu}}}
{\Delta x^6}\!+\!
\frac{\scriptstyle{Ha\gamma^0\gamma^{\mu}\Delta\eta\Delta x_{\mu}}}
{\Delta x^6}\Bigr]\nonumber\\
&&\times\Bigl[(3n^2\!\!-\!2n\!-\!2)\!+\!(n^3\!\!-\!n^2\!\!-\!2n)
\ln(\frac{y}{4})\Bigr]
-\frac{4Ha}{\Delta x^4}
\Bigl[(3n^2\!\!-\!1)\!+\!(n^3\!\!-\!n)
\ln(\frac{y}{4}) \Bigr]\nonumber\\
&&\hspace{2.5cm}
-\frac{2H^2a^2\gamma^0\gamma^{\mu}\Delta x_{\mu}}{\Delta x^4}
\Bigl[(3n^2\!\!+\!4n\!-\!1)\!+\!(n^3\!\!+\!2n^2\!\!-\!n\!-\!2)
\ln(\frac{y}{4})\Bigr]\nonumber\\
&&\hspace{2cm}+\frac{H^2 a^2 Ha'}{\Delta x^4}
\Bigl[(3n^2\!\!+\!6n\!+\!2)\!+\!(n^3\!\!+\!n^2\!\!+\!2n)
\ln(\frac{y}{4})\Bigr]\Biggr\}(\frac{y}{4})^n \,.
\end{eqnarray}
The same pattern here happened again! We should keep it in mind that
this pattern might maintain after summing up each individual contribution.
We also separate temporal terms with spatial ones for our conventional choice
of counternterms and summarized the results in Tables~\ref{cfn11},
\ref{cfn12}, \ref{cfn13} and \ref{cfn14} before many summations are performed.
Table~\ref{cfn12} (Table~\ref{cfn14}) is the partner of Table~\ref{cfn11}
(Table~\ref{cfn13}) with the extra logarithm, $\ln\frac{y}{4}$. From the coefficient
of each individual term at the bottom of each table one might already notice
that the pattern we mentioned above still exists. The benefit for postponing
the infinite summation for each individual contraction not only
because it is a less complicated procedure but also because the pattern
serves us one consistent check whether or not we have done the computation
correctly for such a long and complicated calculation.
\begin{table}

\vbox{\tabskip=0pt \offinterlineskip
\def\tablerule{\noalign{\hrule}}
\halign to390pt {\strut#& \vrule#\tabskip=1em plus2em& \hfil#\hfil&
\vrule#& \hfil#\hfil& \vrule#& \hfil#\hfil& \vrule#& \hfil#\hfil&
\vrule#& \hfil#\hfil& \vrule#& \hfil#\hfil& \vrule#&
\hfil#\hfil& \vrule#\tabskip=0pt\cr
\tablerule
\omit&height4pt&\omit&&\omit&&\omit&&\omit&&\omit&&\omit&&\omit&\cr
&&$\!\!\!\!\!\scriptstyle{(\rm I\!-\!J)_{\rm sub}}\!\!\!\!\!\!\!\!$
&&$\!\!\!\!\!\!\frac{Ha}{\Delta x^4}\!\!\!\!\!\!\!$
&&$\!\!\!\!\!\!\frac{Ha'}{\Delta x^4}\!\!\!\!\!\!\!$
&&$\!\!\!\!\!\!\scriptstyle{\frac{H^2a^2\gamma^0\!\gamma^k\!\Delta x_k}
{\Delta x^4}}\!\!\!\!\!\!\!\!\!\!\!$
&&$\!\!\!\!\!\!\scriptstyle{\frac{H^{2}aa'\gamma^0\!\gamma^k\!\Delta x_k}
{\Delta x^4}}\!\!\!\!\!\!\!\!\!\!\!$
&&$ \!\!\!\!\!\!\frac{H^2a^2\Delta\eta}{\Delta x^4}\!\!\!\!\!\!\!\!$
&&$\!\!\!\!\!\frac{H^2a^2Ha'}{\Delta x^2}\!\!\!\!\!$& \cr
\omit&height4pt&\omit&&\omit&&\omit&&\omit&&\omit&&\omit&&\omit&\cr
\tablerule
\omit&height4pt&\omit&&\omit&&\omit&&\omit&&\omit&&\omit&&\omit&\cr
&&$\!\!\!\!\!\scriptstyle{(\rm 1\!-\!1)}\!\!\!\!\!\!\!\!$
&&$\!\!\!\!\!\!\scriptstyle{-8(3n^2-1)}\!\!\!\!\!\!\!\!\!\!\!$
&&$\!\!\!\!\!\!\scriptstyle{0}\!\!\!\!\!\!\!\!\!\!\!$
&&$\!\!\!\!\!\!\scriptstyle{-4(3n^2+4n-1)}\!\!\!\!\!\!\!\!\!\!\!$
&&$\!\!\!\!\!\!\scriptstyle{0}\!\!\!\!\!\!\!$
&&$ \!\!\!\!\!\!\scriptstyle{4(3n^2+4n-1)}\!\!\!\!\!\!\!\!\!\!$
&&$\!\!\!\!\!\scriptstyle{2(3n^2+6n+2)}\!\!\!\!\!$& \cr
\omit&height4pt&\omit&&\omit&&\omit&&\omit&&\omit&&\omit&&\omit&\cr
\tablerule
\omit&height4pt&\omit&&\omit&&\omit&&\omit&&\omit&&\omit&&\omit&\cr
&&$\!\!\!\!\scriptstyle{(\rm 1\!-\!2)}\!\!\!\!\!\!\!\!$
&&$\!\!\!\!\!\!\scriptstyle{2(3n^2-1)}\!\!\!\!\!\!\!\!\!\!\!$
&&$\!\!\!\!\!\!\scriptstyle{0}\!\!\!\!\!\!\!\!\!\!\!$
&&$\!\!\!\!\!\!\scriptstyle{(3n^2+4n-1)}\!\!\!\!\!\!\!\!\!\!\!$
&&$\!\!\!\!\!\!\scriptstyle{0}\!\!\!\!\!\!\!$
&&$ \!\!\!\!\!\!\scriptstyle{-(3n^2+4n-1)}\!\!\!\!\!\!\!\!\!\!$
&&$\!\!\!\!\!\scriptstyle{-\frac12(3n^2+6n+2)}\!\!\!\!\!$& \cr
\omit&height4pt&\omit&&\omit&&\omit&&\omit&&\omit&&\omit&&\omit&\cr
\tablerule
\omit&height4pt&\omit&&\omit&&\omit&&\omit&&\omit&&\omit&&\omit&\cr
&&$\!\!\!\!\!\scriptstyle{(\rm 1\!-\!3)}\!\!\!\!\!\!\!\!$
&&$\!\!\!\!\!\!\scriptstyle{-3(2n+1)}\!\!\!\!\!\!\!\!\!\!\!$
&&$\!\!\!\!\!\!\scriptstyle{0}\!\!\!\!\!\!\!\!\!\!\!$
&&$\!\!\!\!\!\!\scriptstyle{-\frac32(2n+3)}\!\!\!\!\!\!\!\!\!\!\!$
&&$\!\!\!\!\!\!\scriptstyle{0}\!\!\!\!\!\!\!$
&&$ \!\!\!\!\!\!\scriptstyle{\frac32(2n+3)}\!\!\!\!\!\!\!\!\!\!$
&&$\!\!\!\!\!\scriptstyle{\frac34(2n+3)}\!\!\!\!\!$& \cr
\omit&height4pt&\omit&&\omit&&\omit&&\omit&&\omit&&\omit&&\omit&\cr
\tablerule
\omit&height4pt&\omit&&\omit&&\omit&&\omit&&\omit&&\omit&&\omit&\cr
&&$\!\!\!\!\scriptstyle{(\rm 2\!-\!1)}\!\!\!\!\!\!\!\!$
&&$\!\!\!\!\!\!\scriptstyle{2(3n^2-1)}\!\!\!\!\!\!\!\!\!\!\!$
&&$\!\!\!\!\!\!\scriptstyle{0}\!\!\!\!\!\!\!\!\!\!\!$
&&$\!\!\!\!\!\!\scriptstyle{(3n^2+4n-1)}\!\!\!\!\!\!\!\!\!\!\!$
&&$\!\!\!\!\!\!\scriptstyle{0}\!\!\!\!\!\!\!$
&&$ \!\!\!\!\!\!\scriptstyle{-(3n^2+4n-1)}\!\!\!\!\!\!\!\!\!\!$
&&$\!\!\!\!\!\scriptstyle{-\frac12(3n^2+6n+2)}\!\!\!\!\!$& \cr
\omit&height4pt&\omit&&\omit&&\omit&&\omit&&\omit&&\omit&&\omit&\cr
\tablerule
\omit&height4pt&\omit&&\omit&&\omit&&\omit&&\omit&&\omit&&\omit&\cr
&&$\!\!\!\!\!\scriptstyle{(\rm 2\!-\!2)_{\rm a}}\!\!\!\!\!\!\!\!$
&&$\!\!\!\!\!\!\scriptstyle{-(3n^2+2n)}\!\!\!\!\!\!\!\!\!\!\!$
&&$\!\!\!\!\!\!\scriptstyle{-2(3n^2+n)}\!\!\!\!\!\!\!\!\!\!\!$
&&$\!\!\!\!\!\!\scriptstyle{-\frac12(3n^2+4n-1)}\!\!\!\!\!\!\!\!\!\!\!$
&&$\!\!\!\!\!\!\!\!\!\!\!\scriptstyle{-(3n^2+2n)}\!\!\!\!\!\!\!\!\!\!\!\!\!\!$
&&$ \!\!\!\!\!\!\scriptstyle{-\frac12(3n^2+4n-1)}\!\!\!\!\!\!\!\!\!\!\!$
&&$\!\!\!\!\!\scriptstyle{-\frac14(3n^2+6n+2)}\!\!\!\!\!\!\!$& \cr
\omit&height4pt&\omit&&\omit&&\omit&&\omit&&\omit&&\omit&&\omit&\cr
\tablerule
\omit&height4pt&\omit&&\omit&&\omit&&\omit&&\omit&&\omit&&\omit&\cr
&&$\!\!\!\!\!\scriptstyle{(\rm 2\!-\!2)_{\rm b}}\!\!\!\!\!\!\!\!$
&&$\!\!\!\!\!\!\scriptstyle{-2(2n+1)}\!\!\!\!\!\!\!\!\!\!\!$
&&$\!\!\!\!\!\!\scriptstyle{-2(6n^2+2n)}\!\!\!\!\!\!\!\!\!\!\!$
&&$\!\!\!\!\!\!\scriptstyle{0}\!\!\!\!\!\!\!\!\!\!\!$
&&$\!\!\!\!\!\!\!\!\!\!\!\scriptstyle{0}\!\!\!\!\!\!\!\!\!\!\!\!\!\!$
&&$ \!\!\!\!\!\!\scriptstyle{-2(3n^2+4n-1)}\!\!\!\!\!\!\!\!\!\!\!$
&&$\!\!\!\!\!\scriptstyle{-(3n^2+6n+2)}\!\!\!\!\!\!\!$& \cr
\omit&height4pt&\omit&&\omit&&\omit&&\omit&&\omit&&\omit&&\omit&\cr
\tablerule
\omit&height4pt&\omit&&\omit&&\omit&&\omit&&\omit&&\omit&&\omit&\cr
&&$\!\!\!\!\!\scriptstyle{(\rm 2\!-\!2)_{\rm c}}\!\!\!\!\!\!\!\!$
&&$\!\!\!\!\!\!\scriptstyle{-(3n^2-1)}\!\!\!\!\!\!\!\!\!\!\!$
&&$\!\!\!\!\!\!\scriptstyle{0}\!\!\!\!\!\!\!\!\!\!\!$
&&$\!\!\!\!\!\!\scriptstyle{-\frac12(3n^2+4n-1)}\!\!\!\!\!\!\!\!\!\!\!$
&&$\!\!\!\!\!\!\!\!\!\!\!\scriptstyle{0}\!\!\!\!\!\!\!\!\!\!\!\!\!\!$
&&$ \!\!\!\!\!\!\scriptstyle{\frac12(3n^2+4n-1)}\!\!\!\!\!\!\!\!\!\!\!$
&&$\!\!\!\!\!\scriptstyle{\frac14(3n^2+6n+2)}\!\!\!\!\!\!\!$& \cr
\omit&height4pt&\omit&&\omit&&\omit&&\omit&&\omit&&\omit&&\omit&\cr
\tablerule
\omit&height4pt&\omit&&\omit&&\omit&&\omit&&\omit&&\omit&&\omit&\cr
&&$\!\!\!\!\!\scriptstyle{(\rm 2\!-\!3)_{\rm a}}\!\!\!\!\!\!\!\!$
&&$\!\!\!\!\!\!\scriptstyle{0}\!\!\!\!\!\!\!\!\!\!\!$
&&$\!\!\!\!\!\!\scriptstyle{-(4n+1)}\!\!\!\!\!\!\!\!\!\!\!$
&&$\!\!\!\!\!\!\scriptstyle{0}\!\!\!\!\!\!\!\!\!\!\!$
&&$\!\!\!\!\!\!\!\!\!\!\!\scriptstyle{0}\!\!\!\!\!\!\!\!\!\!\!\!\!\!$
&&$ \!\!\!\!\!\!\scriptstyle{-(2n+3)}\!\!\!\!\!\!\!\!\!\!\!$
&&$\!\!\!\!\!\scriptstyle{-\frac12(2n+3)}\!\!\!\!\!\!\!$& \cr
\omit&height4pt&\omit&&\omit&&\omit&&\omit&&\omit&&\omit&&\omit&\cr
\tablerule
\omit&height4pt&\omit&&\omit&&\omit&&\omit&&\omit&&\omit&&\omit&\cr
&&$\!\!\!\!\!\scriptstyle{(\rm 2\!-\!3)_{\rm b}}\!\!\!\!\!\!\!\!$
&&$\!\!\!\!\!\!\scriptstyle{\frac12(2n+1)}\!\!\!\!\!\!\!\!\!\!\!$
&&$\!\!\!\!\!\!\scriptstyle{0}\!\!\!\!\!\!\!\!\!\!\!$
&&$\!\!\!\!\!\!\scriptstyle{\frac14(2n+3)}\!\!\!\!\!\!\!\!\!\!\!$
&&$\!\!\!\!\!\!\!\!\!\!\!\scriptstyle{0}\!\!\!\!\!\!\!\!\!\!\!\!\!\!$
&&$ \!\!\!\!\!\!\scriptstyle{-\frac14(2n+3)}\!\!\!\!\!\!\!\!\!\!\!$
&&$\!\!\!\!\!\scriptstyle{-\frac18(2n+3)}\!\!\!\!\!\!\!$& \cr
\omit&height4pt&\omit&&\omit&&\omit&&\omit&&\omit&&\omit&&\omit&\cr
\tablerule
\omit&height4pt&\omit&&\omit&&\omit&&\omit&&\omit&&\omit&&\omit&\cr
&&$\!\!\!\!\!\scriptstyle{(\rm 2\!-\!3)_{\rm c}}\!\!\!\!\!\!\!\!$
&&$\!\!\!\!\!\!\scriptstyle{-\frac12(2n+1)}\!\!\!\!\!\!\!\!\!\!\!$
&&$\!\!\!\!\!\!\scriptstyle{-(2n+\frac12)}\!\!\!\!\!\!\!\!\!\!\!$
&&$\!\!\!\!\!\!\scriptstyle{-\frac14(2n+3)}\!\!\!\!\!\!\!\!\!\!\!$
&&$\!\!\!\!\!\!\!\!\!\!\!\scriptstyle{-\frac12(2n+1)}\!\!\!\!\!\!\!\!\!\!\!\!\!\!$
&&$ \!\!\!\!\!\!\scriptstyle{-\frac14(2n+3)}\!\!\!\!\!\!\!\!\!\!\!$
&&$\!\!\!\!\!\scriptstyle{-\frac18(2n+3)}\!\!\!\!\!\!\!$& \cr
\omit&height4pt&\omit&&\omit&&\omit&&\omit&&\omit&&\omit&&\omit&\cr
\tablerule
\omit&height4pt&\omit&&\omit&&\omit&&\omit&&\omit&&\omit&&\omit&\cr
&&$\!\!\!\!\!\scriptstyle{(\rm 3\!-\!1)}\!\!\!\!\!\!\!\!$
&&$\!\!\!\!\!\!\scriptstyle{6(n-\frac34)}\!\!\!\!\!\!\!\!\!\!\!$
&&$\!\!\!\!\!\!\scriptstyle{-\frac32}\!\!\!\!\!\!\!\!\!\!\!$
&&$\!\!\!\!\!\!\scriptstyle{3n}\!\!\!\!\!\!\!\!\!\!\!$
&&$\!\!\!\!\!\!\!\!\!\!\!\scriptstyle{\frac32}\!\!\!\!\!\!\!\!\!\!\!\!\!\!$
&&$ \!\!\!\!\!\!\scriptstyle{-3n}\!\!\!\!\!\!\!\!\!\!\!$
&&$\!\!\!\!\!\scriptstyle{-\frac34(2n+1)}\!\!\!\!\!\!\!$& \cr
\omit&height4pt&\omit&&\omit&&\omit&&\omit&&\omit&&\omit&&\omit&\cr
\tablerule
\omit&height4pt&\omit&&\omit&&\omit&&\omit&&\omit&&\omit&&\omit&\cr
&&$\!\!\!\!\!\scriptstyle{(\rm 3\!-\!2)_{\rm a}}\!\!\!\!\!\!\!\!$
&&$\!\!\!\!\!\!\scriptstyle{\frac14(4n+1)}\!\!\!\!\!\!\!\!\!\!\!$
&&$\!\!\!\!\!\!\scriptstyle{(2n+\frac14)}\!\!\!\!\!\!\!\!\!\!\!$
&&$\!\!\!\!\!\!\scriptstyle{\frac{1}{2}n}\!\!\!\!\!\!\!\!\!\!\!$
&&$\!\!\!\!\!\!\!\!\!\!\!\scriptstyle{\frac14(4n+1)}\!\!\!\!\!\!\!\!\!\!\!\!\!\!$
&&$ \!\!\!\!\!\!\scriptstyle{\frac{1}{2}n}\!\!\!\!\!\!\!\!\!\!\!$
&&$\!\!\!\!\!\scriptstyle{\frac18(2n+1)}\!\!\!\!\!\!\!$& \cr
\omit&height4pt&\omit&&\omit&&\omit&&\omit&&\omit&&\omit&&\omit&\cr
\tablerule
\omit&height4pt&\omit&&\omit&&\omit&&\omit&&\omit&&\omit&&\omit&\cr
&&$\!\!\!\!\!\scriptstyle{(\rm 3\!-\!2)_{\rm b}}\!\!\!\!\!\!\!\!$
&&$\!\!\!\!\!\!\scriptstyle{-(n-\frac34)}\!\!\!\!\!\!\!\!\!\!\!$
&&$\!\!\!\!\!\!\scriptstyle{\frac14}\!\!\!\!\!\!\!\!\!\!\!$
&&$\!\!\!\!\!\!\scriptstyle{-\frac{1}{2}n}\!\!\!\!\!\!\!\!\!\!\!$
&&$\!\!\!\!\!\!\!\!\!\!\!\scriptstyle{-\frac14}\!\!\!\!\!\!\!\!\!\!\!\!\!\!$
&&$ \!\!\!\!\!\!\scriptstyle{\frac{1}{2}n}\!\!\!\!\!\!\!\!\!\!\!$
&&$\!\!\!\!\!\scriptstyle{\frac18(2n+1)}\!\!\!\!\!\!\!$& \cr
\omit&height4pt&\omit&&\omit&&\omit&&\omit&&\omit&&\omit&&\omit&\cr
\tablerule
\omit&height4pt&\omit&&\omit&&\omit&&\omit&&\omit&&\omit&&\omit&\cr
&&$\!\!\!\!\!\scriptstyle{(\rm 3\!-\!2)_{\rm c}}\!\!\!\!\!\!\!\!$
&&$\!\!\!\!\!\!\scriptstyle{2}\!\!\!\!\!\!\!\!\!\!\!$
&&$\!\!\!\!\!\!\scriptstyle{(4n+1)}\!\!\!\!\!\!\!\!\!\!\!$
&&$\!\!\!\!\!\!\scriptstyle{0}\!\!\!\!\!\!\!\!\!\!\!$
&&$\!\!\!\!\!\!\!\!\!\!\!\scriptstyle{0}\!\!\!\!\!\!\!\!\!\!\!\!\!\!$
&&$ \!\!\!\!\!\!\scriptstyle{2n}\!\!\!\!\!\!\!\!\!\!\!$
&&$\!\!\!\!\!\scriptstyle{\frac12(2n+1)}\!\!\!\!\!\!\!$& \cr
\omit&height4pt&\omit&&\omit&&\omit&&\omit&&\omit&&\omit&&\omit&\cr
\tablerule
\omit&height4pt&\omit&&\omit&&\omit&&\omit&&\omit&&\omit&&\omit&\cr
&&$\!\!\!\!\!\scriptstyle{(\rm 3\!-\!3)_{\rm a}}\!\!\!\!\!\!\!\!$
&&$\!\!\!\!\!\!\scriptstyle{-1}\!\!\!\!\!\!\!\!\!\!\!$
&&$\!\!\!\!\!\!\scriptstyle{1}\!\!\!\!\!\!\!\!\!\!\!$
&&$\!\!\!\!\!\!\scriptstyle{0}\!\!\!\!\!\!\!\!\!\!\!$
&&$\!\!\!\!\!\!\!\!\!\!\!\scriptstyle{0}\!\!\!\!\!\!\!\!\!\!\!\!\!\!$
&&$ \!\!\!\!\!\!\scriptstyle{\frac12}\!\!\!\!\!\!\!\!\!\!\!$
&&$\!\!\!\!\!\scriptstyle{\frac14}\!\!\!\!\!\!\!$& \cr
\omit&height4pt&\omit&&\omit&&\omit&&\omit&&\omit&&\omit&&\omit&\cr
\tablerule
\omit&height4pt&\omit&&\omit&&\omit&&\omit&&\omit&&\omit&&\omit&\cr
&&$\!\!\!\!\!\scriptstyle{(\rm 3\!-\!3)_{\rm b}}\!\!\!\!\!\!\!\!$
&&$\!\!\!\!\!\!\scriptstyle{0}\!\!\!\!\!\!\!\!\!\!\!$
&&$\!\!\!\!\!\!\scriptstyle{\frac12}\!\!\!\!\!\!\!\!\!\!\!$
&&$\!\!\!\!\!\!\scriptstyle{\frac18}\!\!\!\!\!\!\!\!\!\!\!$
&&$\!\!\!\!\!\!\!\!\!\!\!\scriptstyle{\frac14}\!\!\!\!\!\!\!\!\!\!\!\!\!\!$
&&$ \!\!\!\!\!\!\scriptstyle{\frac18}\!\!\!\!\!\!\!\!\!\!\!$
&&$\!\!\!\!\!\scriptstyle{\frac{1}{16}}\!\!\!\!\!\!\!$& \cr
\omit&height4pt&\omit&&\omit&&\omit&&\omit&&\omit&&\omit&&\omit&\cr
\tablerule
\omit&height4pt&\omit&&\omit&&\omit&&\omit&&\omit&&\omit&&\omit&\cr
&&$\!\!\!\!\!\scriptstyle{(\rm 3\!-\!3)_{\rm c}}\!\!\!\!\!\!\!\!$
&&$\!\!\!\!\!\!\scriptstyle{\frac52}\!\!\!\!\!\!\!\!\!\!\!$
&&$\!\!\!\!\!\!\scriptstyle{0}\!\!\!\!\!\!\!\!\!\!\!$
&&$\!\!\!\!\!\!\scriptstyle{\frac{5}{8}}\!\!\!\!\!\!\!\!\!\!\!$
&&$\!\!\!\!\!\!\!\!\!\!\!\scriptstyle{0}\!\!\!\!\!\!\!\!\!\!\!\!\!\!$
&&$ \!\!\!\!\!\!\scriptstyle{-\frac{5}{8}}\!\!\!\!\!\!\!\!\!\!\!$
&&$\!\!\!\!\!\scriptstyle{-\frac{5}{16}}\!\!\!\!\!\!\!$& \cr
\omit&height4pt&\omit&&\omit&&\omit&&\omit&&\omit&&\omit&&\omit&\cr
\tablerule
\omit&height4pt&\omit&&\omit&&\omit&&\omit&&\omit&&\omit&&\omit&\cr
&&$\!\!\!\!\!\scriptscriptstyle{\rm Total_{\rm sub}}\!\!\!\!\!\!\!\!\!\!\!\!$
&&$\!\!\!\!\!\!\scriptstyle{-6(3n^2+n)}\!\!\!\!\!\!\!\!\!\!\!$
&&$\!\!\!\!\!\!\scriptstyle{-6(3n^2+n)}\!\!\!\!\!\!\!\!\!\!\!$
&&$\!\!\!\!\!\!\scriptstyle{-(9n^2+12n+\frac34)}\!\!\!\!\!\!\!\!\!\!\!$
&&$\!\!\!\!\!\scriptstyle{(-3n^2-2n+\frac54)}\!\!\!\!\!\!\!\!\!\!\!$
&&$ \!\!\!\!\!\!\scriptstyle{0}\!\!\!\!\!\!\!\!\!\!\!$
&&$\!\!\!\!\!\scriptstyle{0}\!\!\!\!\!\!\!$& \cr
\omit&height4pt&\omit&&\omit&&\omit&&\omit&&\omit&&\omit&&\omit&\cr
\tablerule}}

\caption{$i\Delta_{\rm cf}\times i[S]_{n\geq 1}-{\rm II}$. All contributions
are multiplied by $\frac{\kappa^2 mH}{64\pi^4}\sum_{n\geq 1}^{\infty}
\Bigl(\frac{y}{4}\Bigr)^n$.}

\label{cfn13}

\end{table}
\begin{table}

\vbox{\tabskip=0pt \offinterlineskip
\def\tablerule{\noalign{\hrule}}
\halign to390pt {\strut#& \vrule#\tabskip=1em plus2em& \hfil#\hfil&
\vrule#& \hfil#\hfil& \vrule#& \hfil#\hfil& \vrule#& \hfil#\hfil&
\vrule#& \hfil#\hfil& \vrule#& \hfil#\hfil& \vrule#&
\hfil#\hfil& \vrule#\tabskip=0pt\cr
\tablerule
\omit&height4pt&\omit&&\omit&&\omit&&\omit&&\omit&&\omit&&\omit&\cr
&&$\!\!\!\!\!\scriptstyle{(\rm I\!-\!J)_{\rm sub}}\!\!\!\!\!\!\!\!$
&&$\!\!\!\!\!\!\frac{Ha}{\Delta x^4}\!\!\!\!\!\!\!$
&&$\!\!\!\!\!\!\frac{Ha'}{\Delta x^4}\!\!\!\!\!\!\!$
&&$\!\!\!\!\!\!\scriptstyle{\frac{H^2a^2\gamma^0\!\gamma^k\!\Delta x_k}
{\Delta x^4}}\!\!\!\!\!\!\!\!\!\!\!$
&&$\!\!\!\!\!\!\scriptstyle{\frac{H^{2}aa'\gamma^0\!\gamma^k\!\Delta x_k}
{\Delta x^4}}\!\!\!\!\!\!\!\!\!\!\!$
&&$ \!\!\!\!\!\!\frac{H^2a^2\Delta\eta}{\Delta x^4}\!\!\!\!\!\!\!\!$
&&$\!\!\!\!\!\frac{H^2a^2Ha'}{\Delta x^2}\!\!\!\!\!$& \cr
\omit&height4pt&\omit&&\omit&&\omit&&\omit&&\omit&&\omit&&\omit&\cr
\tablerule
\omit&height4pt&\omit&&\omit&&\omit&&\omit&&\omit&&\omit&&\omit&\cr
&&$\!\!\!\!\!\scriptstyle{(\rm 1\!-\!1)}\!\!\!\!\!\!\!\!$
&&$\!\!\!\!\!\!\scriptstyle{-8(n^3-n)}\!\!\!\!\!\!\!\!\!\!\!$
&&$\!\!\!\!\!\!\scriptstyle{0}\!\!\!\!\!\!\!\!\!\!\!$
&&$\!\!\!\!\!\!\scriptscriptstyle{-4(n^3+2n^2-n-2)}\!\!\!\!\!\!\!\!\!\!\!$
&&$\!\!\!\!\!\!\scriptstyle{0}\!\!\!\!\!\!\!$
&&$ \!\!\!\!\!\!\scriptscriptstyle{4(n^3+2n^2-n-2)}\!\!\!\!\!\!\!\!\!\!$
&&$\!\!\!\!\!\scriptscriptstyle{2(n^3+3n^2+2n)}\!\!\!\!\!$& \cr
\omit&height4pt&\omit&&\omit&&\omit&&\omit&&\omit&&\omit&&\omit&\cr
\tablerule
\omit&height4pt&\omit&&\omit&&\omit&&\omit&&\omit&&\omit&&\omit&\cr
&&$\!\!\!\!\scriptstyle{(\rm 1\!-\!2)}\!\!\!\!\!\!\!\!$
&&$\!\!\!\!\!\!\scriptstyle{2(n^3-n)}\!\!\!\!\!\!\!\!\!\!\!$
&&$\!\!\!\!\!\!\scriptstyle{0}\!\!\!\!\!\!\!\!\!\!\!$
&&$\!\!\!\!\!\!\scriptstyle{(n^3+2n^2-n-2)}\!\!\!\!\!\!\!\!\!\!\!$
&&$\!\!\!\!\!\!\scriptstyle{0}\!\!\!\!\!\!\!$
&&$ \!\!\!\!\!\!\scriptscriptstyle{-(n^3+2n^2-n-2)}\!\!\!\!\!\!\!\!\!\!$
&&$\!\!\!\!\!\scriptscriptstyle{-\frac12(n^3+3n^2+2n)}\!\!\!\!\!$& \cr
\omit&height4pt&\omit&&\omit&&\omit&&\omit&&\omit&&\omit&&\omit&\cr
\tablerule
\omit&height4pt&\omit&&\omit&&\omit&&\omit&&\omit&&\omit&&\omit&\cr
&&$\!\!\!\!\!\scriptstyle{(\rm 1\!-\!3)}\!\!\!\!\!\!\!\!$
&&$\!\!\!\!\!\!\scriptstyle{-3(n^2+n)}\!\!\!\!\!\!\!\!\!\!\!$
&&$\!\!\!\!\!\!\scriptstyle{0}\!\!\!\!\!\!\!\!\!\!\!$
&&$\!\!\!\!\!\!\scriptstyle{-\frac32(n^2+3n+2)}\!\!\!\!\!\!\!\!\!\!\!$
&&$\!\!\!\!\!\!\scriptstyle{0}\!\!\!\!\!\!\!$
&&$ \!\!\!\!\!\!\scriptstyle{\frac32(n^2+3n+2)}\!\!\!\!\!\!\!\!\!\!$
&&$\!\!\!\!\!\scriptscriptstyle{\frac34(n^2+3n+2)}\!\!\!\!\!$& \cr
\omit&height4pt&\omit&&\omit&&\omit&&\omit&&\omit&&\omit&&\omit&\cr
\tablerule
\omit&height4pt&\omit&&\omit&&\omit&&\omit&&\omit&&\omit&&\omit&\cr
&&$\!\!\!\!\scriptstyle{(\rm 2\!-\!1)}\!\!\!\!\!\!\!\!$
&&$\!\!\!\!\!\!\scriptstyle{2(n^3-n)}\!\!\!\!\!\!\!\!\!\!\!$
&&$\!\!\!\!\!\!\scriptstyle{0}\!\!\!\!\!\!\!\!\!\!\!$
&&$\!\!\!\!\!\!\scriptstyle{(n^3+2n^2-n-2)}\!\!\!\!\!\!\!\!\!\!\!$
&&$\!\!\!\!\!\!\scriptstyle{0}\!\!\!\!\!\!\!$
&&$ \!\!\!\!\!\!\scriptscriptstyle{-(n^3+2n^2-n-2)}\!\!\!\!\!\!\!\!\!\!$
&&$\!\!\!\!\!\scriptscriptstyle{-\frac12(n^3+3n^2+2n)}\!\!\!\!\!$& \cr
\omit&height4pt&\omit&&\omit&&\omit&&\omit&&\omit&&\omit&&\omit&\cr
\tablerule
\omit&height4pt&\omit&&\omit&&\omit&&\omit&&\omit&&\omit&&\omit&\cr
&&$\!\!\!\!\!\scriptstyle{(\rm 2\!-\!2)_{\rm a}}\!\!\!\!\!\!\!\!$
&&$\!\!\!\!\!\!\scriptstyle{-(n^3+n^2)}\!\!\!\!\!\!\!\!\!\!\!$
&&$\!\!\!\!\!\!\scriptstyle{-(2n^3+n^2)}\!\!\!\!\!\!\!\!\!\!\!$
&&$\!\!\!\!\!\!\scriptscriptstyle{-\frac12(n^3+2n^2-n-2)}\!\!\!\!\!\!\!\!\!\!\!$
&&$\!\!\!\!\!\!\!\!\!\!\!\scriptstyle{-(n^3+n^2)}\!\!\!\!\!\!\!\!\!\!\!\!\!\!$
&&$ \!\!\!\!\!\!\scriptscriptstyle{-\frac12(n^3+2n^2-n-2)}\!\!\!\!\!\!\!\!\!\!\!$
&&$\!\!\!\!\!\scriptstyle{-\frac14(n^3+3n^2+2n)}\!\!\!\!\!\!\!$& \cr
\omit&height4pt&\omit&&\omit&&\omit&&\omit&&\omit&&\omit&&\omit&\cr
\tablerule
\omit&height4pt&\omit&&\omit&&\omit&&\omit&&\omit&&\omit&&\omit&\cr
&&$\!\!\!\!\!\scriptstyle{(\rm 2\!-\!2)_{\rm b}}\!\!\!\!\!\!\!\!$
&&$\!\!\!\!\!\!\scriptstyle{-2(n^2+n)}\!\!\!\!\!\!\!\!\!\!\!$
&&$\!\!\!\!\!\!\scriptstyle{-2(2n^3+n^2)}\!\!\!\!\!\!\!\!\!\!\!$
&&$\!\!\!\!\!\!\scriptstyle{0}\!\!\!\!\!\!\!\!\!\!\!$
&&$\!\!\!\!\!\!\!\!\!\!\!\scriptstyle{0}\!\!\!\!\!\!\!\!\!\!\!\!\!\!$
&&$ \!\!\!\!\!\!\scriptscriptstyle{-2(n^3+2n^2-n-2)}\!\!\!\!\!\!\!\!\!\!\!$
&&$\!\!\!\!\!\scriptscriptstyle{-(n^3+3n^2+2n)}\!\!\!\!\!\!\!$& \cr
\omit&height4pt&\omit&&\omit&&\omit&&\omit&&\omit&&\omit&&\omit&\cr
\tablerule
\omit&height4pt&\omit&&\omit&&\omit&&\omit&&\omit&&\omit&&\omit&\cr
&&$\!\!\!\!\!\scriptstyle{(\rm 2\!-\!2)_{\rm c}}\!\!\!\!\!\!\!\!$
&&$\!\!\!\!\!\!\scriptstyle{-(n^3-n)}\!\!\!\!\!\!\!\!\!\!\!$
&&$\!\!\!\!\!\!\scriptstyle{0}\!\!\!\!\!\!\!\!\!\!\!$
&&$\!\!\!\!\!\!\scriptscriptstyle{-\frac12(n^3+2n^2-n-2)}\!\!\!\!\!\!\!\!\!\!\!$
&&$\!\!\!\!\!\!\!\!\!\!\!\scriptstyle{0}\!\!\!\!\!\!\!\!\!\!\!\!\!\!$
&&$ \!\!\!\!\!\!\scriptscriptstyle{\frac12(n^3+2n^2-n-2)}\!\!\!\!\!\!\!\!\!\!\!$
&&$\!\!\!\!\!\scriptscriptstyle{\frac14(n^3+3n^2+2n)}\!\!\!\!\!\!\!$& \cr
\omit&height4pt&\omit&&\omit&&\omit&&\omit&&\omit&&\omit&&\omit&\cr
\tablerule
\omit&height4pt&\omit&&\omit&&\omit&&\omit&&\omit&&\omit&&\omit&\cr
&&$\!\!\!\!\!\scriptstyle{(\rm 2\!-\!3)_{\rm a}}\!\!\!\!\!\!\!\!$
&&$\!\!\!\!\!\!\scriptstyle{0}\!\!\!\!\!\!\!\!\!\!\!$
&&$\!\!\!\!\!\!\scriptstyle{-(2n^2+n)}\!\!\!\!\!\!\!\!\!\!\!$
&&$\!\!\!\!\!\!\scriptstyle{0}\!\!\!\!\!\!\!\!\!\!\!$
&&$\!\!\!\!\!\!\!\!\!\!\!\scriptstyle{0}\!\!\!\!\!\!\!\!\!\!\!\!\!\!$
&&$ \!\!\!\!\!\!\scriptstyle{-(n^2+3n+2)}\!\!\!\!\!\!\!\!\!\!\!$
&&$\!\!\!\!\!\scriptscriptstyle{-\frac12(n^2+3n+2)}\!\!\!\!\!\!\!$& \cr
\omit&height4pt&\omit&&\omit&&\omit&&\omit&&\omit&&\omit&&\omit&\cr
\tablerule
\omit&height4pt&\omit&&\omit&&\omit&&\omit&&\omit&&\omit&&\omit&\cr
&&$\!\!\!\!\!\scriptstyle{(\rm 2\!-\!3)_{\rm b}}\!\!\!\!\!\!\!\!$
&&$\!\!\!\!\!\!\scriptstyle{\frac12(n^2+n)}\!\!\!\!\!\!\!\!\!\!\!$
&&$\!\!\!\!\!\!\scriptstyle{0}\!\!\!\!\!\!\!\!\!\!\!$
&&$\!\!\!\!\!\!\scriptstyle{\frac14(n^2+3n+2)}\!\!\!\!\!\!\!\!\!\!\!$
&&$\!\!\!\!\!\!\!\!\!\!\!\scriptstyle{0}\!\!\!\!\!\!\!\!\!\!\!\!\!\!$
&&$ \!\!\!\!\!\!\scriptstyle{-\frac14(n^2+3n+2)}\!\!\!\!\!\!\!\!\!\!\!$
&&$\!\!\!\!\!\scriptscriptstyle{-\frac18(n^2+3n+2)}\!\!\!\!\!\!\!$& \cr
\omit&height4pt&\omit&&\omit&&\omit&&\omit&&\omit&&\omit&&\omit&\cr
\tablerule
\omit&height4pt&\omit&&\omit&&\omit&&\omit&&\omit&&\omit&&\omit&\cr
&&$\!\!\!\!\!\scriptstyle{(\rm 2\!-\!3)_{\rm c}}\!\!\!\!\!\!\!\!$
&&$\!\!\!\!\!\!\scriptstyle{-\frac12(n^2+n)}\!\!\!\!\!\!\!\!\!\!\!$
&&$\!\!\!\!\!\!\scriptstyle{-(n^2+\frac{1}{2}n)}\!\!\!\!\!\!\!\!\!\!\!$
&&$\!\!\!\!\!\!\scriptstyle{-\frac14(n^2+3n+2)}\!\!\!\!\!\!\!\!\!\!\!$
&&$\!\!\!\!\!\!\!\!\!\!\!\scriptstyle{-\frac12(n^2+n)}\!\!\!\!\!\!\!\!\!\!\!\!\!\!$
&&$ \!\!\!\!\!\!\scriptstyle{-\frac14(n^2+3n+2)}\!\!\!\!\!\!\!\!\!\!\!$
&&$\!\!\!\!\!\scriptscriptstyle{-\frac18(n^2+3n+2)}\!\!\!\!\!\!\!$& \cr
\omit&height4pt&\omit&&\omit&&\omit&&\omit&&\omit&&\omit&&\omit&\cr
\tablerule
\omit&height4pt&\omit&&\omit&&\omit&&\omit&&\omit&&\omit&&\omit&\cr
&&$\!\!\!\!\!\scriptstyle{(\rm 3\!-\!1)}\!\!\!\!\!\!\!\!$
&&$\!\!\!\!\!\!\scriptstyle{(3n^2-\frac{9}{2}n)}\!\!\!\!\!\!\!\!\!\!\!$
&&$\!\!\!\!\!\!\scriptstyle{-\frac{3}{2}n}\!\!\!\!\!\!\!\!\!\!\!$
&&$\!\!\!\!\!\!\scriptstyle{\frac32(n^2-1)}\!\!\!\!\!\!\!\!\!\!\!$
&&$\!\!\!\!\!\!\!\!\!\!\!\scriptstyle{\frac{3}{2}n}\!\!\!\!\!\!\!\!\!\!\!\!\!\!$
&&$ \!\!\!\!\!\!\scriptstyle{-\frac32(n^2-1)}\!\!\!\!\!\!\!\!\!\!\!$
&&$\!\!\!\!\!\scriptscriptstyle{-\frac34(n^2+n)}\!\!\!\!\!\!\!$& \cr
\omit&height4pt&\omit&&\omit&&\omit&&\omit&&\omit&&\omit&&\omit&\cr
\tablerule
\omit&height4pt&\omit&&\omit&&\omit&&\omit&&\omit&&\omit&&\omit&\cr
&&$\!\!\!\!\!\scriptstyle{(\rm 3\!-\!2)_{\rm a}}\!\!\!\!\!\!\!\!$
&&$\!\!\!\!\!\!\scriptstyle{\frac14(2n^2+n)}\!\!\!\!\!\!\!\!\!\!\!$
&&$\!\!\!\!\!\!\scriptstyle{(n^2+\frac{1}{4}n)}\!\!\!\!\!\!\!\!\!\!\!$
&&$\!\!\!\!\!\!\scriptstyle{\frac{1}{4}(n^2-1)}\!\!\!\!\!\!\!\!\!\!\!$
&&$\!\!\!\!\!\!\!\!\!\!\!\scriptstyle{\frac14(2n^2+n)}\!\!\!\!\!\!\!\!\!\!\!\!\!\!$
&&$ \!\!\!\!\!\!\scriptstyle{\frac{1}{4}(n^2-1)}\!\!\!\!\!\!\!\!\!\!\!$
&&$\!\!\!\!\!\scriptstyle{\frac18(n^2+n)}\!\!\!\!\!\!\!$& \cr
\omit&height4pt&\omit&&\omit&&\omit&&\omit&&\omit&&\omit&&\omit&\cr
\tablerule
\omit&height4pt&\omit&&\omit&&\omit&&\omit&&\omit&&\omit&&\omit&\cr
&&$\!\!\!\!\!\scriptstyle{(\rm 3\!-\!2)_{\rm b}}\!\!\!\!\!\!\!\!$
&&$\!\!\!\!\!\!\scriptstyle{-(\frac{1}{2}n^2-\frac{3}{4}n)}\!\!\!\!\!\!\!\!\!\!\!$
&&$\!\!\!\!\!\!\scriptstyle{\frac{1}{4}n}\!\!\!\!\!\!\!\!\!\!\!$
&&$\!\!\!\!\!\!\scriptstyle{-\frac{1}{4}(n^2-1)}\!\!\!\!\!\!\!\!\!\!\!$
&&$\!\!\!\!\!\!\!\!\!\!\!\scriptstyle{-\frac{1}{4}n}\!\!\!\!\!\!\!\!\!\!\!\!\!\!$
&&$ \!\!\!\!\!\!\scriptstyle{\frac{1}{4}(n^2-1)}\!\!\!\!\!\!\!\!\!\!\!$
&&$\!\!\!\!\!\scriptstyle{\frac18(n^2+n)}\!\!\!\!\!\!\!$& \cr
\omit&height4pt&\omit&&\omit&&\omit&&\omit&&\omit&&\omit&&\omit&\cr
\tablerule
\omit&height4pt&\omit&&\omit&&\omit&&\omit&&\omit&&\omit&&\omit&\cr
&&$\!\!\!\!\!\scriptstyle{(\rm 3\!-\!2)_{\rm c}}\!\!\!\!\!\!\!\!$
&&$\!\!\!\!\!\!\scriptstyle{2n}\!\!\!\!\!\!\!\!\!\!\!$
&&$\!\!\!\!\!\!\scriptstyle{(2n^2+n)}\!\!\!\!\!\!\!\!\!\!\!$
&&$\!\!\!\!\!\!\scriptstyle{0}\!\!\!\!\!\!\!\!\!\!\!$
&&$\!\!\!\!\!\!\!\!\!\!\!\scriptstyle{0}\!\!\!\!\!\!\!\!\!\!\!\!\!\!$
&&$ \!\!\!\!\!\!\scriptstyle{(n^2-1)}\!\!\!\!\!\!\!\!\!\!\!$
&&$\!\!\!\!\!\scriptstyle{\frac12(n^2+n)}\!\!\!\!\!\!\!$& \cr
\omit&height4pt&\omit&&\omit&&\omit&&\omit&&\omit&&\omit&&\omit&\cr
\tablerule
\omit&height4pt&\omit&&\omit&&\omit&&\omit&&\omit&&\omit&&\omit&\cr
&&$\!\!\!\!\!\scriptstyle{(\rm 3\!-\!3)_{\rm a}}\!\!\!\!\!\!\!\!$
&&$\!\!\!\!\!\!\scriptstyle{-(n+\frac12)}\!\!\!\!\!\!\!\!\!\!\!$
&&$\!\!\!\!\!\!\scriptstyle{(n+\frac12)}\!\!\!\!\!\!\!\!\!\!\!$
&&$\!\!\!\!\!\!\scriptstyle{0}\!\!\!\!\!\!\!\!\!\!\!$
&&$\!\!\!\!\!\!\!\!\!\!\!\scriptstyle{0}\!\!\!\!\!\!\!\!\!\!\!\!\!\!$
&&$ \!\!\!\!\!\!\scriptstyle{\frac12(n+1)}\!\!\!\!\!\!\!\!\!\!\!$
&&$\!\!\!\!\!\scriptstyle{\frac14(n+1)}\!\!\!\!\!\!\!$& \cr
\omit&height4pt&\omit&&\omit&&\omit&&\omit&&\omit&&\omit&&\omit&\cr
\tablerule
\omit&height4pt&\omit&&\omit&&\omit&&\omit&&\omit&&\omit&&\omit&\cr
&&$\!\!\!\!\!\scriptstyle{(\rm 3\!-\!3)_{\rm b}}\!\!\!\!\!\!\!\!$
&&$\!\!\!\!\!\!\scriptstyle{-\frac18}\!\!\!\!\!\!\!\!\!\!\!$
&&$\!\!\!\!\!\!\scriptstyle{\frac{1}{2}n+\frac18}\!\!\!\!\!\!\!\!\!\!\!$
&&$\!\!\!\!\!\!\scriptstyle{\frac18(n+1)}\!\!\!\!\!\!\!\!\!\!\!$
&&$\!\!\!\!\!\!\!\!\!\!\!\scriptstyle{\frac18(2n+1)}\!\!\!\!\!\!\!\!\!\!\!\!\!\!$
&&$ \!\!\!\!\!\!\scriptstyle{\frac18(n+1)}\!\!\!\!\!\!\!\!\!\!\!$
&&$\!\!\!\!\!\scriptstyle{\frac{1}{16}(n+1)}\!\!\!\!\!\!\!$& \cr
\omit&height4pt&\omit&&\omit&&\omit&&\omit&&\omit&&\omit&&\omit&\cr
\tablerule
\omit&height4pt&\omit&&\omit&&\omit&&\omit&&\omit&&\omit&&\omit&\cr
&&$\!\!\!\!\!\scriptstyle{(\rm 3\!-\!3)_{\rm c}}\!\!\!\!\!\!\!\!$
&&$\!\!\!\!\!\!\scriptstyle{\frac{5}{2}n+\frac58}\!\!\!\!\!\!\!\!\!\!\!$
&&$\!\!\!\!\!\!\scriptstyle{-\frac58}\!\!\!\!\!\!\!\!\!\!\!$
&&$\!\!\!\!\!\!\scriptstyle{\frac{5}{8}(n+1)}\!\!\!\!\!\!\!\!\!\!\!$
&&$\!\!\!\!\!\!\!\!\!\!\!\scriptstyle{\frac58}\!\!\!\!\!\!\!\!\!\!\!\!\!\!$
&&$ \!\!\!\!\!\!\scriptstyle{-\frac{5}{8}(n+1)}\!\!\!\!\!\!\!\!\!\!\!$
&&$\!\!\!\!\!\scriptstyle{-\frac{5}{16}(n+1)}\!\!\!\!\!\!\!$& \cr
\omit&height4pt&\omit&&\omit&&\omit&&\omit&&\omit&&\omit&&\omit&\cr
\tablerule
\omit&height4pt&\omit&&\omit&&\omit&&\omit&&\omit&&\omit&&\omit&\cr
&&$\!\!\!\!\!\scriptscriptstyle{\rm Total_{\rm sub}}\!\!\!\!\!\!\!\!\!\!\!\!$
&&$\!\!\!\!\!\!\scriptscriptstyle{-3(2n^3+n^2)}\!\!\!\!\!\!\!\!\!\!\!$
&&$\!\!\!\!\!\!\scriptscriptstyle{-3(3n^3+n^2)}\!\!\!\!\!\!\!\!\!\!\!$
&&$\!\!\!\!\!\!\scriptscriptstyle{\!(\!-3n^3\!-\!6n^2\!-\!\frac{3}{4}n\!+\!\frac94\!)
\!\!\!\!\!\!\!\!\!\!\!}$
&&$\!\!\!\!\!\scriptscriptstyle{\!(\!-n^3\!-\!n^2\!+\!\frac{5}{4}n\!+\!\frac34\!)}
\!\!\!\!\!\!\!\!\!\!\!\!$
&&$ \!\!\!\!\!\!\scriptstyle{0}\!\!\!\!\!\!\!\!\!\!\!$
&&$\!\!\!\!\!\scriptstyle{0}\!\!\!\!\!\!\!$& \cr
\omit&height4pt&\omit&&\omit&&\omit&&\omit&&\omit&&\omit&&\omit&\cr
\tablerule}}

\caption{$i\Delta_{\rm cf}\times i[S]_{n\geq 1}-{\rm II'}$. All contributions
are multiplied by $\frac{\kappa^2 mH}{64\pi^4}\sum_{n\geq 1}^{\infty}
\Bigl(\frac{y}{4}\Bigr)^n\ln(\frac{y}{4})$.}

\label{cfn14}

\end{table}

We can easily read off the contribution from each distinctive term.
For example, the total coefficient of
$\frac{\gamma^0\gamma^k \Delta x_k}{\Delta x^6}$ from Table~\ref{cfn11}
and Table~\ref{cfn12} is,
\begin{eqnarray}
\frac{\kappa^2 m H}{64\pi^4}\sum_{n=1}^{\infty}\Bigl[
8(3n^2\!\!-\!2n\!-\!1)+8(n^3\!\!-\!n^2\!\!-\!n)\ln(\frac{y}{4})\Bigr]
(\frac{y}{4})^n \,.
\end{eqnarray}
After collecting all of terms from the four tables we obtain,
\begin{eqnarray}
&&\frac{\kappa^2 m H}{64\pi^4}
\sum_{n=1}^{\infty}(\frac{y}{4})^n\Biggl\{
\frac{\gamma^0\gamma^k \Delta x_k}
{\Delta x^6}\Bigl[8(3n^2\!\!-\!2n\!-\!1)
+8(n^3\!\!-\!n^2\!\!-\!n)\ln(\frac{y}{4})\Bigr]\nonumber\\
&&+\Bigl[5\frac{\scriptstyle{Ha\gamma^0\Delta\eta\gamma^k\Delta x_k}}
{\Delta x^6}
+\frac{\scriptstyle{Ha'\gamma^0\Delta\eta\gamma^k\Delta x_k}}
{\Delta x^6}
\!-\!3\frac{\scriptstyle{Ha\Delta\eta^2}}{\Delta x^6}\!-\!3
\frac{\scriptstyle{Ha'\Delta\eta^2}}{\Delta x^6}\Bigr]
\Bigl[(6n^2\!\!-\!4n\!-\!\frac32)\nonumber\\
&&+(2n^3\!\!-\!2n^2\!\!-\!\frac{3}{2}n)\ln(\frac{y}{4})\Bigr]
\!-\!3\Bigl[\frac{H\!a}{\Delta x^4}
\!+\!\frac{H\!a'}{\Delta x^4}\Bigr]
\Bigl[(6n^2\!\!+\!2n)\!+\!(2n^3\!\!+\!n^2)\ln(\frac{y}{4})\Bigr]\nonumber\\
&&+\frac{H^2\!a^2\gamma^0\!\gamma^k\!\Delta x_k}
{\Delta x^4}\Bigl[-(9n^2\!\!+\!12n\!+\!\frac34)
\!+\!(-3n^3\!\!-\!6n^2\!\!-\!\frac{3}{4}n\!+\!\frac94)
\ln(\frac{y}{4})\Bigr]\nonumber\\
&&+\frac{H^2\!aa'\gamma^0\!\gamma^k\!\Delta x_k}{\Delta x^4}
\Bigl[(-3n^2\!\!-\!2n\!+\!\frac54)
\!+\!(-n^3\!\!-\!n^2\!\!+\!\frac{5}{4}n\!+\!\frac34)
\ln(\frac{y}{4})\Bigr]\Biggr\}.
\end{eqnarray}
One might notice that the infinite series could be summed easily using the
following identities,
\begin{eqnarray}
&&\sum_{n=1}^{\infty}Y^n=\frac{Y}{1-Y}\,\,\,\,;\,\,\,\,
\sum_{n=1}^{\infty}nY^n=\frac{Y}{(1-Y)^2}\label{sum1} \\
&&\sum_{n=1}^{\infty}n^2Y^n=\frac{Y(Y+1)}{(1-Y)^3}\,\,\,\,;
\,\,\,\,\sum_{n=1}^{\infty}n^3Y^n=
\frac{Y(Y^2+4Y+1)}{(1-Y)^3}\,.\label{sum2}
\end{eqnarray}
Here $Y$ stands for $\frac{y}{4}$.
After the summation, the total contribution from $i\Delta_{\rm cf}~
i[S]_{n\geq 1}$ is,
\begin{eqnarray}
&&-i\Bigl[\Sigma^{\rm cfn1}\Bigr](x;x')\!=\!
\frac{\kappa^2 mH}{64\pi^4}\Biggl\{
8\frac{\gamma^0\gamma^k\Delta x_k}{\Delta x^6}
\Biggl[\frac{(-Y^2\!\!+\!7Y)}{(1\!-\!Y)^3}
+\frac{(Y^3\!\!+\!6Y^2\!\!-\!Y)}{(1\!-\!Y)^4}\ln(Y)\Biggr]\nonumber\\
&&+\Biggl[5\frac{\scriptstyle{Ha\gamma^0\Delta\eta\gamma^k\Delta x_k}}
{\Delta x^6}
\!+\!\frac{\scriptstyle{Ha'\gamma^0\Delta\eta\gamma^k\Delta x_k}}
{\Delta x^6}
\!-\!3\frac{\scriptstyle{Ha\Delta\eta^2}}{\Delta x^6}\!-\!3
\frac{\scriptstyle{Ha'\Delta\eta^2}}{\Delta x^6}\Biggr]
\Biggl[\frac{-3Y^3\!\!+\!26Y^2\!\!+\!Y}{2(1\!-\!Y)^3}\nonumber\\
&&+\frac{3Y^3\!\!+\!22Y^2\!\!-\!3Y}{2(1\!-\!Y)^4}
\ln(Y)\Biggr]\!\!+\!\!\Biggl[\frac{H\!a}{\Delta x^4}\!+\!
\frac{H\!a'}{\Delta x^4}\Biggr]
\Biggl[\frac{-12(Y^2\!+\!2Y)}{(1\!-\!Y)^3}\!+\!
\frac{-3(Y^3\!\!+\!8Y^2\!\!+\!3Y)}{(1\!-\!Y)^4}\ln(Y)\Biggr]\nonumber\\
&&+\frac{H^2\!a^2\gamma^0\!\gamma^k\!\Delta x_k}{\Delta x^4}
\Biggl[\frac{-3(Y^2\!\!-\!6Y\!+\!29)}{4(1\!-\!Y)^3}
\!+\!\frac{-3(3Y^4\!\!-\!12Y^3\!\!+\!23Y^2\!\!+\!10Y)}
{4(1\!-\!Y)^4}\ln(Y)\Biggr]\nonumber\\
&&\hspace{.7cm}+\frac{H^2\!aa'\gamma^0\!\gamma^k\!\Delta x_k}{\Delta x^4}
\Biggl[\frac{(5Y^3\!\!-\!14Y^2\!\!-\!15Y)}{4(1\!-\!Y)^3}
\!+\!\frac{(-3Y^4\!\!+\!14Y^3\!\!-\!35Y^2)}{4(1\!-\!Y)^4}\ln(Y)
\Biggr]\Biggr\}\,.\label{sumcfn1}
\end{eqnarray}

\subsection{Sub-Leading Contributions from $i{\delta \! \Delta}_A$}
In this subsection we compute the contribution from substituting
the residual $A$-type part of the graviton propagator in
Table~\ref{gen3},
\begin{equation}
i\Bigl[{}_{\alpha\beta} \Delta_{\rho\sigma}\Bigr](x;x')
\longrightarrow \Bigl[ \overline{\eta}_{\alpha \rho}
\overline{\eta}_{\sigma \beta} \!+\! \overline{\eta}_{\alpha \sigma}
\overline{\eta}_{\rho \beta} \!-\! \frac2{D\!-\!3}
\overline{\eta}_{\alpha\beta} \overline{\eta}_{\rho\sigma} \Bigr]
i\delta\!\Delta_A(x;x') \; . \label{DApart}
\end{equation}
As with the conformal contributions of the previous section we first
make the requisite contractions and then act the derivatives. The
result of this first step is displayed in Table~\ref{DAcon0} and
Table~\ref{DAcon}. We have sometimes decomposed the result for a single
vertex pair into as many as five terms because the three different tensors
in (\ref{DApart}) can make distinct contributions, and because distinct
contributions also come from breaking up factors of $\gamma^{\alpha}
J^{\beta \mu}$. These distinct contributions are tagged with
subscripts $a$, $b$, $c$, etc.
\begin{table}

\vbox{\tabskip=0pt \offinterlineskip
\def\tablerule{\noalign{\hrule}}
\halign to390pt {\strut#& \vrule#\tabskip=1em plus2em& \hfil#&
\vrule#& \hfil#& \vrule#& \hfil#& \vrule#& \hfil#\hfil&
\vrule#\tabskip=0pt\cr
\tablerule
\omit&height4pt&\omit&&\omit&&\omit&&\omit&\cr &&\hidewidth {\rm I}
&&\hidewidth {\rm J} \hidewidth&& \hidewidth {\rm sub} \hidewidth&&
\hidewidth $iV_I^{\alpha\beta}(x) \, i[S](x;x') \, i
V_J^{\rho\sigma}(x') \, [\mbox{}_{\alpha\beta} T^A_{\rho\sigma}] \,
i\delta\!\Delta_A(x;x')$ \hidewidth&\cr
\omit&height4pt&\omit&&\omit&&\omit&&\omit&\cr
\tablerule
\omit&height2pt&\omit&&\omit&&\omit&&\omit&\cr && 1 && 4 && \omit &&
$-\f{(D-1)}{(D-3)}i\ka^{2}am\hspace{-.1cm}\not{\hspace{-.1cm}\partial}
i[S](x;x')i\d\!\D_{A}(x;x')$ & \cr
\omit&height2pt&\omit&&\omit&&\omit&&\omit&\cr
\tablerule
\omit&height2pt&\omit&&\omit&&\omit&&\omit&\cr && 2 && 4 && \omit &&
$\f{1}{(D-3)}i\ka^{2}am\hspace{-.1cm}\not{\hspace{-.1cm}\bar{\partial}}
i[S](x;x')i\d\!\D_{A}(x;x')$ & \cr
\omit&height2pt&\omit&&\omit&&\omit&&\omit&\cr
\tablerule
\omit&height2pt&\omit&&\omit&&\omit&&\omit&\cr && 3 && 4 && a &&
$-\f{(D-1)}{2(D-3)}i\ka^{2}am \gamma^{0}\partial_{0}
i\d\!\D_{A}(x;x')i[S](x;x')$ & \cr
\omit&height2pt&\omit&&\omit&&\omit&&\omit&\cr
\tablerule
\omit&height2pt&\omit&&\omit&&\omit&&\omit&\cr && 3 && 4 && b &&
$-\f{(D-2)}{2(D-3)}i\ka^{2}am\hspace{-.1cm}\not{\hspace{-.1cm}\bar{\partial}}
i\d\!\D_{A}(x;x')i[S](x;x')$ & \cr
\omit&height2pt&\omit&&\omit&&\omit&&\omit&\cr
\tablerule
\omit&height2pt&\omit&&\omit&&\omit&&\omit&\cr && 4 && 1 && \omit &&
$\f{(D-1)}{(D-3)}i\ka^{2}am \partial_{\mu}'\{
i[S](x;x')\gamma^{\mu}i\d\!\D_{A}(x;x')\}$ & \cr
\omit&height2pt&\omit&&\omit&&\omit&&\omit&\cr
\tablerule
\omit&height2pt&\omit&&\omit&&\omit&&\omit&\cr && 4 && 2 && \omit &&
$\f{1}{(D-3)}i\ka^2 am \partial_{k}\{
i[S](x;x')\gamma_{k}i\d\!\D_{A}(x;x')\}$ & \cr
\omit&height2pt&\omit&&\omit&&\omit&&\omit&\cr
\tablerule
\omit&height2pt&\omit&&\omit&&\omit&&\omit&\cr && 4 && 3 && a &&
$-\f{(D-1)}{2(D-3)}i\ka^2 am i[S](x;x')
\gamma^{0}\partial_{0}i\d\!\D_{A}(x;x')$ & \cr
\omit&height2pt&\omit&&\omit&&\omit&&\omit&\cr
\tablerule
\omit&height2pt&\omit&&\omit&&\omit&&\omit&\cr && 4 && 3 && b &&
$-\f{(D-2)}{2(D-3)}i\ka^2 am i[S](x;x')
\hspace{-.1cm}\not{\hspace{-.1cm}\bar{\del}}i\d\!\D_{A}(x;x')$ & \cr
\omit&height2pt&\omit&&\omit&&\omit&&\omit&\cr
\tablerule}}
\caption{Contractions from the $i\d\!\D_{A}$ part of the graviton
propagator-I}

\label{DAcon0}

\end{table}

The next step is to act the derivatives and it is of course
necessary to have an expression for $i\delta\!\Delta_A(x;x')$.
From (\ref{DeltaA}) one can deduce,
\begin{eqnarray}
\lefteqn{i\delta\!\Delta_A(x;x') = } \nonumber \\
& & \hspace{-.5cm} \frac{H^2}{16 \pi^{\frac{D}2}}
\frac{\Gamma(\frac{D}2 \!+\! 1)}{\frac{D}2 \!-\! 2} \frac{(a a')^{2-
\frac{D}2}}{\Delta x^{D-4}} + \frac{H^{D-2}}{(4\pi)^\frac{D}2}
\frac{\Gamma(D\!-\!1)}{\Gamma( \frac{D}2
)} \Biggl\{- \pi\cot\Bigl(\frac{\pi}2 D\Bigr) + \ln(aa') \Biggr\} \nonumber \\
& & \hspace{-.5cm} + \frac{H^{D-2}}{(4\pi)^{\frac{D}2}} \!
\sum_{n=1}^{\infty} \! \left\{\!\frac1{n} \frac{\Gamma(n \!+\!D\!-\!
1)}{\Gamma(n \!+\! \frac{D}2)} \Bigl(\frac{y}4 \Bigr)^n \!\!\!\! -
\frac1{n \!-\! \frac{D}2 \!+\! 2} \frac{\Gamma(n \!+\!  \frac{D}2
\!+\! 1)}{\Gamma(n \!+\! 2)} \Bigl(\frac{y}4 \Bigr)^{n - \frac{D}2
+2} \!\right\} \! . \quad \label{dA}
\end{eqnarray}
\begin{table}

\vbox{\tabskip=0pt \offinterlineskip
\def\tablerule{\noalign{\hrule}}
\halign to390pt {\strut#& \vrule#\tabskip=1em plus2em& \hfil#&
\vrule#& \hfil#& \vrule#& \hfil#& \vrule#& \hfil#\hfil&
\vrule#\tabskip=0pt\cr
\tablerule
\omit&height4pt&\omit&&\omit&&\omit&&\omit&\cr &&\hidewidth {\rm I}
&&\hidewidth {\rm J} \hidewidth&& \hidewidth {\rm sub} \hidewidth&&
\hidewidth $iV_I^{\alpha\beta}(x) \, i[S](x;x') \, i
V_J^{\rho\sigma}(x') \, [\mbox{}_{\alpha\beta} T^A_{\rho\sigma}] \,
i\delta\!\Delta_A(x;x')$ \hidewidth&\cr
\omit&height4pt&\omit&&\omit&&\omit&&\omit&\cr
\tablerule
\omit&height2pt&\omit&&\omit&&\omit&&\omit&\cr && 1 && 1 && \omit &&
$\f{(D-1)}{(D-3)}\ka^2\del^{'}_{\mu}\{\hspace{-.1cm}\not{\hspace{-.1cm}\del}
i[S](x;x')\g^{\mu}i\d\!\D_{A}(x;x')\}$ & \cr
\omit&height2pt&\omit&&\omit&&\omit&&\omit&\cr
\tablerule
\omit&height2pt&\omit&&\omit&&\omit&&\omit&\cr && 1 && 2 && \omit &&
$\f{1}{(D-3)}\ka^2\del_{k}\{\hspace{-.1cm}\not{\hspace{-.1cm}\del}
i[S](x;x')\g_{k}i\d\!\D_{A}(x;x')\}$ & \cr
\omit&height2pt&\omit&&\omit&&\omit&&\omit&\cr
\tablerule
\omit&height2pt&\omit&&\omit&&\omit&&\omit&\cr && 1 && 3 && a &&
$-\f{(D-1)}{2(D-2)}\ka^2\hspace{-.1cm}\not{\hspace{-.1cm}\del}
i[S](x;x')\g^{0}\partial_{0}' i\d\!\D_{A}(x;x')$ & \cr
\omit&height2pt&\omit&&\omit&&\omit&&\omit&\cr
\tablerule
\omit&height2pt&\omit&&\omit&&\omit&&\omit&\cr && 1 && 3 && b &&
$\f{(D-2)}{2(D-3)}\ka^2\hspace{-.1cm}\not{\hspace{-.1cm}\del}
i[S](x;x')\hspace{-.1cm}\not{\hspace{-.1cm}\bar{\partial}}
i\d\!\D_{A}(x;x')$ & \cr
\omit&height2pt&\omit&&\omit&&\omit&&\omit&\cr
\tablerule
\omit&height2pt&\omit&&\omit&&\omit&&\omit&\cr && 2 && 1 && \omit &&
$-\f{1}{(D-3)}\ka^2\del^{'}_{\mu}\{\hspace{-.1cm}\not{\hspace{-.1cm}\bar{\del}}
i[S](x;x')\g^{\mu}i\d\!\D_{A}(x;x')\}$ & \cr
\omit&height2pt&\omit&&\omit&&\omit&&\omit&\cr
\tablerule
\omit&height2pt&\omit&&\omit&&\omit&&\omit&\cr && 2 && 2 && a && $
\f{1}{4}\ka^2\hspace{-.1cm} \not{\hspace{-.1cm}\bar{\del}}\{\del_{k}
i[S](x;x')\g_{k}i\d\!\D_{A}(x;x')\}$ & \cr
\omit&height2pt&\omit&&\omit&&\omit&&\omit&\cr
\tablerule
\omit&height2pt&\omit&&\omit&&\omit&&\omit&\cr && 2 && 2 && b && $ +
\f{1}{4}\ka^2\del_{\ell}\{\g_{k}\del_{\ell}
i[S](x;x')\g_{k}i\d\!\D_{A}(x;x') \}$ & \cr
\omit&height2pt&\omit&&\omit&&\omit&&\omit&\cr
\tablerule
\omit&height2pt&\omit&&\omit&&\omit&&\omit&\cr && 2 && 2 && c && $-
\f{1}{2 (D-3)}\ka^2\del_{k}\{\hspace{-.1cm}\not{\hspace{-.1cm}\bar{\del}}
i[S](x;x')\g_{k}i\d\!\D_{A}(x;x')\}$ & \cr
\omit&height2pt&\omit&&\omit&&\omit&&\omit&\cr
\tablerule
\omit&height2pt&\omit&&\omit&&\omit&&\omit&\cr && 2 && 3 && a &&
$\f1{2 (D-3)}\ka^2 \hspace{-.1cm}\not{\hspace{-.1cm}\bar{\del}}
i[S](x;x')\gamma^{0}\partial_{0}'i\d\!\D_{A}(x;x') $ & \cr
\omit&height2pt&\omit&&\omit&&\omit&&\omit&\cr
\tablerule
\omit&height2pt&\omit&&\omit&&\omit&&\omit&\cr && 2 && 3 && b &&
$-\f{1}{4}\ka^2\g_{k}\del_{\ell}i[S](x;x')
\g_{(k}\del_{\ell)}i\d\!\D_{A}(x;x') $ & \cr
\omit&height2pt&\omit&&\omit&&\omit&&\omit&\cr
\tablerule
\omit&height2pt&\omit&&\omit&&\omit&&\omit&\cr && 2 && 3 && c &&
$ -\f{1}{4(D-3)}\ka^2\hspace{-.1cm}\not{\hspace{-.1cm}\bar{\del}}
i[S](x;x')\hspace{-.1cm}\not{\hspace{-.1cm}\bar{\del}}
i\d\!\D_{A}(x;x')$&\cr
\omit&height2pt&\omit&&\omit&&\omit&&\omit&\cr
\tablerule
\omit&height2pt&\omit&&\omit&&\omit&&\omit&\cr && 3 && 1 && a &&
$\f{1}{2}(\f{D-1}{D-3})\ka^2\del^{'}_{\mu}\{
\gamma^{0}\partial_{0} i\d\!\D_{A}(x;x')i[S](x;x')\g^{\mu}\} $ &\cr
\omit&height2pt&\omit&&\omit&&\omit&&\omit&\cr
\tablerule
\omit&height2pt&\omit&&\omit&&\omit&&\omit&\cr && 3 && 1 && b &&
$\f{(D-2)}{2(D-3)}\ka^2\del^{'}_{\mu}\{\hspace{-.1cm}\not{\hspace{-.1cm}\bar{\del}}
i\d\!\D_{A}(x;x')i[S](x;x')\g^{\mu}\} $ & \cr
\omit&height2pt&\omit&&\omit&&\omit&&\omit&\cr
\tablerule
\omit&height2pt&\omit&&\omit&&\omit&&\omit&\cr && 3 && 2 && a &&
$\f{1}{2(D-3)}\ka^2\del_{k}\{\gamma^{0}\partial_{0}
i\d\!\D_{A}(x;x')i[S](x;x')\g_{k}\}$ & \cr
\omit&height2pt&\omit&&\omit&&\omit&&\omit&\cr
\tablerule
\omit&height2pt&\omit&&\omit&&\omit&&\omit&\cr && 3 && 2 && b &&
$\f{1}{4(D-3)}\ka^2\del_{k}\{\hspace{-.1cm}\not{\hspace{-.1cm}\bar{\del}}
i\d\!\D_{A}(x;x')i[S](x;x')\g_{k}\} $ & \cr
\omit&height2pt&\omit&&\omit&&\omit&&\omit&\cr
\tablerule
\omit&height2pt&\omit&&\omit&&\omit&&\omit&\cr && 3 && 2 && c &&
$\f{1}{8}\ka^2\hspace{-.1cm}\not{\hspace{-.1cm}\bar{\del}}\{i[S](x;x')
\hspace{-.1cm}\not{\hspace{-.1cm}\bar{\del}}i\d\!\D_{A}(x;x')\}$ &\cr \omit&height2pt&\omit&&\omit&&\omit&&\omit&\cr
\tablerule
\omit&height2pt&\omit&&\omit&&\omit&&\omit&\cr && 3 && 2 && d &&
$\f{1}{8}\ka^2\del_{k}\{\g_{\ell}
i[S](x;x')\g_{\ell}\del_{k}i\d\!\D_{A}(x;x')\}$ & \cr
\omit&height2pt&\omit&&\omit&&\omit&&\omit&\cr
\tablerule
\omit&height2pt&\omit&&\omit&&\omit&&\omit&\cr && 3 && 3 && a &&
$-\f{1}{4}(\f{D-1}{D-3})\ka^2\g^{0}i[S](x;x')\gamma^{0}
\del_{0}\partial_{0}' i\d\!\D_{A}(x;x') $ &\cr
\omit&height2pt&\omit&&\omit&&\omit&&\omit&\cr
\tablerule
\omit&height2pt&\omit&&\omit&&\omit&&\omit&\cr && 3 && 3 && b &&
$\f{(D-2)}{4(D-3)}\ka^2\g^{0}i[S](x;x')\del_{0}\hspace{-.1cm}
\not{\hspace{-.1cm}\bar{\del}}i\d\!\D_{A}(x;x') $ &\cr
\omit&height2pt&\omit&&\omit&&\omit&&\omit&\cr
\tablerule
\omit&height2pt&\omit&&\omit&&\omit&&\omit&\cr && 3 && 3 && c &&
$-\f{(D-2)}{4(D-3)}\ka^2\g_{k}i[S](x;x')\del_{k}
\gamma^{0}\partial_{0}'i\d\!\D_{A}(x;x') $ &\cr
\omit&height2pt&\omit&&\omit&&\omit&&\omit&\cr
\tablerule
\omit&height2pt&\omit&&\omit&&\omit&&\omit&\cr && 3 && 3 && d &&
$\f{3D-7}{16(D-3)}\ka^2\g_{k}i[S](x;x')\del_{k}
\hspace{-.1cm}\not{\hspace{-.1cm}\bar{\del}}i\d\!\D_{A}(x;x')$ & \cr
\omit&height2pt&\omit&&\omit&&\omit&&\omit&\cr
\tablerule
\omit&height2pt&\omit&&\omit&&\omit&&\omit&\cr && 3 && 3 && e &&
$-\f{1}{16}\ka^2\g_{k}i[S](x;x')\g_{k}\nabla^2 i\d\!\D_{A}(x;x')$& \cr \omit&height2pt&\omit&&\omit&&\omit&&\omit&\cr
\tablerule}}
\caption{Contractions from the $i\d\!\D_{A}$ part of the graviton
propagator-II}

\label{DAcon}

\end{table}

In $D\!=\!4$ the most singular contributions to (\ref{3ptloop}) have
the form, $i\delta\!\Delta_A/{\Delta x}^4$. Because the infinite
series terms in (\ref{dA}) behave like positive powers of $\Delta x^2$
these terms make integrable contributions to the quantum-corrected
Dirac equation (\ref{Diraceq1}). We can therefore take $D\!=\!4$ for
those terms, at which point all the infinite series terms drop.
Hence it is only necessary to keep the first line of (\ref{dA}) and
that is all we need to use.

The generic contraction from Table~\ref{DAcon0} which only consists of
one derivative acting on a propagator, the order m contributions
must be and could only be from the most singular part of the
fermion propagator. In reducing these terms the following
derivatives occur many times,
\begin{eqnarray}
\partial_i i \delta\!\Delta_A(x;x') & = & -\frac{H^2}{8 \pi^{\frac{D}2}} \,
\Gamma\Bigl(\frac{D}2\!+\!1\Bigr) (a a')^{2-\frac{D}2} \,
\frac{\Delta x^i}{
\Delta x^{D-2}} = -\partial_i' i \delta\!\Delta_A(x;x'),\label{DAd1}\\
\partial_0 i \delta\!\Delta_A(x;x') & = & \frac{H^2}{8 \pi^{\frac{D}2}}
\, \Gamma\Bigl(\frac{D}2\!+\!1\Bigr) (a a')^{2-\frac{D}2} \Biggl\{
\frac{\Delta \eta}{\Delta x^{D-2}} \!-\! \frac{a H}{2 \Delta
x^{D-4}} \Biggr\}
\nonumber \\
& & \hspace{5cm} + \frac{H^{D-2}}{2^D \pi^{\frac{D}2}} \frac{\Gamma(
D\!-\!1)}{\Gamma(\frac{D}2)} \, a H \; ,\label{DAd2} \qquad \\
\partial_0' i \delta\!\Delta_A(x;x') & = & \frac{H^2}{8 \pi^{\frac{D}2}}
\, \Gamma\Bigl(\frac{D}2\!+\!1\Bigr) (a a')^{2-\frac{D}2} \Biggl\{
-\frac{\Delta \eta}{\Delta x^{D-2}} \!-\! \frac{a' H}{2 \Delta
x^{D-4}} \Biggr\}
\nonumber \\
& & \hspace{5cm} + \frac{H^{D-2}}{2^D \pi^{\frac{D}2}}
\frac{\Gamma(D\!-\!1)}{ \Gamma(\frac{D}2)} \, a' H \; .\label{DAd3} \qquad
\end{eqnarray}
We also make use of following identities to simplify the contributions,
\begin{eqnarray}
&&\frac{\Delta\eta^2}{\Delta x^{2D\!-\!2}}=\frac{1}{4(D\!-\!2)(D\!-\!3)}
\partial_0^2\frac{1}{\Delta x^{2D\!-\!6}}-\frac{1}{2(D\!-\!2)}
\frac{1}{\Delta x^{2D\!-\!4}}\,,\label{id6}\\
&&\frac{\overline{\Delta x}^2}{\Delta x^{2D\!-\!2}}
=\frac{1}{4(D\!-\!2)(D\!-\!3)}\nabla^2\!\frac{1}{\Delta x^{2D\!-\!6}}
+\frac{(D\!-\!1)}{2(D\!-\!2)}\frac{1}{\Delta x^{2D\!-\!4}}\,,\label{id7}\\
&&\frac{\gamma^0\Delta\eta\gamma^k\Delta x_k}{\Delta x ^{2D\!-\!2}}
=\frac{-1}{4(D\!-\!2)(D\!-\!3)}\gamma^0\partial_0\!\!
\not{\hspace{-.1cm}\bar{\partial}}\frac{1}{\Delta x^{2D\!-\!6}}\,,\label{id8}\\
&&\frac{\Delta\eta}{\Delta x^{2D\!-\!4}}=\frac{1}{2(D\!-\!3)}\partial_0
\frac{1}{\Delta x^{2D\!-\!6}}\,\,,\,\,
\frac{\gamma^k\Delta x_k}{\Delta x^{2D\!-\!4}}=\frac{-1}{2(D\!-\!3)}
\!\!\not{\hspace{-.1cm}\bar{\partial}}\frac{1}{\Delta x^{2D\!-\!6}}\,.
\label{id9}
\end{eqnarray}
\begin{table}

\vbox{\tabskip=0pt \offinterlineskip
\def\tablerule{\noalign{\hrule}}
\halign to390pt {\strut#& \vrule#\tabskip=1em plus2em& \hfil#\hfil&
\vrule#& \hfil#\hfil& \vrule#&  \hfil#\hfil&
\vrule#\tabskip=0pt\cr
\tablerule
\omit&height4pt&\omit&&\omit&&\omit&\cr
&&${\rm (I-J)}_{\rm sub}$
&&$\frac{i\kappa^2H^2}{16\pi^{\frac{D}{2}}}
\frac{\Gamma(\frac{D}{2}+1)}{2}\frac{ma}{(D-4)}\delta^{D}(x-x')$
&&$\frac{i\kappa^2H^2}{16\pi^2}ma\delta^4(x-x')$ &\cr
\omit&height4pt&\omit&&\omit&&\omit&\cr
\tablerule
\omit&height4pt&\omit&&\omit&&\omit&\cr
&&${\rm (1-4)}$&&$\frac{32}{\sqrt{\pi}}\frac{H^{D-4}}{(D-3)}$
&&$-12 \ln(a)$ &\cr
\omit&height4pt&\omit&&\omit&&\omit&\cr
\tablerule
\omit&height4pt&\omit&&\omit&&\omit&\cr
&&${\rm (2-4)}$&&$-\mu^{D-4}\frac{(D-1)}{(D-3)^2}$&&$ 3\ln(a)$ &\cr
\omit&height4pt&\omit&&\omit&&\omit&\cr
\tablerule
\omit&height4pt&\omit&&\omit&&\omit&\cr
&&${\rm (3-4)_a}$&&$-\mu^{D-4}\frac{(D-1)}{2(D-3)^2}$&&$\frac32\ln(a)$ &\cr
\omit&height4pt&\omit&&\omit&&\omit&\cr
\tablerule
\omit&height4pt&\omit&&\omit&&\omit&\cr
&&${\rm (3-4)_b}$&&$-\mu^{D-4}\frac{(D-1)(D-2)}{2(D-3)^2}$&&$3\ln(a)$ &\cr
\omit&height4pt&\omit&&\omit&&\omit&\cr
\tablerule
\omit&height4pt&\omit&&\omit&&\omit&\cr
&&${\rm (4-1)}$&&$0$&&$9+6\ln(\frac{H^2}{4\mu^2})$ &\cr
\omit&height4pt&\omit&&\omit&&\omit&\cr
\tablerule
\omit&height4pt&\omit&&\omit&&\omit&\cr
&&${\rm (4-2)}$&&$0$&&$ 0$ &\cr
\omit&height4pt&\omit&&\omit&&\omit&\cr
\tablerule
\omit&height4pt&\omit&&\omit&&\omit&\cr
&&${\rm (4-3)_a}$&&$-\mu^{D-4}\frac{(D-1)}{2(D-3)^2}$&&$\frac32\ln(a)$ &\cr
\omit&height4pt&\omit&&\omit&&\omit&\cr
\tablerule
\omit&height4pt&\omit&&\omit&&\omit&\cr
&&${\rm (4-3)_b}$&&$-\mu^{D-4}\frac{(D-1)(D-2)}{2(D-3)^2}$&&$3\ln(a)$ &\cr
\omit&height4pt&\omit&&\omit&&\omit&\cr
\tablerule
\omit&height4pt&\omit&&\omit&&\omit&\cr
&&${\rm total}$&&$\frac{32}{\sqrt{\pi}}\frac{H^{D-4}}{(D-3)}
-\mu^{D-4}\frac{D(D-1)}{(D-3)^2}$
&&$9+6\ln(\frac{H^2}{4\mu^2})$ &\cr
\omit&height4pt&\omit&&\omit&&\omit&\cr
\tablerule}}

\caption{The local terms from $i\d\!\D_{A}\times i[S]_{\rm cf}$. }

\label{dDA1conf}

\end{table}
Note that the contraction (1-1) produces a delta function through
(\ref{fpeqn}) and picks up one divergent and one finite local term
which do not possesses any dimension-dependent powers of $\Delta x$.
We tabulate this kind of the result in Table~\ref{dDA1conf}. One might
already notice that the first two terms of (\ref{dA})\footnote{
$\pi\cot{(\frac{D\pi}{2})}=\frac{1}{\frac12(D\!-\!4)}$} and the final
two terms of (\ref{DAd2}) and (\ref{DAd3}) tend to cancel in $D=4$.
We indeed encounter this kind of cancelation entirely for the leading
divergent terms in contractions (4-1) and (4-2) and hence we present
the remaining finite results in Table~\ref{dDA2conf}. For the rest of
the contractions, the sum of the leading divergent contributions do
not vanish after we apply (\ref{id6}), (\ref{id7}), (\ref{id8}),
(\ref{id9}), (\ref{id3}), (\ref{id4}), (\ref{id5}), (\ref{ds}) and
(\ref{ds1}). Therefore we give the local terms in Table~\ref{dDA1conf}
and the finite nonlocal terms in Table~\ref{dDA2conf}.
After collecting all contributions from Table~\ref{dDA1conf} and
Table~\ref{dDA2conf} we obtain,
\begin{eqnarray}
&&-i\Bigl[\Sigma^{\rm idAcf}\Bigr](x;\!x')\!=\!
\frac{i\kappa^2H^2}{16\pi^{\frac{D}{2}}}\frac{\Gamma(\frac{D}{2}\!+\!1)}{2}
\frac{ma}{(D\!\!-\!4)}\Biggl[\frac{32}{\sqrt{\pi}}
\frac{H^{D\!-\!4}}{(D\!\!-\!3)}\!-\!\frac{\mu^{D\!-\!4}D(D\!\!-\!1)}
{(D\!\!-\!3)^2}\Biggr]\nonumber\\
&&\times\delta^D\!(x\!\!-\!\!x')\!+\!\frac{i\kappa^2H^2}{16\pi^2}
\Biggl[9+6\ln(\frac{H^2}{4\mu^2})\Biggr]ma\delta^4(x\!-\!x')\!+\!
\frac{\kappa^2H^2}{32\pi^4}ma\Biggl\{\Bigl[6\partial^2
\!\!-\!\!2\nabla^2\Bigr]\nonumber\\
&&\times\Biggl[\frac{\ln(\mu^2\Delta x^2)}{\Delta x^2}\Biggr]
+\Biggl[\frac32\partial_0^2\!-\!\frac12\gamma^0\!\partial_0
\!\!\not{\hspace{-.1cm}\bar{\partial}}-\Bigl(\frac92
+2\ln(\frac{H^2}{4\mu^2})\Bigr)\nabla^2\Biggr]
\frac{1}{\Delta x^2}\Biggr\}\,.\label{idAcf}
\end{eqnarray}
\begin{table}

\vbox{\tabskip=0pt \offinterlineskip
\def\tablerule{\noalign{\hrule}}
\halign to390pt {\strut#& \vrule#\tabskip=1em plus2em& \hfil#\hfil&
\vrule#& \hfil#\hfil& \vrule#& \hfil#\hfil& \vrule#& \hfil#\hfil&
\vrule#& \hfil#\hfil& \vrule#& \hfil#\hfil& \vrule#& \hfil#\hfil&
\vrule#\tabskip=0pt\cr
\tablerule
\omit&height4pt&\omit&&\omit&&\omit&&\omit&&\omit&&\omit&&\omit&\cr
&&$\!\!\!\!\scriptstyle{{\rm (I-J)}_{\rm sub}}\!\!\!\!\!\!\!\!$
&&$\!\!\!\!\partial^2\frac{\ln(\mu\Delta x)^2}{\Delta x^2}\!\!\!\!\!\!\!$
&&$\!\!\!\!\!\gamma^0\!\partial_0\!\!\not{\hspace{-.1cm}\bar{\partial}}
\frac{\ln(\mu\Delta x)^2}{\Delta x^2}\!\!\!\!\!\!\!$
&&$\!\!\!\!\!\nabla^2\frac{\ln(\mu\Delta x)^2}{\Delta x^2}\!\!\!\!\!\!\!$
&&$\!\!\!\!\partial_0^2\frac{1}{\Delta x^2}\!\!\!\!\!\!\!$
&&$\!\!\!\!\!\!\!\gamma^0\!\partial_0\!\!\not{\hspace{-.1cm}\bar{\partial}}
\frac{1}{\Delta x^2}\!\!\!\!\!\!\!\!$
&&$\!\!\!\!\!\!\!\nabla^2\!\frac{1}{\Delta x^2}\!\!\!\!\!\!\!\!$ &\cr
\omit&height4pt&\omit&&\omit&&\omit&&\omit&&\omit&&\omit&&\omit&\cr
\tablerule
\omit&height4pt&\omit&&\omit&&\omit&&\omit&&\omit&&\omit&&\omit&\cr
&&$\!\!\!\!\!\!\scriptstyle{{\rm (1-4)}}\!\!\!\!\!\!\!\!$
&&$\!\!\!\!\!0\!\!\!\!\!\!$
&&$\!\!\!\!\!\!\!\!0\!\!\!\!\!\!\!\!\!$
&&$\!\!\!\!\!\!\!\!0\!\!\!\!$
&&$\!\!\!\!0\!\!\!\!\!\!\!\!$
&&$\!\!\!\!\!\!\!0\!\!\!\!\!\!\!\!$
&&$\!\!\!\!\!\!\!0\!\!\!\!\!\!\!\!$ &\cr
\omit&height4pt&\omit&&\omit&&\omit&&\omit&&\omit&&\omit&&\omit&\cr
\tablerule
\omit&height4pt&\omit&&\omit&&\omit&&\omit&&\omit&&\omit&&\omit&\cr
&&$\!\!\!\!\!\!\scriptstyle{{\rm (2-4)}}\!\!\!\!\!\!\!\!$
&&$\!\!\!\!\!\frac34\!\!\!\!\!\!$
&&$\!\!\!\!\!\!\!\!-1\!\!\!\!\!\!\!\!\!$
&&$\!\!\!\!\!\!\!\!-1\!\!\!\!$
&&$\!\!\!\!0\!\!\!\!\!\!\!\!$
&&$\!\!\!\!\!\!\!-2-\ln(\frac{H^2}{4\mu^2})\!\!\!\!\!\!\!\!$
&&$\!\!\!\!\!\!\!-2-\ln(\frac{H^2}{4\mu^2})\!\!\!\!\!\!\!\!$ &\cr
\omit&height4pt&\omit&&\omit&&\omit&&\omit&&\omit&&\omit&&\omit&\cr
\tablerule
\omit&height4pt&\omit&&\omit&&\omit&&\omit&&\omit&&\omit&&\omit&\cr
&&$\!\!\!\!\!\!\scriptstyle{{\rm (3-4)}_{\rm a}}\!\!\!\!\!\!\!\!$
&&$\!\!\!\!\!\frac38\!\!\!\!\!\!$
&&$\!\!\!\!\!\!\!\!0\!\!\!\!\!\!\!\!\!$
&&$\!\!\!\!\!\!\!\!0\!\!\!\!$
&&$\!\!\!\!\frac34\!\!\!\!\!\!\!\!$
&&$\!\!\!\!\!\!\!\frac34\!\!\!\!\!\!\!\!$
&&$\!\!\!\!\!\!\!0\!\!\!\!\!\!\!\!$ &\cr
\omit&height4pt&\omit&&\omit&&\omit&&\omit&&\omit&&\omit&&\omit&\cr
\tablerule
\omit&height4pt&\omit&&\omit&&\omit&&\omit&&\omit&&\omit&&\omit&\cr
&&$\!\!\!\!\!\!\scriptstyle{{\rm (3-4)}_{\rm b}}\!\!\!\!\!\!\!\!$
&&$\!\!\!\!\!\frac34\!\!\!\!\!\!$
&&$\!\!\!\!\!\!\!\!0\!\!\!\!\!\!\!\!\!$
&&$\!\!\!\!\!\!\!\!0\!\!\!\!$
&&$\!\!\!\!0\!\!\!\!\!\!\!\!$
&&$\!\!\!\!\!\!\!-\frac12\!\!\!\!\!\!\!\!$
&&$\!\!\!\!\!\!\!-\frac12\!\!\!\!\!\!\!\!$ &\cr
\omit&height4pt&\omit&&\omit&&\omit&&\omit&&\omit&&\omit&&\omit&\cr
\tablerule
\omit&height4pt&\omit&&\omit&&\omit&&\omit&&\omit&&\omit&&\omit&\cr
&&$\!\!\!\!\!\!\scriptstyle{{\rm (4-1)}}\!\!\!\!\!\!\!\!$
&&$\!\!\!\!\!3\!\!\!\!\!\!$
&&$\!\!\!\!\!\!\!\!0\!\!\!\!\!\!\!\!\!$
&&$\!\!\!\!\!\!\!\!0\!\!\!\!$
&&$\!\!\!\!0\!\!\!\!\!\!\!\!$
&&$\!\!\!\!\!\!\!0\!\!\!\!\!\!\!\!$
&&$\!\!\!\!\!\!\!0\!\!\!\!\!\!\!\!$ &\cr
\omit&height4pt&\omit&&\omit&&\omit&&\omit&&\omit&&\omit&&\omit&\cr
\tablerule
\omit&height4pt&\omit&&\omit&&\omit&&\omit&&\omit&&\omit&&\omit&\cr
&&$\!\!\!\!\!\!\scriptstyle{{\rm (4-2)}}\!\!\!\!\!\!\!\!$
&&$\!\!\!\!\!0\!\!\!\!\!\!$
&&$\!\!\!\!\!\!\!\!1\!\!\!\!\!\!\!\!\!$
&&$\!\!\!\!\!\!\!\!-1\!\!\!\!$
&&$\!\!\!\!0\!\!\!\!\!\!\!\!$
&&$\!\!\!\!\!\!\!\frac32+\ln(\frac{H^2}{4\mu^2})\!\!\!\!\!\!\!\!$
&&$\!\!\!\!\!\!\!-\frac32-\ln(\frac{H^2}{4\mu^2})\!\!\!\!\!\!\!\!$ &\cr
\omit&height4pt&\omit&&\omit&&\omit&&\omit&&\omit&&\omit&&\omit&\cr
\tablerule
\omit&height4pt&\omit&&\omit&&\omit&&\omit&&\omit&&\omit&&\omit&\cr
&&$\!\!\!\!\!\!\scriptstyle{{\rm (4-3)}_{\rm a}}\!\!\!\!\!\!\!\!$
&&$\!\!\!\!\!\frac38\!\!\!\!\!\!$
&&$\!\!\!\!\!\!\!\!0\!\!\!\!\!\!\!\!\!$
&&$\!\!\!\!\!\!\!\!0\!\!\!\!$
&&$\!\!\!\!\frac34\!\!\!\!\!\!\!\!$
&&$\!\!\!\!\!\!\!-\frac34\!\!\!\!\!\!\!\!$
&&$\!\!\!\!\!\!\!0\!\!\!\!\!\!\!\!$ &\cr
\omit&height4pt&\omit&&\omit&&\omit&&\omit&&\omit&&\omit&&\omit&\cr
\tablerule
\omit&height4pt&\omit&&\omit&&\omit&&\omit&&\omit&&\omit&&\omit&\cr
&&$\!\!\!\!\!\!\scriptstyle{{\rm (4-3)}_{\rm b}}\!\!\!\!\!\!\!\!$
&&$\!\!\!\!\!\frac34\!\!\!\!\!\!$
&&$\!\!\!\!\!\!\!\!0\!\!\!\!\!\!\!\!\!$
&&$\!\!\!\!\!\!\!\!0\!\!\!\!$
&&$\!\!\!\!0\!\!\!\!\!\!\!\!$
&&$\!\!\!\!\!\!\!\frac12\!\!\!\!\!\!\!\!$
&&$\!\!\!\!\!\!\!-\frac12\!\!\!\!\!\!\!\!$ &\cr
\omit&height4pt&\omit&&\omit&&\omit&&\omit&&\omit&&\omit&&\omit&\cr
\tablerule
\omit&height4pt&\omit&&\omit&&\omit&&\omit&&\omit&&\omit&&\omit&\cr
&&$\!\!\!\!\!\!\scriptstyle{{\rm total}}\!\!\!\!\!\!\!\!$
&&$\!\!\!\!\!6\!\!\!\!\!\!$
&&$\!\!\!\!\!\!\!\!0\!\!\!\!\!\!\!\!\!$
&&$\!\!\!\!\!\!\!\!-2\!\!\!\!$
&&$\!\!\!\!\frac32\!\!\!\!\!\!\!\!$
&&$\!\!\!\!\!\!\!-\frac12\!\!\!\!\!\!\!\!$
&&$\!\!\!\!\!\!\!-\frac92-2\ln(\frac{H^2}{4\mu^2})\!\!\!\!\!\!\!\!$ &\cr
\omit&height4pt&\omit&&\omit&&\omit&&\omit&&\omit&&\omit&&\omit&\cr
\tablerule}}

\caption{The non-local terms from $i\d\!\D_{A}\times i[S]_{\rm cf}$.
All contributions are multiplied by $\frac{i\kappa^2 H^2}{32\pi^4}ma$. }

\label{dDA2conf}

\end{table}

\begin{table}

\vbox{\tabskip=0pt \offinterlineskip
\def\tablerule{\noalign{\hrule}}
\halign to390pt {\strut#& \vrule#\tabskip=1em plus2em& \hfil#\hfil&
\vrule#& \hfil#\hfil& \vrule#& \hfil#\hfil& \vrule#& \hfil#\hfil&
\vrule#& \hfil#\hfil& \vrule#& \hfil#\hfil&\vrule#\tabskip=0pt\cr
\tablerule
\omit&height4pt&\omit&&\omit&&\omit&&\omit&&\omit&&\omit&\cr
&&$\!\!\!\!{\rm (I-J)}\!\!\!\!\!\!\!\!$
&&$\!\!\!\!\partial^2\frac{\ln(\mu\Delta x)^2}{\Delta x^2}\!\!\!\!\!\!\!$
&&$\!\!\!\!\!\gamma^0\!\partial_0\!\!\not{\hspace{-.1cm}\bar{\partial}}
\frac{\ln(\mu\Delta x)^2}{\Delta x^2}\!\!\!\!\!\!\!$
&&$\!\!\!\!\!\nabla^2\frac{\ln(\mu\Delta x)^2}{\Delta x^2}\!\!\!\!\!\!\!$
&&$\!\!\!\!\!\!\!H\!a\partial_0\frac{\ln(\mu\Delta x)^2}
{\Delta x^2}\!\!\!\!\!\!\!\!$
&&$\!\!\!\!\!\!\!H\!a\gamma^0\!\!\not{\hspace{-.1cm}\bar{\partial}}
\frac{\ln(\mu\Delta x)^2}{\Delta x^2}\!\!\!\!\!\!\!\!$ &\cr
\omit&height4pt&\omit&&\omit&&\omit&&\omit&&\omit&&\omit&\cr
\tablerule
\omit&height4pt&\omit&&\omit&&\omit&&\omit&&\omit&&\omit&\cr
&&$\!\!\!\!\!\!\rm{(1-1)}\!\!\!\!\!\!\!\!$
&&$\!\!\!\!\!-3\!\!\!\!\!\!$
&&$\!\!\!\!\!\!\!\!0\!\!\!\!\!\!\!\!\!$
&&$\!\!\!\!\!\!\!\!0\!\!\!\!$
&&$\!\!\!\!3\!\!\!\!\!\!\!\!$
&&$\!\!\!\!\!\!\!3\!\!\!\!\!\!\!\!$ &\cr
\omit&height4pt&\omit&&\omit&&\omit&&\omit&&\omit&&\omit&\cr
\tablerule
\omit&height4pt&\omit&&\omit&&\omit&&\omit&&\omit&&\omit&\cr
&&$\!\!\!\!\!\!\rm{(1-2)}\!\!\!\!\!\!\!\!$
&&$\!\!\!\!\!0\!\!\!\!\!\!$
&&$\!\!\!\!\!\!\!\!-1\!\!\!\!\!\!\!\!\!$
&&$\!\!\!\!\!\!\!\!1\!\!\!\!$
&&$\!\!\!\!0\!\!\!\!\!\!\!\!$
&&$\!\!\!\!\!\!\!-1\!\!\!\!\!\!\!\!$ &\cr
\omit&height4pt&\omit&&\omit&&\omit&&\omit&&\omit&&\omit&\cr
\tablerule
\omit&height4pt&\omit&&\omit&&\omit&&\omit&&\omit&&\omit&\cr
&&$\!\!\!\!\!\!\rm{(2-1)}\!\!\!\!\!\!\!\!$
&&$\!\!\!\!\!0\!\!\!\!\!\!$
&&$\!\!\!\!\!\!\!\!1\!\!\!\!\!\!\!\!\!$
&&$\!\!\!\!\!\!\!\!1\!\!\!\!$
&&$\!\!\!\!0\!\!\!\!\!\!\!\!$
&&$\!\!\!\!\!\!\!0\!\!\!\!\!\!\!\!$ &\cr
\omit&height4pt&\omit&&\omit&&\omit&&\omit&&\omit&&\omit&\cr
\tablerule
\omit&height4pt&\omit&&\omit&&\omit&&\omit&&\omit&&\omit&\cr
&&$\!\!\!\!\!\!\rm{(2-2)}\!\!\!\!\!\!\!\!$
&&$\!\!\!\!\!0\!\!\!\!\!\!$
&&$\!\!\!\!\!\!\!\!0\!\!\!\!\!\!\!\!\!$
&&$\!\!\!\!\!\!\!\!\frac12\!\!\!\!$
&&$\!\!\!\!0\!\!\!\!\!\!\!\!$
&&$\!\!\!\!\!\!\!0\!\!\!\!\!\!\!\!$ &\cr
\omit&height4pt&\omit&&\omit&&\omit&&\omit&&\omit&&\omit&\cr
\tablerule
\omit&height4pt&\omit&&\omit&&\omit&&\omit&&\omit&&\omit&\cr
&&$\!\!\!\!\!\!\rm{total}\!\!\!\!\!\!\!\!$
&&$\!\!\!\!\!-3\!\!\!\!\!\!$
&&$\!\!\!\!\!\!\!\!0\!\!\!\!\!\!\!\!\!$
&&$\!\!\!\!\!\!\!\!\frac52\!\!\!\!$
&&$\!\!\!\!3\!\!\!\!\!\!\!\!$
&&$\!\!\!\!\!\!\!2\!\!\!\!\!\!\!\!$ &\cr
\omit&height4pt&\omit&&\omit&&\omit&&\omit&&\omit&&\omit&\cr
\tablerule}}

\caption{$i\d\!\D_{A}\times i[S]_{\rm fm}$.
All contributions are multiplied by $\frac{\kappa^2 H^2}{32\pi^4}ma$. }

\label{dDAfm1}

\end{table}
The generic contractions from Table~\ref{DAcon}, which are comprised
of two derivatives acting on the propagators, the order m contributions
could either come from the flat spacetime mass term or from the
infinite series expansion of the fermion propagator.

We first deal with the contributions from the flat spacetime mass term of
the fermion propagator. At $D=4$ these terms have a dimension
$\frac{i\delta\Delta_A}{\Delta x^4}$ which are not integrable when we substitute
back to the quantum-corrected Dirac equation and working on an arbitrary $D$
is required. The contractions (1-1), (1-2), (2-1) and (2-2) tend to cancel
through (\ref{dA}). It turns out that the sum of divergent terms vanish and
only left the finite terms. Hence we present the results for $D=4$ at the end.
Here we showed (2-1) as an example,
\begin{eqnarray}
&&\frac{-\kappa^2}{(D\!\!-\!\!3)}\partial'_{\mu}\Bigl\{
\not{\hspace{-.1cm}\bar{\partial}}i[S]\gamma^{\mu}
i\delta\Delta_{A}\Bigr\}\!=\!\frac{\kappa^2\!H^2}{32\pi^D}
\frac{\Gamma(\frac{D}{2})\Gamma(\frac{D}{2}\!\!+\!\!1)}{(D\!-\!3)}
ma\partial'_{\mu}\Biggl\{\frac{(aa')^{2-\frac{D}{2}}}
{\frac12(D\!\!-\!\!4)}\nonumber\\
&&\frac{\gamma^k\Delta x_{k}\gamma^{\mu}}{\Delta x^{2D-4}}\Biggr\}
\!+\!\frac{\kappa^2\!H^{D-2}}{2^{D+1}\pi^D}
\frac{\Gamma(D\!\!-\!\!1)}{(D\!\!-\!\!3)}ma\partial'_{\mu}\Biggl\{
\Bigl[\frac{-2}{(D\!\!-\!\!4)}\!+\!\ln(aa')\Bigr]
\frac{\gamma^k\Delta x_{k}\gamma^{\mu}}{\Delta x^D}\Biggr\}\nonumber\\
&&=\!\frac{\kappa^2\!H^2}{32\pi^D}\frac{\mu^{D\!-\!4}\!ma}{(D\!\!-\!\!3)}
\Biggl\{\Biggl[\frac{2}{(D\!\!-\!\!4)}\!-\!\frac{2}{(D\!\!-\!\!4)}
\!-\!\frac12\!\!-\!\ln\!\!\frac{H^2}{4\mu^2}\!-\!1\!\Biggr]\!\!\!
\not{\hspace{-.1cm}\bar{\partial}}\!\!\not{\hspace{-.1cm}\partial}
\frac{1}{\Delta x^2}-\!\not{\hspace{-.1cm}\bar{\partial}}\!\!
\not{\hspace{-.1cm}\partial}\frac{\ln(\mu^2\!\Delta x^2)}
{\Delta x^2}\!\!\Biggr\}\nonumber\\
&&\hspace{2.5cm}=\!\frac{\kappa^2\!H^2}{32\pi^4}ma\Biggl\{
\Biggl[-\frac32\!-\!\ln\!\!\frac{H^2}{4\mu^2}\Biggr]\!\!
\not{\hspace{-.1cm}\bar{\partial}}\!\!\not{\hspace{-.1cm}\partial}
\frac{1}{\Delta x^2}-\not{\hspace{-.1cm}\bar{\partial}}\!\!
\not{\hspace{-.1cm}\partial}\frac{\ln(\mu^2\!\Delta x^2)}
{\Delta x^2}\Biggr\}\,.
\end{eqnarray}
The most singular terms in the contractions ${\rm(2\!\!-\!\!3)_a}$,
${\rm(3\!\!-\!\!1)}$ and ${\rm(3\!\!-\!\!2)}$ are integrable
after exacting out derivatives\footnote{Some less singular terms
in the contractions ${\rm(2\!\!-\!\!3)_a}$, ${\rm(3\!\!-\!\!1)_a}$
and $\rm{(3\!\!-\!\!2)_a}$ get canceled through the property of
(\ref{DAd2}) and (\ref{DAd3}).}. Because the contraction (3-3)
involves various double derivatives directly acting on
$i\delta\Delta_A$ we would give their expressions here,
\begin{eqnarray}
&&\partial_0\partial^{\prime}_0 i\delta\Delta_A\!=\!
\frac{H^2\Gamma(\frac{D}{2}\!+\!1)}
{8\pi^{\frac{D}{2}}(aa')^{\frac{D}{2}-2}}
\Biggl\{\frac{-1}{\Delta x^{D-2}}\!-\!\frac{(D\!-\!2)\Delta\eta^2}
{\Delta x^D}\!+\!\frac12(D\!-\!4)\nonumber\\
&&\hspace{5.5cm}\times\Biggl[\frac{H^2aa'\Delta\eta^2}
{\Delta x^{D-2}}\!+\!\frac12\frac{H^2aa'}{\Delta x^{D-4}}\Biggr]
\Biggr\},\label{DAdd1}\\
&&\hspace{-.5cm}\partial_k\partial_0i\delta\Delta_A\!=\!
\frac{H^2\Gamma(\frac{D}{2}\!+\!1)}
{8\pi^{\frac{D}{2}}(aa')^{\frac{D}{2}-2}}\Biggl\{
\frac{-(D\!-\!2)\Delta\eta\Delta x_k}{\Delta x^{D}}\!+\!\frac12(D\!-\!4)
\frac{Ha\Delta x_k}{\Delta x^{D-2}}\Biggr\},\label{DAdd2}\\
&&\hspace{-.5cm}\partial_k\partial^{\prime}_0 i\delta\Delta_A\!=\!
\frac{H^2\Gamma(\frac{D}{2}\!+\!1)}
{8\pi^{\frac{D}{2}}(aa')^{\frac{D}{2}-2}}\Biggl\{
\frac{(D\!-\!2)\Delta\eta\Delta x_k}{\Delta x^{D}}\!+\!\frac12(D\!-\!4)
\frac{Ha'\Delta x_k}{\Delta x^{D-2}}\Biggr\},\label{DAdd3}\\
&&\partial_k\partial_l i\delta\Delta_A\!=\!
\frac{H^2\Gamma(\frac{D}{2}\!+\!1)}
{8\pi^{\frac{D}{2}}(aa')^{\frac{D}{2}-2}}
\Biggl\{\frac{(D\!-\!2)\Delta x_k\Delta x_l}{\Delta x^{D}}
\!-\!\frac{\delta_{kl}}{\Delta x^{D-2}}\Biggl\}.\label{DAdd4}
\end{eqnarray}
\begin{table}

\vbox{\tabskip=0pt \offinterlineskip
\def\tablerule{\noalign{\hrule}}
\halign to390pt {\strut#& \vrule#\tabskip=1em plus2em& \hfil#\hfil&
\vrule#& \hfil#\hfil& \vrule#& \hfil#\hfil& \vrule#& \hfil#\hfil&
\vrule#& \hfil#\hfil& \vrule#& \hfil#\hfil&\vrule#\tabskip=0pt\cr
\tablerule
\omit&height4pt&\omit&&\omit&&\omit&&\omit&&\omit&&\omit&\cr
&&$\!\!\!\!{\rm (I-J)_{\rm sub}}\!\!\!\!\!\!\!\!$
&&$\!\!\!\!\partial^2\frac{1}{\Delta x^2}\!\!\!\!\!\!\!$
&&$\!\!\!\!\!\gamma^0\!\partial_0\!\!\not{\hspace{-.1cm}\bar{\partial}}
\frac{1}{\Delta x^2}\!\!\!\!\!\!\!$
&&$\!\!\!\!\!\nabla^2\frac{1}{\Delta x^2}\!\!\!\!\!\!\!$
&&$\!\!\!\!\!\!\!H\!a\partial_0\frac{1}{\Delta x^2}\!\!\!\!\!\!\!\!$
&&$\!\!\!\!\!\!\!H\!a\gamma^0\!\!\not{\hspace{-.1cm}\bar{\partial}}
\frac{1}{\Delta x^2}\!\!\!\!\!\!\!\!$ &\cr
\omit&height4pt&\omit&&\omit&&\omit&&\omit&&\omit&&\omit&\cr
\tablerule
\omit&height4pt&\omit&&\omit&&\omit&&\omit&&\omit&&\omit&\cr
&&$\!\!\!\!\!\!\rm{(1-1)}\!\!\!\!\!\!\!\!$
&&$\!\!\!\!\!-\frac32\!-\!3\ln\!\!\frac{H^2}{4\mu^2}\!\!\!\!\!\!$
&&$\!\!\!\!\!\!\!\!0\!\!\!\!\!\!\!\!\!$
&&$\!\!\!\!\!\!\!\!0\!\!\!\!$
&&$\!\!\!\!\!\frac32\!+\!3\!\ln\!\!\frac{H^2}{4\mu^2}\!\!\!\!\!\!\!$
&&$\!\!\!\!\!\!\!\frac32\!+\!3\!\ln\!\!\frac{H^2}{4\mu^2}\!\!\!\!\!\!\!\!$ &\cr
\omit&height4pt&\omit&&\omit&&\omit&&\omit&&\omit&&\omit&\cr
\tablerule
\omit&height4pt&\omit&&\omit&&\omit&&\omit&&\omit&&\omit&\cr
&&$\!\!\!\!\!\!\rm{(1-2)}\!\!\!\!\!\!\!\!$
&&$\!\!\!\!\!0\!\!\!\!\!\!$
&&$\!\!\!\!\!\!\!\!-\frac32\!-\!\ln\!\!\frac{H^2}{4\mu^2}\!\!\!\!\!\!\!\!\!$
&&$\!\!\!\!\!\frac32\!+\!\ln\!\!\frac{H^2}{4\mu^2}\!\!\!\!\!\!\!$
&&$\!\!\!\!0\!\!\!\!\!\!\!\!$
&&$\!\!\!\!\!\!\!-\frac12\!-\!\ln\!\!\frac{H^2}{4\mu^2}\!\!\!\!\!\!\!\!$ &\cr
\omit&height4pt&\omit&&\omit&&\omit&&\omit&&\omit&&\omit&\cr
\tablerule
\omit&height4pt&\omit&&\omit&&\omit&&\omit&&\omit&&\omit&\cr
&&$\!\!\!\!\!\!\rm{(2-1)}\!\!\!\!\!\!\!\!$
&&$\!\!\!\!\!0\!\!\!\!\!\!$
&&$\!\!\!\!\!\!\!\!\frac32\!+\!\ln\!\!\frac{H^2}{4\mu^2}\!\!\!\!\!\!\!\!\!$
&&$\!\!\!\!\!\frac32\!+\!\ln\!\!\frac{H^2}{4\mu^2}\!\!\!\!\!\!\!$
&&$\!\!\!\!0\!\!\!\!\!\!\!\!$
&&$\!\!\!\!\!\!\!0\!\!\!\!\!\!\!\!$ &\cr
\omit&height4pt&\omit&&\omit&&\omit&&\omit&&\omit&&\omit&\cr
\tablerule
\omit&height4pt&\omit&&\omit&&\omit&&\omit&&\omit&&\omit&\cr
&&$\!\!\!\!\!\!\rm{(2-2)}\!\!\!\!\!\!\!\!$
&&$\!\!\!\!\!0\!\!\!\!\!\!$
&&$\!\!\!\!\!\!\!\!0\!\!\!\!\!\!\!\!\!$
&&$\!\!\!\!\!\!\frac34\!+\!\frac12\!\ln\!\!\frac{H^2}{4\mu^2}\!\!\!\!\!\!$
&&$\!\!\!\!0\!\!\!\!\!\!\!\!$
&&$\!\!\!\!\!\!\!0\!\!\!\!\!\!\!\!$ &\cr
\omit&height4pt&\omit&&\omit&&\omit&&\omit&&\omit&&\omit&\cr
\tablerule
\omit&height4pt&\omit&&\omit&&\omit&&\omit&&\omit&&\omit&\cr
&&$\!\!\!\!\!\!{\rm (2-3)_a }\!\!\!\!\!\!\!\!$
&&$\!\!\!\!\!0\!\!\!\!\!\!$
&&$\!\!\!\!\!\!\!\!\frac14\!\!\!\!\!\!\!\!\!$
&&$\!\!\!\!\!\!0\!\!\!\!\!\!$
&&$\!\!\!\!0\!\!\!\!\!\!\!\!$
&&$\!\!\!\!\!\!\!0\!\!\!\!\!\!\!\!$ &\cr
\omit&height4pt&\omit&&\omit&&\omit&&\omit&&\omit&&\omit&\cr
\tablerule
\omit&height4pt&\omit&&\omit&&\omit&&\omit&&\omit&&\omit&\cr
&&$\!\!\!\!\!\!{\rm (3-1)_a }\!\!\!\!\!\!\!\!$
&&$\!\!\!\!\!\frac32\!\!\!\!\!\!$
&&$\!\!\!\!\!\!\!\!-\frac32\!\!\!\!\!\!\!\!\!$
&&$\!\!\!\!\!\!-\frac32\!\!\!\!\!\!$
&&$\!\!\!\!0\!\!\!\!\!\!\!\!$
&&$\!\!\!\!\!\!\!0\!\!\!\!\!\!\!\!$ &\cr
\omit&height4pt&\omit&&\omit&&\omit&&\omit&&\omit&&\omit&\cr
\tablerule
\omit&height4pt&\omit&&\omit&&\omit&&\omit&&\omit&&\omit&\cr
&&$\!\!\!\!\!\!{\rm (3-1)_b }\!\!\!\!\!\!\!\!$
&&$\!\!\!\!\!0\!\!\!\!\!\!$
&&$\!\!\!\!\!\!\!\!-1\!\!\!\!\!\!\!\!\!$
&&$\!\!\!\!\!\!-1\!\!\!\!\!\!$
&&$\!\!\!\!0\!\!\!\!\!\!\!\!$
&&$\!\!\!\!\!\!\!0\!\!\!\!\!\!\!\!$ &\cr
\omit&height4pt&\omit&&\omit&&\omit&&\omit&&\omit&&\omit&\cr
\tablerule
\omit&height4pt&\omit&&\omit&&\omit&&\omit&&\omit&&\omit&\cr
&&$\!\!\!\!\!\!{\rm (3-2) }\!\!\!\!\!\!\!\!$
&&$\!\!\!\!\!0\!\!\!\!\!\!$
&&$\!\!\!\!\!\!\!\!\frac12\!\!\!\!\!\!\!\!\!$
&&$\!\!\!\!\!\!-\frac34\!\!\!\!\!\!$
&&$\!\!\!\!0\!\!\!\!\!\!\!\!$
&&$\!\!\!\!\!\!\!0\!\!\!\!\!\!\!\!$ &\cr
\omit&height4pt&\omit&&\omit&&\omit&&\omit&&\omit&&\omit&\cr
\tablerule
\omit&height4pt&\omit&&\omit&&\omit&&\omit&&\omit&&\omit&\cr
&&$\!\!\!\!\!\!{\rm (3-3)_{b+c} }\!\!\!\!\!\!\!\!$
&&$\!\!\!\!\!0\!\!\!\!\!\!$
&&$\!\!\!\!\!\!\!\!\frac14-\frac14\!\!\!\!\!\!\!\!\!$
&&$\!\!\!\!\!\!0\!\!\!\!\!\!$
&&$\!\!\!\!0\!\!\!\!\!\!\!\!$
&&$\!\!\!\!\!\!\!0\!\!\!\!\!\!\!\!$ &\cr
\omit&height4pt&\omit&&\omit&&\omit&&\omit&&\omit&&\omit&\cr
\tablerule
\omit&height4pt&\omit&&\omit&&\omit&&\omit&&\omit&&\omit&\cr
&&$\!\!\!\!\!\!{\rm total}\!\!\!\!\!\!\!\!$
&&$\!\!\!\!\!-3\!\ln\!\!\frac{H^2}{4\mu^2}\!\!\!\!\!\!$
&&$\!\!\!\!\!\!\!\!-\frac74\!\!\!\!\!\!\!\!\!$
&&$\!\!\!\!\!\!\frac12\!+\!\frac52\!\ln\!\!\frac{H^2}{4\mu^2}\!\!\!\!\!\!$
&&$\!\!\!\!\!\!\frac32\!+\!3\!\ln\!\!\frac{H^2}{4\mu^2}\!\!\!\!\!\!$
&&$\!\!\!\!\!\!\!1\!+\!2\!\ln\!\!\frac{H^2}{4\mu^2}\!\!\!\!\!\!\!\!$ &\cr
\omit&height4pt&\omit&&\omit&&\omit&&\omit&&\omit&&\omit&\cr
\tablerule}}

\caption{$i\d\!\D_{A}\times i[S]_{\rm fm}$.
All contributions are multiplied by $\frac{\kappa^2 H^2}{32\pi^4}ma$. }

\label{dDAfm2}

\end{table}
Note that the terms with a factor of $(D\!-\!4)$ in the equations
(\ref{DAdd1})-(\ref{DAdd3}) have dimensionality, which is either
$\frac{1}{\Delta^{D-4}}$ or $\frac{1}{\Delta x^{D-3}}$. When they
combine with $i[S]_{\rm fm}$ whose dimensionality is $\frac{1}{\Delta x^{D-2}}$,
one can see that those contributions are integrable in four dimensions
and hence could only give the contributions of the order $(D\!-\!4)$.
Therefore we can drop the terms we mentioned above from (\ref{DAdd1})
to (\ref{DAdd3}) when we compute the contraction (3-3). In addition, the
contractions $(3\!-\!3)_{\rm b}$ and $(3\!-\!3)_{\rm c}$ are both finite
in $D=4$ dimensions after performing the partial integration\footnote{
``Partial integration'' is not standard usage in Physics. Here
we mean extracting the derivatives outside the quantum-corrected Dirac
equation.}. The finite contributions mentioned above are displayed
in Table~\ref{dDAfm1} and Table~\ref{dDAfm2}
and they can be read off immediately,
\begin{eqnarray}
&&\frac{\kappa^2\!H^2\!ma}{32\pi^4}\Biggl\{\Bigl[-3\partial^2\!
\!+\!\frac52\!\nabla^2\!+\!3H\!a\partial_0\!+\!2H\!a\gamma^0
\!\!\not{\hspace{-.1cm}\bar{\partial}}\Bigr]
\frac{\ln\mu^2\!\Delta x^2}{\Delta x^2}\!+\!\Biggl[-3\ln\!
\frac{H^2}{4\mu^2}\partial^2\!-\!\frac74\gamma^0\partial_0\!\!
\not{\hspace{-.1cm}\bar{\partial}}\nonumber\\
&&+\Bigl(\frac12\!+\!\frac52\ln\!\frac{H^2}{4\mu^2}\Bigr)\nabla^2\!
\!+\!\Bigl(\frac32\!+\!3\ln\!\frac{H^2}{4\mu^2}\Bigr)H\!a\partial_0
\!+\!\Bigl(1\!+\!2\ln\!\frac{H^2}{4\mu^2}\Bigr)H\!a\gamma^0\!\!
\not{\hspace{-.1cm}\bar{\partial}}\Biggr]\!\frac{1}{\Delta x^2}
\Biggr\}.\label{dDAfm12}
\end{eqnarray}

The rest of the contractions which require further renormalization are
summarized in Table~\ref{dDAfm3}.
We then apply the same formalism to partially integrate, extract
the local divergences and take $D\!=\!4$ for the remaining, integrable
and ultraviolet finite nonlocal terms. The sub-total from
Table~\ref{dDAfm3} could be obtained,
\begin{eqnarray}
\lefteqn{\frac{i\kappa^2\!H^2}{64\pi^{\frac{D}{2}}}\Gamma\!(\frac{D}{2}\!+\!1)
\frac{\mu^{D\!-\!4}(\!D\!-\!1)(D^3\!\!-\!9D^2\!\!+\!20D\!-\!4)}
{8(D\!-\!2)(D\!-\!3)^2(D\!-\!4)}ma\delta^D\!(x\!-\!x')}\nonumber\\
&&+\frac{i\kappa^2\!H^2}{64\pi^2}\frac{3ma}{2}\ln\!a\delta^4\!(x\!-\!x')
+\frac{\kappa^2\!H^2 ma}{32\pi^4}\Biggl\{\frac{3}{32}
\partial^2\frac{\ln\mu^2\!\Delta x^2}{\Delta x^2}\nonumber\\
&&\hspace{2cm}+\Biggl[\frac34\partial_0^2\!+\!
\frac{1}{8}\!\gamma^0\!\partial_0\!\!\not{\hspace{-.1cm}\bar{\partial}}
\!-\!\frac{3}{16}\!\nabla^2\!+\!\frac34\!H\!a\partial_0
\!+\!H\!a\gamma^0\!\!\not{\hspace{-.1cm}\bar{\partial}}\Biggr]
\!\frac{1}{\Delta x^2}\Biggr\}.\label{re-dDAfm3}
\end{eqnarray}
\begin{table}

\vbox{\tabskip=0pt \offinterlineskip
\def\tablerule{\noalign{\hrule}}
\halign to390pt {\strut#& \vrule#\tabskip=1em plus2em& \hfil#\hfil&
\vrule#& \hfil#\hfil& \vrule#& \hfil#\hfil& \vrule#& \hfil#\hfil&
\vrule#& \hfil#\hfil& \vrule#& \hfil#\hfil& \vrule#&
\hfil#\hfil&\vrule#\tabskip=0pt\cr
\tablerule
\omit&height4pt&\omit&&\omit&&\omit&&\omit&&\omit&&\omit&&\omit&\cr
&&$\!\!\!\!{\rm (I\!\!-\!\!J)_{\rm sub}}\!\!\!\!\!\!\!\!$
&&$\!\!\!\!\frac{\Delta\eta^2}{\Delta x^{2D\!-\!2}}\!\!\!\!\!\!\!$
&&$\!\!\!\!\!\frac{\overline{\Delta x}^2}{\Delta x^{2D\!-\!2}}\!\!\!\!\!\!\!$
&&$\!\!\!\!\!\frac{1}{\Delta x^{2D\!-\!4}}\!\!\!\!\!\!\!$
&&$\!\!\!\!\!\!\!\frac{\gamma^0\!\Delta\eta\gamma^k\!\Delta x_k}
{\Delta x^{2D\!-\!2}}\!\!\!\!\!\!\!\!$
&&$\!\!\!\!\!\!\!\frac{H\!a\Delta\eta}{\Delta x^{2D\!-\!4}}\!\!\!\!\!\!\!\!$
&&$\!\!\!\!\!\frac{H\!a\gamma^0\!\gamma^k\!\Delta x_k}
{\Delta x^{2D\!-\!4}}\!\!\!\!\!\!\!$&\cr
\omit&height4pt&\omit&&\omit&&\omit&&\omit&&\omit&&\omit&&\omit&\cr
\tablerule
\omit&height4pt&\omit&&\omit&&\omit&&\omit&&\omit&&\omit&&\omit&\cr
&&$\!\!\!\!{\rm (1\!\!-\!\!3)_{\rm a}}\!\!\!\!\!\!\!\!$
&&$\!\!\!\!-(D\!\!-\!\!1)\!\!\!\!\!\!\!$
&&$\!\!\!\!\!0\!\!\!\!\!\!\!$
&&$\!\!\!\!\!0\!\!\!\!\!\!\!$
&&$\!\!\!\!\!\!\!(D\!\!-\!\!1)\!\!\!\!\!\!\!\!$
&&$\!\!\!\!\!\!\!\frac{(D\!-\!1)}{(D\!-\!2)}\!\!\!\!\!\!\!\!$
&&$\!\!\!\!\!0\!\!\!\!\!\!\!$&\cr
\omit&height4pt&\omit&&\omit&&\omit&&\omit&&\omit&&\omit&&\omit&\cr
\tablerule
\omit&height4pt&\omit&&\omit&&\omit&&\omit&&\omit&&\omit&&\omit&\cr
&&$\!\!\!\!{\rm (1\!\!-\!\!3)_{\rm b}}\!\!\!\!\!\!\!\!$
&&$\!\!\!\!0\!\!\!\!\!\!\!$
&&$\!\!\!\!\!\frac{-(D\!-\!2)^2}{(D\!-\!3)}\!\!\!\!\!\!\!$
&&$\!\!\!\!\!0\!\!\!\!\!\!\!$
&&$\!\!\!\!\!\!\!\frac{-(D\!-\!2)^2}{(D\!-\!3)}\!\!\!\!\!\!\!\!$
&&$\!\!\!\!\!\!\!0\!\!\!\!\!\!\!\!$
&&$\!\!\!\!\!\frac{-(D\!-\!2)}{(D\!-\!3)}\!\!\!\!\!\!\!$&\cr
\omit&height4pt&\omit&&\omit&&\omit&&\omit&&\omit&&\omit&&\omit&\cr
\tablerule
\omit&height4pt&\omit&&\omit&&\omit&&\omit&&\omit&&\omit&&\omit&\cr
&&$\!\!\!\!\!\!\!\!{\rm (2\!\!-\!\!3)_{b+c}}\!\!\!\!\!\!\!\!\!\!$
&&$\!\!\!\!0\!\!\!\!\!\!\!$
&&$\!\!\!\!\!\!\!\!\frac{(D\!-\!1)(D\!-\!2)^2}{4(D\!-\!3)}\!\!\!\!\!\!\!\!\!$
&&$\!\!\!\!\!0\!\!\!\!\!\!\!$
&&$\!\!\!\!\!\!\!0\!\!\!\!\!\!\!\!$
&&$\!\!\!\!\!\!\!0\!\!\!\!\!\!\!\!$
&&$\!\!\!\!\!0\!\!\!\!\!\!\!$&\cr
\omit&height4pt&\omit&&\omit&&\omit&&\omit&&\omit&&\omit&&\omit&\cr
\tablerule
\omit&height4pt&\omit&&\omit&&\omit&&\omit&&\omit&&\omit&&\omit&\cr
&&$\!\!\!\!{\rm (3\!\!-\!\!3)_{a}}\!\!\!\!\!\!\!\!$
&&$\!\!\!\!\!\!\!\frac{(D\!-\!1)(D\!-\!2)}{2(D\!-\!3)}\!\!\!\!\!\!\!\!\!$
&&$\!\!\!\!\!0\!\!\!\!\!\!\!$
&&$\!\!\!\!\!\!\!\!\frac{(D\!-\!1)}{2(D\!-\!3)}\!\!\!\!\!\!\!\!\!\!$
&&$\!\!\!\!\!\!\!0\!\!\!\!\!\!\!\!$
&&$\!\!\!\!\!\!\!0\!\!\!\!\!\!\!\!$
&&$\!\!\!\!\!0\!\!\!\!\!\!\!$&\cr
\omit&height4pt&\omit&&\omit&&\omit&&\omit&&\omit&&\omit&&\omit&\cr
\tablerule
\omit&height4pt&\omit&&\omit&&\omit&&\omit&&\omit&&\omit&&\omit&\cr
&&$\!\!\!\!\!\!\!\!\!{\rm (3\!\!-\!\!3)_{\rm d+e}}\!\!\!\!\!\!\!\!\!\!\!$
&&$\!\!\!\!0\!\!\!\!\!\!\!$
&&$\!\!\!\!\!\!\!\!\!\frac{(D\!-\!2)^2(D\!-\!5)}{8(D\!-\!3)}
\!\!\!\!\!\!\!\!\!\!$
&&$\!\!\!\!\!\!\!\!\frac{-(D\!-\!1)(D\!-\!2)(D\!-\!5)}
{8(D\!-\!3)}\!\!\!\!\!\!\!\!\!\!$
&&$\!\!\!\!\!\!\!0\!\!\!\!\!\!\!\!$
&&$\!\!\!\!\!\!\!0\!\!\!\!\!\!\!\!$
&&$\!\!\!\!\!0\!\!\!\!\!\!\!$&\cr
\omit&height4pt&\omit&&\omit&&\omit&&\omit&&\omit&&\omit&&\omit&\cr
\tablerule
\omit&height4pt&\omit&&\omit&&\omit&&\omit&&\omit&&\omit&&\omit&\cr
&&$\!\!\!\!\!\!\!\!{\rm total}\!\!\!\!\!\!\!\!\!\!\!$
&&$\!\!\!\!\!\!\!\!\frac{(D\!-\!1)(3D\!-\!8)}{2(D\!-\!3)}\!\!\!\!\!\!\!\!\!\!$
&&$\!\!\!\!\!\!\!\!\frac{3(D\!-\!2)^2(D\!-\!5)}
{8(D\!-\!3)}\!\!\!\!\!\!\!\!\!\!$
&&$\!\!\!\!\!\!\!\!\!\!\!\frac{-(D\!-\!1)^2(D\!-\!6)}
{8(D\!-\!3)}\!\!\!\!\!\!\!\!\!\!\!\!$
&&$\!\!\!\!\!\!\!\!\frac{-1}{(D\!-\!3)}\!\!\!\!\!\!\!\!\!\!$
&&$\!\!\!\!\!\!\!\!\!\!\!\frac{(D\!-\!1)}{(D\!-\!2)}\!\!\!\!\!\!\!\!\!\!\!\!$
&&$\!\!\!\!\!\!\!\!\!\frac{-(D\!-\!2)}{(D\!-\!3)}\!\!\!\!\!\!\!\!\!\!$&\cr
\omit&height4pt&\omit&&\omit&&\omit&&\omit&&\omit&&\omit&&\omit&\cr
\tablerule}}

\caption{The contribution from $i\d\!\D_{A}\times i[S]_{\rm fm}$
which are not integrable in four dimensions. Note that all contributions
are multiplied by the factor $\frac{\kappa^2H^2}{64\pi^D}
\Gamma(\frac{D}{2}\!\!-\!\!1)\Gamma(\frac{D}{2}\!\!+\!\!1)
ma(aa')^{2-\frac{D}{2}}$. }

\label{dDAfm3}

\end{table}
Combining (\ref{dDAfm12}) and (\ref{re-dDAfm3}) gives,
\begin{eqnarray}
&&-i\Bigl[\Sigma^{\rm idAfm}\Bigr](x;x')\!=\!\!
\frac{i\kappa^2\!H^2}{64\pi^{\frac{D}{2}}}\Gamma\!(\frac{D}{2}\!+\!1)
\frac{\mu^{D\!-\!4}(\!D\!-\!1)(D^3\!\!-\!9D^2\!\!+\!20D\!-\!4)}
{8(D\!-\!2)(D\!-\!3)^2(D\!-\!4)}\times\nonumber\\
&&ma\delta^D\!(x\!-\!x')
\!+\!\frac{i\kappa^2\!H^2}{64\pi^2}\frac{3ma}{2}\!\ln\!a
\delta^4\!(x\!-\!x')\!+\!\frac{\kappa^2\!H^2 ma}{32\pi^4}
\Biggl\{\Bigl[-\frac{93}{32}\partial^2\!\!+\!
\frac52\!\nabla^2\!\!+\!3H\!a\partial_0\nonumber\\
&&+2H\!a\gamma^0\!\!\not{\hspace{-.1cm}\bar{\partial}}\Bigr]
\Biggl[\frac{\ln(\mu^2\!\Delta x^2)}{\Delta x^2}\Biggr]
+\Biggl[\Bigl(\frac34\!+\!3\ln\!\frac{H^2}{4\mu^2}\Bigr)
\partial_0\!-\!\frac{13}{8}\!\gamma^0\!\partial_0
\!\!\not{\hspace{-.1cm}\bar{\partial}}
\!+\!\Bigl(\!\frac{5}{16}\!-\!\frac12\!
\ln\!\frac{H^2}{4\mu^2}\Bigr)\!\nabla^2\nonumber\\
&&\hspace{2.8cm}+\Bigl(\frac94\!+\!3\ln\!\frac{H^2}{4\mu^2}\Bigr)
H\!a\partial_0\!+\!\Bigl(2\!+\!2\ln\!\frac{H^2}{4\mu^2}\Bigr)
H\!a\gamma^0\!\!\not{\hspace{-.1cm}\bar{\partial}}\Biggr]
\!\frac{1}{\Delta x^2}\Biggr\}\,.\label{idAfm}
\end{eqnarray}

\begin{table}

\vbox{\tabskip=0pt \offinterlineskip
\def\tablerule{\noalign{\hrule}}
\halign to390pt {\strut#& \vrule#\tabskip=1em plus2em& \hfil#\hfil&
\vrule#& \hfil#\hfil& \vrule#& \hfil#\hfil& \vrule#& \hfil#\hfil&
\vrule#& \hfil#\hfil& \vrule#& \hfil#\hfil& \vrule#\tabskip=0pt\cr
\tablerule
\omit&height4pt&\omit&&\omit&&\omit&&\omit&&\omit&&\omit&\cr
&&$\!\!\!\!{\rm I\!-\!J}_{\rm sub}\!\!\!\!\!\!$
&&\omit
&&$\!\!\!\!\frac{\Delta\eta^3}{\Delta x^6}\!\!\!\!\!\!$
&&$\!\!\!\!\!\!\!\frac{\gamma^0\Delta\eta^2\gamma^k\Delta x_k}
{\Delta x^6}\!\!\!\!\!\!\!\!\!$
&&$\!\!\!\!\!\!\!\frac{\Delta\eta}{\Delta x^4}\!\!\!\!\!\!\!\!$
&&$\!\!\!\!\!\!\!\!\!\frac{\gamma^0\gamma^k\Delta x_k}{\Delta x^4}
\!\!\!\!\!\!\!\!$ &\cr
\omit&height4pt&\omit&&\omit&&\omit&&\omit&&\omit&&\omit&\cr
\tablerule
\omit&height2pt&\omit&&\omit&&\omit&&\omit&&\omit&&\omit&\cr
&&$\!\!\!\!\!\!(1\!-\!1)\!\!\!\!\!\!\!$
&&$\!\!\!\!\!1\!\!\!\!\!$ &&$\!\!\!\!\!\!0\!\!\!\!\!\!$
&&$\!\!\!\!\!\!0\!\!\!\!\!\!$ &&
$\!\!\!\!\!\scriptstyle{24(3n^2+4n+1)}\!\!\!\!\!$
&&$\!\!\!\!\!\!\!\!\scriptstyle{-24(3n^2+4n+1)}\!\!\!\!\!\!\!\!$ & \cr
\omit&height2pt&\omit&&\omit&&\omit&&\omit&&\omit&&\omit&\cr
\tablerule
\omit&height2pt&\omit&&\omit&&\omit&&\omit&&\omit&&\omit&\cr
&&$\!\!\!\!\!\!(1\!-\!1)\!\!\!\!\!\!$
&&$\!\!\!\!\!\!\!\scriptstyle{\ln Y}\!\!\!\!\!\!\!$
&&$\!\!\!\!\!0\!\!\!\!\!$ &&$\!\!\!\!\!0\!\!\!\!\!$
&&$\!\!\!\!\!\scriptstyle{24(n^3+2n^2+n)}\!\!\!\!\!$
&&$\!\!\!\!\!\!\!\!\scriptstyle{-24(n^3+2n+n)}\!\!\!\!\!\!\!\!$ & \cr
\omit&height2pt&\omit&&\omit&&\omit&&\omit&&\omit&&\omit&\cr
\tablerule
\omit&height2pt&\omit&&\omit&&\omit&&\omit&&\omit&&\omit&\cr
&&$\!\!\!\!\!\!\!(1\!-\!2)\!\!\!\!\!\!\!$
&&$\!\!\!\!\!\!1\!\!\!\!\!\!$
&&$\!\!\!\!\!\!0\!\!\!\!\!\!$ &&$\!\!\!\!\!\!0\!\!\!\!\!$
&&$\!\!\!\!\!\!\!0\!\!\!\!\!\!\!$
&&$\!\!\!\!\!\scriptstyle{8(3n^2-1)}\!\!\!\!\!$ & \cr
\omit&height2pt&\omit&&\omit&&\omit&&\omit&&\omit&&\omit&\cr
\tablerule
\omit&height2pt&\omit&&\omit&&\omit&&\omit&&\omit&&\omit&\cr
&&$\!\!\!\!\!\!(1\!-\!2)\!\!\!\!\!\!$
&&$\!\!\!\!\!\!\!\scriptstyle{\ln Y}\!\!\!\!\!\!\!$
&&$\!\!\!\!\!\!0\!\!\!\!\!\!$ &&$\!\!\!\!0\!\!\!\!$
&&$\!\!\!\!\!0\!\!\!\!\!$
&&$\!\!\!\!\!\!\!\!\scriptstyle{8(n^3-n)}\!\!\!\!\!\!\!\!$ & \cr
\omit&height2pt&\omit&&\omit&&\omit&&\omit&&\omit&&\omit&\cr
\tablerule
\omit&height2pt&\omit&&\omit&&\omit&&\omit&&\omit&&\omit&\cr
&&$\!\!\!\!\!\!\!(2\!-\!1)\!\!\!\!\!\!\!$
&&$\!\!\!\!\!\!\!1\!\!\!\!\!\!\!$
&&$\!\!\!\!\!\!\!\scriptstyle{-16(3n^2-6n+2)}\!\!\!\!\!\!\!\!$
&&$\!\!\!\!\!\!\!\scriptstyle{16(3n^2-6n+2)}\!\!\!\!\!\!\!$
&&$\!\!\!\!\!\!\!\!\scriptstyle{-8(6n^2-6n+1)}\!\!\!\!\!\!\!\!$
&&$\!\!\!\!\!\!\!\!\scriptstyle{8(3n^2-2n)}\!\!\!\!\!\!\!\!$ & \cr
\omit&height2pt&\omit&&\omit&&\omit&&\omit&&\omit&&\omit&\cr
\tablerule
\omit&height2pt&\omit&&\omit&&\omit&&\omit&&\omit&&\omit&\cr
&&$\!\!\!\!\!\!\!(2\!-\!1)\!\!\!\!\!\!\!$
&&$\!\!\!\!\!\!\!\scriptstyle{\ln Y}\!\!\!\!\!\!\!$
&&$\!\!\!\!\!\!\scriptstyle{-16(n^3-3n^2+2n)}\!\!\!\!\!\!$
&&$\!\!\!\!\!\scriptstyle{16(n^3-3n^2+2n)}\!\!\!\!\!$
&&$\!\!\!\!\!\scriptstyle{-8(2n^3-3n^2+n)}\!\!\!\!\!$
&&$\!\!\!\!\!\!\!\!\scriptstyle{8(n^3-n^2)}\!\!\!\!\!\!\!\!$ & \cr
\omit&height2pt&\omit&&\omit&&\omit&&\omit&&\omit&&\omit&\cr
\tablerule
\omit&height2pt&\omit&&\omit&&\omit&&\omit&&\omit&&\omit&\cr
&&$\!\!\!\!\!\!\!(2\!-\!2)_{\rm a}\!\!\!\!\!\!\!$
&&$\!\!\!\!\!\!\!1\!\!\!\!\!\!\!$
&&$\!\!\!\!\!\!\!\scriptstyle{-2(3n^2-6n+2)}\!\!\!\!\!\!\!\!$
&&$\!\!\!\!\!\!\!\scriptstyle{2(3n^2-6n+2)}\!\!\!\!\!\!\!$
&&$\!\!\!\!\!\!\!\!\scriptstyle{-(6n^2-6n+1)}\!\!\!\!\!\!\!\!$
&&$\!\!\!\!\!\!\!\!\scriptstyle{(6n^2-2n-1)}\!\!\!\!\!\!\!\!$ & \cr
\omit&height2pt&\omit&&\omit&&\omit&&\omit&&\omit&&\omit&\cr
\tablerule
\omit&height2pt&\omit&&\omit&&\omit&&\omit&&\omit&&\omit&\cr
&&$\!\!\!\!\!\!(2\!-\!2)_{\rm a}\!\!\!\!\!\!$
&&$\!\!\!\!\!\!\!\scriptstyle{\ln Y}\!\!\!\!\!\!\!$
&&$\!\!\!\!\!\!\scriptstyle{-2(n^3-3n^2+2n)}\!\!\!\!\!$
&&$\!\!\!\!\!\scriptstyle{2(n^3-3n^2+2n)}\!\!\!\!\!$
&&$\!\!\!\!\!\scriptstyle{-(2n^3-3n^2+n)}\!\!\!\!\!$
&&$\!\!\!\!\!\!\!\!\scriptstyle{(2n^3-n^2-n)}\!\!\!\!\!\!\!\!$ & \cr
\omit&height2pt&\omit&&\omit&&\omit&&\omit&&\omit&&\omit&\cr
\tablerule
\omit&height2pt&\omit&&\omit&&\omit&&\omit&&\omit&&\omit&\cr
&&$\!\!\!\!\!\!\!\!(2\!-\!2)_{\rm b}\!\!\!\!\!\!\!\!$
&&$\!\!\!\!\!\!\!1\!\!\!\!\!\!\!$
&&$\!\!\!\!\!\!\!\scriptstyle{-6(3n^2-6n+2)}\!\!\!\!\!\!\!\!$
&&$\!\!\!\!\!\!\!\scriptstyle{-2(3n^2-6n+2)}\!\!\!\!\!\!\!$
&&$\!\!\!\!\!\!\!\!\scriptstyle{-3(6n^2-6n+1)}\!\!\!\!\!\!\!\!$
&&$\!\!\!\!\!\!\!\!\scriptstyle{-(6n^2-2n-1)}\!\!\!\!\!\!\!\!$ & \cr
\omit&height2pt&\omit&&\omit&&\omit&&\omit&&\omit&&\omit&\cr
\tablerule
\omit&height2pt&\omit&&\omit&&\omit&&\omit&&\omit&&\omit&\cr
&&$\!\!\!\!\!\!\!(2\!-\!2)_{\rm b}\!\!\!\!\!\!\!$
&&$\!\!\!\!\!\!\!\scriptstyle{\ln Y}\!\!\!\!\!\!\!$
&&$\!\!\!\!\!\!\scriptstyle{-6(n^3-3n^2+2n)}\!\!\!\!\!\!$
&&$\!\!\!\!\!\scriptstyle{-2(n^3-3n^2+2n)}\!\!\!\!\!$
&&$\!\!\!\!\!\scriptstyle{-3(2n^3-3n^2+n)}\!\!\!\!\!$
&&$\!\!\!\!\!\!\!\!\scriptstyle{-(2n^3-n^2-n)}\!\!\!\!\!\!\!\!$ & \cr
\omit&height2pt&\omit&&\omit&&\omit&&\omit&&\omit&&\omit&\cr
\tablerule
\omit&height2pt&\omit&&\omit&&\omit&&\omit&&\omit&&\omit&\cr
&&$\!\!\!\!\!\!\!(2\!-\!2)_{\rm c}\!\!\!\!\!\!\!$
&&$\!\!\!\!\!\!\!\!1\!\!\!\!\!\!\!\!$
&&$\!\!\!\!\!\!\!\scriptstyle{4(3n^2-6n+2)}\!\!\!\!\!\!\!\!$
&&$\!\!\!\!\!\!\!\scriptstyle{-4(3n^2-6n+2)}\!\!\!\!\!\!\!$
&&$\!\!\!\!\!\!\!\!\scriptstyle{2(6n^2-6n+1)}\!\!\!\!\!\!\!\!$
&&$\!\!\!\!\!\!\!\!\scriptstyle{-2(6n^2-2n-1)}\!\!\!\!\!\!\!\!$ & \cr
\omit&height2pt&\omit&&\omit&&\omit&&\omit&&\omit&&\omit&\cr
\tablerule
\omit&height2pt&\omit&&\omit&&\omit&&\omit&&\omit&&\omit&\cr
&&$\!\!\!\!\!\!\!(2\!-\!2)_{\rm c}\!\!\!\!\!\!$
&&$\!\!\!\!\!\!\!\scriptstyle{\ln Y}\!\!\!\!\!\!\!$
&&$\!\!\!\!\!\!\scriptstyle{-4(n^3-3n^2+2n)}\!\!\!\!\!\!$
&&$\!\!\!\!\!\scriptstyle{-4(n^3-3n^2+2n)}\!\!\!\!\!$
&&$\!\!\!\!\!\scriptstyle{2(2n^3-3n^2+n)}\!\!\!\!\!$
&&$\!\!\!\!\!\!\!\!\scriptstyle{-2(2n^3-n^2-n)}\!\!\!\!\!\!\!\!$ & \cr
\omit&height2pt&\omit&&\omit&&\omit&&\omit&&\omit&&\omit&\cr
\tablerule
\omit&height2pt&\omit&&\omit&&\omit&&\omit&&\omit&&\omit&\cr
&&$\!\!\!\!\!\!\!\scriptstyle{\rm{total}}\!\!\!\!\!\!\!\!\!\!$
&&$\!\!\!\!\!\!\!1\!\!\!\!\!\!\!$
&&$\!\!\!\!\!\!\!\!\scriptstyle{-20(3n^2-6n+2)}\!\!\!\!\!\!\!\!$
&&$\!\!\!\!\!\!\!\scriptstyle{12(3n^2-6n+2)}\!\!\!\!\!\!\!$
&&$\!\!\!\!\!\!\!\!\scriptstyle{2(6n^2+78n+7)}\!\!\!\!\!\!\!\!$
&&$\!\!\!\!\!\!\!\!\scriptstyle{-6(6n^2+18n+5)}\!\!\!\!\!\!\!\!$ & \cr
\omit&height2pt&\omit&&\omit&&\omit&&\omit&&\omit&&\omit&\cr
\tablerule
\omit&height2pt&\omit&&\omit&&\omit&&\omit&&\omit&&\omit&\cr
&&$\!\!\!\!\!\!\!\scriptstyle{\rm{total}}\!\!\!\!\!\!\!\!\!\!$
&&$\!\!\!\!\!\!\!\scriptstyle{\ln Y}\!\!\!\!\!\!\!$
&&$\!\!\!\!\!\!\scriptstyle{-20(n^3-3n^2+2n)}\!\!\!\!\!\!$
&&$\!\!\!\!\!\scriptstyle{12(n^3-3n^2+2n)}\!\!\!\!\!$
&&$\!\!\!\!\!\scriptstyle{2(2n^3+39n^2+7n)}\!\!\!\!\!\!\!\!\!\!$
&&$\!\!\!\!\!\!\!\scriptstyle{-6(2n^3+9n^2+5n)}\!\!\!\!\!\!\!\!$ & \cr
\omit&height2pt&\omit&&\omit&&\omit&&\omit&&\omit&&\omit&\cr
\tablerule}}

\caption{$i\delta\!\D_{A}\times i[S]_{n\geq 0}-{\rm IA}$. The factor
$\frac{i \kappa^2H^2}{2^6 \pi^4}\frac{mHaa'}{2}\Bigl(
\ln{\frac{H^2\Delta x^2}{4}}\Bigr)\sum_{n=0}^{\infty}Y^{n}$
multiplies all contributions. Here $Y=\frac{y}{4}$; $\ln Y$
and $1$ are the multiplicative factors for the each individual row.}

\label{dAn0-1}

\end{table}

\begin{table}

\vbox{\tabskip=0pt \offinterlineskip
\def\tablerule{\noalign{\hrule}}
\halign to390pt {\strut#& \vrule#\tabskip=1em plus2em& \hfil#\hfil&
\vrule#& \hfil#\hfil& \vrule#& \hfil#\hfil& \vrule#& \hfil#\hfil&
\vrule#& \hfil#\hfil& \vrule#& \hfil#\hfil& \vrule#&
\hfil#\hfil& \vrule#\tabskip=0pt\cr
\tablerule
\omit&height4pt&\omit&&\omit&&\omit&&\omit&&\omit&&\omit&&\omit&\cr
&&$\!\!\!\!\scriptstyle{{\rm I\!-\!J}_{\rm sub}}\!\!\!\!\!\!$
&&\omit
&&$\!\!\!\!\scriptstyle{\frac{Ha\Delta\eta^2}{\Delta x^4}}\!\!\!\!\!\!$
&&$\!\!\!\!\!\!\!\scriptstyle{\frac{Ha'\Delta\eta^2}
{\Delta x^4}}\!\!\!\!\!\!\!\!\!$
&&$\!\!\!\!\!\!\!\!\!\!\scriptstyle{\frac{Ha\gamma^0\!\Delta\eta
\gamma^k\!\Delta x_k}{\Delta x^4}}\!\!\!\!\!\!\!\!\!\!\!$
&&$\!\!\!\!\!\!\!\scriptstyle{\frac{Ha'\gamma^0\!\Delta\eta
\gamma^k\!\Delta x_k}{\Delta x^4}}\!\!\!\!\!\!\!\!\!\!\!$
&&$\!\!\!\!\!\!\frac{Ha}{\Delta x^2}\!\!\!\!\!\!$&\cr
\omit&height4pt&\omit&&\omit&&\omit&&\omit&&\omit&&\omit&&\omit&\cr
\tablerule
\omit&height2pt&\omit&&\omit&&\omit&&\omit&&\omit&&\omit&&\omit&\cr
&&$\!\!\!\!\!\!\!\scriptstyle{(1-1)}\!\!\!\!\!\!\!\!\!\!$
&&$\!\!\!\!\!\!\!\scriptstyle{1}\!\!\!\!\!\!\!\!\!\!\!$
&&$\!\!\!\!\!\!\scriptstyle{24(3n^2+4n+1)}\!\!\!\!\!\!\!\!\!$
&&$\!\!\!\!\!\scriptstyle{0}\!\!\!\!\!$
&&$\!\!\!\!\!\!\!\!\scriptstyle{-24(3n^2+4n+1)}\!\!\!\!\!\!\!\!\!\!$
&&$\!\!\!\!\!\!\!\scriptstyle{0}\!\!\!\!\!\!\!\!$
&&$\!\!\!\!\!\!\!\scriptstyle{12(3n^2+6n+2)}\!\!\!\!\!\!\!\!$& \cr
\omit&height2pt&\omit&&\omit&&\omit&&\omit&&\omit&&\omit&&\omit&\cr
\tablerule
\omit&height2pt&\omit&&\omit&&\omit&&\omit&&\omit&&\omit&&\omit&\cr
&&$\!\!\!\!\!\!\!\scriptstyle{(1-1)}\!\!\!\!\!\!\!\!\!\!$
&&$\!\!\!\!\!\!\!\scriptstyle{\ln Y}\!\!\!\!\!\!\!\!\!\!\!$
&&$\!\!\!\!\!\!\scriptstyle{24(n^3+2n^2+n)}\!\!\!\!\!\!\!\!\!$
&&$\!\!\!\!\!\scriptstyle{0}\!\!\!\!\!$
&&$\!\!\!\!\!\!\!\!\scriptstyle{-24(n^3+2n^2+n)}\!\!\!\!\!\!\!\!\!\!$
&&$\!\!\!\!\!\!\!\scriptstyle{0}\!\!\!\!\!\!\!\!$
&&$\!\!\!\!\!\!\!\scriptstyle{12(n^3+3n^2+2n)}\!\!\!\!\!\!\!\!$& \cr
\omit&height2pt&\omit&&\omit&&\omit&&\omit&&\omit&&\omit&&\omit&\cr
\tablerule
\omit&height2pt&\omit&&\omit&&\omit&&\omit&&\omit&&\omit&&\omit&\cr
&&$\!\!\!\!\!\!\!\scriptstyle{(1-2)}\!\!\!\!\!\!\!\!\!\!$
&&$\!\!\!\!\!\!\!\scriptstyle{1}\!\!\!\!\!\!\!\!\!\!\!$
&&$\!\!\!\!\!\!\scriptstyle{0}\!\!\!\!\!\!$
&&$\!\!\!\!\!\scriptstyle{0}\!\!\!\!\!$
&&$\!\!\!\!\!\scriptstyle{8(3n^2-1)}\!\!\!\!\!\!\!\!\!\!$
&&$\!\!\!\!\!\!\!\scriptstyle{0}\!\!\!\!\!\!\!\!$
&&$\!\!\!\!\!\!\!\scriptstyle{0}\!\!\!\!\!\!\!\!$& \cr
\omit&height2pt&\omit&&\omit&&\omit&&\omit&&\omit&&\omit&&\omit&\cr
\tablerule
\omit&height2pt&\omit&&\omit&&\omit&&\omit&&\omit&&\omit&&\omit&\cr
&&$\!\!\!\!\!\!\!\scriptstyle{(1-2)}\!\!\!\!\!\!\!\!\!\!$
&&$\!\!\!\!\!\!\!\scriptstyle{\ln Y}\!\!\!\!\!\!\!\!\!\!\!$
&&$\!\!\!\!\!\!\scriptstyle{0}\!\!\!\!\!\!$
&&$\!\!\!\!\!\scriptstyle{0}\!\!\!\!\!$
&&$\!\!\!\!\!\scriptstyle{8(n^3-n)}\!\!\!\!\!\!\!\!\!\!$
&&$\!\!\!\!\!\!\!\scriptstyle{0}\!\!\!\!\!\!\!\!$
&&$\!\!\!\!\!\!\!\scriptstyle{0}\!\!\!\!\!\!\!\!$& \cr
\omit&height2pt&\omit&&\omit&&\omit&&\omit&&\omit&&\omit&&\omit&\cr
\tablerule
\omit&height2pt&\omit&&\omit&&\omit&&\omit&&\omit&&\omit&&\omit&\cr
&&$\!\!\!\!\!\!\!\scriptstyle{(2-1)}\!\!\!\!\!\!\!\!\!\!$
&&$\!\!\!\!\!\!\!\scriptstyle{1}\!\!\!\!\!\!\!\!\!\!\!$
&&$\!\!\!\!\!\!\scriptstyle{4(3n^2-1)}\!\!\!\!\!\!$
&&$\!\!\!\!\!\scriptstyle{-4(3n^2-1)}\!\!\!\!\!$
&&$\!\!\!\!\!\scriptstyle{-4(3n^2-1)}\!\!\!\!\!\!\!\!\!\!$
&&$\!\!\!\!\!\!\!\scriptstyle{4(3n^2-1)}\!\!\!\!\!\!\!\!$
&&$\!\!\!\!\!\!\!\scriptstyle{2(6n^2+6n+1)}\!\!\!\!\!\!\!\!$& \cr
\omit&height2pt&\omit&&\omit&&\omit&&\omit&&\omit&&\omit&&\omit&\cr
\tablerule
\omit&height2pt&\omit&&\omit&&\omit&&\omit&&\omit&&\omit&&\omit&\cr
&&$\!\!\!\!\!\!\!\scriptstyle{(2-1)}\!\!\!\!\!\!\!\!\!\!$
&&$\!\!\!\!\!\!\!\scriptstyle{\ln Y}\!\!\!\!\!\!\!\!\!\!\!$
&&$\!\!\!\!\!\!\scriptstyle{4(n^3-n)}\!\!\!\!\!\!$
&&$\!\!\!\!\!\scriptstyle{-4(n^3-n)}\!\!\!\!\!$
&&$\!\!\!\!\!\scriptstyle{-4(n^3-n)}\!\!\!\!\!\!\!\!\!\!$
&&$\!\!\!\!\!\!\!\scriptstyle{4(n^3-n)}\!\!\!\!\!\!\!\!$
&&$\!\!\!\!\!\!\!\scriptstyle{2(2n^3+3n^2+n)}\!\!\!\!\!\!\!\!$& \cr
\omit&height2pt&\omit&&\omit&&\omit&&\omit&&\omit&&\omit&&\omit&\cr
\tablerule
\omit&height2pt&\omit&&\omit&&\omit&&\omit&&\omit&&\omit&&\omit&\cr
&&$\!\!\!\!\!\!\!\scriptstyle{(2-2)_{\rm a}}\!\!\!\!\!\!\!\!\!\!$
&&$\!\!\!\!\!\!\!\scriptstyle{1}\!\!\!\!\!\!\!\!\!\!\!$
&&$\!\!\!\!\!\!\scriptstyle{(3n^2-1)}\!\!\!\!\!\!$
&&$\!\!\!\!\!\scriptstyle{0}\!\!\!\!\!$
&&$\!\!\!\!\!\scriptstyle{0}\!\!\!\!\!\!\!\!\!\!$
&&$\!\!\!\!\!\!\!\scriptstyle{0}\!\!\!\!\!\!\!\!$
&&$\!\!\!\!\!\!\!\scriptstyle{\frac12(6n^2+6n+1)}\!\!\!\!\!\!\!\!$&\cr
\omit&height2pt&\omit&&\omit&&\omit&&\omit&&\omit&&\omit&&\omit&\cr
\tablerule
\omit&height2pt&\omit&&\omit&&\omit&&\omit&&\omit&&\omit&&\omit&\cr
&&$\!\!\!\!\!\!\!\scriptstyle{(2-2)}_{\rm a}\!\!\!\!\!\!\!\!\!\!$
&&$\!\!\!\!\!\!\!\scriptstyle{\ln Y}\!\!\!\!\!\!\!\!\!\!\!$
&&$\!\!\!\!\!\!\scriptstyle{(n^3-n)}\!\!\!\!\!\!$
&&$\!\!\!\!\!\scriptstyle{0}\!\!\!\!\!$
&&$\!\!\!\!\!\scriptstyle{0}\!\!\!\!\!\!\!\!\!\!$
&&$\!\!\!\!\!\!\!\scriptstyle{0}\!\!\!\!\!\!\!\!$
&&$\!\!\!\!\!\!\!\scriptstyle{\frac12(2n^3+3n^2+n)}\!\!\!\!\!\!\!\!$&\cr
\omit&height2pt&\omit&&\omit&&\omit&&\omit&&\omit&&\omit&&\omit&\cr
\tablerule
\omit&height2pt&\omit&&\omit&&\omit&&\omit&&\omit&&\omit&&\omit&\cr
&&$\!\!\!\!\!\!\!\scriptstyle{(2-2)_{\rm b}}\!\!\!\!\!\!\!\!\!\!$
&&$\!\!\!\!\!\!\!\scriptstyle{1}\!\!\!\!\!\!\!\!\!\!\!$
&&$\!\!\!\!\!\!\scriptstyle{3(3n^2-1)}\!\!\!\!\!\!$
&&$\!\!\!\!\!\scriptstyle{0}\!\!\!\!\!$
&&$\!\!\!\!\!\scriptstyle{0}\!\!\!\!\!\!\!\!\!\!$
&&$\!\!\!\!\!\!\!\scriptstyle{0}\!\!\!\!\!\!\!\!$
&&$\!\!\!\!\!\!\!\scriptstyle{\frac32(6n^2+6n+1)}\!\!\!\!\!\!\!\!$& \cr
\omit&height2pt&\omit&&\omit&&\omit&&\omit&&\omit&&\omit&&\omit&\cr
\tablerule
\omit&height2pt&\omit&&\omit&&\omit&&\omit&&\omit&&\omit&&\omit&\cr
&&$\!\!\!\!\!\!\!\scriptstyle{(2-2)}_{\rm b}\!\!\!\!\!\!\!\!\!\!$
&&$\!\!\!\!\!\!\!\scriptstyle{\ln Y}\!\!\!\!\!\!\!\!\!\!\!$
&&$\!\!\!\!\!\!\scriptstyle{3(n^3-n)}\!\!\!\!\!\!$
&&$\!\!\!\!\!\scriptstyle{0}\!\!\!\!\!$
&&$\!\!\!\!\!\scriptstyle{0}\!\!\!\!\!\!\!\!\!\!$
&&$\!\!\!\!\!\!\!\scriptstyle{0}\!\!\!\!\!\!\!\!$
&&$\!\!\!\!\!\!\!\scriptstyle{\frac32(2n^3+3n^2+n)}\!\!\!\!\!\!\!\!$& \cr
\omit&height2pt&\omit&&\omit&&\omit&&\omit&&\omit&&\omit&&\omit&\cr
\tablerule
\omit&height2pt&\omit&&\omit&&\omit&&\omit&&\omit&&\omit&&\omit&\cr
&&$\!\!\!\!\!\!\!\scriptstyle{(2-2)_{\rm c}}\!\!\!\!\!\!\!\!\!\!$
&&$\!\!\!\!\!\!\!\scriptstyle{1}\!\!\!\!\!\!\!\!\!\!\!$
&&$\!\!\!\!\!\!\scriptstyle{-2(3n^2-1)}\!\!\!\!\!\!$
&&$\!\!\!\!\!\scriptstyle{0}\!\!\!\!\!$
&&$\!\!\!\!\!\scriptstyle{0}\!\!\!\!\!\!\!\!\!\!$
&&$\!\!\!\!\!\!\!\scriptstyle{0}\!\!\!\!\!\!\!\!$
&&$\!\!\!\!\!\!\!\scriptstyle{-(6n^2+6n+1)}\!\!\!\!\!\!\!\!$& \cr
\omit&height2pt&\omit&&\omit&&\omit&&\omit&&\omit&&\omit&&\omit&\cr
\tablerule
\omit&height2pt&\omit&&\omit&&\omit&&\omit&&\omit&&\omit&&\omit&\cr
&&$\!\!\!\!\!\!\!\scriptstyle{(2-2)_{\rm c}}\!\!\!\!\!\!\!\!\!\!$
&&$\!\!\!\!\!\!\!\scriptstyle{\ln Y}\!\!\!\!\!\!\!\!\!\!\!$
&&$\!\!\!\!\!\!\scriptstyle{-2(n^3-n)}\!\!\!\!\!\!$
&&$\!\!\!\!\!\scriptstyle{0}\!\!\!\!\!$
&&$\!\!\!\!\!\scriptstyle{0}\!\!\!\!\!\!\!\!\!\!$
&&$\!\!\!\!\!\!\!\scriptstyle{0}\!\!\!\!\!\!\!\!$
&&$\!\!\!\!\!\!\!\scriptstyle{-(2n^3+3n^2+n)}\!\!\!\!\!\!\!\!$& \cr
\omit&height2pt&\omit&&\omit&&\omit&&\omit&&\omit&&\omit&&\omit&\cr
\tablerule
\omit&height2pt&\omit&&\omit&&\omit&&\omit&&\omit&&\omit&&\omit&\cr
&&$\!\!\!\!\!\!\!\scriptstyle{\rm{total}}\!\!\!\!\!\!\!\!\!\!$
&&$\!\!\!\!\!\!\!\scriptstyle{1}\!\!\!\!\!\!\!\!\!\!\!$
&&$\!\!\!\!\!\!\scriptstyle{6(15n^2+16n+3)}\!\!\!\!\!\!\!\!\!$
&&$\!\!\!\!\!\scriptstyle{-4(3n^2-1)}\!\!\!\!\!\!\!\!$
&&$\!\!\!\!\!\!\!\scriptstyle{-4(15n^2+24n+7)}\!\!\!\!\!\!\!\!\!\!$
&&$\!\!\!\!\!\!\!\scriptstyle{4(3n^2-1)}\!\!\!\!\!\!\!\!$
&&$\!\!\!\!\!\!\!\scriptstyle{9(6n^2+10n+3)}\!\!\!\!\!\!\!\!$& \cr
\omit&height2pt&\omit&&\omit&&\omit&&\omit&&\omit&&\omit&&\omit&\cr
\tablerule
\omit&height2pt&\omit&&\omit&&\omit&&\omit&&\omit&&\omit&&\omit&\cr
&&$\!\!\!\!\!\!\!\scriptstyle{\rm total}\!\!\!\!\!\!\!\!\!\!$
&&$\!\!\!\!\!\!\!\scriptstyle{\ln Y}\!\!\!\!\!\!\!\!\!\!\!$
&&$\!\!\!\!\!\!\scriptstyle{6(5n^3+8n^2+3n)}\!\!\!\!\!\!\!\!\!$
&&$\!\!\!\!\!\!\!\scriptstyle{-4(n^3-n)}\!\!\!\!\!\!\!\!$
&&$\!\!\!\!\!\!\scriptstyle{-4(5n^3+12n^2+7n)}\!\!\!\!\!\!\!\!\!\!$
&&$\!\!\!\!\!\!\!\scriptstyle{4(n^3-n)}\!\!\!\!\!\!\!\!\!\!\!$
&&$\!\!\!\!\!\!\scriptstyle{9(2n^3+5n^2+3n)}\!\!\!\!\!\!\!\!$& \cr
\omit&height2pt&\omit&&\omit&&\omit&&\omit&&\omit&&\omit&&\omit&\cr
\tablerule}}

\caption{$i\delta\!\D_{A}\times i[S]_{n\geq 0}-{\rm IB}$. The factor
$\frac{i \kappa^2H^2}{2^6 \pi^4}\frac{mHaa'}{2}\Bigl(
\ln{\frac{H^2\Delta x^2}{4}}\Bigr)\sum_{n=0}^{\infty}Y^{n}$
multiplies all contributions. Here $Y=\frac{y}{4}$; $\ln Y$
and $1$ are the multiplicative factors for the each individual row.}

\label{dAn0-2}

\end{table}

\begin{table}

\vbox{\tabskip=0pt \offinterlineskip
\def\tablerule{\noalign{\hrule}}
\halign to390pt {\strut#& \vrule#\tabskip=1em plus2em& \hfil#\hfil&
\vrule#& \hfil#\hfil& \vrule#& \hfil#\hfil& \vrule#& \hfil#\hfil&
\vrule#& \hfil#\hfil& \vrule#& \hfil#\hfil& \vrule#&
\hfil#\hfil& \vrule#\tabskip=0pt\cr
\tablerule
\omit&height4pt&\omit&&\omit&&\omit&&\omit&&\omit&&\omit&&\omit&\cr
&&$\!\!\!\!\scriptstyle{{\rm I\!-\!J}_{\rm sub}}\!\!\!\!\!\!$
&&\omit
&&$\!\!\!\!\!\!\!\!\frac{Ha'}{\Delta x^2}\!\!\!\!\!\!\!\!\!$
&&$\!\!\!\!\!\!\!\scriptstyle{\frac{H^2a^2\Delta\eta}
{\Delta x^2}}\!\!\!\!\!\!\!\!\!$
&&$\!\!\!\!\!\!\!\!\!\!\scriptstyle{\frac{H^2\!a^2\gamma^0\!
\!\gamma^k\!\Delta x_k}{\Delta x^2}}\!\!\!\!\!\!\!\!\!\!\!$
&&$\!\!\!\!\!\!\!\scriptstyle{\frac{H^{2}\!aa'\gamma^0\!
\gamma^k\!\Delta x_k}{\Delta x^2}}\!\!\!\!\!\!\!\!\!\!\!$
&&$\!\!\!\!\!\!\scriptstyle{H^{3}a^{2}a'}\!\!\!\!\!\!$&\cr
\omit&height4pt&\omit&&\omit&&\omit&&\omit&&\omit&&\omit&&\omit&\cr
\tablerule
\omit&height2pt&\omit&&\omit&&\omit&&\omit&&\omit&&\omit&&\omit&\cr
&&$\!\!\!\!\!\!\!\scriptstyle{(1-1)}\!\!\!\!\!\!\!\!\!\!$
&&$\!\!\!\!\!\!\!\scriptstyle{1}\!\!\!\!\!\!\!\!\!\!\!$
&&$\!\!\!\!\!\!\scriptstyle{0}\!\!\!\!\!\!\!\!\!$
&&$\!\!\!\!\!\scriptstyle{-6(3n^2+6n+2)}\!\!\!\!\!\!\!\!\!$
&&$\!\!\!\!\!\!\!\scriptstyle{6(3n^2+6n+2)}\!\!\!\!\!\!\!\!\!\!$
&&$\!\!\!\!\!\!\!\scriptstyle{0}\!\!\!\!\!\!\!\!$
&&$\!\!\!\!\!\!\!\scriptstyle{-3(3n^2+8n+5)}\!\!\!\!\!\!\!\!$& \cr
\omit&height2pt&\omit&&\omit&&\omit&&\omit&&\omit&&\omit&&\omit&\cr
\tablerule
\omit&height2pt&\omit&&\omit&&\omit&&\omit&&\omit&&\omit&&\omit&\cr
&&$\!\!\!\!\!\!\!\scriptstyle{(1-1)}\!\!\!\!\!\!\!\!\!\!$
&&$\!\!\!\!\!\!\!\scriptstyle{\ln Y}\!\!\!\!\!\!\!\!\!\!\!$
&&$\!\!\!\!\!\!\scriptstyle{0}\!\!\!\!\!\!\!\!\!\!\!\!$
&&$\!\!\!\!\!\!\!\scriptstyle{-6(n^3+3n^2+2n)}\!\!\!\!\!\!\!\!\!\!\!$
&&$\!\!\!\!\!\!\!\!\scriptstyle{6(n^3+3n^2+2n)}\!\!\!\!\!\!\!\!\!\!\!$
&&$\!\!\!\!\!\!\!\scriptstyle{0}\!\!\!\!\!\!\!\!$
&&$\!\!\!\!\!\!\!\scriptstyle{-3(n^3\!+4n^2\!+5n\!+\!2)}
\!\!\!\!\!\!\!\!$& \cr
\omit&height2pt&\omit&&\omit&&\omit&&\omit&&\omit&&\omit&&\omit&\cr
\tablerule
\omit&height2pt&\omit&&\omit&&\omit&&\omit&&\omit&&\omit&&\omit&\cr
&&$\!\!\!\!\!\!\!\scriptstyle{(1-2)}\!\!\!\!\!\!\!\!\!\!$
&&$\!\!\!\!\!\!\!\scriptstyle{1}\!\!\!\!\!\!\!\!\!\!\!$
&&$\!\!\!\!\!\!\scriptstyle{0}\!\!\!\!\!\!$
&&$\!\!\!\!\!\scriptstyle{0}\!\!\!\!\!$
&&$\!\!\!\!\!\scriptstyle{-2(3n^2\!+\!6n\!+\!2)}\!\!\!\!\!\!\!\!\!\!$
&&$\!\!\!\!\!\!\!\scriptstyle{0}\!\!\!\!\!\!\!\!$
&&$\!\!\!\!\!\!\!\scriptstyle{0}\!\!\!\!\!\!\!\!$& \cr
\omit&height2pt&\omit&&\omit&&\omit&&\omit&&\omit&&\omit&&\omit&\cr
\tablerule
\omit&height2pt&\omit&&\omit&&\omit&&\omit&&\omit&&\omit&&\omit&\cr
&&$\!\!\!\!\!\!\!\scriptstyle{(1-2)}\!\!\!\!\!\!\!\!\!\!$
&&$\!\!\!\!\!\!\!\scriptstyle{\ln Y}\!\!\!\!\!\!\!\!\!\!\!$
&&$\!\!\!\!\!\!\scriptstyle{0}\!\!\!\!\!\!$
&&$\!\!\!\!\!\scriptstyle{0}\!\!\!\!\!$
&&$\!\!\!\!\!\scriptstyle{-2(n^3\!+\!3n^2\!+\!2n)}\!\!\!\!\!\!\!\!\!\!$
&&$\!\!\!\!\!\!\!\scriptstyle{0}\!\!\!\!\!\!\!\!$
&&$\!\!\!\!\!\!\!\scriptstyle{0}\!\!\!\!\!\!\!\!$& \cr
\omit&height2pt&\omit&&\omit&&\omit&&\omit&&\omit&&\omit&&\omit&\cr
\tablerule
\omit&height2pt&\omit&&\omit&&\omit&&\omit&&\omit&&\omit&&\omit&\cr
&&$\!\!\!\!\!\!\!\scriptstyle{(2-1)}\!\!\!\!\!\!\!\!\!\!$
&&$\!\!\!\!\!\!\!\scriptstyle{1}\!\!\!\!\!\!\!\!\!\!\!$
&&$\!\!\!\!\!\!\scriptstyle{-2(6n^2\!+6n+1)}\!\!\!\!\!\!\!\!\!\!$
&&$\!\!\!\!\!\scriptstyle{0}\!\!\!\!\!$
&&$\!\!\!\!\!\scriptstyle{0}\!\!\!\!\!\!\!\!\!\!$
&&$\!\!\!\!\!\!\!\scriptstyle{-2(3n^2\!+4n+1)}\!\!\!\!\!\!\!\!$
&&$\!\!\!\!\!\!\!\scriptstyle{0}\!\!\!\!\!\!\!\!$& \cr
\omit&height2pt&\omit&&\omit&&\omit&&\omit&&\omit&&\omit&&\omit&\cr
\tablerule
\omit&height2pt&\omit&&\omit&&\omit&&\omit&&\omit&&\omit&&\omit&\cr
&&$\!\!\!\!\!\!\!\scriptstyle{(2-1)}\!\!\!\!\!\!\!\!\!\!$
&&$\!\!\!\!\!\!\!\scriptstyle{\ln Y}\!\!\!\!\!\!\!\!\!\!\!$
&&$\!\!\!\!\!\!\scriptstyle{-2(2n^3\!+3n^2\!+n)}\!\!\!\!\!\!\!\!\!\!$
&&$\!\!\!\!\!\scriptstyle{0}\!\!\!\!\!$
&&$\!\!\!\!\!\scriptstyle{0}\!\!\!\!\!\!\!\!\!\!$
&&$\!\!\!\!\!\!\!\scriptstyle{-2(n^3\!+2n^2\!+n)}\!\!\!\!\!\!\!\!$
&&$\!\!\!\!\!\!\!\scriptstyle{0}\!\!\!\!\!\!\!\!$& \cr
\omit&height2pt&\omit&&\omit&&\omit&&\omit&&\omit&&\omit&&\omit&\cr
\tablerule
\omit&height2pt&\omit&&\omit&&\omit&&\omit&&\omit&&\omit&&\omit&\cr
&&$\!\!\!\!\!\!\!\scriptstyle{(2-2)_{\rm a}}\!\!\!\!\!\!\!\!\!\!$
&&$\!\!\!\!\!\!\!\scriptstyle{1}\!\!\!\!\!\!\!\!\!\!\!$
&&$\!\!\!\!\!\!\scriptstyle{0}\!\!\!\!\!\!$
&&$\!\!\!\!\!\scriptstyle{0}\!\!\!\!\!$
&&$\!\!\!\!\!\scriptstyle{0}\!\!\!\!\!\!\!\!\!\!$
&&$\!\!\!\!\!\!\!\scriptstyle{0}\!\!\!\!\!\!\!\!$
&&$\!\!\!\!\!\!\!\scriptstyle{0}\!\!\!\!\!\!\!\!$&\cr
\omit&height2pt&\omit&&\omit&&\omit&&\omit&&\omit&&\omit&&\omit&\cr
\tablerule
\omit&height2pt&\omit&&\omit&&\omit&&\omit&&\omit&&\omit&&\omit&\cr
&&$\!\!\!\!\!\!\!\scriptstyle{(2-2)}_{\rm a}\!\!\!\!\!\!\!\!\!\!$
&&$\!\!\!\!\!\!\!\scriptstyle{\ln Y}\!\!\!\!\!\!\!\!\!\!\!$
&&$\!\!\!\!\!\!\scriptstyle{0}\!\!\!\!\!\!$
&&$\!\!\!\!\!\scriptstyle{0}\!\!\!\!\!$
&&$\!\!\!\!\!\scriptstyle{0}\!\!\!\!\!\!\!\!\!\!$
&&$\!\!\!\!\!\!\!\scriptstyle{0}\!\!\!\!\!\!\!\!$
&&$\!\!\!\!\!\!\!\scriptstyle{0}\!\!\!\!\!\!\!\!$&\cr
\omit&height2pt&\omit&&\omit&&\omit&&\omit&&\omit&&\omit&&\omit&\cr
\tablerule
\omit&height2pt&\omit&&\omit&&\omit&&\omit&&\omit&&\omit&&\omit&\cr
&&$\!\!\!\!\!\!\!\scriptstyle{(2-2)_{\rm b}}\!\!\!\!\!\!\!\!\!\!$
&&$\!\!\!\!\!\!\!\scriptstyle{1}\!\!\!\!\!\!\!\!\!\!\!$
&&$\!\!\!\!\!\!\scriptstyle{0}\!\!\!\!\!\!$
&&$\!\!\!\!\!\scriptstyle{0}\!\!\!\!\!$
&&$\!\!\!\!\!\scriptstyle{0}\!\!\!\!\!\!\!\!\!\!$
&&$\!\!\!\!\!\!\!\scriptstyle{0}\!\!\!\!\!\!\!\!$
&&$\!\!\!\!\!\!\!\scriptstyle{0}\!\!\!\!\!\!\!\!$& \cr
\omit&height2pt&\omit&&\omit&&\omit&&\omit&&\omit&&\omit&&\omit&\cr
\tablerule
\omit&height2pt&\omit&&\omit&&\omit&&\omit&&\omit&&\omit&&\omit&\cr
&&$\!\!\!\!\!\!\!\scriptstyle{(2-2)}_{\rm b}\!\!\!\!\!\!\!\!\!\!$
&&$\!\!\!\!\!\!\!\scriptstyle{\ln Y}\!\!\!\!\!\!\!\!\!\!\!$
&&$\!\!\!\!\!\!\scriptstyle{0}\!\!\!\!\!\!$
&&$\!\!\!\!\!\scriptstyle{0}\!\!\!\!\!$
&&$\!\!\!\!\!\scriptstyle{0}\!\!\!\!\!\!\!\!\!\!$
&&$\!\!\!\!\!\!\!\scriptstyle{0}\!\!\!\!\!\!\!\!$
&&$\!\!\!\!\!\!\!\scriptstyle{0}\!\!\!\!\!\!\!\!$& \cr
\omit&height2pt&\omit&&\omit&&\omit&&\omit&&\omit&&\omit&&\omit&\cr
\tablerule
\omit&height2pt&\omit&&\omit&&\omit&&\omit&&\omit&&\omit&&\omit&\cr
&&$\!\!\!\!\!\!\!\scriptstyle{(2-2)_{\rm c}}\!\!\!\!\!\!\!\!\!\!$
&&$\!\!\!\!\!\!\!\scriptstyle{1}\!\!\!\!\!\!\!\!\!\!\!$
&&$\!\!\!\!\!\!\scriptstyle{0}\!\!\!\!\!\!$
&&$\!\!\!\!\!\scriptstyle{0}\!\!\!\!\!$
&&$\!\!\!\!\!\scriptstyle{0}\!\!\!\!\!\!\!\!\!\!$
&&$\!\!\!\!\!\!\!\scriptstyle{0}\!\!\!\!\!\!\!\!$
&&$\!\!\!\!\!\!\!\scriptstyle{0}\!\!\!\!\!\!\!\!$& \cr
\omit&height2pt&\omit&&\omit&&\omit&&\omit&&\omit&&\omit&&\omit&\cr
\tablerule
\omit&height2pt&\omit&&\omit&&\omit&&\omit&&\omit&&\omit&&\omit&\cr
&&$\!\!\!\!\!\!\!\scriptstyle{(2-2)_{\rm c}}\!\!\!\!\!\!\!\!\!\!$
&&$\!\!\!\!\!\!\!\scriptstyle{\ln Y}\!\!\!\!\!\!\!\!\!\!\!$
&&$\!\!\!\!\!\!\scriptstyle{0}\!\!\!\!\!\!$
&&$\!\!\!\!\!\scriptstyle{0}\!\!\!\!\!$
&&$\!\!\!\!\!\scriptstyle{0}\!\!\!\!\!\!\!\!\!\!$
&&$\!\!\!\!\!\!\!\scriptstyle{0}\!\!\!\!\!\!\!\!$
&&$\!\!\!\!\!\!\!\scriptstyle{0}\!\!\!\!\!\!\!\!$& \cr
\omit&height2pt&\omit&&\omit&&\omit&&\omit&&\omit&&\omit&&\omit&\cr
\tablerule
\omit&height2pt&\omit&&\omit&&\omit&&\omit&&\omit&&\omit&&\omit&\cr
&&$\!\!\!\!\!\!\!\scriptstyle{\rm{total}}\!\!\!\!\!\!\!\!\!\!$
&&$\!\!\!\!\!\!\!\scriptstyle{1}\!\!\!\!\!\!\!\!\!\!\!$
&&$\!\!\!\!\!\!\scriptstyle{-2(6n^2\!+6n+1)}\!\!\!\!\!\!\!\!\!$
&&$\!\!\!\!\!\scriptstyle{-6(3n^2\!+6n+2)}\!\!\!\!\!\!\!\!\!\!\!$
&&$\!\!\!\!\!\!\!\scriptstyle{4(3n^2\!+6n+2)}\!\!\!\!\!\!\!\!\!\!\!\!\!$
&&$\!\!\!\!\!\!\scriptstyle{-2(3n^2\!+4n+1)}\!\!\!\!\!\!\!\!$
&&$\!\!\!\!\!\!\!\scriptstyle{-3(3n^2\!+8n+5)}\!\!\!\!\!\!\!\!$& \cr
\omit&height2pt&\omit&&\omit&&\omit&&\omit&&\omit&&\omit&&\omit&\cr
\tablerule
\omit&height2pt&\omit&&\omit&&\omit&&\omit&&\omit&&\omit&&\omit&\cr
&&$\!\!\!\!\!\!\!\scriptstyle{\rm total}\!\!\!\!\!\!\!\!\!\!$
&&$\!\!\!\!\!\!\!\scriptstyle{\ln Y}\!\!\!\!\!\!\!\!\!\!\!$
&&$\!\!\!\!\!\!\scriptstyle{-2(2n^3\!+3n^2\!+n)}\!\!\!\!\!\!\!\!\!$
&&$\!\!\!\!\!\!\scriptstyle{-6(n^3\!+3n^2\!+2n)}\!\!\!\!\!\!\!\!\!\!$
&&$\!\!\!\!\!\!\scriptstyle{4(n^3\!+3n^2\!+2n)}\!\!\!\!\!\!\!\!\!\!$
&&$\!\!\!\!\!\!\!\scriptstyle{-2(n^3\!+2n^2\!+n)}\!\!\!\!\!\!\!\!\!\!\!$
&&$\!\!\!\!\!\!\scriptstyle{-3(n^3\!+4n^2\!+5n+2)}\!\!\!\!\!\!\!\!$& \cr
\omit&height2pt&\omit&&\omit&&\omit&&\omit&&\omit&&\omit&&\omit&\cr
\tablerule}}

\caption{$i\delta\!\D_{A}\times i[S]_{n\geq 0}-{\rm IC}$. The factor
$\frac{i \kappa^2H^2}{2^6 \pi^4}\frac{mHaa'}{2}
\Bigl(\ln\frac{H^2\Delta x^2}{4}\Bigr)\sum_{n=0}^{\infty}Y^{n}$
multiplies all contributions. Here $Y=\frac{y}{4}$; $\ln Y$
and $1$ are the multiplicative factors for the each individual row.}

\label{dAn0-3}

\end{table}
\begin{table}

\vbox{\tabskip=0pt \offinterlineskip
\def\tablerule{\noalign{\hrule}}
\halign to390pt {\strut#& \vrule#\tabskip=1em plus2em& \hfil#\hfil&
\vrule#& \hfil#\hfil& \vrule#& \hfil#\hfil& \vrule#& \hfil#\hfil&
\vrule#& \hfil#\hfil& \vrule#& \hfil#\hfil& \vrule#\tabskip=0pt\cr
\tablerule
\omit&height4pt&\omit&&\omit&&\omit&&\omit&&\omit&&\omit&\cr
&&$\!\!\!\!{\rm I\!-\!J}\!\!\!\!\!\!$
&&\omit
&&$\!\!\!\!\frac{\Delta\eta^3}{\Delta x^6}\!\!\!\!\!\!$
&&$\!\!\!\!\!\!\!\frac{\gamma^0\Delta\eta^2\gamma^k\Delta x_k}
{\Delta x^6}\!\!\!\!\!\!\!\!\!$
&&$\!\!\!\!\!\!\!\frac{\Delta\eta}{\Delta x^4}\!\!\!\!\!\!\!\!$
&&$\!\!\!\!\!\!\!\!\!\frac{\gamma^0\gamma^k\Delta x_k}{\Delta x^4}
\!\!\!\!\!\!\!\!$ &\cr
\omit&height4pt&\omit&&\omit&&\omit&&\omit&&\omit&&\omit&\cr
\tablerule
\omit&height2pt&\omit&&\omit&&\omit&&\omit&&\omit&&\omit&\cr
&&$\!\!\!\!\!\!(1\!-\!1)\!\!\!\!\!\!\!$
&&$\!\!\!\!\!\scriptstyle{1}\!\!\!\!\!$
&&$\!\!\!\!\!\!\scriptstyle{0}\!\!\!\!\!\!$
&&$\!\!\!\!\!\!\scriptstyle{0}\!\!\!\!\!\!$ &&
$\!\!\!\!\!\scriptstyle{12(3n^2+8n+3)}\!\!\!\!\!$
&&$\!\!\!\!\!\!\!\!\scriptstyle{-12(3n^2+8n+3)}\!\!\!\!\!\!\!\!$ & \cr
\omit&height2pt&\omit&&\omit&&\omit&&\omit&&\omit&&\omit&\cr
\tablerule
\omit&height2pt&\omit&&\omit&&\omit&&\omit&&\omit&&\omit&\cr
&&$\!\!\!\!\!\!(1\!-\!1)\!\!\!\!\!\!$
&&$\!\!\!\!\!\!\!\scriptstyle{\ln Y}\!\!\!\!\!\!\!$
&&$\!\!\!\!\!\scriptstyle{0}\!\!\!\!\!$
&&$\!\!\!\!\!\scriptstyle{0}\!\!\!\!\!$
&&$\!\!\!\!\!\scriptstyle{12(n^3+4n^2+3n)}\!\!\!\!\!$
&&$\!\!\!\!\!\!\!\!\scriptstyle{-12(n^3+4n^2+3n)}\!\!\!\!\!\!\!\!$ & \cr
\omit&height2pt&\omit&&\omit&&\omit&&\omit&&\omit&&\omit&\cr
\tablerule
\omit&height2pt&\omit&&\omit&&\omit&&\omit&&\omit&&\omit&\cr
&&$\!\!\!\!\!\!\!(1\!-\!2)\!\!\!\!\!\!\!$
&&$\!\!\!\!\!\!\scriptstyle{1}\!\!\!\!\!\!$
&&$\!\!\!\!\!\!\scriptstyle{0}\!\!\!\!\!\!$
&&$\!\!\!\!\!\!\scriptstyle{0}\!\!\!\!\!$
&&$\!\!\!\!\!\!\!\scriptstyle{0}\!\!\!\!\!\!\!$
&&$\!\!\!\!\!\scriptstyle{4(3n^2+4n+1)}\!\!\!\!\!$ & \cr
\omit&height2pt&\omit&&\omit&&\omit&&\omit&&\omit&&\omit&\cr
\tablerule
\omit&height2pt&\omit&&\omit&&\omit&&\omit&&\omit&&\omit&\cr
&&$\!\!\!\!\!\!(1\!-\!2)\!\!\!\!\!\!$
&&$\!\!\!\!\!\!\!\scriptstyle{\ln Y}\!\!\!\!\!\!\!$
&&$\!\!\!\!\!\!\scriptstyle{0}\!\!\!\!\!\!$
&&$\!\!\!\!\!\scriptstyle{0}\!\!\!\!\!\!$
&&$\!\!\!\!\!\scriptstyle{0}\!\!\!\!\!$
&&$\!\!\!\!\!\!\!\!\scriptstyle{4(n^3+2n^2+n)}\!\!\!\!\!\!\!\!$ & \cr
\omit&height2pt&\omit&&\omit&&\omit&&\omit&&\omit&&\omit&\cr
\tablerule
\omit&height2pt&\omit&&\omit&&\omit&&\omit&&\omit&&\omit&\cr
&&$\!\!\!\!\!\!\!(1\!-\!3)\!\!\!\!\!\!\!$
&&$\!\!\!\!\!\!\!\scriptstyle{1}\!\!\!\!\!\!\!$
&&$\!\!\!\!\!\!\!\scriptstyle{0}\!\!\!\!\!\!\!\!$
&&$\!\!\!\!\!\!\!\scriptstyle{0}\!\!\!\!\!\!\!$
&&$\!\!\!\!\!\!\!\!\scriptstyle{-6(2n+1)}\!\!\!\!\!\!\!\!$
&&$\!\!\!\!\!\!\!\!\scriptstyle{8(2n+1)}\!\!\!\!\!\!\!\!$ & \cr
\omit&height2pt&\omit&&\omit&&\omit&&\omit&&\omit&&\omit&\cr
\tablerule
\omit&height2pt&\omit&&\omit&&\omit&&\omit&&\omit&&\omit&\cr
&&$\!\!\!\!\!\!\!(1\!-\!3)\!\!\!\!\!\!\!$
&&$\!\!\!\!\!\!\!\scriptstyle{\ln Y}\!\!\!\!\!\!\!$
&&$\!\!\!\!\!\!\scriptstyle{0}\!\!\!\!\!\!$
&&$\!\!\!\!\!\scriptstyle{0}\!\!\!\!\!$
&&$\!\!\!\!\!\scriptstyle{-6(n^2+n)}\!\!\!\!\!$
&&$\!\!\!\!\!\!\!\!\scriptstyle{8(n^2+n)}\!\!\!\!\!\!\!\!$ & \cr
\omit&height2pt&\omit&&\omit&&\omit&&\omit&&\omit&&\omit&\cr
\tablerule
\omit&height2pt&\omit&&\omit&&\omit&&\omit&&\omit&&\omit&\cr
&&$\!\!\!\!\!\!\!(2\!-\!1)\!\!\!\!\!\!\!$
&&$\!\!\!\!\!\!\!\scriptstyle{1}\!\!\!\!\!\!\!$
&&$\!\!\!\!\!\!\!\scriptstyle{-8(3n^2-2n)}\!\!\!\!\!\!\!\!$
&&$\!\!\!\!\!\!\!\scriptstyle{8(3n^2-2n)}\!\!\!\!\!\!\!$
&&$\!\!\!\!\!\!\!\!\scriptstyle{-4(6n^2+2n)}\!\!\!\!\!\!\!\!$
&&$\!\!\!\!\!\!\!\!\scriptstyle{4(3n^2+2n+1)}\!\!\!\!\!\!\!\!$ & \cr
\omit&height2pt&\omit&&\omit&&\omit&&\omit&&\omit&&\omit&\cr
\tablerule
\omit&height2pt&\omit&&\omit&&\omit&&\omit&&\omit&&\omit&\cr
&&$\!\!\!\!\!\!(2\!-\!1)\!\!\!\!\!\!$
&&$\!\!\!\!\!\!\!\scriptstyle{\ln Y}\!\!\!\!\!\!\!$
&&$\!\!\!\!\!\!\scriptstyle{-8(n^3-n^2)}\!\!\!\!\!$
&&$\!\!\!\!\!\scriptstyle{8(n^3-n^2)}\!\!\!\!\!$
&&$\!\!\!\!\!\scriptstyle{-4(2n^3+n^2)}\!\!\!\!\!$
&&$\!\!\!\!\!\!\!\!\scriptstyle{4(n^3+n^2+n)}\!\!\!\!\!\!\!\!$ & \cr
\omit&height2pt&\omit&&\omit&&\omit&&\omit&&\omit&&\omit&\cr
\tablerule
\omit&height2pt&\omit&&\omit&&\omit&&\omit&&\omit&&\omit&\cr
&&$\!\!\!\!\!\!\!\!(2\!-\!2)\!\!\!\!\!\!\!\!$
&&$\!\!\!\!\!\!\!\scriptstyle{1}\!\!\!\!\!\!\!$
&&$\!\!\!\!\!\!\!\scriptstyle{-2(3n^2-2n)}\!\!\!\!\!\!\!\!$
&&$\!\!\!\!\!\!\!\scriptstyle{-2(3n^2-2n)}\!\!\!\!\!\!\!$
&&$\!\!\!\!\!\!\!\!\scriptstyle{-(6n^2+2n-3)}\!\!\!\!\!\!\!\!$
&&$\!\!\!\!\!\!\!\!\scriptstyle{-(6n^2+6n-1)}\!\!\!\!\!\!\!\!$ & \cr
\omit&height2pt&\omit&&\omit&&\omit&&\omit&&\omit&&\omit&\cr
\tablerule
\omit&height2pt&\omit&&\omit&&\omit&&\omit&&\omit&&\omit&\cr
&&$\!\!\!\!\!\!\!(2\!-\!2)\!\!\!\!\!\!\!$
&&$\!\!\!\!\!\!\!\scriptstyle{\ln Y}\!\!\!\!\!\!\!$
&&$\!\!\!\!\!\!\scriptstyle{-2(n^3-n^2)}\!\!\!\!\!\!$
&&$\!\!\!\!\!\scriptstyle{-2(n^3-n^2)}\!\!\!\!\!$
&&$\!\!\!\!\!\scriptstyle{-(2n^3+n^2-3n)}\!\!\!\!\!$
&&$\!\!\!\!\!\!\!\!\scriptstyle{-(2n^3+3n^2-n)}\!\!\!\!\!\!\!\!$ & \cr
\omit&height2pt&\omit&&\omit&&\omit&&\omit&&\omit&&\omit&\cr
\tablerule
\omit&height2pt&\omit&&\omit&&\omit&&\omit&&\omit&&\omit&\cr
&&$\!\!\!\!\!\!\!(2\!-\!3)\!\!\!\!\!\!\!$
&&$\!\!\!\!\!\!\!\!\scriptstyle{1}\!\!\!\!\!\!\!\!$
&&$\!\!\!\!\!\!\!\scriptstyle{10(2n-1)}\!\!\!\!\!\!\!\!$
&&$\!\!\!\!\!\!\!\scriptstyle{-6(2n-1)}\!\!\!\!\!\!\!$
&&$\!\!\!\!\!\!\!\!\scriptstyle{2(10n-2)}\!\!\!\!\!\!\!\!$
&&$\!\!\!\!\!\!\!\!\scriptstyle{-2(2n+1)}\!\!\!\!\!\!\!\!$ & \cr
\omit&height2pt&\omit&&\omit&&\omit&&\omit&&\omit&&\omit&\cr
\tablerule
\omit&height2pt&\omit&&\omit&&\omit&&\omit&&\omit&&\omit&\cr
&&$\!\!\!\!\!\!\!(2\!-\!3)\!\!\!\!\!\!$
&&$\!\!\!\!\!\!\!\scriptstyle{\ln Y}\!\!\!\!\!\!\!$
&&$\!\!\!\!\!\!\scriptstyle{10(n^2-n)}\!\!\!\!\!\!$
&&$\!\!\!\!\!\scriptstyle{-6(n^2-n)}\!\!\!\!\!$
&&$\!\!\!\!\!\scriptstyle{2(5n^2-2n)}\!\!\!\!\!$
&&$\!\!\!\!\!\!\!\!\scriptstyle{-2(n^2+n)}\!\!\!\!\!\!\!\!$ & \cr
\omit&height2pt&\omit&&\omit&&\omit&&\omit&&\omit&&\omit&\cr
\tablerule
\omit&height2pt&\omit&&\omit&&\omit&&\omit&&\omit&&\omit&\cr
&&$\!\!\!\!\!\!\!(3\!-\!1)\!\!\!\!\!\!\!$
&&$\!\!\!\!\!\!\!\!\scriptstyle{1}\!\!\!\!\!\!\!\!$
&&$\!\!\!\!\!\!\!\scriptstyle{-8(2n-2)}\!\!\!\!\!\!\!\!$
&&$\!\!\!\!\!\!\!\scriptstyle{8(2n-2)}\!\!\!\!\!\!\!$
&&$\!\!\!\!\!\!\!\!\scriptstyle{2(4n-1)}\!\!\!\!\!\!\!\!$
&&$\!\!\!\!\!\!\!\!\scriptstyle{-2(8n-3)}\!\!\!\!\!\!\!\!$ & \cr
\omit&height2pt&\omit&&\omit&&\omit&&\omit&&\omit&&\omit&\cr
\tablerule
\omit&height2pt&\omit&&\omit&&\omit&&\omit&&\omit&&\omit&\cr
&&$\!\!\!\!\!\!\!(3\!-\!1)\!\!\!\!\!\!\!$
&&$\!\!\!\!\!\!\!\!\scriptstyle{\ln Y}\!\!\!\!\!\!\!\!$
&&$\!\!\!\!\!\!\!\scriptstyle{-8(n^2-2n)}\!\!\!\!\!\!\!\!$
&&$\!\!\!\!\!\!\!\scriptstyle{8(n^2-2n)}\!\!\!\!\!\!\!$
&&$\!\!\!\!\!\!\!\!\scriptstyle{2(2n^2-n)}\!\!\!\!\!\!\!\!$
&&$\!\!\!\!\!\!\!\!\scriptstyle{-2(4n^2-3n)}\!\!\!\!\!\!\!\!$ & \cr
\omit&height2pt&\omit&&\omit&&\omit&&\omit&&\omit&&\omit&\cr
\tablerule
\omit&height2pt&\omit&&\omit&&\omit&&\omit&&\omit&&\omit&\cr
&&$\!\!\!\!\!\!\!(3\!-\!2)\!\!\!\!\!\!\!$
&&$\!\!\!\!\!\!\!\!\scriptstyle{1}\!\!\!\!\!\!\!\!$
&&$\!\!\!\!\!\!\!\scriptstyle{-2(2n-2)}\!\!\!\!\!\!\!\!$
&&$\!\!\!\!\!\!\!\scriptstyle{-2(2n-2)}\!\!\!\!\!\!\!$
&&$\!\!\!\!\!\!\!\!\scriptstyle{-(4n-1)}\!\!\!\!\!\!\!\!$
&&$\!\!\!\!\!\!\!\!\scriptstyle{(4n-3)}\!\!\!\!\!\!\!\!$ & \cr
\omit&height2pt&\omit&&\omit&&\omit&&\omit&&\omit&&\omit&\cr
\tablerule
\omit&height2pt&\omit&&\omit&&\omit&&\omit&&\omit&&\omit&\cr
&&$\!\!\!\!\!\!\!(3\!-\!2)\!\!\!\!\!\!\!$
&&$\!\!\!\!\!\!\!\!\scriptstyle{\ln Y}\!\!\!\!\!\!\!\!$
&&$\!\!\!\!\!\!\!\scriptstyle{-2(n^2-2n)}\!\!\!\!\!\!\!\!$
&&$\!\!\!\!\!\!\!\scriptstyle{-2(n^2-2n)}\!\!\!\!\!\!\!$
&&$\!\!\!\!\!\!\!\!\scriptstyle{-(2n^2-n)}\!\!\!\!\!\!\!\!$
&&$\!\!\!\!\!\!\!\!\scriptstyle{(2n^2-3n)}\!\!\!\!\!\!\!\!$ & \cr
\omit&height2pt&\omit&&\omit&&\omit&&\omit&&\omit&&\omit&\cr
\tablerule
\omit&height2pt&\omit&&\omit&&\omit&&\omit&&\omit&&\omit&\cr
&&$\!\!\!\!\!\!\!(3\!-\!3)\!\!\!\!\!\!\!$
&&$\!\!\!\!\!\!\!\!\scriptstyle{1}\!\!\!\!\!\!\!\!$
&&$\!\!\!\!\!\!\!\scriptstyle{-5}\!\!\!\!\!\!\!\!$
&&$\!\!\!\!\!\!\!\scriptstyle{3}\!\!\!\!\!\!\!$
&&$\!\!\!\!\!\!\!\!\scriptstyle{-\frac72}\!\!\!\!\!\!\!\!$
&&$\!\!\!\!\!\!\!\!\scriptstyle{-\frac12}\!\!\!\!\!\!\!\!$ & \cr
\omit&height2pt&\omit&&\omit&&\omit&&\omit&&\omit&&\omit&\cr
\tablerule
\omit&height2pt&\omit&&\omit&&\omit&&\omit&&\omit&&\omit&\cr
&&$\!\!\!\!\!\!\!(3\!-\!3)\!\!\!\!\!\!\!$
&&$\!\!\!\!\!\!\!\!\scriptstyle{\ln Y}\!\!\!\!\!\!\!\!$
&&$\!\!\!\!\!\!\!\scriptstyle{-5n}\!\!\!\!\!\!\!\!$
&&$\!\!\!\!\!\!\!\scriptstyle{3n}\!\!\!\!\!\!\!$
&&$\!\!\!\!\!\!\!\!\scriptstyle{-\frac{7}{2}n}\!\!\!\!\!\!\!\!$
&&$\!\!\!\!\!\!\!\!\scriptstyle{-\frac{1}{2}n}\!\!\!\!\!\!\!\!$ & \cr
\omit&height2pt&\omit&&\omit&&\omit&&\omit&&\omit&&\omit&\cr
\tablerule
\omit&height2pt&\omit&&\omit&&\omit&&\omit&&\omit&&\omit&\cr
&&$\!\!\!\!\!\!\!\scriptstyle{\rm{total}}\!\!\!\!\!\!\!\!\!\!$
&&$\!\!\!\!\!\!\!\scriptstyle{1}\!\!\!\!\!\!\!$
&&$\!\!\!\!\!\!\!\!\scriptstyle{-5(6n^2\!-4n-1)}\!\!\!\!\!\!\!\!$
&&$\!\!\!\!\!\!\!\scriptstyle{3(6n^2\!-4n-1)}\!\!\!\!\!\!\!$
&&$\!\!\!\!\!\!\!\!\scriptstyle{(6n^2\!+98n+\frac{49}{2})}\!\!\!\!\!\!\!\!$
&&$\!\!\!\!\!\!\!\!\scriptstyle{-(18n^2\!+78n+\frac{37}{2})}\!\!\!\!\!\!\!\!$ & \cr
\omit&height2pt&\omit&&\omit&&\omit&&\omit&&\omit&&\omit&\cr
\tablerule
\omit&height2pt&\omit&&\omit&&\omit&&\omit&&\omit&&\omit&\cr
&&$\!\!\!\!\!\!\!\scriptstyle{\rm{total}}\!\!\!\!\!\!\!\!\!\!$
&&$\!\!\!\!\!\!\!\scriptstyle{\ln Y}\!\!\!\!\!\!\!$
&&$\!\!\!\!\!\!\scriptstyle{-5(2n^3\!-2n^2\!-n)}\!\!\!\!\!\!$
&&$\!\!\!\!\!\scriptstyle{3(2n^3\!-2n^2\!-n)}\!\!\!\!\!$
&&$\!\!\!\!\!\scriptstyle{(2n^3\!+49n^2\!+\frac{49}{2}n)}\!\!\!\!\!\!\!\!\!\!$
&&$\!\!\!\!\!\!\!\scriptstyle{-(6n^3\!+39n^2\!+\frac{37}{2}n)}\!\!\!\!\!\!\!\!$ & \cr
\omit&height2pt&\omit&&\omit&&\omit&&\omit&&\omit&&\omit&\cr
\tablerule}}

\caption{$i\delta\!\D_{A}\times i[S]_{n\geq 0}-{\rm IIA}$. The factor
$\frac{i \kappa^2H^2}{2^6 \pi^4}\frac{mHaa'}{2}\sum_{n=0}^{\infty}Y^{n}$
multiplies all contributions. Here $Y=\frac{y}{4}$; $\ln Y$
and $1$ are the multiplicative factors for the each individual row.}

\label{dAn0-4}

\end{table}

\begin{table}

\vbox{\tabskip=0pt \offinterlineskip
\def\tablerule{\noalign{\hrule}}
\halign to390pt {\strut#& \vrule#\tabskip=1em plus2em& \hfil#\hfil&
\vrule#& \hfil#\hfil& \vrule#& \hfil#\hfil& \vrule#& \hfil#\hfil&
\vrule#& \hfil#\hfil& \vrule#& \hfil#\hfil& \vrule#&
\hfil#\hfil& \vrule#\tabskip=0pt\cr
\tablerule
\omit&height4pt&\omit&&\omit&&\omit&&\omit&&\omit&&\omit&&\omit&\cr
&&$\!\!\!\!\scriptstyle{{\rm I\!-\!J}}\!\!\!\!\!\!$
&&\omit
&&$\!\!\!\!\scriptstyle{\frac{Ha\Delta\eta^2}{\Delta x^4}}\!\!\!\!\!\!$
&&$\!\!\!\!\!\!\!\scriptstyle{\frac{Ha'\Delta\eta^2}
{\Delta x^4}}\!\!\!\!\!\!\!\!\!$
&&$\!\!\!\!\!\!\!\!\!\!\scriptstyle{\frac{Ha\gamma^0\!\Delta\eta
\gamma^k\!\Delta x_k}{\Delta x^4}}\!\!\!\!\!\!\!\!\!\!\!$
&&$\!\!\!\!\!\!\scriptstyle{\frac{Ha'\!\gamma^0\!\Delta\eta
\gamma^k\!\Delta x_k}{\Delta x^4}}\!\!\!\!\!\!\!\!\!\!\!$
&&$\!\!\!\!\!\!\frac{Ha}{\Delta x^2}\!\!\!\!\!\!$&\cr
\omit&height4pt&\omit&&\omit&&\omit&&\omit&&\omit&&\omit&&\omit&\cr
\tablerule
\omit&height2pt&\omit&&\omit&&\omit&&\omit&&\omit&&\omit&&\omit&\cr
&&$\!\!\!\!\!\scriptstyle{(1-1)}\!\!\!\!\!\!\!\!\!\!\!\!$
&&$\!\!\!\!\!\!\!\scriptstyle{1}\!\!\!\!\!\!\!\!\!\!\!$
&&$\!\!\!\!\!\!\scriptstyle{12(3n^2+8n+3)}\!\!\!\!\!\!\!\!\!$
&&$\!\!\!\!\!\scriptstyle{0}\!\!\!\!\!$
&&$\!\!\!\!\!\!\!\!\scriptstyle{-12(3n^2+8n+3)}\!\!\!\!\!\!\!\!\!\!$
&&$\!\!\!\!\!\!\!\scriptstyle{0}\!\!\!\!\!\!\!\!$
&&$\!\!\!\!\!\!\!\scriptstyle{6(3n^2+6n+2)}\!\!\!\!\!\!\!\!$& \cr
\omit&height2pt&\omit&&\omit&&\omit&&\omit&&\omit&&\omit&&\omit&\cr
\tablerule
\omit&height2pt&\omit&&\omit&&\omit&&\omit&&\omit&&\omit&&\omit&\cr
&&$\!\!\!\!\!\!\scriptstyle{(1-1)}\!\!\!\!\!\!\!\!\!\!\!\!$
&&$\!\!\!\!\!\!\!\!\!\!\scriptstyle{\ln Y}\!\!\!\!\!\!\!\!\!\!\!\!\!\!\!$
&&$\!\!\!\!\!\!\scriptstyle{12(n^3+4n^2+3n)}\!\!\!\!\!\!\!\!\!$
&&$\!\!\!\!\!\scriptstyle{0}\!\!\!\!\!$
&&$\!\!\!\!\!\!\!\!\scriptstyle{-12(n^3+4n^2+3n)}\!\!\!\!\!\!\!\!\!\!$
&&$\!\!\!\!\!\!\!\scriptstyle{0}\!\!\!\!\!\!\!\!$
&&$\!\!\!\!\!\!\!\scriptstyle{6(n^3+3n^2+2n)}\!\!\!\!\!\!\!\!$& \cr
\omit&height2pt&\omit&&\omit&&\omit&&\omit&&\omit&&\omit&&\omit&\cr
\tablerule
\omit&height2pt&\omit&&\omit&&\omit&&\omit&&\omit&&\omit&&\omit&\cr
&&$\!\!\!\!\!\!\scriptstyle{(1-2)}\!\!\!\!\!\!\!\!\!\!\!\!$
&&$\!\!\!\!\!\!\!\scriptstyle{1}\!\!\!\!\!\!\!\!\!\!\!$
&&$\!\!\!\!\!\!\scriptstyle{0}\!\!\!\!\!\!$
&&$\!\!\!\!\!\scriptstyle{0}\!\!\!\!\!$
&&$\!\!\!\!\!\scriptstyle{4(3n^2+4n+1)}\!\!\!\!\!\!\!\!\!\!$
&&$\!\!\!\!\!\!\!\scriptstyle{0}\!\!\!\!\!\!\!\!$
&&$\!\!\!\!\!\!\!\scriptstyle{0}\!\!\!\!\!\!\!\!$& \cr
\omit&height2pt&\omit&&\omit&&\omit&&\omit&&\omit&&\omit&&\omit&\cr
\tablerule
\omit&height2pt&\omit&&\omit&&\omit&&\omit&&\omit&&\omit&&\omit&\cr
&&$\!\!\!\!\!\!\scriptstyle{(1-2)}\!\!\!\!\!\!\!\!\!\!\!\!$
&&$\!\!\!\!\!\!\!\!\!\scriptstyle{\ln Y}\!\!\!\!\!\!\!\!\!\!\!\!\!\!$
&&$\!\!\!\!\!\!\scriptstyle{0}\!\!\!\!\!\!$
&&$\!\!\!\!\!\scriptstyle{0}\!\!\!\!\!$
&&$\!\!\!\!\!\scriptstyle{4(n^3+2n^2+n)}\!\!\!\!\!\!\!\!\!\!$
&&$\!\!\!\!\!\!\!\scriptstyle{0}\!\!\!\!\!\!\!\!$
&&$\!\!\!\!\!\!\!\scriptstyle{0}\!\!\!\!\!\!\!\!$& \cr
\omit&height2pt&\omit&&\omit&&\omit&&\omit&&\omit&&\omit&&\omit&\cr
\tablerule
\omit&height2pt&\omit&&\omit&&\omit&&\omit&&\omit&&\omit&&\omit&\cr
&&$\!\!\!\!\!\!\scriptstyle{(1-3)}\!\!\!\!\!\!\!\!\!\!\!\!$
&&$\!\!\!\!\!\!\!\scriptstyle{1}\!\!\!\!\!\!\!\!\!\!\!$
&&$\!\!\!\!\!\!\scriptstyle{-6(2n+1)}\!\!\!\!\!\!$
&&$\!\!\!\!\!\scriptstyle{0}\!\!\!\!\!$
&&$\!\!\!\!\!\scriptstyle{8(2n+1)}\!\!\!\!\!\!\!\!\!\!$
&&$\!\!\!\!\!\!\!\scriptstyle{0}\!\!\!\!\!\!\!\!$
&&$\!\!\!\!\!\!\!\scriptstyle{0}\!\!\!\!\!\!\!\!$& \cr
\omit&height2pt&\omit&&\omit&&\omit&&\omit&&\omit&&\omit&&\omit&\cr
\tablerule
\omit&height2pt&\omit&&\omit&&\omit&&\omit&&\omit&&\omit&&\omit&\cr
&&$\!\!\!\!\!\!\scriptstyle{(1-3)}\!\!\!\!\!\!\!\!\!\!\!\!$
&&$\!\!\!\!\!\!\!\!\!\scriptstyle{\ln Y}\!\!\!\!\!\!\!\!\!\!\!\!\!\!$
&&$\!\!\!\!\!\!\scriptstyle{-6(n^2+n)}\!\!\!\!\!\!$
&&$\!\!\!\!\!\scriptstyle{0}\!\!\!\!\!$
&&$\!\!\!\!\!\scriptstyle{8(n^2+n)}\!\!\!\!\!\!\!\!\!\!$
&&$\!\!\!\!\!\!\!\scriptstyle{0}\!\!\!\!\!\!\!\!$
&&$\!\!\!\!\!\!\!\scriptstyle{0}\!\!\!\!\!\!\!\!$& \cr
\omit&height2pt&\omit&&\omit&&\omit&&\omit&&\omit&&\omit&&\omit&\cr
\tablerule
\omit&height2pt&\omit&&\omit&&\omit&&\omit&&\omit&&\omit&&\omit&\cr
&&$\!\!\!\!\!\!\scriptstyle{(2-1)}\!\!\!\!\!\!\!\!\!\!\!\!$
&&$\!\!\!\!\!\!\!\scriptstyle{1}\!\!\!\!\!\!\!\!\!\!\!$
&&$\!\!\!\!\!\!\scriptstyle{2(3n^2+4n+1)}\!\!\!\!\!\!$
&&$\!\!\!\!\!\scriptstyle{-2(3n^2-1)}\!\!\!\!\!$
&&$\!\!\!\!\!\scriptstyle{-2(3n^2+4n+1)}\!\!\!\!\!\!\!\!\!\!$
&&$\!\!\!\!\!\!\!\scriptstyle{2(3n^2-1)}\!\!\!\!\!\!\!\!$
&&$\!\!\!\!\!\!\!\scriptstyle{(6n^2+14n+5)}\!\!\!\!\!\!\!\!$&\cr
\omit&height2pt&\omit&&\omit&&\omit&&\omit&&\omit&&\omit&&\omit&\cr
\tablerule
\omit&height2pt&\omit&&\omit&&\omit&&\omit&&\omit&&\omit&&\omit&\cr
&&$\!\!\!\!\!\!\scriptstyle{(2-1)}\!\!\!\!\!\!\!\!\!\!\!\!$
&&$\!\!\!\!\!\!\!\!\!\scriptstyle{\ln Y}\!\!\!\!\!\!\!\!\!\!\!\!\!\!$
&&$\!\!\!\!\!\!\scriptstyle{2(n^3+2n^2+n)}\!\!\!\!\!\!$
&&$\!\!\!\!\!\scriptstyle{-2(n^3-n)}\!\!\!\!\!$
&&$\!\!\!\!\!\scriptstyle{-2(n^3+2n^2+n)}\!\!\!\!\!\!\!\!\!\!$
&&$\!\!\!\!\!\!\!\scriptstyle{2(n^3-n)}\!\!\!\!\!\!\!\!$
&&$\!\!\!\!\!\!\!\scriptstyle{(2n^3+7n^2+5n)}\!\!\!\!\!\!\!\!$&\cr
\omit&height2pt&\omit&&\omit&&\omit&&\omit&&\omit&&\omit&&\omit&\cr
\tablerule
\omit&height2pt&\omit&&\omit&&\omit&&\omit&&\omit&&\omit&&\omit&\cr
&&$\!\!\!\!\!\!\scriptstyle{(2-2)}\!\!\!\!\!\!\!\!\!\!\!\!\!$
&&$\!\!\!\!\!\!\!\scriptstyle{1}\!\!\!\!\!\!\!\!\!\!\!$
&&$\!\!\!\!\!\!\scriptstyle{(3n^2+4n+1)}\!\!\!\!\!\!$
&&$\!\!\!\!\!\scriptstyle{0}\!\!\!\!\!$
&&$\!\!\!\!\!\scriptstyle{0}\!\!\!\!\!\!\!\!\!\!$
&&$\!\!\!\!\!\!\!\scriptstyle{0}\!\!\!\!\!\!\!\!$
&&$\!\!\!\!\!\!\!\scriptstyle{\frac12(6n^2\!+14n+5)}\!\!\!\!\!\!\!\!$& \cr
\omit&height2pt&\omit&&\omit&&\omit&&\omit&&\omit&&\omit&&\omit&\cr
\tablerule
\omit&height2pt&\omit&&\omit&&\omit&&\omit&&\omit&&\omit&&\omit&\cr
&&$\!\!\!\!\!\!\scriptstyle{(2-2)}\!\!\!\!\!\!\!\!\!\!\!$
&&$\!\!\!\!\!\!\!\!\!\!\scriptstyle{\ln Y}\!\!\!\!\!\!\!\!\!\!\!\!\!\!$
&&$\!\!\!\!\!\!\scriptstyle{(n^3+2n^2+n)}\!\!\!\!\!\!$
&&$\!\!\!\!\!\scriptstyle{0}\!\!\!\!\!$
&&$\!\!\!\!\!\scriptstyle{0}\!\!\!\!\!\!\!\!\!\!$
&&$\!\!\!\!\!\!\!\scriptstyle{0}\!\!\!\!\!\!\!\!$
&&$\!\!\!\!\!\!\!\scriptstyle{\frac12(2n^3\!+7n^2\!+5n)}
\!\!\!\!\!\!\!\!$& \cr
\omit&height2pt&\omit&&\omit&&\omit&&\omit&&\omit&&\omit&&\omit&\cr
\tablerule
\omit&height2pt&\omit&&\omit&&\omit&&\omit&&\omit&&\omit&&\omit&\cr
&&$\!\!\!\!\!\!\scriptstyle{(2-3)}\!\!\!\!\!\!\!\!\!\!\!$
&&$\!\!\!\!\!\!\!\scriptstyle{1}\!\!\!\!\!\!\!\!\!\!\!$
&&$\!\!\!\!\!\!\scriptstyle{-3(2n+1)}\!\!\!\!\!\!$
&&$\!\!\!\!\!\scriptstyle{0}\!\!\!\!\!$
&&$\!\!\!\!\!\scriptstyle{2(2n+1)}\!\!\!\!\!\!\!\!\!\!$
&&$\!\!\!\!\!\!\!\scriptstyle{0}\!\!\!\!\!\!\!\!$
&&$\!\!\!\!\!\!\!\scriptstyle{-3(2n+1)}\!\!\!\!\!\!\!\!$& \cr
\omit&height2pt&\omit&&\omit&&\omit&&\omit&&\omit&&\omit&&\omit&\cr
\tablerule
\omit&height2pt&\omit&&\omit&&\omit&&\omit&&\omit&&\omit&&\omit&\cr
&&$\!\!\!\!\!\!\scriptstyle{(2-3)}\!\!\!\!\!\!\!\!\!\!\!$
&&$\!\!\!\!\!\!\!\!\!\scriptstyle{\ln Y}\!\!\!\!\!\!\!\!\!\!\!\!\!\!\!$
&&$\!\!\!\!\!\!\scriptstyle{-3(n^2+n)}\!\!\!\!\!\!$
&&$\!\!\!\!\!\scriptstyle{0}\!\!\!\!\!$
&&$\!\!\!\!\!\scriptstyle{2(n^2+n)}\!\!\!\!\!\!\!\!\!\!$
&&$\!\!\!\!\!\!\!\scriptstyle{0}\!\!\!\!\!\!\!\!$
&&$\!\!\!\!\!\!\!\scriptstyle{-3(n^2+n)}\!\!\!\!\!\!\!\!$& \cr
\omit&height2pt&\omit&&\omit&&\omit&&\omit&&\omit&&\omit&&\omit&\cr
\tablerule
\omit&height2pt&\omit&&\omit&&\omit&&\omit&&\omit&&\omit&&\omit&\cr
&&$\!\!\!\!\!\!\scriptstyle{(3-1)}\!\!\!\!\!\!\!\!\!\!\!$
&&$\!\!\!\!\!\!\!\scriptstyle{1}\!\!\!\!\!\!\!\!\!\!\!$
&&$\!\!\!\!\!\!\scriptstyle{4n}\!\!\!\!\!\!\!\!\!$
&&$\!\!\!\!\!\scriptstyle{-2(2n+1)}\!\!\!\!\!\!\!\!$
&&$\!\!\!\!\!\!\!\scriptstyle{-2(2n)}\!\!\!\!\!\!\!\!\!\!$
&&$\!\!\!\!\!\!\!\scriptstyle{2(2n+1)}\!\!\!\!\!\!\!\!$
&&$\!\!\!\!\!\!\!\scriptstyle{-(2n+3)}\!\!\!\!\!\!\!\!$& \cr
\omit&height2pt&\omit&&\omit&&\omit&&\omit&&\omit&&\omit&&\omit&\cr
\tablerule
\omit&height2pt&\omit&&\omit&&\omit&&\omit&&\omit&&\omit&&\omit&\cr
&&$\!\!\!\!\!\!\scriptstyle{(3-1)}\!\!\!\!\!\!\!\!\!\!\!$
&&$\!\!\!\!\!\!\!\!\!\scriptstyle{\ln Y}\!\!\!\!\!\!\!\!\!\!\!\!\!\!$
&&$\!\!\!\!\!\!\scriptstyle{(2n^2-2)}\!\!\!\!\!\!\!\!\!$
&&$\!\!\!\!\!\scriptstyle{-2(n^2+n)}\!\!\!\!\!\!\!\!$
&&$\!\!\!\!\!\!\!\scriptstyle{-2(n^2-1)}\!\!\!\!\!\!\!\!\!\!$
&&$\!\!\!\!\!\!\!\scriptstyle{2(n^2+n)}\!\!\!\!\!\!\!\!$
&&$\!\!\!\!\!\!\!\scriptstyle{-(n^2+3n+2)}\!\!\!\!\!\!\!\!$& \cr
\omit&height2pt&\omit&&\omit&&\omit&&\omit&&\omit&&\omit&&\omit&\cr
\tablerule
\omit&height2pt&\omit&&\omit&&\omit&&\omit&&\omit&&\omit&&\omit&\cr
&&$\!\!\!\!\!\!\scriptstyle{(3-2)}\!\!\!\!\!\!\!\!\!\!\!$
&&$\!\!\!\!\!\!\!\scriptstyle{1}\!\!\!\!\!\!\!\!\!\!\!$
&&$\!\!\!\!\!\!\scriptstyle{6n}\!\!\!\!\!\!\!\!\!$
&&$\!\!\!\!\!\scriptstyle{0}\!\!\!\!\!\!\!\!$
&&$\!\!\!\!\!\!\!\scriptstyle{2(2n)}\!\!\!\!\!\!\!\!\!\!$
&&$\!\!\!\!\!\!\!\scriptstyle{0}\!\!\!\!\!\!\!\!$
&&$\!\!\!\!\!\!\!\scriptstyle{6n+\frac92}\!\!\!\!\!\!\!\!$& \cr
\omit&height2pt&\omit&&\omit&&\omit&&\omit&&\omit&&\omit&&\omit&\cr
\tablerule
\omit&height2pt&\omit&&\omit&&\omit&&\omit&&\omit&&\omit&&\omit&\cr
&&$\!\!\!\!\!\!\scriptstyle{(3-2)}\!\!\!\!\!\!\!\!\!\!\!$
&&$\!\!\!\!\!\!\!\!\scriptstyle{\ln Y}\!\!\!\!\!\!\!\!\!\!\!\!\!\!$
&&$\!\!\!\!\!\!\scriptstyle{3(n^2-1)}\!\!\!\!\!\!\!\!\!$
&&$\!\!\!\!\!\scriptstyle{0}\!\!\!\!\!\!\!\!$
&&$\!\!\!\!\!\!\!\scriptstyle{2(n^2-1)}\!\!\!\!\!\!\!\!\!\!$
&&$\!\!\!\!\!\!\!\scriptstyle{0}\!\!\!\!\!\!\!\!$
&&$\!\!\!\!\!\!\!\scriptstyle{3n^2+\frac{9}{2}n+\frac32}\!\!\!\!\!\!\!\!$& \cr
\omit&height2pt&\omit&&\omit&&\omit&&\omit&&\omit&&\omit&&\omit&\cr
\tablerule
\omit&height2pt&\omit&&\omit&&\omit&&\omit&&\omit&&\omit&&\omit&\cr
&&$\!\!\!\!\!\!\scriptstyle{(3-3)}\!\!\!\!\!\!\!\!\!\!\!$
&&$\!\!\!\!\!\!\!\scriptstyle{1}\!\!\!\!\!\!\!\!\!\!\!$
&&$\!\!\!\!\!\!\scriptstyle{\frac52}\!\!\!\!\!\!\!\!\!$
&&$\!\!\!\!\!\scriptstyle{0}\!\!\!\!\!\!\!\!$
&&$\!\!\!\!\!\!\!\scriptstyle{0}\!\!\!\!\!\!\!\!\!\!$
&&$\!\!\!\!\!\!\!\scriptstyle{0}\!\!\!\!\!\!\!\!$
&&$\!\!\!\!\!\!\!\scriptstyle{\frac74}\!\!\!\!\!\!\!\!$& \cr
\omit&height2pt&\omit&&\omit&&\omit&&\omit&&\omit&&\omit&&\omit&\cr
\tablerule
\omit&height2pt&\omit&&\omit&&\omit&&\omit&&\omit&&\omit&&\omit&\cr
&&$\!\!\!\!\!\!\scriptstyle{(3-3)}\!\!\!\!\!\!\!\!\!\!\!$
&&$\!\!\!\!\!\!\!\!\scriptstyle{\ln Y}\!\!\!\!\!\!\!\!\!\!\!\!\!\!\!$
&&$\!\!\!\!\!\!\scriptstyle{\frac52(n+1)}\!\!\!\!\!\!\!\!\!$
&&$\!\!\!\!\!\scriptstyle{0}\!\!\!\!\!\!\!\!$
&&$\!\!\!\!\!\!\!\scriptstyle{0}\!\!\!\!\!\!\!\!\!\!$
&&$\!\!\!\!\!\!\!\scriptstyle{0}\!\!\!\!\!\!\!\!$
&&$\!\!\!\!\!\!\!\scriptstyle{\frac74(n+1)}\!\!\!\!\!\!\!\!$& \cr
\omit&height2pt&\omit&&\omit&&\omit&&\omit&&\omit&&\omit&&\omit&\cr
\tablerule
\omit&height2pt&\omit&&\omit&&\omit&&\omit&&\omit&&\omit&&\omit&\cr
&&$\!\!\!\!\!\!\!\scriptstyle{\rm{total}}\!\!\!\!\!\!\!\!\!\!$
&&$\!\!\!\!\!\!\!\scriptstyle{1}\!\!\!\!\!\!\!\!\!\!\!$
&&$\!\!\!\!\!\!\scriptstyle{5(9n^2\!+\!20n\!+\!\frac{13}{2})}
\!\!\!\!\!\!\!\!\!$
&&$\!\!\!\!\!\scriptstyle{-2(3n^2+2n)}\!\!\!\!\!\!\!\!$
&&$\!\!\!\!\!\!\!\scriptstyle{-2(15n^2\!+\!34n\!+\!12)}\!\!\!\!\!\!\!\!\!\!$
&&$\!\!\!\!\!\!\!\scriptstyle{2(3n^2+2n)}\!\!\!\!\!\!\!\!$
&&$\!\!\!\!\!\!\!\scriptstyle{27n^2\!+55n\!+\!\frac{79}{4}}\!\!\!\!\!\!\!\!$& \cr
\omit&height2pt&\omit&&\omit&&\omit&&\omit&&\omit&&\omit&&\omit&\cr
\tablerule
\omit&height2pt&\omit&&\omit&&\omit&&\omit&&\omit&&\omit&&\omit&\cr
&&$\!\!\!\!\!\!\!\scriptstyle{\rm total}\!\!\!\!\!\!\!\!\!\!$
&&$\!\!\!\!\!\!\!\!\!\scriptstyle{\ln Y}\!\!\!\!\!\!\!\!\!\!\!\!\!\!\!\!$
&&$\!\!\!\!\!\!\scriptstyle{5(3n^3\!+\!10n^2\!+\!\frac{13}{2}n\!-\!\frac12)}
\!\!\!\!\!\!\!\!\!$
&&$\!\!\!\!\!\!\!\scriptstyle{-2(n^3+n^2)}\!\!\!\!\!\!\!\!$
&&$\!\!\!\!\!\!\scriptstyle{-2(5n^3\!+\!17n^2\!+\!12n)}\!\!\!\!\!\!\!\!\!\!$
&&$\!\!\!\!\!\!\!\scriptstyle{2(n^3+n^2)}\!\!\!\!\!\!\!\!\!\!\!$
&&$\!\!\!\!\!\!\scriptstyle{9n^3\!+\!\frac{55}{2}n^2\!+\!\frac{77}{4}n
\!+\!\frac54}\!\!\!\!\!\!\!\!$& \cr
\omit&height2pt&\omit&&\omit&&\omit&&\omit&&\omit&&\omit&&\omit&\cr
\tablerule}}

\caption{$i\delta\!\D_{A}\times i[S]_{n\geq 0}-{\rm IIB}$. The factor
$\frac{i \kappa^2H^2}{2^6 \pi^4}\frac{mHaa'}{2}\sum_{n=0}^{\infty}Y^{n}$
multiplies all contributions. Here $Y=\frac{y}{4}$; $\ln Y$
and $1$ are the multiplicative factors for the each individual row.}

\label{dAn0-5}
\end{table}

\begin{table}

\vbox{\tabskip=0pt \offinterlineskip
\def\tablerule{\noalign{\hrule}}
\halign to390pt {\strut#& \vrule#\tabskip=1em plus2em& \hfil#\hfil&
\vrule#& \hfil#\hfil& \vrule#& \hfil#\hfil& \vrule#& \hfil#\hfil&
\vrule#& \hfil#\hfil& \vrule#& \hfil#\hfil& \vrule#&
\hfil#\hfil& \vrule#\tabskip=0pt\cr
\tablerule
\omit&height4pt&\omit&&\omit&&\omit&&\omit&&\omit&&\omit&&\omit&\cr
&&$\!\!\!\!\scriptstyle{{\rm I\!-\!J}}\!\!\!\!\!\!$
&&\omit
&&$\!\!\!\!\!\!\!\!\frac{Ha'}{\Delta x^2}\!\!\!\!\!\!\!\!\!$
&&$\!\!\!\!\!\!\!\scriptstyle{\frac{H^2a^2\Delta\eta}
{\Delta x^2}}\!\!\!\!\!\!\!\!\!$
&&$\!\!\!\!\!\!\!\!\!\!\scriptstyle{\frac{H^2\!a^2\gamma^0\!
\!\gamma^k\!\Delta x_k}{\Delta x^2}}\!\!\!\!\!\!\!\!\!\!\!$
&&$\!\!\!\!\!\!\!\scriptstyle{\frac{H^{2}\!aa'\gamma^0\!
\gamma^k\!\Delta x_k}{\Delta x^2}}\!\!\!\!\!\!\!\!\!\!\!$
&&$\!\!\!\!\!\!\scriptstyle{H^{3}a^{2}a'}\!\!\!\!\!\!$&\cr
\omit&height4pt&\omit&&\omit&&\omit&&\omit&&\omit&&\omit&&\omit&\cr
\tablerule
\omit&height2pt&\omit&&\omit&&\omit&&\omit&&\omit&&\omit&&\omit&\cr
&&$\!\!\!\!\!\!\!\scriptstyle{(1-1)}\!\!\!\!\!\!\!\!\!\!$
&&$\!\!\!\!\!\!\!\scriptstyle{1}\!\!\!\!\!\!\!\!\!\!\!$
&&$\!\!\!\!\!\scriptstyle{0}\!\!\!\!\!\!\!\!\!$
&&$\!\!\!\!\!\scriptstyle{-3(3n^2\!+\!10n\!+\!8)}\!\!\!\!\!\!\!\!\!$
&&$\!\!\!\!\!\!\!\scriptstyle{3(3n^2\!+\!10n\!+\!8)}\!\!\!\!\!\!\!\!\!\!$
&&$\!\!\!\!\!\!\!\scriptstyle{0}\!\!\!\!\!\!\!\!$
&&$\!\!\!\!\!\!\!\scriptstyle{-\frac32(3n^2\!+\!8n\!+\!5)}\!\!\!\!\!\!\!\!$& \cr
\omit&height2pt&\omit&&\omit&&\omit&&\omit&&\omit&&\omit&&\omit&\cr
\tablerule
\omit&height2pt&\omit&&\omit&&\omit&&\omit&&\omit&&\omit&&\omit&\cr
&&$\!\!\!\!\!\!\!\scriptstyle{(1-1)}\!\!\!\!\!\!\!\!\!\!$
&&$\!\!\!\!\!\!\!\scriptstyle{\ln Y}\!\!\!\!\!\!\!\!\!\!\!$
&&$\!\!\!\!\!\!\scriptstyle{0}\!\!\!\!\!\!\!\!\!\!\!$
&&$\!\!\!\!\!\!\!\scriptstyle{-3(n^3\!+\!5n^2\!+\!8n\!+\!4)}
\!\!\!\!\!\!\!\!\!\!\!$
&&$\!\!\!\!\!\!\!\!\scriptstyle{3(n^3\!+\!5n^2\!+\!8n\!+\!4)}
\!\!\!\!\!\!\!\!\!\!\!$
&&$\!\!\!\!\!\!\!\scriptstyle{0}\!\!\!\!\!\!\!\!$
&&$\!\!\!\!\!\!\!\scriptstyle{-\frac32(n^3\!+\!4n^2\!+\!5n\!+\!2)}
\!\!\!\!\!\!\!\!\!$& \cr
\omit&height2pt&\omit&&\omit&&\omit&&\omit&&\omit&&\omit&&\omit&\cr
\tablerule
\omit&height2pt&\omit&&\omit&&\omit&&\omit&&\omit&&\omit&&\omit&\cr
&&$\!\!\!\!\!\!\!\scriptstyle{(1-2)}\!\!\!\!\!\!\!\!\!\!$
&&$\!\!\!\!\!\!\!\scriptstyle{1}\!\!\!\!\!\!\!\!\!\!\!$
&&$\!\!\!\!\!\scriptstyle{0}\!\!\!\!\!\!\!\!\!$
&&$\!\!\!\!\!\scriptstyle{0}\!\!\!\!\!$
&&$\!\!\!\!\!\scriptstyle{-(3n^2\!+\!10n\!+\!8)}\!\!\!\!\!\!\!\!\!\!$
&&$\!\!\!\!\!\!\!\scriptstyle{0}\!\!\!\!\!\!\!\!$
&&$\!\!\!\!\!\!\!\scriptstyle{0}\!\!\!\!\!\!\!\!$& \cr
\omit&height2pt&\omit&&\omit&&\omit&&\omit&&\omit&&\omit&&\omit&\cr
\tablerule
\omit&height2pt&\omit&&\omit&&\omit&&\omit&&\omit&&\omit&&\omit&\cr
&&$\!\!\!\!\!\!\!\scriptstyle{(1-2)}\!\!\!\!\!\!\!\!\!\!$
&&$\!\!\!\!\!\!\!\scriptstyle{\ln Y}\!\!\!\!\!\!\!\!\!\!\!$
&&$\!\!\!\!\scriptstyle{0}\!\!\!\!\!\!\!\!\!$
&&$\!\!\!\!\!\scriptstyle{0}\!\!\!\!\!$
&&$\!\!\!\!\!\scriptstyle{-(n^3\!+\!5n^2\!+\!8n\!+\!4)}
\!\!\!\!\!\!\!\!\!\!$
&&$\!\!\!\!\!\!\!\scriptstyle{0}\!\!\!\!\!\!\!\!$
&&$\!\!\!\!\!\!\!\scriptstyle{0}\!\!\!\!\!\!\!\!$& \cr
\omit&height2pt&\omit&&\omit&&\omit&&\omit&&\omit&&\omit&&\omit&\cr
\tablerule
\omit&height2pt&\omit&&\omit&&\omit&&\omit&&\omit&&\omit&&\omit&\cr
&&$\!\!\!\!\!\!\!\scriptstyle{(1-3)}\!\!\!\!\!\!\!\!\!\!$
&&$\!\!\!\!\!\!\!\scriptstyle{1}\!\!\!\!\!\!\!\!\!\!\!$
&&$\!\!\!\!\!\scriptstyle{0}\!\!\!\!\!\!\!\!\!\!$
&&$\!\!\!\!\!\scriptstyle{\frac32(2n+3)}\!\!\!\!\!$
&&$\!\!\!\!\!\scriptstyle{-2(2n+3)}\!\!\!\!\!\!\!\!\!\!$
&&$\!\!\!\!\!\!\!\scriptstyle{0}\!\!\!\!\!\!\!\!$
&&$\!\!\!\!\!\!\!\scriptstyle{0}\!\!\!\!\!\!\!\!$& \cr
\omit&height2pt&\omit&&\omit&&\omit&&\omit&&\omit&&\omit&&\omit&\cr
\tablerule
\omit&height2pt&\omit&&\omit&&\omit&&\omit&&\omit&&\omit&&\omit&\cr
&&$\!\!\!\!\!\!\!\scriptstyle{(1-3)}\!\!\!\!\!\!\!\!\!\!$
&&$\!\!\!\!\!\!\!\scriptstyle{\ln Y}\!\!\!\!\!\!\!\!\!\!\!$
&&$\!\!\!\!\!\!\scriptstyle{0}\!\!\!\!\!\!\!\!\!\!$
&&$\!\!\!\!\!\scriptstyle{\frac32(n^2\!+\!3n\!+\!2)}\!\!\!\!\!$
&&$\!\!\!\!\!\scriptstyle{-2(n^2+3n+2)}\!\!\!\!\!\!\!\!\!\!$
&&$\!\!\!\!\!\!\!\scriptstyle{0}\!\!\!\!\!\!\!\!$
&&$\!\!\!\!\!\!\!\scriptstyle{0}\!\!\!\!\!\!\!\!$& \cr
\omit&height2pt&\omit&&\omit&&\omit&&\omit&&\omit&&\omit&&\omit&\cr
\tablerule
\omit&height2pt&\omit&&\omit&&\omit&&\omit&&\omit&&\omit&&\omit&\cr
&&$\!\!\!\!\!\!\!\scriptstyle{(2-1)}\!\!\!\!\!\!\!\!\!\!$
&&$\!\!\!\!\!\!\!\scriptstyle{1}\!\!\!\!\!\!\!\!\!\!\!$
&&$\!\!\!\!\!\!\scriptstyle{-(6n^2\!+\!6n\!+\!1)}\!\!\!\!\!\!$
&&$\!\!\!\!\!\scriptstyle{0}\!\!\!\!\!$
&&$\!\!\!\!\!\scriptstyle{0}\!\!\!\!\!\!\!\!\!\!$
&&$\!\!\!\!\!\!\!\scriptstyle{-(3n^2\!+\!4n\!+\!1)}\!\!\!\!\!\!\!\!\!\!$
&&$\!\!\!\!\!\!\!\scriptstyle{0}\!\!\!\!\!\!\!\!$&\cr
\omit&height2pt&\omit&&\omit&&\omit&&\omit&&\omit&&\omit&&\omit&\cr
\tablerule
\omit&height2pt&\omit&&\omit&&\omit&&\omit&&\omit&&\omit&&\omit&\cr
&&$\!\!\!\!\!\!\!\scriptstyle{(2-1)}\!\!\!\!\!\!\!\!\!\!$
&&$\!\!\!\!\!\!\!\scriptstyle{\ln Y}\!\!\!\!\!\!\!\!\!\!\!$
&&$\!\!\!\!\!\!\scriptstyle{-(2n^3\!+\!3n^2\!+\!n)}\!\!\!\!\!\!\!\!\!$
&&$\!\!\!\!\!\scriptstyle{0}\!\!\!\!\!$
&&$\!\!\!\!\!\scriptstyle{0}\!\!\!\!\!\!\!\!\!\!$
&&$\!\!\!\!\!\!\scriptstyle{-(n^3\!+\!2n^2\!+\!n)}\!\!\!\!\!\!\!\!\!\!$
&&$\!\!\!\!\!\!\!\scriptstyle{0}\!\!\!\!\!\!\!\!$&\cr
\omit&height2pt&\omit&&\omit&&\omit&&\omit&&\omit&&\omit&&\omit&\cr
\tablerule
\omit&height2pt&\omit&&\omit&&\omit&&\omit&&\omit&&\omit&&\omit&\cr
&&$\!\!\!\!\!\!\!\scriptstyle{(2-2)}\!\!\!\!\!\!\!\!\!\!$
&&$\!\!\!\!\!\!\!\scriptstyle{1}\!\!\!\!\!\!\!\!\!\!\!$
&&$\!\!\!\!\!\!\scriptstyle{0}\!\!\!\!\!\!$
&&$\!\!\!\!\!\scriptstyle{0}\!\!\!\!\!$
&&$\!\!\!\!\!\scriptstyle{0}\!\!\!\!\!\!\!\!\!\!$
&&$\!\!\!\!\!\!\!\scriptstyle{0}\!\!\!\!\!\!\!\!$
&&$\!\!\!\!\!\!\!\scriptstyle{0}\!\!\!\!\!\!\!\!$& \cr
\omit&height2pt&\omit&&\omit&&\omit&&\omit&&\omit&&\omit&&\omit&\cr
\tablerule
\omit&height2pt&\omit&&\omit&&\omit&&\omit&&\omit&&\omit&&\omit&\cr
&&$\!\!\!\!\!\!\!\scriptstyle{(2-2)}\!\!\!\!\!\!\!\!\!\!$
&&$\!\!\!\!\!\!\!\scriptstyle{\ln Y}\!\!\!\!\!\!\!\!\!\!\!$
&&$\!\!\!\!\!\!\scriptstyle{0}\!\!\!\!\!\!$
&&$\!\!\!\!\!\scriptstyle{0}\!\!\!\!\!$
&&$\!\!\!\!\!\scriptstyle{0}\!\!\!\!\!\!\!\!\!\!$
&&$\!\!\!\!\!\!\!\scriptstyle{0}\!\!\!\!\!\!\!\!$
&&$\!\!\!\!\!\!\!\scriptstyle{0}\!\!\!\!\!\!\!\!$& \cr
\omit&height2pt&\omit&&\omit&&\omit&&\omit&&\omit&&\omit&&\omit&\cr
\tablerule
\omit&height2pt&\omit&&\omit&&\omit&&\omit&&\omit&&\omit&&\omit&\cr
&&$\!\!\!\!\!\!\!\scriptstyle{(2-3)}\!\!\!\!\!\!\!\!\!\!$
&&$\!\!\!\!\!\!\!\scriptstyle{1}\!\!\!\!\!\!\!\!\!\!\!$
&&$\!\!\!\!\!\!\scriptstyle{0}\!\!\!\!\!\!$
&&$\!\!\!\!\!\scriptstyle{0}\!\!\!\!\!$
&&$\!\!\!\!\!\scriptstyle{0}\!\!\!\!\!\!\!\!\!\!$
&&$\!\!\!\!\!\!\!\scriptstyle{0}\!\!\!\!\!\!\!\!$
&&$\!\!\!\!\!\!\!\scriptstyle{0}\!\!\!\!\!\!\!\!$& \cr
\omit&height2pt&\omit&&\omit&&\omit&&\omit&&\omit&&\omit&&\omit&\cr
\tablerule
\omit&height2pt&\omit&&\omit&&\omit&&\omit&&\omit&&\omit&&\omit&\cr
&&$\!\!\!\!\!\!\!\scriptstyle{(2-3)}\!\!\!\!\!\!\!\!\!\!$
&&$\!\!\!\!\!\!\!\scriptstyle{\ln Y}\!\!\!\!\!\!\!\!\!\!\!$
&&$\!\!\!\!\!\!\scriptstyle{0}\!\!\!\!\!\!$
&&$\!\!\!\!\!\scriptstyle{0}\!\!\!\!\!$
&&$\!\!\!\!\!\scriptstyle{0}\!\!\!\!\!\!\!\!\!\!$
&&$\!\!\!\!\!\!\!\scriptstyle{0}\!\!\!\!\!\!\!\!$
&&$\!\!\!\!\!\!\!\scriptstyle{0}\!\!\!\!\!\!\!\!$& \cr
\omit&height2pt&\omit&&\omit&&\omit&&\omit&&\omit&&\omit&&\omit&\cr
\tablerule
\omit&height2pt&\omit&&\omit&&\omit&&\omit&&\omit&&\omit&&\omit&\cr
&&$\!\!\!\!\!\!\!\scriptstyle{(3-1)}\!\!\!\!\!\!\!\!\!\!$
&&$\!\!\!\!\!\!\!\scriptstyle{1}\!\!\!\!\!\!\!\!\!\!\!$
&&$\!\!\!\!\!\!\scriptstyle{(2n-2)}\!\!\!\!\!\!$
&&$\!\!\!\!\!\scriptstyle{0}\!\!\!\!\!$
&&$\!\!\!\!\!\scriptstyle{0}\!\!\!\!\!\!\!\!\!\!$
&&$\!\!\!\!\!\!\!\scriptstyle{2(2n+2)}\!\!\!\!\!\!\!\!$
&&$\!\!\!\!\!\!\!\scriptstyle{0}\!\!\!\!\!\!\!\!$& \cr
\omit&height2pt&\omit&&\omit&&\omit&&\omit&&\omit&&\omit&&\omit&\cr
\tablerule
\omit&height2pt&\omit&&\omit&&\omit&&\omit&&\omit&&\omit&&\omit&\cr
&&$\!\!\!\!\!\!\!\scriptstyle{(3-1)}\!\!\!\!\!\!\!\!\!\!$
&&$\!\!\!\!\!\!\!\scriptstyle{\ln Y}\!\!\!\!\!\!\!\!\!\!\!$
&&$\!\!\!\!\!\!\scriptstyle{(n^2\!-\!2n\!-\!3)}\!\!\!\!\!\!$
&&$\!\!\!\!\!\scriptstyle{0}\!\!\!\!\!$
&&$\!\!\!\!\!\scriptstyle{0}\!\!\!\!\!\!\!\!\!\!$
&&$\!\!\!\!\!\!\!\scriptstyle{2(n^2\!+\!2n\!+\!1)}\!\!\!\!\!\!\!\!$
&&$\!\!\!\!\!\!\!\scriptstyle{0}\!\!\!\!\!\!\!\!$& \cr
\omit&height2pt&\omit&&\omit&&\omit&&\omit&&\omit&&\omit&&\omit&\cr
\tablerule
\omit&height2pt&\omit&&\omit&&\omit&&\omit&&\omit&&\omit&&\omit&\cr
&&$\!\!\!\!\!\!\!\scriptstyle{(3-2)}\!\!\!\!\!\!\!\!\!\!$
&&$\!\!\!\!\!\!\!\scriptstyle{1}\!\!\!\!\!\!\!\!\!\!\!$
&&$\!\!\!\!\!\!\scriptstyle{0}\!\!\!\!\!\!$
&&$\!\!\!\!\!\scriptstyle{0}\!\!\!\!\!$
&&$\!\!\!\!\!\scriptstyle{0}\!\!\!\!\!\!\!\!\!\!$
&&$\!\!\!\!\!\!\!\scriptstyle{0}\!\!\!\!\!\!\!\!$
&&$\!\!\!\!\!\!\!\scriptstyle{0}\!\!\!\!\!\!\!\!$& \cr
\omit&height2pt&\omit&&\omit&&\omit&&\omit&&\omit&&\omit&&\omit&\cr
\tablerule
\omit&height2pt&\omit&&\omit&&\omit&&\omit&&\omit&&\omit&&\omit&\cr
&&$\!\!\!\!\!\!\!\scriptstyle{(3-2)}\!\!\!\!\!\!\!\!\!\!$
&&$\!\!\!\!\!\!\!\scriptstyle{\ln Y}\!\!\!\!\!\!\!\!\!\!\!$
&&$\!\!\!\!\!\!\scriptstyle{0}\!\!\!\!\!\!$
&&$\!\!\!\!\!\scriptstyle{0}\!\!\!\!\!$
&&$\!\!\!\!\!\scriptstyle{0}\!\!\!\!\!\!\!\!\!\!$
&&$\!\!\!\!\!\!\!\scriptstyle{0}\!\!\!\!\!\!\!\!$
&&$\!\!\!\!\!\!\!\scriptstyle{0}\!\!\!\!\!\!\!\!$& \cr
\omit&height2pt&\omit&&\omit&&\omit&&\omit&&\omit&&\omit&&\omit&\cr
\tablerule
\omit&height2pt&\omit&&\omit&&\omit&&\omit&&\omit&&\omit&&\omit&\cr
&&$\!\!\!\!\!\!\!\scriptstyle{(3-3)}\!\!\!\!\!\!\!\!\!\!$
&&$\!\!\!\!\!\!\!\scriptstyle{1}\!\!\!\!\!\!\!\!\!\!\!$
&&$\!\!\!\!\!\!\scriptstyle{0}\!\!\!\!\!\!$
&&$\!\!\!\!\!\scriptstyle{0}\!\!\!\!\!$
&&$\!\!\!\!\!\scriptstyle{0}\!\!\!\!\!\!\!\!\!\!$
&&$\!\!\!\!\!\!\!\scriptstyle{0}\!\!\!\!\!\!\!\!$
&&$\!\!\!\!\!\!\!\scriptstyle{0}\!\!\!\!\!\!\!\!$& \cr
\omit&height2pt&\omit&&\omit&&\omit&&\omit&&\omit&&\omit&&\omit&\cr
\tablerule
\omit&height2pt&\omit&&\omit&&\omit&&\omit&&\omit&&\omit&&\omit&\cr
&&$\!\!\!\!\!\!\!\scriptstyle{(3-3)}\!\!\!\!\!\!\!\!\!\!$
&&$\!\!\!\!\!\!\!\scriptstyle{\ln Y}\!\!\!\!\!\!\!\!\!\!\!$
&&$\!\!\!\!\!\!\scriptstyle{0}\!\!\!\!\!\!$
&&$\!\!\!\!\!\scriptstyle{0}\!\!\!\!\!$
&&$\!\!\!\!\!\scriptstyle{0}\!\!\!\!\!\!\!\!\!\!$
&&$\!\!\!\!\!\!\!\scriptstyle{0}\!\!\!\!\!\!\!\!$
&&$\!\!\!\!\!\!\!\scriptstyle{0}\!\!\!\!\!\!\!\!$& \cr
\omit&height2pt&\omit&&\omit&&\omit&&\omit&&\omit&&\omit&&\omit&\cr
\tablerule
\omit&height2pt&\omit&&\omit&&\omit&&\omit&&\omit&&\omit&&\omit&\cr
&&$\!\!\!\!\!\!\!\scriptstyle{\rm{total}}\!\!\!\!\!\!\!\!\!\!$
&&$\!\!\!\!\!\!\!\scriptstyle{1}\!\!\!\!\!\!\!\!\!\!\!$
&&$\!\!\!\!\!\!\scriptstyle{-(6n^2\!+\!4n\!+\!3)}\!\!\!\!\!\!\!\!\!$
&&$\!\!\!\!\!\scriptstyle{-3(3n^2\!+\!9n\!+\!\frac{13}{2})}
\!\!\!\!\!\!\!\!\!\!\!$
&&$\!\!\!\!\!\!\!\scriptstyle{2(3n^2\!+\!8n\!+\!5)}
\!\!\!\!\!\!\!\!\!\!\!\!\!$
&&$\!\!\!\!\!\!\scriptstyle{-(3n^2\!-\!3)}\!\!\!\!\!\!\!\!$
&&$\!\!\!\!\!\!\!\scriptstyle{-\frac32(3n^2\!+\!8n\!+\!5)}
\!\!\!\!\!\!\!\!$& \cr
\omit&height2pt&\omit&&\omit&&\omit&&\omit&&\omit&&\omit&&\omit&\cr
\tablerule
\omit&height2pt&\omit&&\omit&&\omit&&\omit&&\omit&&\omit&&\omit&\cr
&&$\!\!\!\!\!\!\!\scriptstyle{\rm total}\!\!\!\!\!\!\!\!\!\!$
&&$\!\!\!\!\!\!\!\scriptstyle{\ln Y}\!\!\!\!\!\!\!\!\!\!\!$
&&$\!\!\!\!\!\!\scriptstyle{-(2n^3\!+\!2n^2\!+\!3n\!+\!3)}\!\!\!\!\!\!\!\!\!$
&&$\!\!\!\!\!\!\scriptstyle{-3(n^3\!+\!\frac{9}{2}n^2\!
+\!\frac{13}{2}n\!+\!3)}\!\!\!\!\!\!\!\!\!\!$
&&$\!\!\!\!\!\!\scriptstyle{2(n^3\!+\!4n^2\!+\!5n\!+\!2)}
\!\!\!\!\!\!\!\!\!\!$
&&$\!\!\!\!\!\!\!\scriptstyle{-(n^3\!-\!3n\!-\!2)}\!\!\!\!\!\!\!\!\!\!\!$
&&$\!\!\!\!\!\!\scriptstyle{-\frac32(n^3\!+\!4n^2\!+\!5n\!+\!2)}
\!\!\!\!\!\!\!\!$& \cr
\omit&height2pt&\omit&&\omit&&\omit&&\omit&&\omit&&\omit&&\omit&\cr
\tablerule}}

\caption{$i\delta\!\D_{A}\times i[S]_{n\geq 0}-{\rm IIC}$. The factor
$\frac{i \kappa^2H^2}{2^6 \pi^4}\frac{mHaa'}{2}
\sum_{n=0}^{\infty}Y^{n}$
multiplies all contributions. Here $Y=\frac{y}{4}$; $\ln Y$
and $1$ are the multiplicative factors for the each individual row.}

\label{dAn0-6}

\end{table}
The final class is comprised of terms in which comes from the infinite
series expansion. Theses contributions are integrable in $D=4$ and do not
require any further renormalizations so we could set $D=4$ right at the
beginning without alternating the final result. Therefore we shall apply
the infinite series expansion of the fermion propagator for $D=4$ to this
calculation. It is actually the same expression as (\ref{4Dfpropn1}).
In addition, one can also use (\ref{4Ddfpropn}) and all related identities
in Appendix \ref{fermionprop} but remember that the series is summed up
from $n=0$ instead of $n=1$. The residual A-type graviton propagator
and its various derivatives in four dimensions occurred very frequently,
\begin{eqnarray}
&&i\delta\Delta_A(x;x')=\frac{-H^2}{8\pi^2}\Bigl[\ln
(\frac{H^2\Delta x^2}{4})+\frac12\Bigr]\,,\\
&&\partial_{i}i\delta\Delta_A(x;x')=
-\partial_{i}i\delta\Delta_A(x;x')=\frac{-H^2}{4\pi^2}
\frac{\Delta x_i}{\Delta x^2}\,,\\
&&\partial_{0}i\delta\Delta_A(x;x')=-\partial^{\prime}_{0}
i\delta\Delta_A(x;x')=\frac{H^2}{4\pi^2}\frac{\Delta\eta}
{\Delta x^2}\,\\
&&\partial_0\partial^{\prime}_{0}i\delta\Delta_A(x;x')=
\frac{-H^2}{4\pi^2}\Bigl[\frac{1}{\Delta x^2}
+\frac{2\Delta\eta^2}{\Delta x^4}\Bigr]\,,\\
&&\partial_0\partial_{i}i\delta\Delta_A(x;x')=
-\partial^{\prime}_0\partial_{i}i\delta\Delta_A(x;x')=
\frac{H^2}{4\pi^2}\frac{-2\Delta\eta\Delta x_i}{\Delta x^4}\,,\\
&&\partial_{i}\partial_{l}i\delta\Delta_A(x;x')=
\frac{H^2}{4\pi^2}\Bigl[\frac{-\delta_{il}}{\Delta x^2}
+\frac{2\Delta x_i\Delta x_l}{\Delta x^4}\Bigr]\,.
\end{eqnarray}
\begin{table}

\vbox{\tabskip=0pt \offinterlineskip
\def\tablerule{\noalign{\hrule}}
\halign to390pt {\strut#& \vrule#\tabskip=1em plus2em& \hfil#\hfil&
\vrule#& \hfil#\hfil& \vrule#&
\hfil#\hfil& \vrule#\tabskip=0pt\cr
\tablerule
\omit&height4pt&\omit&&\omit&&\omit&\cr
&&\omit&&$\!\!\!\!\!\ln \frac{H^2 \Delta x^2}{4}\!\!\!\!\!$
&&$\!\!\!\!\!1\!\!\!\!\!$&\cr
\omit&height4pt&\omit&&\omit&&\omit&\cr
\tablerule
\omit&height4pt&\omit&&\omit&&\omit&\cr
&&$\!\!\!\!\!\frac{\Delta\eta^3}{\Delta x^6}\!\!\!\!\!$
&&$\!\!\!\!\!-20f_1(Y)-20f_2(Y)\ln Y\!\!\!\!\!$
&&$\!\!\!\!\!-5g_1(Y)-5g_2(Y)\ln Y\!\!\!\!\!$&\cr
\omit&height4pt&\omit&&\omit&&\omit&\cr
\tablerule
\omit&height4pt&\omit&&\omit&&\omit&\cr
&&$\frac{\gamma^0\Delta\eta^2\gamma^k\Delta x_k}{\Delta x^6}$
&&$12f_1(Y)+12f_2(Y)\ln Y$&&$3g_1(Y)+3g_2(Y)\ln Y$&\cr
\omit&height4pt&\omit&&\omit&&\omit&\cr
\tablerule
\omit&height4pt&\omit&&\omit&&\omit&\cr
&&$\frac{\Delta\eta}{\Delta x^4}$
&&$2f_3(Y)+2f_4(Y)\ln Y$&&$g_3(Y)+g_4(Y)\ln Y$&\cr
\omit&height4pt&\omit&&\omit&&\omit&\cr
\tablerule
\omit&height4pt&\omit&&\omit&&\omit&\cr
&&$\frac{\gamma^0\gamma^k\Delta x_k}{\Delta x^4}$
&&$-6f_5(Y)-6f_6(Y)\ln Y$&&$-g_5(Y)-g_6(Y)\ln Y$&\cr
\omit&height4pt&\omit&&\omit&&\omit&\cr
\tablerule
\omit&height4pt&\omit&&\omit&&\omit&\cr
&&$\frac{Ha\Delta\eta^2}{\Delta x^4}$
&&$6f_7(Y)+6f_8(Y)\ln Y$&&$5g_7(Y)+5g_8(Y)\ln Y$&\cr
\omit&height4pt&\omit&&\omit&&\omit&\cr
\tablerule
\omit&height4pt&\omit&&\omit&&\omit&\cr
&&$\frac{Ha'\Delta\eta^2}{\Delta x^4}$
&&$-4f_9(Y)-4f_{10}(Y)\ln Y$&&$-2g_9(Y)-2g_{10}(Y)\ln Y$&\cr
\omit&height4pt&\omit&&\omit&&\omit&\cr
\tablerule
\omit&height4pt&\omit&&\omit&&\omit&\cr
&&$\!\!\!\!\!\frac{Ha\gamma^0\!\Delta\eta
\gamma^k\!\Delta x_k}{\Delta x^4}\!\!\!\!\!$
&&$\!\!\!\!\!-4f_{11}(Y)-4f_{12}(Y)\ln Y\!\!\!\!\!$
&&$\!\!\!\!\!-2g_{11}(Y)-2g_{12}(Y)\ln Y\!\!\!\!\!$&\cr
\omit&height4pt&\omit&&\omit&&\omit&\cr
\tablerule
\omit&height4pt&\omit&&\omit&&\omit&\cr
&&$\!\!\!\!\!\frac{Ha'\!\gamma^0\!\Delta\eta
\gamma^k\!\Delta x_k}{\Delta x^4}\!\!\!\!\!$
&&$\!\!\!\!\!4f_9(Y)+4f_{10}(Y)\ln Y\!\!\!\!\!$
&&$\!\!\!\!\!2g_9(Y)+2g_{10}(Y)\ln Y\!\!\!\!\!$&\cr
\omit&height4pt&\omit&&\omit&&\omit&\cr
\tablerule
\omit&height4pt&\omit&&\omit&&\omit&\cr
&&$\frac{Ha}{\Delta x^2}$
&&$9f_{13}(Y)+9f_{14}(Y)\ln Y$&&$g_{13}(Y)+g_{14}(Y)\ln Y$&\cr
\omit&height4pt&\omit&&\omit&&\omit&\cr
\tablerule
\omit&height4pt&\omit&&\omit&&\omit&\cr
&&$\frac{Ha'}{\Delta x^2}$
&&$-2f_{15}(Y)-2f_{16}(Y)\ln Y$&&$-g_{15}(Y)-g_{16}(Y)\ln Y$&\cr
\omit&height4pt&\omit&&\omit&&\omit&\cr
\tablerule
\omit&height4pt&\omit&&\omit&&\omit&\cr
&&$\frac{H^2\!a^2\Delta\eta}{\Delta x^2}$
&&$-6f_{17}(Y)-6f_{18}(Y)\ln Y$&&$-3g_{17}(Y)-3g_{18}(Y)\ln Y$&\cr
\omit&height4pt&\omit&&\omit&&\omit&\cr
\tablerule
\omit&height4pt&\omit&&\omit&&\omit&\cr
&&$\frac{H^2\!a^2\gamma^0\gamma^k\Delta x_k}{\Delta x^2}$
&&$4f_{17}(Y)+4f_{18}(Y)\ln Y$&&$2g_{19}(Y)+2g_{20}(Y)\ln Y$&\cr
\omit&height4pt&\omit&&\omit&&\omit&\cr
\tablerule
\omit&height4pt&\omit&&\omit&&\omit&\cr
&&$\frac{H^2\!aa'\gamma^0\gamma^k\Delta x_k}{\Delta x^2}$
&&$-2f_{19}(Y)-2f_{20}(Y)\ln Y$&&$-g_{21}(Y)-g_{22}(Y)\ln Y$&\cr
\omit&height4pt&\omit&&\omit&&\omit&\cr
\tablerule
\omit&height4pt&\omit&&\omit&&\omit&\cr
&&$\!\!\!\!\!H^3\!a^2\!a'\!\!\!\!\!$
&&$\!\!\!\!\!-3f_{21}(Y)-3f_{22}(Y)\ln Y\!\!\!\!\!$
&&$\!\!\!\!\!-\frac{3}{2}g_{19}(Y)
-\frac{3}{2}g_{20}(Y)\ln Y\!\!\!\!\!$&\cr
\omit&height4pt&\omit&&\omit&&\omit&\cr
\tablerule}}

\caption{The total result for $i\delta\!\D_{A}\times i[S]_{n\geq 0}$.
The factor $\frac{i \kappa^2H^2}{2^6 \pi^4}\frac{mHaa'}{2}$
multiplies all contributions. Here $Y=\frac{y}{4}$;
$\ln\frac{H^2\Delta x^2}{4} $ and $1$ are the multiplicative factors
for the each individual column. The various functions $f_i(Y)$ and
$g_i(Y)$ are presented in Table~\ref{dAn0-Coeff} }

\label{dAn0-tot}

\end{table}

\begin{table}

\vbox{\tabskip=0pt \offinterlineskip
\def\tablerule{\noalign{\hrule}}
\halign to390pt {\strut#& \vrule#\tabskip=1em plus2em&
\hfil#\hfil& \vrule#& \hfil#\hfil& \vrule#&
\hfil#\hfil& \vrule#& \hfil#\hfil& \vrule#\tabskip=0pt\cr
\tablerule
\omit&height2pt&\omit&&\omit&&\omit&&\omit&\cr
&&$\!\!\!\!\!f_i(Y)\!\!\!\!\!$&&\omit
&&$\!\!\!\!\!g_i(Y)\!\!\!\!\!$&&\omit&\cr
\omit&height3pt&\omit&&\omit&&\omit&&\omit&\cr
\tablerule
\omit&height2pt&\omit&&\omit&&\omit&&\omit&\cr
&&$f_1(Y)$&&$\frac{Y(2Y^2+5Y-1)}{(1-Y)^3}\!+\!2$
&&$g_1(Y)$&&$\frac{-Y(Y^2-12Y-1)}{(1-Y)^3}\!-\!1$&\cr
\omit&height2pt&\omit&&\omit&&\omit&&\omit&\cr
\tablerule
\omit&height2pt&\omit&&\omit&&\omit&&\omit&\cr
&&$f_2(Y)$ &&$\frac{6Y^3}{(1-Y)^4}$
&&$g_2(Y)$ &&$\frac{Y(3Y^2+10Y-1)}{(1-Y)^4}$&\cr
\omit&height2pt&\omit&&\omit&&\omit&&\omit&\cr
\tablerule
\omit&height2pt&\omit&&\omit&&\omit&&\omit&\cr
&&$f_3(Y)$ &&$\frac{Y(7Y^2-86Y+91)}{(1-Y)^3}\!+\!7$
&&$g_3(Y)$
&&$\frac{Y(49Y^2-282Y+257)}{2(1-Y)^3}\!+\!\frac{49}{2}$&\cr
\omit&height2pt&\omit&&\omit&&\omit&&\omit&\cr
\tablerule
\omit&height2pt&\omit&&\omit&&\omit&&\omit&\cr
&&$f_4(Y)$ &&$\frac{-6Y(5Y^2+Y-8)}{(1-Y)^4}$
&&$g_4(Y)$ &&$\frac{Y(-45Y^2-82Y+151)}{2(1-Y)^4}$&\cr
\omit&height2pt&\omit&&\omit&&\omit&&\omit&\cr
\tablerule
\omit&height2pt&\omit&&\omit&&\omit&&\omit&\cr
&&$f_5(Y)$ &&$\frac{Y(5Y^2-22Y+29)}{(1-Y)^3}\!+\!5$
&&$g_5(Y)$
&&$\frac{Y(37Y^2-194Y+229)}{2(1-Y)^3}\!+\!\frac{37}{2} $&\cr
\omit&height2pt&\omit&&\omit&&\omit&&\omit&\cr
\tablerule
\omit&height2pt&\omit&&\omit&&\omit&&\omit&\cr
&&$f_6(Y)$ &&$\frac{2Y(-Y^2-Y+8)}{(1-Y)^4}$
&&$g_6(Y)$ &&$\frac{Y(-29Y^2-26Y+127)}{2(1-Y)^4}$&\cr
\omit&height2pt&\omit&&\omit&&\omit&&\omit&\cr
\tablerule
\omit&height2pt&\omit&&\omit&&\omit&&\omit&\cr
&&$f_7(Y)$ &&$\frac{Y(3Y^2-7Y+34)}{(1-Y)^3}\!+\!3$
&&$g_7(Y)$
&&$\frac{Y(13Y^2-48Y+71)}{2(1-Y)^3}\!+\!\frac{13}{2}$&\cr
\omit&height2pt&\omit&&\omit&&\omit&&\omit&\cr
\tablerule
\omit&height2pt&\omit&&\omit&&\omit&&\omit&\cr
&&$f_8(Y)$ &&$\frac{2Y(7Y+8)}{(1-Y)^4}$
&&$g_8(Y)$
&&$\frac{Y(Y^3-4Y^2+Y+38)}{2(1-Y)^4}\!-\!\frac12$&\cr
\omit&height2pt&\omit&&\omit&&\omit&&\omit&\cr
\tablerule
\omit&height2pt&\omit&&\omit&&\omit&&\omit&\cr
&&$f_9(Y)$ &&$\frac{Y(-Y^2+5Y+2)}{(1-Y)^3}\!-\!1$
&&$g_9(Y)$&&$\frac{Y(Y+5)}{(1-Y)^3}$&\cr
\omit&height2pt&\omit&&\omit&&\omit&&\omit&\cr
\tablerule
\omit&height2pt&\omit&&\omit&&\omit&&\omit&\cr
&&$\!\!\!\!\!f_{10}(Y)\!\!\!\!\!$ &&$\frac{6Y^2}{(1-Y)^4}$
&&$\!\!\!\!\!g_{10}(Y)\!\!\!\!\!$&&$\frac{2Y(2Y+1)}{(1-Y)^4}$&\cr
\omit&height2pt&\omit&&\omit&&\omit&&\omit&\cr
\tablerule
\omit&height2pt&\omit&&\omit&&\omit&&\omit&\cr
&&$f_{11}(Y)$ &&$\frac{Y(7Y^2-23Y+46)}{(1-Y)^3}\!+\!7$
&&$g_{11}(Y)$
&&$\frac{Y(12Y^2-43Y+61)}{(1-Y)^3}\!+\!12$&\cr
\omit&height2pt&\omit&&\omit&&\omit&&\omit&\cr
\tablerule
\omit&height2pt&\omit&&\omit&&\omit&&\omit&\cr
&&$f_{12}(Y)$ &&$\frac{6Y(Y+4)}{(1-Y)^4}$
&&$g_{12}(Y)$&&$\frac{-2Y(2Y-17)}{(1-Y)^4}$&\cr
\omit&height2pt&\omit&&\omit&&\omit&&\omit&\cr
\tablerule
\omit&height2pt&\omit&&\omit&&\omit&&\omit&\cr
&&$f_{13}(Y)$ &&$\frac{Y(3Y^2-10Y+19)}{(1-Y)^3}\!+\!3$
&&$g_{13}(Y)$
&&$\frac{Y(79Y^2-270Y+407)}{4(1-Y)^3}\!+\!\frac{79}{4}$&\cr
\omit&height2pt&\omit&&\omit&&\omit&&\omit&\cr
\tablerule
\omit&height2pt&\omit&&\omit&&\omit&&\omit&\cr
&&$f_{14}(Y)$ &&$\frac{2Y(Y+5)}{(1-Y)^4}$
&&$g_{14}(Y)$
&&$\frac{Y(-5Y^3+20Y^2-29Y+230)}{4(1-Y)^4}\!+\!\frac{5}{4}$&\cr
\omit&height2pt&\omit&&\omit&&\omit&&\omit&\cr
\tablerule
\omit&height2pt&\omit&&\omit&&\omit&&\omit&\cr
&&$f_{15}(Y)$ &&$\frac{Y(Y^2-2Y+13)}{(1-Y)^3}\!+\!1$
&&$g_{15}(Y)$
&&$\frac{Y(3Y^2-4Y+13)}{(1-Y)^3}\!+\!3$&\cr
\omit&height2pt&\omit&&\omit&&\omit&&\omit&\cr
\tablerule
\omit&height2pt&\omit&&\omit&&\omit&&\omit&\cr
&&$f_{16}(Y)$ &&$\frac{6Y(Y+1)}{(1-Y)^4}$
&&$g_{16}(Y)$
&&$\frac{Y(-3Y^3+12Y^2-7Y+10)}{(1-Y)^4}\!+\!3$ &\cr
\omit&height2pt&\omit&&\omit&&\omit&&\omit&\cr
\tablerule
\omit&height2pt&\omit&&\omit&&\omit&&\omit&\cr
&&$f_{17}(Y)$ &&$\frac{Y(2Y^2-7Y+11)}{(1-Y)^3}\!+\!2$
&&$g_{17}(Y)$
&&$\frac{Y(13Y^2-38Y+37)}{2(1-Y)^3}\!+\!\frac{13}{2}$ &\cr
\omit&height2pt&\omit&&\omit&&\omit&&\omit&\cr
\tablerule
\omit&height2pt&\omit&&\omit&&\omit&&\omit&\cr
&&$f_{18}(Y)$ &&$\frac{6Y}{(1-Y)^4}$
&&$g_{18}(Y)$
&&$\frac{3Y(-Y^3+4Y^2-6Y+5)}{(1-Y)^4}\!+\!3$ &\cr
\omit&height2pt&\omit&&\omit&&\omit&&\omit&\cr
\tablerule
\omit&height2pt&\omit&&\omit&&\omit&&\omit&\cr
&&$f_{19}(Y)$ &&$\frac{Y(Y^2-3Y+8)}{(1-Y)^3}\!+\!1$
&&$g_{19}(Y)$
&&$\frac{Y(5Y^2-15Y+16)}{(1-Y)^3}\!+\!5$ &\cr
\omit&height2pt&\omit&&\omit&&\omit&&\omit&\cr
\tablerule
\omit&height2pt&\omit&&\omit&&\omit&&\omit&\cr
&&$f_{20}(Y)$ &&$\frac{2Y(Y+2)}{(1-Y)^4}$
&&$g_{20}(Y)$
&&$\frac{-2Y(Y^3-4Y^2+6Y-6)}{(1-Y)^4}\!+\!2$ &\cr
\omit&height2pt&\omit&&\omit&&\omit&&\omit&\cr
\tablerule
\omit&height2pt&\omit&&\omit&&\omit&&\omit&\cr
&&$f_{21}(Y)$ &&$\frac{Y(5Y^2-15Y+16)}{(1-Y)^3}\!+\!5$
&&$g_{21}(Y)$
&&$\frac{-3Y^2(Y-3)}{(1-Y)^3}\!-\!3$ &\cr
\omit&height2pt&\omit&&\omit&&\omit&&\omit&\cr
\tablerule
\omit&height2pt&\omit&&\omit&&\omit&&\omit&\cr
&&$\!\!\!\!\!\!f_{22}(Y)\!\!\!\!\!$ &&$\!\!\!\!\!\frac{-2Y(Y^3-4Y^2+6Y-6)}{(1-Y)^4}\!+\!2\!\!\!\!\!$
&&$\!\!\!\!\!g_{22}(Y)\!\!\!\!\!$
&&$\!\!\!\!\!\frac{2Y(Y^3-4Y^2+8Y-2)}{(1-Y)^4}\!-\!2\!\!\!\!\!$ &\cr
\omit&height2pt&\omit&&\omit&&\omit&&\omit&\cr
\tablerule}}

\caption{The coefficient functions for the table~\ref{dAn0-tot}}

\label{dAn0-Coeff}

\end{table}
We also make use of the following identities to facilitate our computation
more effectively,
\begin{eqnarray}
&&\gamma^{l}\gamma^{k}\gamma_{l}=\gamma^{k}\,\,\,\,,\,\,\,\,
\gamma^{\mu}\gamma^0\gamma_{\mu}=2\gamma^0\,,\\
&&\gamma^l\gamma^k\gamma^0\gamma^l\Delta x_k
=\gamma^0\gamma^k\Delta x_k\,\,\,,\,\,\,
\gamma^k\gamma^l\gamma^0\gamma^l\Delta x_k=
-3\gamma^0\gamma^k\Delta x_k\,,\\
&&\gamma^j\gamma^{\nu}\gamma^0\gamma^j\Delta x_{\nu}=
3\Delta\eta+\gamma^0\gamma^k\Delta x_k\,,\\
&&\gamma^0\gamma^{\mu}\gamma^0\gamma^{\nu}
\Delta x_{\mu}\Delta x _{\nu}=\Delta x^2-2\gamma^0\Delta\eta
\gamma^{\mu}\Delta x_{\mu}\,,\\
&&\gamma^k\gamma^{\mu}\gamma^0\gamma^{\nu}\Delta x_k
\Delta x_{\mu}\Delta x_{\nu}\!=\!-\gamma^0\gamma^k\Delta x_k
\Delta x^2\!-2\Delta\eta\gamma^k\Delta x_k
\gamma^{\mu}\Delta x_{\mu}\,,\\
&&\gamma^0\gamma^{\nu}\gamma^0\gamma^k\Delta\eta
\Delta x_{\nu}\Delta x_k\!=\!\Delta\eta\overline{\Delta x}^2
-\gamma^0\Delta\eta^2\gamma^k\Delta x_k\,,\\
&&\gamma^k\gamma^{\nu}\gamma^0\gamma^l\Delta x_k\Delta x_l
\Delta x_{\nu}\!=\!\Delta\eta\overline{\Delta x}^2
-\gamma^0\gamma^k\Delta x_k\overline{\Delta x}^2\,.
\end{eqnarray}

Note that any derivatives acting on $i\delta\Delta_{A}$ would eliminate
$\ln(\frac{H^2\Delta x^2}{4})$ and that the exceptions are the generic
contractions (1-1), (1-2), (2-1) and (2-2). We list theses results
in Table~\ref{dAn0-1}, \ref{dAn0-2} and \ref{dAn0-3}. The rest of the
contributions without $\ln\frac{H^2\Delta x^2}{4}$ are summarized in the
Table~\ref{dAn0-4}, \ref{dAn0-5} and \ref{dAn0-6}. From these tables
one can see that the derivative of the coefficient with $\ln\frac{y}{4}$
is the coefficient without $\ln\frac{y}{4}$. Based on the characteristic
of (\ref{4Dfpropn1}) we should not be too surprised at this peculiar pattern
occurring here again as in the previous sub-section. The final result for
$-i[\Sigma^{\rm dAn0}](x;x')$ could be computed using (\ref{sum1}) and
(\ref{sum2}) and then adding the lowest order constant to each distinct contribution
because the summation here starts from $n=0$ rather than $n=1$.
Finally we tabulate the lengthy results in Table~\ref{dAn0-tot} and
Table~\ref{dAn0-Coeff}.

\subsection{Sub-Leading Contributions from $i{\delta \! \Delta}_B$}

In this subsection we evaluate the contribution from substituting
the residual $B$-type part of the graviton propagator in
Table~\ref{gen3},
\begin{equation}
i\Bigl[{}_{\alpha\beta} \Delta_{\rho\sigma}\Bigr] \longrightarrow
-\Bigl[ \delta^0_{\alpha} \delta^0_{\sigma} \overline{\eta}_{\beta
\rho} + \delta^0_{\alpha} \delta^0_{\rho} \overline{\eta}_{\beta
\sigma} + \delta^0_{\beta} \delta^0_{\sigma} \overline{\eta}_{\alpha
\rho} + \delta^0_{\beta} \delta^0_{\rho} \overline{\eta}_{\alpha
\sigma} \Bigr] i\delta\!\Delta_B \; . \label{DBpart}
\end{equation}
As in the two previous sub-sections we first make the requisite
contractions and then act the derivatives. The result of this first
step is presented in Table~\ref{DBcon}. Because the four different
tensors in (\ref{DBpart}) can make distinct contributions,
and because distinct contributions also come from breaking up
factors of $\gamma^{\alpha} J^{\beta \mu}$, we have sometimes
decomposed the result for a single vertex pair into parts.
These distinct parts in Table~\ref{DBcon} are subsequently
labeled by subscripts $a$, $b$, $c$, etc.

\begin{table}

\vbox{\tabskip=0pt \offinterlineskip
\def\tablerule{\noalign{\hrule}}
\halign to390pt {\strut#& \vrule#\tabskip=1em plus2em& \hfil#\hfil&
\vrule#& \hfil#\hfil& \vrule#& \hfil#\hfil& \vrule#& \hfil#\hfil&
\vrule#\tabskip=0pt\cr
\tablerule
\omit&height4pt&\omit&&\omit&&\omit&&\omit&\cr &&$\!\!\!\!{\rm
I}\!\!\!\!$ && $\!\!\!\!{\rm J} \!\!\!\!$ && $\!\!\!\! {\rm sub}
\!\!\!\!$ && $iV_I^{\alpha\beta}(x) \, i[S](x;x') \, i
V_J^{\rho\sigma}(x') \, [\mbox{}_{\alpha\beta} T^B_{\rho\sigma}] \,
i\delta\!\Delta_B(x;x')$ &\cr
\omit&height4pt&\omit&&\omit&&\omit&&\omit&\cr
\tablerule
\omit&height2pt&\omit&&\omit&&\omit&&\omit&\cr
&& 2 && 1 && \omit &&$0$ & \cr
\omit&height2pt&\omit&&\omit&&\omit&&\omit&\cr
\tablerule
\omit&height2pt&\omit&&\omit&&\omit&&\omit&\cr
&& 2 && 2 && a &&$-\f{1}{2}\ka^2\del^{'}_0 \{
 \g^{(0}\del^{k)} i[S](x;x')\g_{k} i\d\!\D_{B}(x;x') \} $ & \cr
\omit&height2pt&\omit&&\omit&&\omit&&\omit&\cr
\tablerule
\omit&height2pt&\omit&&\omit&&\omit&&\omit&\cr && 2 && 2 && b &&
$-\f{1}{2}\ka^2\del_k \{ \g^{(0}\del^{k)} i[S](x;x')\g^{0}
i\d\!\D_{B}(x;x')\}$ & \cr
\omit&height2pt&\omit&&\omit&&\omit&&\omit&\cr
\tablerule
\omit&height2pt&\omit&&\omit&&\omit&&\omit&\cr && 2 && 3 && a &&
$-\f{1}{8}\ka^2\g_k\del_{0}i[S](x;x') \g^k \,
\del^{'}_{0}i\d\!\D_{B}(x;x')$ & \cr
\omit&height2pt&\omit&&\omit&&\omit&&\omit&\cr
\tablerule
\omit&height2pt&\omit&&\omit&&\omit&&\omit&\cr && 2 && 3 && b &&
$\f{1}{8}\ka^2\g ^0\del^{'}_{0} i\d\!\D_{B}(x;x') \,
\del_{k}i[S](x;x')\g^{k} $ & \cr
\omit&height2pt&\omit&&\omit&&\omit&&\omit&\cr
\tablerule
\omit&height2pt&\omit&&\omit&&\omit&&\omit&\cr && 2 && 3 && c &&
$-\f{1}{8}\ka^2\g^k\del_{k} i\d\!\D_{B}(x;x') \,
\del_{0}i[S](x;x')\g^{0} $ & \cr
\omit&height2pt&\omit&&\omit&&\omit&&\omit&\cr
\tablerule
\omit&height2pt&\omit&&\omit&&\omit&&\omit&\cr && 2 && 3 && d &&
$\f{1}{8}\ka^2\g^0\del^{k}i[S](x;x') \g^0 \,
\del_{k}i\d\!\D_{B}(x;x') $ & \cr
\omit&height2pt&\omit&&\omit&&\omit&&\omit&\cr
\tablerule
\omit&height2pt&\omit&&\omit&&\omit&&\omit&\cr && 3 && 1 && \omit &&
$0$ & \cr \omit&height2pt&\omit&&\omit&&\omit&&\omit&\cr
\tablerule
\omit&height2pt&\omit&&\omit&&\omit&&\omit&\cr && 3 && 2 && a &&
$\f{1}{8}\ka^2\del^{'}_{0}\{\g^{k} i[S](x;x') \g_k \,
\del_{0}i\d\!\D_{B}(x;x')\}$ & \cr
\omit&height2pt&\omit&&\omit&&\omit&&\omit&\cr
\tablerule
\omit&height2pt&\omit&&\omit&&\omit&&\omit&\cr && 3 && 2 && b &&
$\f{1}{8}\ka^2\g^k\del_{k}\{ i[S](x;x') \g^0 \,
\del_{0}i\d\!\D_{B}(x;x')\} $ & \cr
\omit&height2pt&\omit&&\omit&&\omit&&\omit&\cr
\tablerule
\omit&height2pt&\omit&&\omit&&\omit&&\omit&\cr && 3 && 2 && c &&
$-\f{1}{8}\ka^2\g^0\del^{'}_{0}\{ i[S](x;x') \g^k \,
\del_{k}i\d\!\D_{B}(x;x')\} $ & \cr
\omit&height2pt&\omit&&\omit&&\omit&&\omit&\cr
\tablerule
\omit&height2pt&\omit&&\omit&&\omit&&\omit&\cr && 3 && 2 && d &&
$-\f{1}{8}\ka^2\del_{k}\{\g^{0} i[S](x;x') \g^0 \,
\del^{k}i\d\!\D_{B}(x;x')\}$ & \cr
\omit&height2pt&\omit&&\omit&&\omit&&\omit&\cr
\tablerule
\omit&height2pt&\omit&&\omit&&\omit&&\omit&\cr && 3 && 3 && a &&
$-\f{1}{16}\ka^2\g_{k}i[S](x;x')\g^{k}
\del_0\del^{'}_{0}i\d\!\D_{B}(x;x') $ & \cr
\omit&height2pt&\omit&&\omit&&\omit&&\omit&\cr
\tablerule
\omit&height2pt&\omit&&\omit&&\omit&&\omit&\cr && 3 && 3 && b &&
$\f{1}{16}\ka^2\g^{0}i[S](x;x')\g^k \,
\del_k\del^{'}_{0}i\d\!\D_{B}(x;x') $ & \cr
\omit&height2pt&\omit&&\omit&&\omit&&\omit&\cr
\tablerule
\omit&height2pt&\omit&&\omit&&\omit&&\omit&\cr && 3 && 3 && c &&
$-\f{1}{16}\ka^2\g^{k}i[S](x;x')\g^0 \,
\del_0\del_{k}i\d\!\D_{B}(x;x') $ & \cr
\omit&height2pt&\omit&&\omit&&\omit&&\omit&\cr
\tablerule
\omit&height2pt&\omit&&\omit&&\omit&&\omit&\cr && 3 && 3 && d &&
$\f{1}{16}\ka^2\g^{0}i[S](x;x')\g^0 \nabla^2 i\d\!\D_{B}(x;x') $ &
\cr \omit&height2pt&\omit&&\omit&&\omit&&\omit&\cr
\tablerule}}

\caption{Contractions from the $i\d\!\D_B$ part of the graviton
propagator. The contributions from (1-1)--(1-4), (4-1)--(4-4),
(2-4) and (3-4) are zero.}

\label{DBcon}

\end{table}

After the conformal contribution has been subtracted,
$i\delta\!\Delta_{\rm B}(x;x')$ is the residual of the
$B$-type propagator (\ref{DeltaB}) ,
\begin{eqnarray}
\lefteqn{i\delta\!\Delta_B(x;x') = \frac{H^2 \Gamma(\frac{D}2)}{16
\pi^{\frac{D}2}} \frac{(a a')^{2-\frac{D}2}}{\Delta x^{D-4}}
-\frac{H^{D-2}}{(4\pi)^\frac{D}2}
\frac{\Gamma(D\!-\!2)}{\Gamma\Bigl(\frac{D}2
\Bigr)} } \nonumber \\
& & \hspace{2cm} + \frac{H^{D-2}}{(4 \pi)^{\frac{D}2}}
\sum_{n=1}^{\infty} \left\{ \frac{\Gamma(n \!+\!
\frac{D}2)}{\Gamma(n \!+\! 2)} \Bigl( \frac{y}4 \Bigr)^{n -
\frac{D}2 +2} - \frac{\Gamma(n \!+\! D \!-\! 2)}{\Gamma(n \!+\!
\frac{D}2)} \Bigl(\frac{y}4 \Bigr)^n \right\} \; . \qquad \label{dB}
\end{eqnarray}
As was the case for the $i\delta\!\Delta_A(x;x')$ contributions
considered in the previous sub-section, for the infinite series terms
from $i\delta\!\Delta_B(x;x')$ this diagram is not sufficiently
singular enough to make a nonzero contribution in the $D\!=\!4$ limit.
Unlike $i\delta\!\Delta_A(x;x')$, even the $n\!=\!0$ terms of
$i\delta\!\Delta_B(x;x')$ vanish for $D\!=\!4$. This means that they
only survive when multiplied by a singular term.

Because most of the contractions involve at least one derivative of $i\delta\!\Delta_B$, it worth working out its various derivatives and
observing their behaviors in the $D=4$ limit,
\begin{eqnarray}
&&\partial_i i \delta\!\Delta_B(x;x')\!=\!-\frac{H^2 \Gamma(\frac{D}2)}{16
\pi^{\frac{D}2}} (D\!-\!4) (a a')^{2-\frac{D}2} \, \frac{\Delta
x^i}{
\Delta x^{D-2}}\!=\! -\partial_i' \delta\!\Delta_B(x;x'), \qquad \\
&&\partial_0 i \delta\!\Delta_B(x;x')= \frac{H^2 \Gamma(\frac{D}2)}{16
\pi^{\frac{D}2}} (D\!-\!4) (a a')^{2-\frac{D}2} \Biggl\{
\frac{\Delta \eta}{
\Delta x^{D-2}} \!-\! \frac{a H}{2 \Delta x^{D-4}} \Biggr\} , \\
&&\partial_0' i \delta\!\Delta_B(x;x')= \frac{H^2 \Gamma(\frac{D}2)}{16
\pi^{\frac{D}2}} (D\!-\!4) (a a')^{2-\frac{D}2}
\Biggl\{-\frac{\Delta \eta}{\Delta x^{D-2}} \!-\! \frac{a' H}{2
\Delta x^{D-4}} \Biggr\} \;,\\
&&\partial_0\partial_0'i\delta\Delta_B(x;x')\!=\!
\frac{H^2\Gamma(\frac{D}{2})}{16\pi^{\frac{D}{2}}}
\frac{(D\!-\!4)}{(aa')^{\frac{D}{2}\!-\!2}}\Biggl\{
\frac{-(D\!-\!2)\Delta\eta^2}{\Delta x^D}
\!-\!\frac{1}{\Delta x^{D-2}}\Biggr\}\nonumber\\
&&\hspace{8cm}+\mathcal{O}[(D\!-\!4)^2]\,,\\
&&\partial_0\partial_ki\delta\Delta_B(x;x')=
-\partial_0'\partial_k\delta\Delta_B(x;x')\nonumber\\
&&\hspace{2cm}=\frac{H^2\Gamma(\frac{D}{2})}{16\pi^{\frac{D}{2}}}
\frac{(D\!-\!4)(D\!-\!2)}{(aa')^{\frac{D}{2}\!-\!2}}\Biggl\{
\frac{-\Delta\eta\Delta x_k}{\Delta x^{D}}\Biggr\}
\!+\!\mathcal{O}[(D-4)^2]\;,\\
&&\nabla^{2}i\delta\Delta_B(x;x')=\frac{H^2\Gamma(D){2}}
{16\pi^{\frac{D}{2}}}\frac{(D\!-\!4)}{(aa')^{\frac{D}{2}\!-\!2}}
\Biggl\{\frac{(D\!-\!2)\overline{\Delta x^2}}{\Delta x^D}
-\frac{(D\!-\!1)}{\Delta x^{D-2}}\Biggr\}\;.
\end{eqnarray}
\begin{table}

\vbox{\tabskip=0pt \offinterlineskip
\def\tablerule{\noalign{\hrule}}
\halign to390pt {\strut#& \vrule#\tabskip=1em plus2em&
\hfil#\hfil& \vrule#& \hfil#\hfil& \vrule#&
\hfil#\hfil&\vrule#\tabskip=0pt\cr
\tablerule
\omit&height4pt&\omit&&\omit&&\omit&\cr
&&$\!\!\!\!\!\!{\rm pre\!-\!factor}\!\!\!\!\!\!$
&&$\frac{\kappa^2H^2}{2^8\pi^D}
\Gamma(\frac{D}{2})\Gamma(\frac{D}{2}\!-\!1)\
\!\frac{ma}{(aa')^{\frac{D}{2}-2}}$
&&$\frac{\kappa^2H^{D\!-\!2}}{2^{D\!+\!4}\pi^D}
\frac{\Gamma(D\!-\!2)\Gamma(\frac{D}{2}\!-\!1)}
{\Gamma(\frac{D}{2})}ma$ &\cr
\omit&height4pt&\omit&&\omit&&\omit&\cr
\tablerule
\omit&height2pt&\omit&&\omit&&\omit&\cr
&& $(2\!-\!2)_{\rm a_1}$ && ${\scriptstyle(D\!-\!1)}
[\frac{(D\!-\!2)}{2(D\!-\!3)}\partial^2_0\!+\!
Ha\partial_0]\frac{1}{\Delta x^{2D-6}}$
&&${\scriptstyle-(D\!-\!1)}[\partial^2_0\!
+\!Ha\partial_0]\frac{1}{\Delta x^{D-2}}$&\cr
\omit&height2pt&\omit&&\omit&&\omit&\cr
\tablerule
\omit&height2pt&\omit&&\omit&&\omit&\cr
&& $(2\!-\!2)_{\rm a_2}$ && $[\frac{-(D\!-\!2)}{2(D\!-\!3)}
\gamma^0\!\partial_0\!\!\not{\hspace{-.1cm}\bar{\partial}}\!-\!
Ha\gamma^0\!\!\not{\hspace{-.1cm}\bar{\partial}}]
\frac{1}{\Delta x^{2D-6}}$ &&$[\gamma^0\!\partial_0
\!\!\not{\hspace{-.1cm}\bar{\partial}}\!+\!Ha\gamma^0\!
\!\not{\hspace{-.1cm}\bar{\partial}}]\frac{1}{\Delta x^{D-2}}$&\cr
\omit&height2pt&\omit&&\omit&&\omit&\cr
\tablerule
\omit&height2pt&\omit&&\omit&&\omit&\cr
&& $(2\!-\!2)_{\rm b_1}$ && $\frac{(D\!-\!2)}{2(D\!-\!3)}
\gamma^0\!\partial_0\!\not{\hspace{-.1cm}\bar{\partial}}
\frac{1}{\Delta x^{2D-6}}$ &&$-\gamma^0\!\partial_0
\!\not{\hspace{-.1cm}\bar{\partial}}\frac{1}{\Delta x^{D-2}}$&\cr
\omit&height2pt&\omit&&\omit&&\omit&\cr
\tablerule
\omit&height2pt&\omit&&\omit&&\omit&\cr
&& $(2\!-\!2)_{\rm b_2}$ && $\frac{-(D\!-\!2)}{2(D\!-\!3)}
\nabla^2\!\frac{1}{\Delta x^{2D-6}}$
&&$\nabla^2\!\frac{1}{\Delta x^{D-2}}$&\cr
\omit&height2pt&\omit&&\omit&&\omit&\cr
\tablerule}}

\caption{The contributions for the contraction (2-2)
from $i\delta\!\Delta_B\times i[S]_{\rm fm}$. }

\label{dBfm2-2}

\end{table}

\begin{table}

\vbox{\tabskip=0pt \offinterlineskip
\def\tablerule{\noalign{\hrule}}
\halign to390pt {\strut#& \vrule#\tabskip=1em plus2em& \hfil#\hfil&
\vrule#&  \hfil#\hfil&\vrule#\tabskip=0pt\cr
\tablerule
\omit&height4pt&\omit&&\omit&\cr
&& ${\rm I\!-\!J}_{\rm sub}$ && ${\rm contributions}$ &\cr
\omit&height4pt&\omit&&\omit&\cr
\tablerule
\omit&height2pt&\omit&&\omit&\cr
&& $(2\!-\!3)_{\rm a}$ && $\frac{(D\!-\!1)}{2(D\!-\!3)}[\partial^2
\!-\!{\scriptstyle(D\!-\!4)}\partial^2_0]
\frac{1}{\Delta x^{2D-6}}$ &\cr
\omit&height2pt&\omit&&\omit&\cr
\tablerule
\omit&height2pt&\omit&&\omit&\cr
&& $(2\!-\!3)_{\rm b}$ && $-\frac{(D\!-\!4)}{2(D\!-\!3)}
\gamma^0\partial_0\!\not{\hspace{-.1cm}\bar{\partial}}
\frac{1}{\Delta x^{2D-6}}$ &\cr
\omit&height2pt&\omit&&\omit&\cr
\tablerule
\omit&height2pt&\omit&&\omit&\cr
&& $(2\!-\!3)_{\rm c}$ && $\frac{(D\!-\!4)}{2(D\!-\!3)}
\gamma^0\partial_0\!\not{\hspace{-.1cm}\bar{\partial}}
\frac{1}{\Delta x^{2D-6}}$ &\cr
\omit&height2pt&\omit&&\omit&\cr
\tablerule
\omit&height2pt&\omit&&\omit&\cr
&& $(2\!-\!3)_{\rm d}$ && $\frac{(D\!-\!1)}{2(D\!-\!3)}
[\partial^2\!+\!{\scriptstyle(D\!-\!4)(D\!-\!1)}
\nabla^2]\frac{1}{\Delta x^{2D-6}}$ &\cr
\omit&height2pt&\omit&&\omit&\cr
\tablerule
\omit&height2pt&\omit&&\omit&\cr
&& $(3\!-\!2)_{\rm a}$ && $\frac{(D\!-\!4)(D\!-\!1)}{(D\!-\!3)}
\partial^2_0\frac{1}{\Delta x^{2D-6}}$ &\cr
\omit&height2pt&\omit&&\omit&\cr
\tablerule
\omit&height2pt&\omit&&\omit&\cr
&& $(3\!-\!2)_{\rm b}$ && $-\frac{(D\!-\!4)}{(D\!-\!3)}
\gamma^0\partial_0\!\not{\hspace{-.1cm}\bar{\partial}}
\frac{1}{\Delta x^{2D-6}}$ &\cr
\omit&height2pt&\omit&&\omit&\cr
\tablerule
\omit&height2pt&\omit&&\omit&\cr
&& $(3\!-\!2)_{\rm c}$ && $\frac{(D\!-\!4)}{(D\!-\!3)}
\gamma^0\partial_0\!\not{\hspace{-.1cm}\bar{\partial}}
\frac{1}{\Delta x^{2D-6}}$ &\cr
\omit&height2pt&\omit&&\omit&\cr
\tablerule
\omit&height2pt&\omit&&\omit&\cr
&& $(3\!-\!2)_{\rm d}$ && $-\frac{(D\!-\!4)}{(D\!-\!3)}
\nabla^2\frac{1}{\Delta x^{2D-6}}$ &\cr
\omit&height2pt&\omit&&\omit&\cr
\tablerule
\omit&height2pt&\omit&&\omit&\cr
&& $(3\!-\!3)_{\rm a}$ && $\frac{-(D\!-\!1)}{4(D\!-\!3)}
\partial^2\frac{1}{\Delta x^{2D-6}}$ &\cr
\omit&height2pt&\omit&&\omit&\cr
\tablerule
\omit&height2pt&\omit&&\omit&\cr
&& $(3\!-\!3)_{\rm b}$ && $\frac{-(D\!-\!4)}{4(D\!-\!3)}
\gamma^0\partial_0\!\not{\hspace{-.1cm}\bar{\partial}}
\frac{1}{\Delta x^{2D-6}}$ &\cr
\omit&height2pt&\omit&&\omit&\cr
\tablerule
\omit&height2pt&\omit&&\omit&\cr
&& $(3\!-\!3)_{\rm c}$ && $\frac{(D\!-\!4)}{4(D\!-\!3)}
\gamma^0\partial_0\!\not{\hspace{-.1cm}\bar{\partial}}
\frac{1}{\Delta x^{2D-6}}$  &\cr
\omit&height2pt&\omit&&\omit&\cr
\tablerule
\omit&height2pt&\omit&&\omit&\cr
&& $(3\!-\!3)_{\rm d}$ && $[\frac{-(D\!-\!1)}{4(D\!-\!3)}
\partial^2\!+\!\frac{(D\!-\!4)}{4(D\!-\!3)}\nabla^2]
\frac{1}{\Delta x^{2D-6}}$ &\cr
\omit&height2pt&\omit&&\omit&\cr
\tablerule
\omit&height2pt&\omit&&\omit&\cr
&& total && $\Bigl\{\frac{(D\!-\!1)}{2(D\!-\!3)}\partial^2\!+\!
{\scriptstyle(D\!-\!4)}[\frac{(D\!-\!1)}{(D\!-\!3)}\partial^2_0]\!-\!
\frac{3}{4(D\!-\!3)}\nabla^2\Bigr\}\frac{1}{\Delta x^{2D-6}}$ &\cr
\omit&height2pt&\omit&&\omit&\cr
\tablerule}}

\caption{$i\delta\!\Delta_B\times i[S]_{\rm fm}$ terms.
All contributions are multiplied by
$\frac{\kappa^2 H^2}{2^{10} \pi^D}ma\Gamma(\frac{D}{2})
\Gamma(\frac{D}{2}-1)(aa')^{2-\frac{D}{2}}$. }

\label{dBfmmost}

\end{table}
The fact that the first line of $i\delta\Delta_B(x;x')$ and its various
derivatives are all of the order $(D\!-\!4)$ means that they would only
contribute when they are multiplied by a divergence. Note that the
contractions consisting of mass interaction vertices all vanish through
(\ref{DBpart}), so no order m contributions come from
$i\delta\Delta_{\rm B}\times i[S]_{\rm cf}$. The potential non-zero,
order m contributions could either come from the flat spacetime mass term
or from the infinite series expansion of the fermion propagator.
Remember that the only term in $i[S]_{\rm fm}$ behaves like
$\frac{1}{\Delta x^{D-2}}$ and that the most singular one in $i[S]_{n\geq 0}$
goes like $\frac{1}{\Delta x^{D-3}}$. In addition, the generic contractions
in Table~\ref{DBcon} are comprised of two derivatives. Therefore one can
count that the dimensionality of most singular contribution from $i[S]_{\rm fm}$
is $\frac{1}{\Delta x^{2D-4}}$ whereas the one from $i[S]_{n\geq 0}$ is
$\frac{1}{\Delta x^{2D-5}}$. The former is logarithmically divergent in
$D=4$ before performing the partial integration, so one still needs to
keep arbitrary $D$ for the computation; the latter is entirely integrable
in $D=4$ so that one could compute this part in four dimensions and
the result turns out to be zero owning to the cancelation of the first
two series of $i\delta\Delta_B(x;x')$ and owning to $(D\!-\!4)$ factor
from its various derivatives. Therefore the only class we need to work out
in this sub-suction is $i\delta\Delta_B(x;x')\times i[S]_{\rm fm}(x;x')$.

We take special care of the contraction (2-2) because it is the only
contraction in Table~\ref{DBcon} which derivatives might have a chance
not to act upon $i\delta\Delta_B$. We also break up $\gamma^{(0}\partial^{k)}$
into $\frac12\gamma^0\partial^k$ and $\frac12\gamma^k\partial^0$
for each sub-contraction and tabulate the results in Table~\ref{dBfm2-2}.
Theses expressions are integrable in four dimensions and the contributions
from the left column cancel out exactly with that from the right column in $D=4$.
Table~\ref{dBfmmost} gives the rest of our results for the most singular
contributions, those in which all derivatives act upon the conformal
coordinate separation. There is no net contribution when one or more of
the derivatives acts upon a scale factor. Those expressions are also
integrable in $D\!=\!4$ dimensions, at which point we can take $D\!=\!4$
and the result vanishes on account of the overall factor of $(D\!-\!4)$
or $(D\!-\!4)^2$. Finally we read off the net contribution from
Table~\ref{dBfmmost} and take $D=4$,
\begin{eqnarray}
-i\Bigl[\Sigma^{\rm idBfm}\Bigr](x;x')=\frac{\kappa^2H^2}
{2^{10}\pi^4}ma\Biggl\{\frac32\partial^2
\frac{1}{\Delta x^2}\Biggr\}=\frac{\kappa^2H^2}
{2^{8}\pi^4}ma\Biggl\{\frac38\partial^2
\frac{1}{\Delta x^2}\Biggr\}\;.\label{idBfm}
\end{eqnarray}

\subsection{Sub-Leading Contributions from $i{\delta \! \Delta}_C$}

The point of this subsection is to work out the contribution from
replacing the graviton propagator in Table~\ref{gen3} by its
residual $C$-type part,
\begin{equation}
i\Bigl[{}_{\alpha\beta} \Delta_{\rho\sigma}\Bigr] \!\rightarrow \!2
\Biggl[ \frac{\bar{\eta}_{\alpha\beta}
\bar{\eta}_{\rho\sigma}}{(D\!-\!2)(D\!-\!3)} \!+\! \frac{\delta^0_{\alpha}
\delta^0_{\beta} \bar{\eta}_{\rho\sigma} \!+\! \bar{\eta}_{\alpha\beta}
\delta^0_{\rho} \delta^0_{\sigma}}{D\!-\!2} \!+\!
\Bigl(\frac{D\!-\!3}{D\!-\!2}\Bigr) \delta^0_{\alpha}
\delta^0_{\beta} \delta^0_{\rho} \delta^0_{\sigma} \Biggr]
i\delta\!\Delta_C . \label{DCpart}
\end{equation}
As in the previous sub-sections we first make the requisite
contractions and then act the derivatives. The result of this first
step is displayed in Table~\ref{DCcon1} and Table~\ref{DCcon2}.
The four different tensors in (\ref{DCpart})can make distinct
contributions and distinct contributions also come from breaking up
factors of $\gamma^{\alpha} J^{\beta \mu}$, so we have sometimes
decomposed the result for a single vertex pair into parts.
These distinct contributions are tagged with subscripts $a$, $b$, $c$, etc.
\begin{table}

\vbox{\tabskip=0pt \offinterlineskip
\def\tablerule{\noalign{\hrule}}
\halign to390pt {\strut#& \vrule#\tabskip=1em plus2em& \hfil#\hfil&
\vrule#& \hfil#\hfil& \vrule#& \hfil#\hfil& \vrule#& \hfil#\hfil&
\vrule#\tabskip=0pt\cr
\tablerule
\omit&height4pt&\omit&&\omit&&\omit&&\omit&\cr &&$\!\!\!\! {\rm
I}\!\!\!\!$ && $\!\!\!\! {\rm J} \!\!\!\!$ && $\!\!\!\! {\rm sub}
\!\!\!\!$ && $iV_I^{\alpha\beta}(x) \, i[S](x;x') \, i
V_J^{\rho\sigma}(x') \, [\mbox{}_{\alpha\beta} T^C_{\rho\sigma}] \,
i\delta\!\Delta_C(x;x')$ &\cr
\omit&height4pt&\omit&&\omit&&\omit&&\omit&\cr
\tablerule
\omit&height2pt&\omit&&\omit&&\omit&&\omit&\cr
&& 1 && 4 && \omit
&&$\frac2{(D-3)(D-2)}\kappa^{2}iam\!\not{\hspace{-.1cm}\partial}
i[S](x;x')i\delta\!\Delta_C(x;x')$ &\cr
\omit&height2pt&\omit&&\omit&&\omit&&\omit&\cr
\tablerule
\omit&height2pt&\omit&&\omit&&\omit&&\omit&\cr
&& 2 && 4 && a
&&$\frac1{(D-2)}\kappa^{2}iam\gamma^0\partial_0
i[S](x;x')i\delta\!\Delta_C(x;x')$ &\cr
\omit&height2pt&\omit&&\omit&&\omit&&\omit&\cr
\tablerule
\omit&height2pt&\omit&&\omit&&\omit&&\omit&\cr
&& 2 && 4 && b
&&$-\frac1{(D-3)(D-2)}\kappa^{2}iam
\!\not{\hspace{-.1cm}\bar{\partial}}i[S](x;x')
i\delta\!\Delta_C(x;x')$ &\cr
\omit&height2pt&\omit&&\omit&&\omit&&\omit&\cr
\tablerule
\omit&height2pt&\omit&&\omit&&\omit&&\omit&\cr
&& 3 && 4 && a
&&$\frac{(D-1)}{2(D-3)(D-2)}\kappa^{2}iam\gamma^0\partial_0
i\delta\!\Delta_C(x;x')i[S](x;x')$ &\cr
\omit&height2pt&\omit&&\omit&&\omit&&\omit&\cr
\tablerule
\omit&height2pt&\omit&&\omit&&\omit&&\omit&\cr
&& 3 && 4 && b
&&$\frac{1}{2(D-3)(D-2)}\kappa^{2}iam
\!\not{\hspace{-.1cm}\bar{\partial}}i\delta\!\Delta_C(x;x')
i[S](x;x')$ &\cr
\omit&height2pt&\omit&&\omit&&\omit&&\omit&\cr
\tablerule
\omit&height2pt&\omit&&\omit&&\omit&&\omit&\cr
&& 4 && 1 && \omit
&&$-\frac{2}{(D-3)(D-2)}\kappa^{2}iam\partial'_{\mu}\{
i[S](x;x')\gamma^{\mu}i\delta\!\Delta_C(x;x')\}$ &\cr
\omit&height2pt&\omit&&\omit&&\omit&&\omit&\cr
\tablerule
\omit&height2pt&\omit&&\omit&&\omit&&\omit&\cr
&& 4 && 2 && a
&&$-\frac{1}{(D-2)}\kappa^{2}iam\partial'_0\{
i[S](x;x')\gamma^0 i\delta\!\Delta_C(x;x')\}$ &\cr
\omit&height2pt&\omit&&\omit&&\omit&&\omit&\cr
\tablerule
\omit&height2pt&\omit&&\omit&&\omit&&\omit&\cr
&& 4 && 2 && b
&&$-\frac{1}{(D-3)(D-2)}\kappa^{2}iam\partial_k\{
i[S](x;x')\gamma^k i\delta\!\Delta_C(x;x')\}$ &\cr
\omit&height2pt&\omit&&\omit&&\omit&&\omit&\cr
\tablerule
\omit&height2pt&\omit&&\omit&&\omit&&\omit&\cr
&& 4 && 3 && a
&&$\frac{(D-1)}{2(D-3)(D-2)}\kappa^{2}iam\,i[S](x;x')
\gamma^0\partial_0 i\delta\!\Delta_C(x;x')$ &\cr
\omit&height2pt&\omit&&\omit&&\omit&&\omit&\cr
\tablerule
\omit&height2pt&\omit&&\omit&&\omit&&\omit&\cr
&& 4 && 3 && b
&&$\frac{1}{2(D-3)(D-2)}\kappa^{2}iam\,i[S](x;x')
\!\not{\!\bar{\partial}}\delta\!\Delta_C(x;x')$ &\cr
\omit&height2pt&\omit&&\omit&&\omit&&\omit&\cr
\tablerule}}

\caption{Contractions from the $i\delta\!\Delta_C$ part of the
graviton propagator-I.}

\label{DCcon1}

\end{table}
\begin{table}

\vbox{\tabskip=0pt \offinterlineskip
\def\tablerule{\noalign{\hrule}}
\halign to390pt {\strut#& \vrule#\tabskip=1em plus2em& \hfil#\hfil&
\vrule#& \hfil#\hfil& \vrule#& \hfil#\hfil& \vrule#& \hfil#\hfil&
\vrule#\tabskip=0pt\cr
\tablerule
\omit&height1.5pt&\omit&&\omit&&\omit&&\omit&\cr &&$\!\!\!\! {\rm
I}\!\!\!\!$ && $\!\!\!\! {\rm J} \!\!\!\!$ && $\!\!\!\! {\rm sub}
\!\!\!\!$ && $iV_I^{\alpha\beta}(x) \, i[S](x;x') \, i
V_J^{\rho\sigma}(x') \, [\mbox{}_{\alpha\beta} T^C_{\rho\sigma}] \,
i\delta\!\Delta_C(x;x')$ &\cr
\omit&height1.5pt&\omit&&\omit&&\omit&&\omit&\cr
\tablerule
\omit&height1.5pt&\omit&&\omit&&\omit&&\omit&\cr
&& 1 && 1 && \omit
&&$-\frac{2}{(D-3)(D-2)}\kappa^2\partial'_{\mu}\{
\not{\hspace{-.1cm}\partial}i[S](x;x')\gamma^{\mu}
i\delta\!\Delta_C(x;x')\}$&\cr
\omit&height1.5pt&\omit&&\omit&&\omit&&\omit&\cr
\tablerule
\omit&height1.5pt&\omit&&\omit&&\omit&&\omit&\cr
&& 1 && 2 && a
&&$-\frac{1}{(D-2)}\kappa^2\partial'_0\{
\not{\hspace{-.1cm}\partial}i[S](x;x')\gamma^0
i\delta\!\Delta_C(x;x')\}$&\cr
\omit&height1.5pt&\omit&&\omit&&\omit&&\omit&\cr
\tablerule
\omit&height1.5pt&\omit&&\omit&&\omit&&\omit&\cr
&& 1 && 2 && b
&&$-\frac{1}{(D-3)(D-2)}\kappa^2\partial_k\{
\not{\hspace{-.1cm}\partial}i[S](x;x')\gamma^k
i\delta\!\Delta_C(x;x')\}$&\cr
\omit&height1.5pt&\omit&&\omit&&\omit&&\omit&\cr
\tablerule
\omit&height1.5pt&\omit&&\omit&&\omit&&\omit&\cr
&& 1 && 3 && a
&&$\frac{(D-1)}{2(D-3)(D-2)}\kappa^2
\!\not{\hspace{-.1cm}\partial}i[S](x;x')\gamma^0\partial'_0
i\delta\!\Delta_C(x;x')$&\cr
\omit&height1.5pt&\omit&&\omit&&\omit&&\omit&\cr
\tablerule
\omit&height1.5pt&\omit&&\omit&&\omit&&\omit&\cr
&& 1 && 3 && b
&&$-\frac{1}{2(D-3)(D-2)}\kappa^2\!\not{\hspace{-.1cm}\partial}
i[S](x;x')\!\not{\hspace{-.1cm}\bar{\partial}}
i\delta\!\Delta_C(x;x')$&\cr
\omit&height1.5pt&\omit&&\omit&&\omit&&\omit&\cr
\tablerule
\omit&height1.5pt&\omit&&\omit&&\omit&&\omit&\cr
&& 2 && 1 && a
&&$-\frac1{(D-2)}\kappa^2\partial'_{\mu}\{\gamma^0\partial_0
i[S](x;x')\gamma^{\mu}i\delta\!\Delta_C(x;x)\}$ &\cr
\omit&height1.5pt&\omit&&\omit&&\omit&&\omit&\cr
\tablerule
\omit&height1.5pt&\omit&&\omit&&\omit&&\omit&\cr
&& 2 && 1 && b
&&$\frac1{(D-3)(D-2)}\kappa^2\partial'_{\mu}\{
\!\not{\hspace{-.1cm}\bar{\partial}}i[S](x;x')
\gamma^{\mu}i\delta\!\Delta_C(x;x)\}$ &\cr
\omit&height1.5pt&\omit&&\omit&&\omit&&\omit&\cr
\tablerule
\omit&height1.5pt&\omit&&\omit&&\omit&&\omit&\cr
&& 2 && 2 && a
&&$-\frac{(D-3)}{2(D-2)}\kappa^2\partial'_0\{\gamma^0
\partial_0i[S](x;x')\gamma^0 i\delta\!\Delta_C(x;x')\}$ & \cr
\omit&height2pt&\omit&&\omit&&\omit&&\omit&\cr
\tablerule
\omit&height1.5pt&\omit&&\omit&&\omit&&\omit&\cr
&& 2 && 2 && b
&&$-\frac{1}{2(D-2)}\kappa^2\partial_k \{\gamma^0
\partial_0 i[S](x;x')\gamma^k i\delta\!\Delta_C(x;x')\}$ &\cr
\omit&height1.5pt&\omit&&\omit&&\omit&&\omit&\cr
\tablerule
\omit&height1.5pt&\omit&&\omit&&\omit&&\omit&\cr
&& 2 && 2 && c
&&$\frac1{2(D-2)}\kappa^2\partial'_0\{
\not{\hspace{-.1cm}\bar{\partial}}i[S](x;x')\gamma^0
i\delta\!\Delta_C(x;x)\}$ &\cr
\omit&height1.5pt&\omit&&\omit&&\omit&&\omit&\cr
\tablerule
\omit&height1.5pt&\omit&&\omit&&\omit&&\omit&\cr
&& 2 && 2 && d
&&$\frac1{2(D-3)(D-2)}\kappa^2\partial_k\{
\not{\hspace{-.1cm}\bar{\partial}}i[S](x;x')\gamma^k
i\delta\!\Delta_C(x;x)\}$ &\cr
\omit&height1.5pt&\omit&&\omit&&\omit&&\omit&\cr
\tablerule
\omit&height1.5pt&\omit&&\omit&&\omit&&\omit&\cr
&& 2 && 3 && a
&&$\frac14(\frac{D-1}{D-2})\kappa^2\gamma^0\partial_0
i[S](x;x')\gamma^0\partial'_0 i\delta\!\Delta_C(x;x')$ &\cr
\omit&height1.5pt&\omit&&\omit&&\omit&&\omit&\cr
\tablerule
\omit&height1.5pt&\omit&&\omit&&\omit&&\omit&\cr
&& 2 && 3 && b
&&$-\frac{1}{4(D-2)}\kappa^2\gamma^0\partial_0i[S](x;x')
\!\not{\hspace{-.1cm}\bar{\partial}}i\delta\!\Delta_C(x;x')$ &\cr
\omit&height1.5pt&\omit&&\omit&&\omit&&\omit&\cr
\tablerule
\omit&height1.5pt&\omit&&\omit&&\omit&&\omit&\cr
&& 2 && 3 && c
&&$-\frac{(D-1)}{4(D-3)(D-2)}\kappa^2
\!\not{\hspace{-.1cm}\bar{\partial}}i[S](x;x')
\gamma^0\partial'_0 i\delta\!\Delta_C(x;x')$ &\cr
\omit&height1.5pt&\omit&&\omit&&\omit&&\omit&\cr
\tablerule
\omit&height1.5pt&\omit&&\omit&&\omit&&\omit&\cr
&& 2 && 3 && d
&&$\frac{1}{4(D-3)(D-2)}\kappa^2
\!\not{\hspace{-.1cm}\bar{\partial}}i[S](x;x')
\!\not{\hspace{-.1cm}\bar{\partial}}i\delta\!\Delta_C(x;x')$ &\cr
\omit&height1.5pt&\omit&&\omit&&\omit&&\omit&\cr
\tablerule
\omit&height1.5pt&\omit&&\omit&&\omit&&\omit&\cr
&& 3 && 1 && a
&&$-\frac{(D-1)}{2(D-3)(D-2)}\kappa^2\partial'_{\mu}\{\gamma^0
\partial_0\,i\delta\!\Delta_C(x;x')i[S](x;x')\gamma^{\mu}\}$ &\cr
\omit&height1.5pt&\omit&&\omit&&\omit&&\omit&\cr
\tablerule
\omit&height1.5pt&\omit&&\omit&&\omit&&\omit&\cr
&& 3 && 1 && b
&&$-\frac1{2(D-3)(D-2)}\kappa^2\partial'_{\mu}\{
\not{\hspace{-.1cm}\bar{\partial}}i\delta\!\Delta_C(x;x')
i[S](x;x')\gamma^{\mu}\}$ &\cr
\omit&height1.5pt&\omit&&\omit&&\omit&&\omit&\cr
\tablerule
\omit&height1.5pt&\omit&&\omit&&\omit&&\omit&\cr
&& 3 && 2 && a
&&$-\frac14(\frac{D-1}{D-2})\kappa^2\partial'_0\{\gamma^0
\partial_0 i\delta\!\Delta_C(x;x')i[S](x;x')\gamma^0\}$ &\cr
\omit&height1.5pt&\omit&&\omit&&\omit&&\omit&\cr
\tablerule
\omit&height1.5pt&\omit&&\omit&&\omit&&\omit&\cr && 3 && 2 && b &&
$-\frac{1}{4(D-2)}\kappa^2\partial'_0\{
\not{\hspace{-.1cm}\bar{\partial}}i\delta\!\Delta_C(x;x')
i[S](x;x')\gamma^0 \}$ &\cr
\omit&height1.5pt&\omit&&\omit&&\omit&&\omit&\cr
\tablerule
\omit&height1.5pt&\omit&&\omit&&\omit&&\omit&\cr
&& 3 && 2 && c
&&$-\frac{(D-1)}{4(D-3)(D-2)}\kappa^2\partial_k\{\gamma^0
\partial_0\,i\delta\!\Delta_C(x;x')i[S](x;x')\gamma^k\}$ &\cr
\omit&height1.5pt&\omit&&\omit&&\omit&&\omit&\cr
\tablerule
\omit&height1.5pt&\omit&&\omit&&\omit&&\omit&\cr
&& 3 && 2 && d
&&$-\frac{1}{4(D-3)(D-2)}\kappa^2\partial_k \{
\not{\hspace{-.1cm}\bar{\partial}}\delta\!\Delta_C(x;x')
i[S](x;x')\gamma^k \}$ &\cr
\omit&height1.5pt&\omit&&\omit&&\omit&&\omit&\cr
\tablerule
\omit&height1.5pt&\omit&&\omit&&\omit&&\omit&\cr
&& 3 && 3 && a
&&$\frac{(D-1)^2}{8(D-3)(D-2)}\kappa^2\gamma^0 i[S](x;x')
\gamma^0\partial_0\partial'_0i\delta\!\Delta_C(x;x')$ &\cr
\omit&height1.5pt&\omit&&\omit&&\omit&&\omit&\cr
\tablerule
\omit&height1.5pt&\omit&&\omit&&\omit&&\omit&\cr
&& 3 && 3 && b
&&$-\frac{(D-1)}{8(D-3)(D-2)}\kappa^2\gamma^0 i[S](x;x')
\gamma^k\partial_0\partial_k i\delta\!\Delta_C(x;x')$ &\cr
\omit&height1.5pt&\omit&&\omit&&\omit&&\omit&\cr
\tablerule
\omit&height1.5pt&\omit&&\omit&&\omit&&\omit&\cr
&& 3 && 3 && c
&&$\frac{(D-1)}{8(D-3)(D-2)}\kappa^2\gamma^k i[S](x;x')
\gamma^0\partial_k\partial'_0 i\delta\!\Delta_C(x;x')$ &\cr
\omit&height1.5pt&\omit&&\omit&&\omit&&\omit&\cr
\tablerule
\omit&height1.5pt&\omit&&\omit&&\omit&&\omit&\cr
&& 3 && 3 && d
&&$-\frac{1}{8(D-3)(D-2)}\kappa^2\gamma^k i[S](x;x')
\gamma^l \partial_k\partial_l i\delta\!\Delta_C(x;x')$ &\cr
\omit&height1.5pt&\omit&&\omit&&\omit&&\omit&\cr
\tablerule}}

\caption{Contractions from the $i\delta\!\Delta_C$ part of the
graviton propagator-II.}

\label{DCcon2}

\end{table}

Here $i\delta\!\Delta_C(x;x')$ is the residual of the $C$-type
propagator (\ref{DeltaC}) after the conformal contribution has been
subtracted,
\begin{eqnarray}
\lefteqn{i \delta\!\Delta_C(x;x') = \frac{H^2}{16 \pi^{\frac{D}2}}
\Bigl( \frac{D}2 \!-\! 3\Bigr) \Gamma\Bigl(\frac{D}2 \!-\! 1\Bigr)
\frac{(a a')^{2-\frac{D}2}}{\Delta x^{D-4}}+
\frac{H^{D-2}}{(4\pi)^{\frac{D}2}}
\frac{\Gamma(D \!-\! 3)}{\Gamma(\frac{D}2)} } \nonumber \\
& & \hspace{-.7cm} - \frac{H^{D-2}}{(4\pi)^{\frac{D}2}}
\!\!\sum_{n=1}^{\infty} \!\!\left\{ \!\!\Bigl(n \!-\! \frac{D}2
\!+\! 3\Bigr) \frac{\Gamma(n \!+\! \frac{D}2 \!-\! 1)}{\Gamma(n
\!+\! 2)} \Bigl(\frac{y}4 \Bigr)^{n -\frac{D}2 +2} \!\!\!\!\!\!\! -
(n\!+\!1) \frac{\Gamma(n \!+\! D \!-\! 3)}{\Gamma(n \!+\!
\frac{D}2)} \Bigl(\frac{y}4 \Bigr)^n \!\right\} \!. \qquad
\label{dC}
\end{eqnarray}
The only way $i\delta\!\Delta_C(x;x')$ can give a nonzero contribution
in $D\!=\!4$ dimensions is for it to multiply a divergence as with
the contributions from $i\delta\!\Delta_B(x;x')$ considered
in the previous sub-section.  That means only the $n\!=\!0$ term can
possibly contribute. Even for the $n\!=\!0$ term, both derivatives
must act upon the coordinate separation to make a nonzero contribution
in $D\!=\!4$ dimensions.

Because all the vertex pairs involve one or more derivatives of
$i\delta\!\Delta_C$, here we list them as follows,
\begin{eqnarray}
&&\partial_i i\delta\!\Delta_C \!=\!\frac{H^2\Gamma(\frac{D}2\!-\!1)}
{16\pi^{\frac{D}2}}(\frac{D}{2}\!-\!3)(D\!-\!4)(aa')^{2-\frac{D}2}\,
\frac{-\Delta x^i}{\Delta x^{D-2}}=-\partial'_i
i\delta\!\Delta_C \;,\qquad \\
&&\partial_0 i\delta\!\Delta_C \!=\!\frac{H^2\Gamma(\frac{D}2\!-\!1)}
{16\pi^{\frac{D}2}}(\frac{D}{2}\!-\!3)(D\!-\!4)(aa')^{2-\frac{D}2}
\Biggl\{\frac{\Delta\eta}{\Delta x^{D-2}} \!-\!
\frac{aH}{2\Delta x^{D-4}}\Biggr\},\qquad \\
&&\partial_0' i\delta\!\Delta_C \!=\!\frac{H^2\Gamma(\frac{D}2\!-\!1)}
{16\pi^{\frac{D}2}}(\frac{D}{2}\!-\!3)(D\!-\!4)(aa')^{2-\frac{D}2}
\Biggl\{-\frac{\Delta\eta}{\Delta x^{D-2}}\!-\!
\frac{a'H}{2\Delta x^{D-4}}\Biggr\},\qquad\\
&&\partial_0\partial'_0 i\delta\!\Delta_C=
\frac{H^2\Gamma(\frac{D}{2}\!-\!1)}{16\pi^{\frac{D}{2}}}
(\frac{D}{2}\!-\!3)\frac{(D\!-\!4)}{(aa')^{\frac{D}{2}-2}}
\Biggl\{\frac{-(D\!-\!2)\Delta\eta^2}{\Delta x^D}
-\frac{1}{\Delta x^{D-2}}\Biggr\}\nonumber\\
&&\hspace{9cm}+\mathcal{O}[(D\!-\!4)^2],\\
&&\partial_k\partial_0 i\delta\!\Delta_C\!=\!
-\partial_k\partial'_0i\delta\!\Delta_C\!\nonumber\\
&&\hspace{1cm}=\frac{H^2\Gamma(\frac{D}{2}\!-\!1)}{16\pi^{\frac{D}{2}}}
(\frac{D}{2}\!-\!3)\frac{(D\!-\!2)(D\!-\!4)}{(aa')^{\frac{D}{2}-2}}
\frac{-\Delta\eta\Delta x_k}{\Delta x^D}
+\mathcal{O}[(D\!-\!4)^2],\\
&&\partial_k\partial_l i\delta\!\Delta_C\!=\!
\frac{H^2\Gamma(\frac{D}{2}\!-\!1)}{16\pi^{\frac{D}{2}}}
(\frac{D}{2}\!-\!3)\frac{(D\!-\!4)}{(aa')^{\frac{D}{2}-2}}
\Biggl\{\frac{{\scriptstyle(D\!-\!2)}\Delta x_k\Delta x_l}{\Delta x^D}
-\frac{\delta_{kl}}{\Delta x^{D-2}}\Biggr\}.
\end{eqnarray}
Note that $i\delta\!\Delta_C$ and its various derivatives have
the same behaviors as $i\delta\!\Delta_B$. The propagator itself
tends to cancel in $D=4$ dimensions and its various derivatives
all carry $(D\!-\!4)$ factor. This means that they could give the
non-zero contributions only when they are multiplied by the singular
terms. For the generic contraction I in Table~\ref{DCcon1}, the only
order m contribution must come from the conformal part of
the fermion propagator whereas the generic contraction II in
Table~\ref{DCcon2} the order m contribution could be either from
the flat spacetime mass term or from the infinite series expansion
of the fermion propagator. The fortunate thing is that the contribution
from the infinite series expansion of the fermion propagator vanishes.
The most singular terms from this particular contribution have
dimensionality $\frac{1}{\Delta x^{2D-5}}$, which are integrable in $D=4$.
Therefore they completely vanish in $D=4$ dimensions owing to the
behaviors of the residual part of the C-type graviton propagator
we mentioned above.

We first work out the contributions from Table~\ref{DCcon1}, which contain
a logarithmic divergence. The contraction $(1\!-\!4)$ and $(2\!-\!4)_{\rm a}$
are simple owning to the special property of the conformal part of
the fermion propagator (\ref{fpeqn}) and hence they only pick up
the constant part of $i\delta\!\Delta_C$,
\begin{eqnarray}
&&[1\!-\!4]=\frac{\kappa^2H^{D\!-\!2}}{2^D\pi^{\frac{D}{2}}}
\frac{\Gamma(D\!-\!3)}{\Gamma(\frac{D}{2})}\frac{2}
{(D\!-\!2)(D\!-\!3)}ima\delta^D\!(x-x')\,,\label{1-4}\\
&&[2\!-\!4]_{\rm a}=\frac{\kappa^2H^{D\!-\!2}}{2^D\pi^{\frac{D}{2}}}
\frac{\Gamma(D\!-\!3)}{\Gamma(\frac{D}{2})}\frac{1}
{(D\!-\!2)}ima\delta^D\!(x-x')\,.\label{2-4a}
\end{eqnarray}
The same procedure is employed to make the expressions integrable in $D=4$.
We summarized the contractions $(4\!-\!1)$, $(4\!-\!2)_{\rm a}$ and
$(4\!-\!2)_{\rm b}$ in Table~\ref{dCcf1} and listed the rest of this
category in Table~\ref{dCcf2}. The two series in Table~\ref{dCcf1}
from each contraction cancel out with each other precisely in four
dimensions. Finally the total summation from (\ref{1-4}), (\ref{2-4a}),
Table~\ref{dCcf1} and Table~\ref{dCcf2} is quite simple in $D=4$,
\begin{eqnarray}
-i[\Sigma^{\rm idCcf}](x;x')\!=\!\frac{i\kappa^2H^2}{16\pi^2}
\frac32 ma\delta^4\!(x-x')\!+\!\frac{\kappa^2H^2}{32\pi^4}
ma\Bigl\{\frac{-9}{16}\partial^2\frac{1}{\Delta x^2}
\Bigr\}\,.\label{idCcf}
\end{eqnarray}

\begin{table}

\vbox{\tabskip=0pt \offinterlineskip
\def\tablerule{\noalign{\hrule}}
\halign to390pt {\strut#& \vrule#\tabskip=1em plus2em& \hfil#\hfil&
\vrule#& \hfil#\hfil& \vrule#&  \hfil#\hfil&
\vrule#\tabskip=0pt\cr
\tablerule
\omit&height2pt&\omit&&\omit&&\omit&\cr
&&${\rm I\!-\!J}_{\rm sub}$
&&$\frac{\kappa^2H^2}{2^5\pi^D}\Gamma(\frac{D}{2})
\Gamma(\frac{D}{2}\!-\!1)\frac{(3\!-\!\frac{D}{2})}{(D\!-\!3)}
\frac{ma}{(aa')^{\frac{D}{2}\!-\!2}}$
&&$\frac{\kappa^2 H^{D\!-\!2}}{2^{D\!+\!1}\pi^D}
\Gamma(D\!-\!3)ma$ &\cr
\omit&height2pt&\omit&&\omit&&\omit&\cr
\tablerule
\omit&height2pt&\omit&&\omit&&\omit&\cr
&&$(4\!-\!1)$ &&$\frac{-1}{(D\!-\!2)(D\!-\!3)}
\partial^2\frac{1}{\Delta x^{2D-6}}$
&&$\frac{2}{(D\!-\!2)^2(D\!-\!3)}\partial^2\frac{1}{\Delta x^{D-2}}$ &\cr
\omit&height2pt&\omit&&\omit&&\omit&\cr
\tablerule
\omit&height2pt&\omit&&\omit&&\omit&\cr
&&$(4\!-\!2)_{\rm a}$ &&$\frac{1}{2(D\!-\!2)}\!\not{\!\partial}
\gamma^0\partial_0\frac{1}{\Delta x^{2D-6}}$
&&$\frac{-1}{(D\!-\!2)^2}\!\not{\!\partial}\gamma^0\!\partial_0
\frac{1}{\Delta x^{D-2}}$ &\cr
\omit&height2pt&\omit&&\omit&&\omit&\cr
\tablerule
\omit&height2pt&\omit&&\omit&&\omit&\cr
&&$(4\!-\!2)_{\rm b}$ &&$\frac{-1}{2(D\!-\!2)(D\!-\!3)}
\!\not{\!\partial}\!\!\not{\!\bar{\partial}}\frac{1}{\Delta x^{2D-6}}$
&&$\frac{1}{(D\!-\!2)^2(D\!-\!3)}\!\not{\!\partial}
\!\!\not{\!\bar{\partial}}\frac{1}{\Delta x^{D-2}}$ &\cr
\omit&height2pt&\omit&&\omit&&\omit&\cr
\tablerule}}

\caption{Contributions from Table~\ref{DCcon1}
for $i\delta\!\Delta_C\times i[S]_{\rm cf}(x;x')$-I.}

\label{dCcf1}

\end{table}

\begin{table}

\vbox{\tabskip=0pt \offinterlineskip
\def\tablerule{\noalign{\hrule}}
\halign to390pt {\strut#& \vrule#\tabskip=1em plus2em&
\hfil#\hfil& \vrule#&  \hfil#\hfil& \vrule#\tabskip=0pt\cr
\tablerule
\omit&height2pt&\omit&&\omit&\cr
&&${\rm I\!-\!J}_{\rm sub}$
&&$\partial^2\frac{1}{\Delta x^{2D-6}}$ &\cr
\omit&height2pt&\omit&&\omit&\cr
\tablerule
\omit&height2pt&\omit&&\omit&\cr
&&$(2\!-\!4)_{\rm b}$ &&$\frac{-1}{4}$&\cr
\omit&height2pt&\omit&&\omit&\cr
\tablerule
\omit&height2pt&\omit&&\omit&\cr
&&$(3\!-\!4)_{\rm a}$ &&$\frac{-1}{8}$&\cr
\omit&height2pt&\omit&&\omit&\cr
\tablerule
\omit&height2pt&\omit&&\omit&\cr
&&$(3\!-\!4)_{\rm b}$ &&$\frac{-1}{8}$&\cr
\omit&height2pt&\omit&&\omit&\cr
\tablerule
\omit&height2pt&\omit&&\omit&\cr
&&$(4\!-\!3)_{\rm a}$ &&$\frac{-1}{8}$&\cr
\omit&height2pt&\omit&&\omit&\cr
\tablerule
\omit&height2pt&\omit&&\omit&\cr
&&$(4\!-\!3)_{\rm b}$ &&$\frac{-1}{8}$&\cr
\omit&height2pt&\omit&&\omit&\cr
\tablerule}}

\caption{Contributions from Table~\ref{DCcon1}
for $i\delta\!\Delta_C\times i[S]_{\rm cf}(x;x')$-II.
All the terms are multiplied by
$\frac{\kappa^2H^2}{2^5\pi^D}\Gamma(\frac{D}{2})
\Gamma(\frac{D}{2}\!-\!1)\frac{(3\!-\!\frac{D}{2})(D\!-\!1)}
{(D\!-\!2)^2(D\!-\!3)^2}\frac{ma}{(aa')^{\frac{D}{2}\!-\!2}}$.}

\label{dCcf2}

\end{table}

\begin{table}

\vbox{\tabskip=0pt \offinterlineskip
\def\tablerule{\noalign{\hrule}}
\halign to390pt {\strut#& \vrule#\tabskip=1em plus2em&
\hfil#\hfil& \vrule#& \hfil#\hfil& \vrule#&  \hfil#\hfil&
\vrule#\tabskip=0pt\cr
\tablerule
\omit&height2pt&\omit&&\omit&&\omit&\cr
&&$\!\!\!\!\!\!{\rm I\!-\!J}_{\rm sub}\!\!\!\!\!\!$
&&$\frac{\kappa^2H^2}{2^6\pi^D}\Gamma^2(\frac{D}{2}\!-\!1)
(3\!-\!\frac{D}{2})\frac{ma}{(aa')^{\frac{D}{2}\!-\!2}}$
&&$\frac{\kappa^2 H^{D\!-\!2}}{2^{D\!+\!2}\pi^D}
\Gamma(D\!-\!3)ma\frac{2}{(D\!-\!2)}$ &\cr
\omit&height2pt&\omit&&\omit&&\omit&\cr
\tablerule
\omit&height2pt&\omit&&\omit&&\omit&\cr
&&$\!\!\!\!\!\!(1\!-\!1)\!\!\!\!\!\!$
&&$[\frac{1}{(D\!-\!3)^2}\partial^2
\!-\!\frac{2Ha\gamma^0}{(D\!-\!2)(D\!-\!3)}
\!\!\not{\!\partial}]\frac{1}{\Delta x^{2D-6}}$
&&$\frac{2[-\partial^2+Ha\gamma^0\!\not{\hspace{.05cm}\partial}]}
{(D\!-\!2)(D\!-\!3)}\frac{1}{\Delta x^{D-2}}$ &\cr
\omit&height2pt&\omit&&\omit&&\omit&\cr
\tablerule
\omit&height2pt&\omit&&\omit&&\omit&\cr
&&$(1\!-\!2)_{\rm a}$ &&$[\frac{-1}{2(D\!-\!3)}
\!\not{\!\partial}\gamma^0\!\partial_0\!-\!\frac{Ha}{(D\!-\!2)}
\partial_0]\frac{1}{\Delta x^{2D-6}}$
&&$\frac{[\not{\hspace{.05cm}\partial\gamma^0\partial_0}
+Ha\partial_0]}{(D\!-\!2)}\frac{1}{\Delta x^{D-2}}$ &\cr
\omit&height2pt&\omit&&\omit&&\omit&\cr
\tablerule
\omit&height2pt&\omit&&\omit&&\omit&\cr
&&$(1\!-\!2)_{\rm b}$
&&$[\frac{1}{2(D\!-\!3)^2}\!\!\not{\!\partial}
\!+\!\frac{Ha\gamma^0}{(D\!-\!2)(D\!-\!3)}
]\!\!\not{\!\bar{\partial}}\frac{1}{\Delta x^{2D-6}}$
&&$\frac{-[\not{\partial}+Ha\gamma^0]}{(D\!-\!2)(D\!-\!3)}
\!\!\not{\!\bar{\partial}}\frac{1}{\Delta x^{D-2}}$ &\cr
\omit&height2pt&\omit&&\omit&&\omit&\cr
\tablerule
\omit&height2pt&\omit&&\omit&&\omit&\cr
&&$(2\!-\!1)_{\rm a}$
&&$[\frac{-1}{2(D\!-\!3)}\gamma^0\partial_0
\!-\!\frac{Ha\gamma^0}{(D\!-\!2)}
]\!\!\not{\!\partial}\frac{1}{\Delta x^{2D-6}}$
&&$\frac{[\gamma^0\partial_0+Ha\gamma^0]}{(D\!-\!2)}
\!\!\not{\!\partial}\frac{1}{\Delta x^{D-2}}$ &\cr
\omit&height2pt&\omit&&\omit&&\omit&\cr
\tablerule
\omit&height2pt&\omit&&\omit&&\omit&\cr
&&$(2\!-\!1)_{\rm b}$
&&$\frac{1}{2(D\!-\!3)^2}\!\!\not{\!\bar{\partial}}
\!\!\not{\!\partial}\frac{1}{\Delta x^{2D-6}}$
&&$\frac{-1}{(D\!-\!2)(D\!-\!3)}\!\!\not{\!\bar{\partial}}
\!\!\not{\!\partial}\frac{1}{\Delta x^{D-2}}$ &\cr
\omit&height2pt&\omit&&\omit&&\omit&\cr
\tablerule
\omit&height2pt&\omit&&\omit&&\omit&\cr
&&$(2\!-\!2)_{\rm a}$
&&$[-\frac14\partial^2_0\!-\!\frac{(D\!-\!3)}{2(D\!-\!2)}
Ha\partial_0]\frac{1}{\Delta x^{2D-6}}$
&&$\frac{(D\!-\!3)}{2(D\!-\!2)}[\partial^2_0\!+\!Ha\partial_0]
\frac{1}{\Delta x^{D-2}}$ &\cr
\omit&height2pt&\omit&&\omit&&\omit&\cr
\tablerule
\omit&height2pt&\omit&&\omit&&\omit&\cr
&&$(2\!-\!2)_{\rm b}$
&&$[\frac{1}{4(D\!-\!3)}\gamma^0\partial_0
\!+\!\frac{Ha\gamma^0}{2(D\!-\!2)}
]\!\!\not{\!\bar{\partial}}\frac{1}{\Delta x^{2D-6}}$
&&$\frac{-[\gamma^0\partial_0+Ha\gamma^0]}{2(D\!-\!2)}
\!\!\not{\!\bar{\partial}}\frac{1}{\Delta x^{D-2}}$ &\cr
\omit&height2pt&\omit&&\omit&&\omit&\cr
\tablerule
\omit&height2pt&\omit&&\omit&&\omit&\cr
&&$(2\!-\!2)_{\rm c}$
&&$\frac{-1}{4}\gamma^0\partial_0\!\!
\not{\!\bar{\partial}}\frac{1}{\Delta x^{2D-6}}$
&&$\frac{1}{2(D\!-\!2)}\gamma^0\partial_0\!\!
\not{\!\bar{\partial}}\frac{1}{\Delta x^{D-2}}$ &\cr
\omit&height2pt&\omit&&\omit&&\omit&\cr
\tablerule
\omit&height2pt&\omit&&\omit&&\omit&\cr
&&$(2\!-\!2)_{\rm d}$
&&$\frac{1}{4(D\!-\!3)^2}\nabla^2\!\frac{1}{\Delta x^{2D-6}}$
&&$\frac{-1}{2(D\!-\!2)(D\!-\!3)}\nabla^2
\!\frac{1}{\Delta x^{D-2}}$ &\cr
\omit&height2pt&\omit&&\omit&&\omit&\cr
\tablerule}}

\caption{Contributions from Table~\ref{DCcon2} for
$i\delta\!\Delta_C\times i[S]_{\rm fm}(x;x')$-I.}

\label{dCfm1}

\end{table}

\begin{table}

\vbox{\tabskip=0pt \offinterlineskip
\def\tablerule{\noalign{\hrule}}
\halign to390pt {\strut#& \vrule#\tabskip=1em plus2em&
\hfil#\hfil& \vrule#&  \hfil#\hfil& \vrule#\tabskip=0pt\cr
\tablerule
\omit&height2pt&\omit&&\omit&\cr
&&${\rm I\!-\!J}_{\rm sub}$
&&$\partial^2\frac{1}{\Delta x^{2D-6}}$ &\cr
\omit&height2pt&\omit&&\omit&\cr
\tablerule
\omit&height2pt&\omit&&\omit&\cr
&&$(1\!-\!3)_{\rm a}$ &&$\frac{-1}{2(D\!-\!3)}$&\cr
\omit&height2pt&\omit&&\omit&\cr
\tablerule
\omit&height2pt&\omit&&\omit&\cr
&&$(1\!-\!3)_{\rm b}$ &&$\frac{-1}{2(D\!-\!3)}$&\cr
\omit&height2pt&\omit&&\omit&\cr
\tablerule
\omit&height2pt&\omit&&\omit&\cr
&&$(2\!-\!3)_{\rm a}$ &&$\frac{-1}{4}$&\cr
\omit&height2pt&\omit&&\omit&\cr
\tablerule
\omit&height2pt&\omit&&\omit&\cr
&&$(2\!-\!3)_{\rm d}$ &&$\frac{1}{4(D\!-\!3)}$&\cr
\omit&height2pt&\omit&&\omit&\cr
\tablerule
\omit&height2pt&\omit&&\omit&\cr
&&$(3\!-\!3)_{\rm a}$ &&$\frac18(\frac{D\!-\!1}{D\!-\!3})$&\cr
\omit&height2pt&\omit&&\omit&\cr
\tablerule
\omit&height2pt&\omit&&\omit&\cr
&&$(3\!-\!3)_{\rm d}$ &&$\frac{1}{8(D\!-\!3)}$&\cr
\omit&height2pt&\omit&&\omit&\cr
\tablerule
\omit&height2pt&\omit&&\omit&\cr
&&${\rm total}$ &&$\frac{(D\!-\!8)}{8(D\!-\!3)}
\!-\!\frac{(D\!-\!4)}{4(D\!-\!3)}$&\cr
\omit&height2pt&\omit&&\omit&\cr
\tablerule}}

\caption{Contribution from Table~\ref{DCcon2} for
$i\delta\!\Delta_C\times i[S]_{\rm fm}(x;x')$-II.
All the terms are multiplied by
$\frac{\kappa^2H^2}{2^8\pi^D}\Gamma^2(\frac{D}{2}\!-\!1)
\frac{(3\!-\!\frac{D}{2})(D\!-\!1)}
{(D\!-\!2)(D\!-\!3)}\frac{ma}{(aa')^{\frac{D}{2}\!-\!2}}$.}

\label{dCfm2}

\end{table}
The final class we need to complete is the contributions from
Table~\ref{DCcon2}. Very similarly to what happened with
$i\delta\!\Delta_B$, the contractions from (1-1), (1-2), (2-1),
(2-2) tend to cancel and we summarized them in Table~\ref{dCfm1}.
We also tabulate the rest of the contributions which do not vanish
in $D=4$ in Table~\ref{dCfm2}. As already explained, terms for which
one or more derivative acts upon a scale factor make no contribution
in $D\!=\!4$ dimensions, so the final nonzero contribution come from
the derivatives only acting upon the coordinate separation,
$\Delta x^2$. The net contributions from Table~\ref{dCfm1} do vanish
completely in $D=4$ dimensions and the only non-zero contributions of
this class come from Table~\ref{dCfm2},
\begin{eqnarray}
-i[\Sigma^{\rm idCfm}](x;x')\!=\!\frac{\kappa^2H^2}{2^8\pi^4}ma
\Biggl\{\frac{-3}{4}\partial^2\frac{1}{\Delta x^2}
\Biggr\}\,.\label{idCfm}
\end{eqnarray}

\section{Renormalization}

Except for the finite results from (\ref{sumcfn1}), and the results
from Table~\ref{dAn0-Coeff}-\ref{dAn0-tot}, each of which possesses
distinctive expressions, the rest of regulated result we
have worked so hard to compute derives from summing expressions
(\ref{tot4pt}), (\ref{re-cfcf}), (\ref{re-cffm}), (\ref{n012}),
(\ref{n03}), (\ref{n0456}), (\ref{idAcf}), (\ref{idAfm}), (\ref{idBfm}),
(\ref{idCcf}) and (\ref{idCfm})
\begin{eqnarray}
&&i\kappa^2\Biggl\{\beta_1\frac{m}{a}\partial^2\!+\!\beta_2 mH\partial_0
\!+\!\beta_3mH\gamma^0\!\!\!\not{\!\bar{\partial}}\!+\!\beta_4 mH^2\!a
\Biggr\}\delta^D\!(x\!-\!x')\!+\!\frac{i\kappa^2}{16\pi^2}\Biggl\{
\Bigl(3\ln a\!-\!\frac38\Bigr)\frac{m}{a}\partial^2\nonumber\\
&&+\Bigl(\frac{97}{16}\ln a\!-\!\frac{63}{16}\Bigr)mH\partial_0
\!+\!\Bigl(\frac{9}{16}\ln a\!+\!\frac18\Bigr)mH\gamma^0\!\!\!
\not{\!\bar{\partial}}\!+\!\Bigl(\frac{95}{8}\ln a\!+\!\frac{195}{32}
\Bigr)H^2\!ma\Biggr\}\delta^4\!(x\!-\!x')\nonumber\\
&&+\frac{\kappa^2}{64\pi^4}\Biggl\{\Biggl[\frac32\frac{m}{a'}
\partial^2\!+\!\Bigl(\frac78\frac{a}{a'}\!-\!\frac{27}{32}\Bigr)
mH\partial_0\!+\!\Bigl(\frac{9}{16}\frac{a}{a'}\!-\!
\frac{9}{32}\Bigr)mH\gamma^0\!\!\!\not{\!\bar{\partial}}\!+\!
H^2m\Bigl(\frac{215}{32}a+\nonumber\\
&&\frac{9}{32}a'\Bigr)\Biggr]\partial^2\!+\!H^2ma\Bigl[\nabla^2
\!+\!6Ha\partial_0\!+\!4Ha\gamma^0\!\!\!\not{\!\bar{\partial}}\Bigr]
\Biggr\}\Biggl[\frac{\ln(\mu^2\Delta x^2)}{\Delta x^2}\Biggr]
\!+\!\frac{\kappa^2H^2}{64\pi^4}\Biggl\{
\frac{9}{16}ma'\partial_0^2\nonumber\\
&&+\Bigl(\frac{-53}{16}ma\!+\!\frac{3}{16}ma'\Bigr)
\gamma^0\partial_0\!\!\not{\!\bar{\partial}}\!+\!\Bigl(
\frac{-49}{16}\!+\!\ln\frac{H^2}{4\mu^2}\Bigr)ma\nabla^2
\!+\!\Bigl(\frac92\!+\!6\ln\frac{H^2}{4\mu^2}
\Bigr)Hma^2\partial_0\nonumber\\
&&+\Bigl(\frac{35}{8}\!-\!\frac{11}{8}\ln\frac{y}{4}
\!+\!4\ln\frac{H^2}{4\mu^2}\Bigr)Hma^2\gamma^0\!\!\!
\not{\!\bar{\partial}}+\Bigl(\frac{-5}{8}\!-\!\frac38
\ln\frac{y}{4}\Bigr)Hmaa'\gamma^0\!\!\!\not{\!\bar{\partial}}
\Biggr\}\frac{1}{\Delta x^2}\,.\label{regres}
\end{eqnarray}
The various $D$-dependent constants in (\ref{regres}) are,
\begin{eqnarray}
&&\hspace{-1cm}\beta_1\!=\!\frac{\mu^{D-4}}{16\pi^{\frac{D}2}}
\frac{\Gamma(\frac{D}2\!-\!1)}{(D\!-\!3)(D\!-\!4)}\Biggl\{
\frac{-2(D\!-\!1)}{(D\!-\!2)} \Biggr\} , \\
&&\hspace{-1cm}\beta_2\!=\!\frac{\mu^{D-4}}{16\pi^{\frac{D}2}}
\frac{\Gamma(\frac{D}2\!-\!1)}{(D\!-\!3)(D\!-\!4)}
\Biggl\{\frac{-4(D\!-\!1)}{(D\!-\!2)}\!+\!
(D\!-\!2)(b_2\!+\!b_3)\Biggr\} , \qquad \\
&&\hspace{-1cm}\beta_3\!=\!\frac{\mu^{D-4}}{16\pi^{\frac{D}2}}
\frac{\Gamma(\frac{D}2\!-\!1)}{(D\!-\!3)(D\!-\!4)}
\Biggl\{\frac{-(D\!-\!1)}{2}\!+\!(D\!-\!2)(b_{2a}\!+\!b_{3a})
\!+\!\frac{(D\!-\!2)}{2}d_1\Biggr\} , \\
&&\hspace{-1cm}\beta_4\!=\!\frac{\mu^{D-4}}{16\pi^{\frac{D}2}}
\frac{\Gamma(\frac{D}{2}\!-\!1)}{(D\!-\!3)(D\!-\!4)}\Biggl\{
\frac{(D\!-\!1)}{2}\!+\!(D\!-\!2)(b_4\!-\!b_2)\!+\!
\frac{(D\!-\!2)(d_2\!+\!d_3\!+\!d_4)}{2} \nonumber\\
&&+\frac{D(D\!-\!1)}{8(D\!-\!3)}\Bigl[-D(D\!-\!2)\!-\!\frac14
\!-\!\frac14(D\!-\!4)\Bigr]\Biggr\}\!+\!\frac{H^{D-4}}
{2^D\pi^{\frac{D}{2}}(D\!-\!4)}\Biggl\{\frac{-\Gamma(D\!+\!1)}
{2\Gamma(\frac{D}{2})}\nonumber\\
&&+\frac{(2D\!-\!3)}{4}\Gamma(\frac{D}{2})\Gamma(\frac{D}{2}\!-\!1)
\Gamma(3\!-\!\frac{D}{2})+\frac{\Gamma(D)}
{\Gamma(\frac{D}{2})}\frac{2}{(D\!-\!3)}\Biggr\}\,.
\end{eqnarray}
Here $b_2$, $b_{\rm 2a}$, $b_3$, $b_{\rm 3a}$, $b_4$, $d_1$,
$d_2$, $d_3$ and $d_4$ are defined in (\ref{cffmcoef}) and
(\ref{d1234}). In obtaining these expressions we have always chosen
to convert finite, $D\!=\!4$ terms with $\partial^2$ acting on
$1/\Delta x^2$, into delta functions,
\begin{equation}
\partial^2 \Bigl[\frac1{\Delta x^2}\Bigr] = i 4 \pi^2 \delta^4(x\!-\!x') \; .
\end{equation}
All such terms have then been included in those which are proportional
to $\delta^4\!(x\!-\!x')$.

\begin{figure}
\begin{center}
\includegraphics[width=4.0cm]{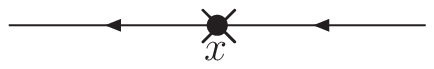}
\\{\rm Fig.~3: Contribution from counterterms.}
\label{Fig3}
\end{center}
\end{figure}

The local divergences in this expression can be absorbed by the BPHZ
counterterms enumerated at the end of section 3. The generic diagram
topology is depicted in Fig.~3, and the analytic form is,
\begin{eqnarray}
\lefteqn{-i\Bigl[\Sigma^{\rm ctm}\Bigr](x;x') = \sum_{I=1}^4 i
C_{Iij}\,\delta^D(x-x')\;,} \\
&&=i\kappa^2\Bigl\{\alpha_1 \frac{m}{a}\partial^2\!+\!\alpha_2
mH\partial_0\!+\!\alpha_3 mH\gamma^0\!\!\!\not{\!\bar{\partial}}
\!+\!\alpha_4 H^2ma\Bigr\}\delta^D(x\!-\!x')\;.\label{genctm}
\end{eqnarray}
In comparing (\ref{regres}) and (\ref{genctm}) it would seem that
the simplest choice for the coefficients $\alpha_i$ is,
\begin{eqnarray}
&&\hspace{-1cm}\alpha_1=3F_1\,\,\, ,\,\,\,
\alpha_2=\frac{97}{16}F_1 \,\,\,,\,\,\,
\alpha_3=\frac{9}{16}F_1 \,\,\,,{\rm and}\,\,\,
\alpha_4=\frac{85}{8}F_2\; . \label{sim}\\
&&{\rm Here}\quad F_2=\frac{\mu^{D-4}}{16\pi^{\frac{D}{2}}(D\!-\!4)}
\quad,\quad F_1=\frac{\Gamma(\frac{D}{2}\!-\!1)}{(D\!-\!3)}\times
F_2\,.
\end{eqnarray}
Except for the distinctive contributions from (\ref{sumcfn1}),
Table~\ref{dAn0-Coeff} and Table~\ref{dAn0-tot}, the rest of
our final result for the renormalized self-energy is,
\begin{eqnarray}
&&-i\Bigl[\Sigma^{\rm ren}\Bigr](x;x')\!=\!\!
\frac{i\kappa^2}{16\pi^2}\Biggl\{\Biggl[3\ln a\!+\!\frac18\Biggr]
\frac{m}{a}\partial^2+\Biggl[\frac{97}{16}\ln a\!-\!\frac{39}{16}\Biggr]
mH\partial_0\!+\!\Biggl[\frac{9}{16}\ln a\nonumber\\
&&+\frac{5}{16}\Biggr]mH\gamma^0\!\!\!\not{\!\bar{\partial}}\!+\!
\Biggl[\frac{95}{8}\ln a\!-\!\frac{29}{32}\!-\!\frac{85}{16}
\psi(1)\!+\!\frac58\ln\frac{H^2}{4\mu^2}\Biggr]H^2\!ma\Biggr\}
\delta^4\!(x\!-\!x')\!+\!\frac{\kappa^2}{64\pi^4}
\Biggl\{\nonumber\\
&&\Biggl[\frac32\frac{m}{a'}\partial^2+\Bigl(\frac78\frac{a}{a'}
\!-\!\frac{27}{32}\Bigr)mH\partial_0\!+\!\Bigl(\frac{9}{16}
\frac{a}{a'}\!-\!\frac{9}{32}\Bigr)mH\gamma^0\!\!\!
\not{\!\bar{\partial}}\!+\!H^2m\Bigl(\frac{215}{32}a\!+\!
\frac{9}{32}a'\Bigr)\Biggr]\partial^2\nonumber\\
&&\hspace{1cm}+H^2ma\Bigl[\nabla^2\!+\!6Ha\partial_0\!+\!4Ha\gamma^0
\!\!\!\not{\!\bar{\partial}}\Bigr]\Biggr\}\Biggl[
\frac{\ln(\mu^2\Delta x^2)}{\Delta x^2}\Biggr]
\!+\!\frac{\kappa^2H^2}{64\pi^4}\Biggl\{
\frac{9}{16}ma'\partial_0^2\nonumber\\
&&+\Biggl[\frac{-53}{16}ma\!+\!\frac{3}{16}ma'\Biggr]
\gamma^0\partial_0\!\!\not{\!\bar{\partial}}\!+\!\Biggl[
\frac{-49}{16}\!+\!\ln\frac{H^2}{4\mu^2}\Biggr]ma\nabla^2
\!+\!\Biggl[\frac92\!+\!6\ln\frac{H^2}{4\mu^2}
\Biggr]Hma^2\partial_0\nonumber\\
&&+\Biggl[\frac{35}{8}\!-\!\frac{11}{8}\ln\frac{y}{4}
\!+\!4\ln\frac{H^2}{4\mu^2}\Biggr]Hma^2\gamma^0\!\!\!
\not{\!\bar{\partial}}\!+\!\Biggl[\frac{-5}{8}\!-\!\frac38
\ln\frac{y}{4}\Biggr]Hmaa'\gamma^0\!\!\!\not{\!\bar{\partial}}
\Biggr\}\frac{1}{\Delta x^2}.
\label{ren}
\end{eqnarray}

\section{Discussion}

Dimensional regularization has been used to compute quantum
gravitational corrections to the fermion self-energy at one loop
order in a locally de Sitter background. Our regulated result is
(\ref{regres}). Although Dirac $+$ Einstein is not perturbatively
renormalizable \cite{DVN} a finite result (\ref{ren}) is obtained
by absorbing the divergences with BPHZ counterterms.

At first order in $m$ only four counterterms are necessary for this
one-loop 1PI function. None of them represents redefinitions of terms
in the Lagrangian of Dirac $+$ Einstein. Two de Sitter invariant
counterterm operators (\ref{inv12}) come from generally coordinate
invariant fermion bilinears (\ref{invctms}). The other two counterterm
operators (\ref{noninv34}) are from other fermion bilinears (\ref{nictm})
which respect the symmetries of our de Sitter noninvariant gauge (\ref{GR}).

BPHZ renormalization does not yield a complete theory because
no physical principle fixs the finite part of these counterterms.
Hence our renormalized result could be changed by altering the
finite parts of the four BPHZ counterterms. It is simple to be
quantitative about this. Were we to make finite shifts $\Delta
\alpha_i$ in our counterterms (\ref{sim}) the induced change
in the renormalized self-energy would be,
\begin{eqnarray}
&&\hspace{-1cm}-i\Bigl[\Delta \Sigma^{\rm ren}\Bigr](x;x')\!=\!i\kappa^2
\Biggl\{\Delta\alpha_1 \frac{m}{a}\partial^2\!+\!\Delta\alpha_2
mH\partial_0\!+\!\Delta\alpha_3mH\gamma^0\!\!\!\not{\!\bar{\partial}}
\!+\!\Delta\alpha_4 H^2ma\Biggr\}\delta^4(x\!-\!x'). \label{arb}\nonumber\\
\end{eqnarray}
However, at late times (which accesses the far infrared because
all momenta are redshifted by $a(t) = e^{Ht}$) the local part of
the renormalized self-energy (\ref{ren}) is dominated by the large
logarithms,
\begin{eqnarray}
&&\hspace{-1.5cm}\frac{i\kappa^2}{16\pi^2}\Biggl\{3\ln\!a
\frac{m}{a}\partial^2\!+\!\frac{97\ln\!a}{16}mH\partial_0\!+\!
\frac{9\ln\!a}{16}mH\gamma^0\!\!\!\not{\!\bar{\partial}}\!+\!
\frac{95\ln\!a}{8}H^2\!ma\Biggr\}\delta^4\!(x\!-\!x').
\label{fixed}
\end{eqnarray}
The coefficients of these logarithms are finite and completely fixed
by our calculation. As long as the shifts $\Delta \alpha_i$ are
finite, their impact (\ref{arb}) must eventually be dwarfed by the
large logarithms (\ref{fixed}).

It does not seem too surprising that the leading behavior in the far
infrared cannot be disturbed by the nonrenormalizability of quantum
gravity. Loops of massless particles make finite, nonanalytic
contributions which cannot be changed by local counterterms and which
dominate the far infrared. Further, no matter how general relativity
is corrected to fix the ultraviolet problem, it cannot involve any
new massless particle or else we should have seen new long range
force. In addition, the correction also cannot change how the existing
massless particle interact at low energy, otherwise we should have
detected classical violation of general relativity. Therefore these
effects must occur as well, \emph{with precisely the same numerical
values}, in whatever fundamental theory ultimately resolves the
ultraviolet problem of quantum gravity. The concept we have just
emphasized is known as low energy effective field theory and has a very
old and distinguished pedigree
\cite{BN,SW,FS,HS,CDH,CD,DMC1,DL,JFD1,JFD2,MV,HL,ABS,oldKK1,oldKK2}.

So we can use (\ref{ren}) reliably in the far infrared. The point
of this exercise has been to study the effect of breaking conformal
invariance with a small fermion mass. Obtaining (\ref{ren}) completes
the first part of our study. What remains is to use our result to
solve the quantum-corrected Dirac equation (\ref{Diraceq1}). We shall
undertake that in a subsequent paper. However, it seems clear that the
dominant effect must come from the terms which possess large
logarithms in local terms and in (\ref{sumcfn1}),
Table~\ref{dAn0-tot} and Table~\ref{dAn0-Coeff}\footnote{Without an
explicit calculation we cannot determine whether or not the
$\ln(\frac{y}{4})$ in the non-local terms will produce infrared
enhancements because the same term occurring in \cite{DW} fails to
give them.}.

As predicted in the Introduction, these terms are enhanced by
a factor of $\ln(a)$ relative to the classical mass term and
$a\ln(a)$ relative to the classical kinetic term. When the classical
mass term is much smaller than the Hubble Parameter, classical
dynamics are dominated by the kinetic term. Therefore a larger
enhancement of the fermion field strength $a\ln(a)$ is expected
in contrast with the only $\ln(a)$ enhancement which soft virtual
gravitons induce on massless fermions \cite{MW1, MW3}.

Loop corrections from massless, minimally coupled scalars and
gravitons during inflation have attracted more and more attention
recently. The interesting time-growing effects of infrared logarithms
might have a chance to eventually overcome the smallness of the loop
counting parameter of $G H^2 \ltwid 10^{-10}$ and yield significant
results. It is not even possible to exclude the possibility that infrared
logarithms can contaminate the power spectrum of cosmological density
perturbations \cite{SW2, IRstudies, MW5, KOW}! However, the logarithms
would only start to grow at horizon crossing, and must cease growing
when the mode reenters the horizon after inflation. Hence the largest
enhancement for a currently observable mode would be $\ln(a) \ltwid 100$
which must be set against $G H^2 \ltwid 10^{-10}$. Therefore
the proportional correction in theses studies are still too small to
be detected by current measuring technique.

The enormously super-horizon modes which have not experienced the
second horizon crossing would give more significant corrections.
Although they are also down by the constant $GH^2$, the time-dependent
enhancement factor $\ln(a)$ could be arbitrarily big so that
perturbation theory eventually breaks down. One must develop the
non-perturbative technique to follow what happens.
Starobinski\u{\i} has advocated gaining quantitative control over
this regime by summing the leading infrared logarithms at each order
\cite{AAS}. With Yokoyama he has given a complete solution for the
case of a minimally coupled scalar with arbitrary potential which is
a spectator to de Sitter inflation \cite{SY}. This powerful
non-perturbative technique has been successfully generalized to Yukawa
theory \cite{MW2}, which showed that the system decays in a Big Rip
singularity, and to SQED \cite{PTW2}, which confirmed the conjecture
by Davis, Dimopoulos, Prokopec and Tornkvist that super-horizon photons
acquire mass during inflation \cite{DDPT}.

The asymptotic late time effect is small in the simple scalar models,
and in SQED for which the series of leading infrared logarithms has
been summed. However, the same kind of effect from Yukawa is huge.
Therefore, it is by no means clear what might be the outcome for more
complicated theories\footnote{By which we mean theories which possess
derivative interactions that cannot be avoided by imposing a special
gauge as was done in SQED.} that also show infrared logarithms
such as quantum gravity \cite{TW4,TW5,TW6}. Another application of
our result (\ref{ren}) is to serve as ``data'' in checking the validity
of the new, more general rule \cite{TW8, MW4} for reproducing the leading
logarithms of massive Dirac $+$ Einstein. This might serve as an important
intermediate point in the difficult task of generalizing Starobinski\u{\i}'s
techniques to full blown quantum gravity.


It is well to close with a comment on whether or not the infrared
logarithm which appears in (\ref{ren}) is a gauge artifact. One obvious
way of checking this is to re-do the computation in a different gauge.
Recently two graviton propagators have been constructed respectively by
imposing the exact de Donder gauge \cite{MTW2} and a general
one parameter gauge \cite{MoraTW}. Both gauges respect de Sitter invariance
but the same de Sitter breaking factor $\ln(a)$ shows up even in the 
transverse-traceless sector of the propagator \cite{MoraTW,KMW}.
\emph{Hence the infrared logarithm must be universal
because the spin two part of the graviton propagator cannot be altered by
changing gauges.} We therefore conjecture that the leading infrared
logarithm at the one loop order in quantum gravity might be
gauge-independent.

\vskip 1cm

\centerline{\bf Acknowledgements}
The author would like to thank R. P. Woodard for useful discussions.
This work was supported by NWO Veni Project $\sharp$ 680-47-406
, by the Institute for Theoretical Physics at the University of Utrecht
and by the Department of Physics at the University of Crete.
\appendix
\section{The reduced fermion propagator and its related identities}
\label{fermionprop}
Here we list some identities we have used for various gamma functions
contracted with the first derivative of the $n=0$ part of the fermion
propagator (\ref{n0mfd1}).
\begin{eqnarray}
&&\not{\!\partial}i[S]_{n=0}=\Gamma(\frac{D}{2}\!-\!1)\Biggl\{
\frac{mHaa'}{4\pi^{\frac{D}{2}}}\Biggl[\frac{-\gamma^0}{\Delta x^{D-2}}
\!-\!\frac{Ha\gamma^0\Delta\eta}{\Delta x^{D-2}}\!-\!
\frac{H^2a^2\gamma^0}{(D\!-\!4)\Delta x^{D-4}}\Biggr]\nonumber\\
&&\hspace{2.5cm}-\frac{mH^{D-3}}{(4\pi)^{\frac{D}{2}}}(aa')^{\frac{D}{2}-1}
\Gamma(\frac{D}{2}\!+\!1)\Gamma(2\!-\!\frac{D}{2})
H^2a^2\gamma^0\Biggr\}\,,\\
&&\partial_{\mu}i[S]\gamma^{\mu}\!=\!\Gamma(\frac{D}{2}\!-\!1)
\Biggl\{\frac{mHaa'}{4\pi^{\frac{D}{2}}}\Biggl[
\frac{(D\!-\!2)\Delta\eta\gamma^{\mu}\Delta x_{\mu}}{\Delta x^D}
\!+\!\frac{Ha\gamma^{\mu}\Delta x_{\mu}}{\Delta x^{D-2}}\nonumber\\
&&-\frac{H^2a^2\gamma^0}{(D\!-\!4)\Delta x^{D-4}}\Biggr]
\!-\!\frac{mH^{D-3}}{(4\pi)^{\frac{D}{2}}}(aa')^{\frac{D}{2}-1}
\Gamma(\frac{D}{2}\!+\!1)\Gamma(2\!-\!\frac{D}{2})
H^2a^2\gamma^0\Biggr\}.
\end{eqnarray}

To facilitate the calculation from the infinite series expansion
of the fermion propagator for $D=4$, we might employ the following
identities,
\begin{eqnarray}
&&\not{\!\partial}i[S]\!=\!\frac{mHaa'}{16\pi^2}\Sigma_{n=0}^{\infty}
(\frac{y}{4})^n\Biggl\{\Bigl[\frac{-4\gamma^0}{\Delta x^2}
\!-\!\frac{4Ha\gamma^0\!\Delta\eta}{\Delta x^2}\Bigr]\Bigl[
(2n\!+\!1)\!+\!(n^2\!+\!n)\ln\frac{y}{4}\Bigr]\nonumber\\
&&\hspace{4cm}+H^2a^2\delta^0_{\mu}\Bigl[(2n\!+\!3)
\!+\!(n^2\!+\!3n\!+\!2)\ln\frac{y}{4}\Bigr]\Biggr\}\,, \\
&&\partial_{\mu}i[S]\gamma^{\mu}\!=\!\frac{mHaa'}{16\pi^2}
\Sigma_{n=0}^{\infty}(\frac{y}{4})^n\Biggl\{\frac{4\gamma^0}{\Delta x^2}
\Bigl[2n\!+\!n^2\ln\frac{y}{4}\Bigr]\!-\!
\frac{8\Delta\eta\gamma^{\mu}\Delta x_{\mu}}{\Delta x^4}\Bigl[
(2n\!-\!1)\!+\!(n^2\!-\!n)\ln\frac{y}{4}\Bigr]\nonumber\\
&&+\frac{4Ha\gamma^{\mu}\Delta x_{\mu}}{\Delta x^2}\Bigl[
(2n\!+\!1)\!+\!(n^2\!+n\!)\ln\frac{y}{4}\Bigr]
\!+\!H^2a^2\delta^0_{\mu}\Bigl[(2n\!+\!3)\!+\!(n^2\!+\!3n\!+\!2)
\ln\frac{y}{4}\Bigr]\Biggr\}\,,\\
&&\not{\!\bar{\partial}}i[S]\!=\!\frac{mHaa'}{16\pi^2}
\Sigma_{n=0}^{\infty}(\frac{y}{4})^n\Biggl\{-4\Bigl[
\frac{\Delta\eta\gamma^k\Delta x_k}{\Delta x^4}\!+\!
\frac{\gamma^0\overline{\Delta x}^2}{\Delta x^4}\Bigr]\Bigl[
(2n\!-\!1)\!+\!(n^2\!-\!n)\ln\frac{y}{4}\Bigr]\nonumber\\
&&\hspace{4cm}-\frac{6\gamma^0}{\Delta x^2}\Bigl[
1\!+\!n\ln\frac{y}{4}\Bigr]\!+\!\frac{2Ha\gamma^k\Delta x_k}
{\Delta x^2}\Bigl[(2n\!+\!1)\!+\!(n^2\!+\!n)\ln\frac{y}{4}
\Bigr]\Biggr\},\\
&&\partial_{k}i[S]\gamma^k\!=\!\frac{mHaa'}{16\pi^2}
\Sigma_{n=0}^{\infty}(\frac{y}{4})^n\Biggl\{4\Bigl[
\frac{-\Delta\eta\gamma^k\Delta x_k}{\Delta x^4}\!+\!
\frac{\gamma^0\overline{\Delta x}^2}{\Delta x^4}\Bigr]\Bigl[
(2n\!-\!1)\!+\!(n^2\!-\!n)\ln\frac{y}{4}\Bigr]\nonumber\\
&&\hspace{4cm}+\frac{6\gamma^0}{\Delta x^2}\Bigl[
1\!+\!n\ln\frac{y}{4}\Bigr]\!+\!\frac{2Ha\gamma^k\Delta x_k}
{\Delta x^2}\Bigl[(2n\!+\!1)\!+\!(n^2\!+\!n)\ln\frac{y}{4}
\Bigr]\Biggr\}.
\end{eqnarray}
The formulae keep the same form for the summation starting from $n=1$
as long as we are working in $D=4$.

\end{document}